\documentclass[prb,showpacs,preprintnumbers,amsmath,amssymb]{revtex4}

\usepackage{epsfig}
\usepackage{graphicx}
\usepackage{dcolumn}
\usepackage{bm}

\newcommand{\EQ}{\begin{equation}}
\newcommand{\EN}{\end{equation}}
\newcommand{\ea}{\end{eqnarray}}
\newcommand{\ba}{\begin{eqnarray}}

\newcommand{\bear}{\begin{eqnarray}}
\newcommand{\ear}{\end{eqnarray}}

\begin{document}

\title{A description of the Hubbard model on a square lattice consistent
with its global $SO(3)\times SO(3)\times U(1)$ symmetry}
\author{J. M. P. Carmelo and M. J. Sampaio} 
\affiliation{GCEP-Centre of Physics, University of Minho, Campus Gualtar, 
P-4710-057 Braga, Portugal}



\begin{abstract}
In this paper a description of the Hubbard model on the square lattice with nearest-neighbor transfer 
integral $t$, on-site repulsion $U$, and $N_a^2\gg 1$ sites consistent with its exact global 
$SO(3)\times SO(3)\times U(1)$ symmetry is constructed. Our studies profit from the interplay of 
that recently found global symmetry of the model on any bipartite lattice with the transformation 
laws under a suitable electron - rotated-electron unitary transformation of
a well-defined set of operators and quantum objects. For $U/4t>0$ 
the occupancy configurations of these objects generate the energy eigenstates that span the 
one- and two-electron subspace. Such a subspace as defined in this paper contains nearly the whole 
spectral weight of the excitations generated by application onto the zero-spin-density ground
state of one- and two-electron operators. Our description involves
three basic objects: charge $c$ fermions, spin-$1/2$ 
spinons, and $\eta$-spin-$1/2$ $\eta$-spinons. 
Independent spinons and independent $\eta$-spinons
are invariant under the above unitary transformation.
Alike in chromodynamics the quarks have 
color but all quark-composite physical particles are 
color-neutral, the $\eta$-spinon (and spinons) that
are not invariant under that transformation 
have $\eta$ spin $1/2$ (and spin $1/2$) but are part of
$\eta$-spin-neutral (and spin-neutral) 
$2\nu$-$\eta$-spinon (and $2\nu$-spinon) composite 
$\eta\nu$ fermions (and $s\nu$ fermions) where $\nu=1,2,...$
is the number of $\eta$-spinon (and spinon) pairs. The occupancy 
configurations of the $c$ fermions, independent spinons and $2\nu$-spinon composite $s\nu$ fermions,
and independent $\eta$-spinons and $2\nu$-$\eta$-spinon composite $\eta\nu$ fermions
correspond to the state representations of the $U(1)$, 
spin $SU(2)$, and $\eta$-spin $SU(2)$ symmetries, 
respectively, associated with the model
$SO(3)\times SO(3)\times U(1) =[SU(2)\times SU(2)\times U(1)]/Z_2^2$
global symmetry. The components of the $\alpha\nu$ fermion discrete momentum values 
$\vec{q}_j=[{q_j}_{x1},{q_j}_{x2}]$ are eigenvalues of the corresponding set of $\alpha\nu$ translation generators
in the presence of fictitious magnetic fields ${\vec{B}}_{\alpha\nu}$. Our operator description has
been constructed to inherently the $\alpha\nu$ translation generators ${\hat{{\vec{q}}}}_{\alpha\nu}$ 
in the presence of the fictitious magnetic field ${\vec{B}}_{\alpha\nu}$
commuting with the momentum operator, consistently with their component operators
$\hat{q}_{\alpha\nu\,x_1}$ and $\hat{q}_{\alpha\nu\,x_1}$ commuting with each other.
In turn, unlike for the 1D model such generators not commute in general with the Hamiltonian, except 
for the Hubbard model on the square lattice in the one- and two-electron subspace.
Concerning one- and two-electron excitations,
the picture that emerges is that of a two-component quantum liquid
of charge $c$ fermions and spin-neutral two-spinon $s1$ fermions.
The description introduced here is consistent with a Mott-Hubbard insulating
ground state with antiferromagnetic 
long-range order for half filling at $x=0$ hole concentration and a ground state with 
short-range spin order for a well-defined range of finite hole concentrations $x>0$. 
For $0<x\ll1$ the latter short-range spin order has an incomensurate-spiral character. 
Our results are of interest for studies of ultra-cold fermionic atoms on optical lattices
and elsewhere evidence is provided that upon addition of a small three-dimensional
anisotropy plane-coupling perturbation to the square-lattice quantum liquid considered
here its short-range spin order coexists for low temperatures and a well-defined range of 
hole concentrations with a long-range superconducting order so that the use of the general description
introduced in this paper contributes to the further understanding
of the role of electronic correlations in the unusual properties of the hole-doped cuprates. 
\end{abstract}

\maketitle

\section{Introduction}

There is some consensus that the Hubbard model on a square lattice 
is the simplest toy model for describing the effects of electronics 
correlations in high-$T_c$ superconductors \cite{companion,bipartite,general,sqql,ARPES-review,2D-MIT} 
and their Mott-Hubbard insulators parent compounds \cite{LCO-neutr-scatt}.
In addition, the model can be experimentally realized with unprecedented 
precision in systems of ultra-cold fermionic atoms on an optical lattice 
and one may expect very detailed experimental results over a wide range of 
parameters to be available \cite{Zoller}. That includes recent studies on
systems of ultra-cold fermionic atoms describing the Mott-Hubbard insulating 
phase of the the Hubbard model on a cubic lattice \cite{cubic}.
Unfortunately, the model has no exact solution and
many open questions about its properties remain 
unsolved. A recent exact result, which 
applies to the model on any bipartite lattice \cite{bipartite}, 
is that for on-site repulsion $U>0$ its 
global symmetry is $SO(3)\times SO(3)\times U(1) =[SO(4)\times U(1)]/Z_2$. That
is an extension of the model well-known $SO(4)$ symmetry 
\cite{Zhang}, which becomes explicit
provided that one describes the problem in
terms of the rotated electrons obtained from
the electrons by any of the unitary transformations
of the type considered in Refs. \cite{bipartite,Stein}. 

In this paper a general description is introduced for the Hubbard model on 
the two-dimensional (2D) square lattice with spacing $a$, 
$N_a^D\equiv [N_a]^D$ sites, $D=2$, $N_a$ even, and 
lattice edge length $L=N_a\,a$. (Due to editorial decisions, the results
presented in this paper were distributed over three articles, specifically 
Refs. \cite{bipartite,general,sqql}. We thus forward the reader to these references, 
in which the presentation of the results was further improved, yet the
scientific message remains the same.)

The representation introduced here refers to suitable quantum objects whose occupancy configurations
generate the state representations of the group 
$SO(3)\times SO(3)\times U(1)$. Addition of 
chemical-potential and magnetic-field operator 
terms to the Hamiltonian lowers its symmetry. 
Such terms commute though with it and the momentum and energy eigenstates
correspond to representations of that group.
It is shown in Appendix A that all the physics of the model in the
whole Hilbert space can be obtained from that of the model
in the subspace spanned by the lowest-weight states (LWSs)
of both the $\eta$-spin $SU(2)$ and spin $SU(2)$ algebras.
Hence often our results refer to that large subspace.
Another subspace that plays a key role in our description
is the one- and two-electron subspace defined in Section V.
For 2D, the states generated by the momentum occupancy configurations
of the above objects belonging to such a subspace are for $U/4t>0$
energy eigenstates.

Furthermore, often we discuss the use of our description for the model on an
one-dimensional (1D) lattice as well, so that the coefficient $D$ 
appearing in several quantities considered in this paper
may have the values $D=1,2$. Due to the integrability of the 1D model,
the states generated by the momentum occupancy configurations
of the above-mentioned objects are exact energy eigenstates 
for the whole Hilbert space and all values of $U/4t$. Our results refer to a 
very large system whose lattice has $N_a^D\gg 1$ sites and for
the square (and 1D) lattice torus periodic 
boundary conditions (and periodic boundary 
conditions) are used. The description involves the 
generalization to all finite values of the on-site repulsion $U>0$
of the exact transformation for separation of spin-$1/2$ 
fermions without constraints introduced for large $U$ values
in Ref. \cite{Ostlund-06}. 

The global $SO(3)\times SO(3)\times U(1)$ 
symmetry refers to the Hubbard model on any
bipartite lattice. Hence it applies to the $D$-dimensional cubic
lattice \cite{2D-RA,2D-DA}. However, some of the results obtained below
are valid for the square lattice only. In spite of our study 
focusing onto the model on that lattice, 
the reason why often our analysis refers to 
the corresponding 1D problem as well is that in contrast to real-space 
dimensions $D>1$ there is an exact solution for the 
model on a 1D lattice \cite{Lieb,Martins,Takahashi}.
Our quantum-object description also applies to 1D, yet
due to the integrability of the model often it then leads to a 
different physics. In turn, one of the procedures used to control the validity of the approximations used in 
the construction of our description profits both from identifying
the common proprieties of the model on the square and 1D lattices and checking 
whether they are valid for the model on the latter lattice by means of its exact solution. 
Some of these common properties are related to the model commuting with the generators of the group 
$SO(3)\times SO(3)\times U(1)$ both for the square and 1D bipartite lattices.

For such two lattices the fulfillment of our program reveals
the emergence of three basic quantum objects:
spin-less and $\eta$-spin-less charge $c$ fermions, 
spin-$1/2$ spinons, and $\eta$-spin-$1/2$ $\eta$-spinons. 
In principle there are some occupancy configurations of 
such objects that generate the exact momentum and energy eigenstates  
for $U/4t>0$, yet the detailed structure of such configurations
is a complex problem. It simplifies in the one- and two-electron subspace 
constructed in this paper, where for the model on the square lattice and 
$U/4t>0$ the occupancy configurations of the above objects generate exact
ground states and excited energy eigenstates.
Such a subspace contains nearly the whole spectral weight generated by application
onto the zero-spin-density ground state of one- and two-electron operators.
Throughout this paper we often call that subspace {\it one- and two-electron
subspace}. In reference \cite{companion0} the spinon occupancy configurations of 
the states that span that subspace are studied. 

The composite $2\nu$-$\eta$-spinon $\eta\nu$ fermions (and
$2\nu$-spinon $s\nu$ fermions) emerge from $\eta$-spin
(and spin) neutral $\eta\nu$ bond particles (and 
$s\nu$ bond particles) through extended Jordan-Wigner transformations \cite{J-W,Wang,Feng}.
For the model on the square lattice the two-spinon 
$s1$ bond particles are found in Refs. \cite{companion,cuprates0} to be closely
related to the spin-singlet bonds of Ref. \cite{duality}. 
It is confirmed in Appendix B that for 1D the discrete momentum
values of the $c$ fermions and $\alpha\nu$ fermions
are good quantum numbers for all $U/4t$ values and their
momentum occupancy configurations generate the exact energy eigenstates. 
Evidence that for the one- and two-electron
subspace the states generated by the occupancy
configurations of such discrete momentum values 
are for the model on the square lattice and $U/4t>0$ 
energy eigenstates is given in this paper and in Ref. \cite{companion}. 

The paper is organized as follows. The model, the uniquely defined 
rotated-electron description used in our studies, its limitations
and usefulness, and the global $SO(3)\times SO(3)\times U(1)$ symmetry are the
subjects of Section II. In Section III the $c$ fermion, $\eta$-spin-$1/2$ $\eta$-spinon, and
spin-$1/2$ spinon and corresponding $c$, $\eta$-spin, and spin 
effective lattices are introduced. The vacua of the theory, 
their transformation laws under the electron - rotated-electron unitary
transformation, and the subspaces they refer to are issues also addressed
in that section. The composite $\alpha\nu$ bond particles and $\alpha\nu$ fermions, 
corresponding $\alpha\nu$ effective lattices, ground-state occupancy configurations,
and a suitable complete set of momentum eigenstates
are the subjects of Section IV. 
In Section V the one- and two-electron subspace where only the $c$ fermions and
$s1$ fermions play an active role is introduced. The corresponding $c$ and $s1$ 
effective lattices and elementary excitations are discussed. In Section VI it is shown that the description 
introduced in this paper is consistent with a Mott-Hubbard insulating 
ground state with long-range antiferromagnetic order for half filling and a ground state with 
short-range spin order for a well-defined range of finite hole concentrations. 
Strong evidence is given that for $0<x\ll1$ and intermediate and large values of $U/4t$ the latter order 
has an incommensurate-spiral character. 
Section VII contains the concluding remarks.
Moreover, that the whole physics can be extracted from the Hubbard model in the 
LWS subspace spanned by the $\eta$-spin and spin LWSs is confirmed in Appendix A.
In Appendix B it is confirmed that the general operator description introduced in this 
paper for the Hubbard model on the square and 1D lattices is consistent with the exact 
solution of the 1D problem. For instance, it is confirmed that the discrete momentum values 
of the $c$ fermions and composite $\alpha\nu$ fermions are good quantum numbers. Furthermore, 
in that Appendix the relation of the creation and annihilation fields of the charge ABCDF algebra  
\cite{Martins} and more traditional spin ABCD Faddeev-Zamolodchikov algebra \cite{ISM} of the 
algebraic formulation of the 1D exact solution of Ref. \cite{Martins} to the $c$
and $\alpha\nu$ fermion operators is discussed and the consistency between the two corresponding 
operational representations confirmed.     

\section{The model, a suitable rotated-electron description, 
and the global $SO(3)\times SO(3)\times U(1)$ symmetry}

\subsection{The model, a suitable rotated-electron description, and its limitations
and usefulness}

The Hubbard model is given by,
\begin{equation}
\hat{H} = t\,\hat{T} + U\,[N_a^D-\hat{Q}]/2 
\, ; \hspace{0.25cm}
\hat{T} = -\sum_{\langle\vec{r}_j\vec{r}_{j'}\rangle}\sum_{\sigma}[c_{\vec{r}_j,\sigma}^{\dag}\,c_{\vec{r}_{j'},\sigma}+h.c.] \, ;
\hspace{0.25cm}
{\hat{Q}} = \sum_{j=1}^{N_a^D}\sum_{\sigma =\uparrow
,\downarrow}\,n_{\vec{r}_j,\sigma}\,(1- n_{\vec{r}_j, -\sigma}) \, .
\label{H}
\end{equation}
Here $D=1$ and $D=2$ for the 1D and square lattices, respectively,
$t$ is the nearest-neighbor transfer integral, $\hat{T}$ the kinetic-energy operator 
in units of $t$, ${\hat{Q}}$ the operator that counts the number of electron 
singly occupied sites so that the operator ${\hat{D}}=[{\hat{N}}-{\hat{Q}}]/2$
counts the number of electron doubly
occupied sites, $n_{{\vec{r}}_j,\sigma} = c_{\vec{r}_j,\sigma}^{\dag} c_{\vec{r}_j,\sigma}$
where $-\sigma=\uparrow$ (and $-\sigma=\downarrow$)
for $\sigma =\downarrow$ (and $\sigma =\uparrow$), ${\hat{N}} = \sum_{\sigma}
{\hat{N}}_{\sigma}$, and ${\hat{N}}_{\sigma}=\sum_{j=1}^{N_a^D}
n_{{\vec{r}}_j,\sigma}$. The kinetic-energy operator $\hat{T}$
can be expressed in terms of the operators, 
\begin{eqnarray}
\hat{T}_0 & = & -\sum_{\langle\vec{r}_j\vec{r}_{j'}\rangle}\sum_{\sigma}[n_{\vec{r}_{j},-\sigma}\,c_{\vec{r}_j,\sigma}^{\dag}\,
c_{\vec{r}_{j'},\sigma}\,n_{\vec{r}_{j'},-\sigma} +
(1-n_{\vec{r}_{j},-\sigma})\,c_{\vec{r}_j,\sigma}^{\dag}\,
c_{\vec{r}_{j'},\sigma}\,(1-n_{\vec{r}_{j'},-\sigma})+h.c.] \, ,
\nonumber \\
\hat{T}_{+1} & = & -\sum_{\langle\vec{r}_j\vec{r}_{j'}\rangle}\sum_{\sigma}
[n_{\vec{r}_{j},-\sigma}\,c_{\vec{r}_j,\sigma}^{\dag}\,c_{\vec{r}_{j'},\sigma}\,(1-n_{\vec{r}_{j'},-\sigma})+h.c.] \, ,
\nonumber \\
\hat{T}_{-1} & = & -\sum_{\langle\vec{r}_j\vec{r}_{j'}\rangle}\sum_{\sigma}
[(1-n_{\vec{r}_{j},-\sigma})\,c_{\vec{r}_j,\sigma}^{\dag}\,
c_{\vec{r}_{j'},\sigma}\,n_{\vec{r}_{j'},-\sigma} + h.c.] \, ,
\label{T-op}
\end{eqnarray}
as $\hat{T}= \hat{T}_0 + \hat{T}_{+1} + \hat{T}_{-1}$. These three kinetic operators play an 
important role in the physics. The operator $\hat{T}_0$ does not change electron double 
occupancy whereas the operators $\hat{T}_{+1}$ and $\hat{T}_{-1}$ do it by $+1$ 
and $-1$, respectively. 

The studies of Ref. \cite{bipartite} consider unitary operators
$\hat{V}=\hat{V}(U/4t)$ and corresponding rotated-electron 
operators ${\tilde{c}}_{\vec{r}_j,\sigma}^{\dag} =
{\hat{V}}^{\dag}\,c_{\vec{r}_j,\sigma}^{\dag}\,{\hat{V}}$
such that rotated-electron
single and double occupancy are good quantum
numbers for $U/4t>0$. The generator ${\tilde{S}}_c$
of the global $U(1)$ symmetry given below and
in that reference has eigenvalue $S_c$ with $2S_c$ being the number of
rotated-electron singly occupied sites. 
Most choices of the unitary operators $\hat{V}$ correspond 
to choices of $U/4t\rightarrow\infty$ sets
$\{\vert\Psi_{\infty}\rangle\}$ of $4^{N_a^D}$
energy eigenstates such that the states 
$\vert \Psi_{U/4t}\rangle={\hat{V}}^{\dag}\vert\Psi_{\infty}\rangle$
are not energy and momentum eigenstates for finite $U/4t$ values yet belong
to a subspace with constant and well-defined values of $S_c$, $S_{\eta}$, 
and $S_s$. Here $S_{\eta}$ (and $S_s$) denotes the $\eta$-spin (and spin) 
value of the energy eigenstates. We call
$S^z_{\eta}= -[N_a^D-N]/2$ (and $S^z_s= -[N_{\uparrow}-
N_{\downarrow}]/2$) the corresponding 
projection. We focus our attention on ground states with 
hole concentration $x= [N_a^D-N]/N_a^D\geq 0$ and spin
density $m= [N_{\uparrow}-N_{\downarrow}]/N_a^D=0$ and are particularly interested in the
{\it LWS-subspace} spanned by the LWSs of both the $\eta$-spin and
spin algebras. Such energy eigenstates refer to values
of $S_{\alpha}$ and $S^z_{\alpha}$ such that 
$S_{\alpha}=-S^z_{\alpha}$ for $\alpha =\eta,s$. 
Indeed, since the off-diagonal generators that generate the
non-LWSs from the LWSs commute with the operator $\hat{V}$
\cite{bipartite}, the whole physics can be extracted 
from the model (\ref{H}) in the LWS-subspace,
as confirmed in Appendix A.

Let $\{\vert \Psi_{U/4t}\rangle\}$ be a
complete set of $4^{N_a^D}$ energy, momentum, $\eta$-spin,
$\eta$-spin projection, spin, and spin-projection eigenstates for
$U/4t>0$. In the limit $U/4t\rightarrow\infty$ such states correspond
to one of the many choices of sets $\{\vert\Psi_{\infty}\rangle\}$ of 
$4^{N_a^D}$ $U/4t$-infinite energy eigenstates. 
For this choice there exists exactly one unitary operator ${\hat{V}}={\hat{V}}(U/4t)$
such that $\vert \Psi_{U/4t}\rangle ={\hat{V}}^{\dag}\vert\Psi_{\infty}\rangle$.
Here we consider such a unitary operator and corresponding rotated-electron
operators ${\tilde{c}}_{\vec{r}_j,\sigma}^{\dag} =
{\hat{V}}^{\dag}\,c_{\vec{r}_j,\sigma}^{\dag}\,{\hat{V}}$.
The states $\vert \Psi_{U/4t}\rangle ={\hat{V}}^{\dag}\vert\Psi_{\infty}\rangle$
(one for each value of $U/4t>0$) that are 
generated from the same initial state $\vert\Psi_{\infty}\rangle$ 
belong to the same {\it $V$ tower}. 

We associate with any operator ${\hat{O}}$ a rotated
operator ${\tilde{O}}={\hat{V}}^{\dag}\,{\hat{O}}\,{\hat{V}}$ that
has the same expression in terms of rotated-electron creation and 
annihilation operators as ${\hat{O}}$ of electron creation and 
annihilation operators, respectively. Any operator ${\hat{O}}$ can 
be written in terms of rotated-electron creation and annihilation
operators as,
\begin{eqnarray}
{\hat{O}} & = & {\hat{V}}\,{\tilde{O}}\,{\hat{V}}^{\dag}
= {\tilde{O}}+ [{\tilde{O}},\,{\hat{S}}\,] + {1\over
2}\,[[{\tilde{O}},\,{\hat{S}}\,],\,{\hat{S}}\,] + ... 
=  {\tilde{O}}+ [{\tilde{O}},\,{\tilde{S}}\,] + {1\over
2}\,[[{\tilde{O}},\,{\tilde{S}}\,],\,{\tilde{S}}\,] + ... \, ,
\nonumber \\
{\hat{S}} & = & {t\over U}\,\left[\hat{T}_{+1} -\hat{T}_{-1}\right] 
+ {\cal{O}} (t^2/U^2) \, ; \hspace{0.25cm}
{\tilde{S}} = {t\over U}\,\left[\tilde{T}_{+1} -\tilde{T}_{-1}\right] 
+ {\cal{O}} (t^2/U^2) \, .
\label{OOr}
\end{eqnarray}
The operator $\hat{S}$ appearing in this equation 
is related to the unitary operator as ${\hat{V}}^{\dag} = e^{{\hat{S}}}$ and
${\hat{V}} = e^{-{\hat{S}}}$. The general unknown expression of ${\hat{S}}$ involves only the 
kinetic operators $\hat{T}_0$, $\hat{T}_{+1}$, and $\hat{T}_{-1}$
of Eq. (\ref{T-op}) and numerical $U/4t$ dependent coefficients and 
for $U/4t>0$ can be expanded in a series of $t/U$. The corresponding first-order
term has the universal form given in Eq. (\ref{OOr}). To arrive
to the expression in terms of the operator
${\tilde{S}}$ also given in Eq. (\ref{OOr}), the property 
that the operator ${\hat{V}}$ commutes with itself
so that ${\hat{V}} = e^{-{\hat{S}}}={\tilde{V}} = e^{-{\tilde{S}}}$ and
${\hat{S}}={\tilde{S}}$ and both the operators
${\hat{V}}$ and ${\hat{S}}$ have the same expression in terms of
electron and rotated-electron creation and annihilation operators
is used. This is behind the expansion 
${\tilde{S}} = (t/U)\,[\tilde{T}_{+1} -\tilde{T}_{-1}] + {\cal{O}} (t^2/U^2)$
given in that equation for the operator ${\tilde{S}}$. Its
higher-order terms can be written as a product of operator factors whose expressions involve the rotated
kinetic operators $\tilde{T}_0$, $\tilde{T}_{+1}$, and $\tilde{T}_{-1}$. 
The full expression of the operator $\hat{S}={\tilde{S}}$ can 
for $U/4t>0$ be written as 
${\hat{S}}={\hat{S}}(\infty)+\delta {\hat{S}}={\tilde{S}}(\infty)+\delta {\tilde{S}}$ 
where ${\hat{S}}(\infty)={\tilde{S}}(\infty)$
corresponds to the operator $S(l)$ at $l=\infty$ defined in Eq. (61) 
of Ref. \cite{Stein} and $\delta {\hat{S}}=\delta {\tilde{S}}$ has the general form
provided in its Eq. (64). The unitary operator ${\hat{V}} = e^{-{\hat{S}}}= e^{-{\tilde{S}}}$
considered here corresponds to exactly one choice of the coefficients $D^{(k)}(\bf{m})$ of 
that equation, where for the operator ${\tilde{V}}$ the index $k=1,2,...$ refers to the number 
of rotated-electron doubly occupied sites. The problem of finding the explicit form 
of the operators ${\hat{V}} ={\tilde{V}}$ and ${\hat{S}}={\tilde{S}}$ is equivalent
to finding all coefficients $D^{(k)}(\bf{m})$ associated with the electron - rotated-electron
unitary transformation as defined in this paper.

Since for finite $U/4t$ values 
the Hamiltonian $\hat{H}$ of Eq. (\ref{H}) does not commute with 
the unitary operator ${\hat{V}} = e^{-{\hat{S}}}$, when expressed in terms 
of rotated-electron creation and annihilation operators it has an infinite 
number of terms and according to Eq. (\ref{OOr}) reads, 
\begin{equation}
{\hat{H}} = {\hat{V}}\,{\tilde{H}}\,{\hat{V}}^{\dag}
= {\tilde{H}} + [{\tilde{H}},\,{\tilde{S}}\,] + {1\over
2}\,[[{\tilde{H}},\,{\tilde{S}}\,],\,{\tilde{S}}\,] + ... \, .
\label{HHr}
\end{equation}
The commutator $[{\tilde{H}},\,{\tilde{S}}\,]$ does not vanish
except for $U/4t\rightarrow\infty$ so that ${\hat{H}} \neq {\tilde{H}}$ for finite values of $U/4t$. 
For $U/4t$ very large the Hamiltonian of Eq. (\ref{HHr}) 
corresponds in terms of rotated-electron creation and annihilation
operators to a simple rotated-electron $t-J$ model. In turn, the higher-order 
$t/U$ terms, which become increasingly important upon decreasing $U/4t$,
generate effective rotated-electron hopping between second, 
third, and more distant neighboring sites. 
Indeed, the products of the kinetic operators
$\tilde{T}_0$, $\tilde{T}_{+1}$, and $\tilde{T}_{-1}$
contained in the higher-order terms of ${\tilde{S}} = (t/U)\,[\tilde{T}_{+1} -\tilde{T}_{-1}] + {\cal{O}} (t^2/U^2)$
also appear in the Hamiltonian expression (\ref{HHr}) in terms of rotated-electron 
creation and annihilation operators. In spite of the operators 
$\tilde{T}_0$, $\tilde{T}_{+1}$, and $\tilde{T}_{-1}$ generating 
only rotated-electron hoping between nearest-neighboring sites,
their products generate effective hoping between for instance second and
third neighboring sites whose real-space distance in units of
the lattice spacing $a$ is for the model on the square lattice $\sqrt{2}\,a$ and $2\,a$, 
usually associated with transfer integrals $t'$ and $t''$, respectively \cite{Tiago}. 

It follows that when expressed in terms of the
rotated-electron operators emerging from the specific
unitary transformation considered above the simple Hubbard model
(\ref{H}) as given in Eq. (\ref{HHr}) contains Hamiltonian terms associated with
higher-order contributions which can be effectively described by transfer 
integrals $t'$, $t''$, and of higher order. It is found in the following
that for hole concentration equal to or larger than zero 
both the model ground state and the excited states that 
span the one- and two-electron subspace as defined in this
paper have vanishing rotated-electron double occupancy. It follows from a 
Renormalization Group analysis based on the results of Refs. \cite{Wieg,Mura}
that for intermediate and large values of $U/4t$ obeying approximately 
the inequality $U/4t\geq u_0\approx 1.302$, 
besides the original nearest-neighboring hoping processes  
only those involving second and third neighboring sites are
relevant for the square-lattice quantum liquid 
described by the Hamiltonian of Eqs. (\ref{H}) and (\ref{HHr}) in the
one- and two-electron subspace. The value $U/4t=u_0\approx 1.302$
is that where according to the results of Section VI the $s1$ fermion spinon
pairing energy reaches its maximum magnitude for small hole concentration
$0<x\ll 1$. Such a pairing energy is found to vanish both in the
limits $U/4t\rightarrow 0$ and $U/4t\rightarrow\infty$. In contrast, for the 1D
model it vanishes for all $U/4t$ values.

Hence for approximately $U/4t\geq u_0$, out of the infinite terms on the right-hand-side
of Eq. (\ref{HHr}) only the first few Hamiltonian terms play
an active role in the physics of the Hubbard model on the square lattice
in the one- and two-electron subspace. Therefore, for intermediate and large values of $U/4t$ 
such a square-lattice quantum liquid can be mapped onto an effective $t-J$ model on
a square lattice with $t$, $t'=t'(U/4t)$, and $t''=t''(U/4t)$ transfer integrals \cite{Tiago}
for which the role of the processes associated with 
$t'=t'(U/4t)$ and $t''=t''(U/4t)$ becomes increasingly important
upon decreasing the $U/4t$ value. For that $U/4t$ range
the latter model is equivalent to the Hubbard model on the square lattice
(\ref{HHr}) in the subspace under consideration and  
expressed in terms of rotated-electron creation and annihilation operators. 
Indeed, the $t-J$ model constraint against double occupancy is in that subspace
equivalent to expressing the Hubbard model in terms of rotated-electron 
creation and annihilation operators.

However, the rotated electrons are not the ultimate objects whose occupancy configurations
generate the exact ground states and excited energy eigenstates. For $U/4t>0$ suitable 
occupancy configurations of the charge $c$ fermions and spin-neutral two-spinon $s1$ fermions 
simpler than those of the rotated electrons generate such states: for the square-lattice model in the one- and
two-electron subspace the excited states
generated by the occupancy configurations of such objects are energy eigenstates. (Actually, for
1D they refer to exact energy eigenstates for the whole Hilbert space provided that $U/4t>0$.) Another advantage relative to the rotated
electrons is that the processes associated with the above effective transfer integrals $t'=t'(U/4t)$ and 
$t''=t''(U/4t)$ are implicitly contained in the $U/4t$ dependence of the energy scales
associated with the $c$ and $s1$ fermions spectra derived in Ref. \cite{companion}.
 
The general operator description introduced in this paper for the Hubbard
model on the square lattice has two main limitations: 
\vspace{0.25cm}

I) For small and intermediate values of $U/4t$ the explicit form of the unitary operator $\hat{V}$ 
associated with the rotated-electron operators as defined in this paper
remains an open problem. It is known that such a unitary operator has for $U/4t>0$ the general
form ${\hat{V}} = e^{-{\hat{S}}}$ where the expression of $\hat{S}$ involves only the three kinetic 
operators $\hat{T}_0$, $\hat{T}_{+1}$, and $\hat{T}_{-1}$ of Eq. (\ref{T-op}). However, the finding 
of the explicit form of the $U/4t$-dependent functional of $\hat{S}$ in terms of the latter three operators,
valid for the whole range of finite $U/4t$ values, is a very 
involved and unsolved quantum problem, beyond the reach of the present status of our scheme. 
Indeed, the operator description introduced in this paper has been constructed to inherently the solution 
of that problem being equivalent to the solution of the Hubbard model on the square lattice. 
\vspace{0.25cm}

II) It turns out that the quantum problem under consideration is non-perturbative
in terms of electron operators so that, in contrast to a three-dimensional (3D) isotropic
Fermi liquid \cite{Pines}, rewriting the square-lattice quantum-liquid theory emerging
from the general description introduced in this paper for the
model in the one- and two-electron subspace in terms of
the standard formalism of many-electron physics is in general an extremely
complex problem. Fortunately, such a quantum liquid 
dramatically simplifies when expressed in terms of the $c$ fermion and $s1$ fermion
operators and one can rich limited but valuable information on
its physics in spite of the lack of explicit information about the
matrix elements between energy eigenstates. The point is that as justified below in the above
subspace the $c$ and $s1$ fermion discrete momentum values have been constructed to 
inherently being good quantum numbers so that the interactions of these objects are residual 
\cite{companion,cuprates0,cuprates}.
\vspace{0.25cm}

Concerning the Hamiltonian, the microscopic processes 
corresponding to the effective transfer integrals $t'=t'(U/4t)$ and $t''=t''(U/4t)$ are important to characterize
the type of order associated with the phases of the square-lattice quantum liquid as for 
instance the short-range incomensurate-spiral spin order considered in 
Subsection VI-B. Fortunately, for small finite hole concentrations, it is confirmed in
Subsection VI-A that for simple one- and two-electron operators ${\hat{O}}$ 
the leading operator term ${\tilde{O}}$ on the right-hand side of Eq. (\ref{OOr}) 
generates nearly the whole spectral weight. Hence in spite of the limitations I and II,
our description provides useful information 
about the physics contained in the model on the square lattice.
Indeed, there are several reasons why, in spite of both the explicit form
of the unitary operator $\hat{V}$ being known only for
large values of $U/4t$ and the difficulties in rewriting 
the theory emerging from the description introduced here
in terms of the standard formalism of many-electron physics, that 
description is rather useful for finite values of $U/4t$ and particularly
for approximately $U/4t\geq u_0$. This is confirmed
in Ref. \cite{companion} from comparison of results obtained by our description
and the standard formalism of many-electron physics concerning
one of the few problems where controlled and reliable approximations 
exist within the latter formalism for the Hubbard model on the square
lattice.  

For one- and two-electron operators ${\hat{O}}$ the terms of the general expression 
(\ref{OOr}) containing commutators involving the related operator $\hat{S}={\tilde{S}}$
are found in Section V-B to generate very little spectral weight.
Hence one can reach a quite faithful representation of
such operators by expressing them in terms of the
$c$ and $s1$ fermion operators. In Subsection III-A we provide
the expression of the rotated-electron creation and annihilation
operators in terms of the operators of the $c$ fermions, $\eta$-spinons, and
spinons. The construction of the spin-neutral two-spinon composite $s1$
bond-particle operators and corresponding $s1$ fermion operators is a
more involved problem addressed in Refs. \cite{companion0,companion}.
Furthermore, in spite of the explicit form of the unitary operator $\hat{V}$
being unknown except for $U/4t\gg 1$, we are able to access the transformation laws and/or 
invariance of several operators and quantum objects under the
corresponding electron - rotated-electron unitary transformation, 
what provides valuable information about the physics of the Hubbard model
on the square lattice for $U/4t>0$. 

The physical reason why the operator terms of the general expression (\ref{OOr}) containing
commutators involving the operator $\hat{S}={\tilde{S}}$ generate very little
one- and two-electron spectral weight is that the rotated electrons as defined in this paper are
directly related to objects whose interactions are residual so that the leading elementary 
processes in terms of them generate nearly the whole spectral weight. Expression of the operator 
terms of Eq. (\ref{OOr}) containing
commutators involving the operator $\hat{S}={\tilde{S}}$ in terms of
the operators of such objects reveals that those
generate higher-order processes, whereas the elementary and leading
processes are generated by the operator ${\tilde{O}}$. 

In Subsection VI-B we study the $U/4t$ dependence of the maximum
magnitude $2\Delta_0$ of the $s1$ fermion spinon pairing energy of
the Hubbard model on the square lattice for small hole concentrations $0<x\ll 1$. 
Such an energy scale is also well defined for a larger range of finite
hole concentrations yet then has a different physical meaning and
plays an important role in square-lattice quantum liquid. 
For $0<x\ll 1$ it plays the role of short-range incommensurate-spiral spin order
parameter. That order is absent in the 1D model for which $2\Delta_0=0$. For
the square-lattice model it occurs both for
$0<x\ll 1$ and temperatures $0\leq T<T_0^*$ and at half filling $x=0$ for $0< T<T_0^*$, 
where $T_0^*\approx \Delta_0/k_B$.
Our study of the $U/4t$ dependence of the energy parameter $2\Delta_0$ 
profits from combination of the description introduced 
here with results of the low-temperature approach of Ref. \cite{Hubbard-T*-x=0} to the half-filled model. 
Such an approach uses a CP$^1$ representation and
is valid below the temperature $T_0^*$ called in that reference $T_x$, which marks the 
onset of the short-range spin order. For the specific value $U/4t\approx 1.525$ the
magnitude of $2\Delta_0$ provided in Subsection VI-B is obtained in Ref.
\cite{companion} from combination of the description introduced
in this paper with the results of Ref. \cite{LCO-Hubbard-NuMi}
for the spin-wave spectrum obtained by summing up an infinite set of ladder diagrams for 
the half-filled model in a spin-density-wave-broken symmetry ground state.
Moreover, in the same subsection the spiral-incommensurate character
of such a short-range spin order is studied by combining results obtained from
the use of our description with the action and corresponding
CP$^1$ field theory of Refs. \cite{Wieg,Mura}.

A realistic program to be carried out here and in following papers includes the fulfillment of two main tasks:
(i) Introducing the objects whose occupancy configurations generate the energy eigenstates 
that span the one- and two-electron subspace and writing the rotated-electron creation and/or annihilation operators 
in the expression of the one- and two-electron operators ${\hat{O}}$ of Eq. (\ref{OOr})
in terms of the operators of such objects; (ii) Finding the specific
occupancy configurations of the latter objects that generate such states.

As discussed above, the expression of the operator ${\tilde{O}}$
in terms of the $c$ fermion, $\eta$-spinon, and spinon operators
provides a rather good representation of the corresponding one-
or two-electron operator ${\hat{O}}$ itself. In this paper we fulfill task (i)
and start solving the problems needed for the fulfillment of task (ii). The
latter task is concluded in Refs. \cite{companion0,companion}, 
whose studies profit from the general description introduced here. 
Furthermore, the further construction of the square-lattice quantum liquid 
is fulfilled in Refs. \cite{companion,cuprates0}. The 
lack for $U/4t>0$ of the explicit general form of the unitary operator $\hat{V}$  
prevents the derivation of matrix elements between energy eigenstates. It is
confirmed elsewhere that the information about the Hubbard model on the
square lattice in the one- and two-electron subspace reached by combining
different methods and approximations with
the general operator description introduced in this paper clarifies important open issues 
of the physics contained in the model. That includes its usefulness and suitability
to describe the physics of real materials \cite{companion,cuprates0,cuprates}.    
       
\subsection{The model global $SO(3)\times SO(3)\times U(1)$ symmetry}

The studies of Ref. \cite{U(1)-NL} reveal that for $U/4t\rightarrow\infty$ the 
Hamiltonian (\ref{H}) has a local $SU(2)\times SU(2) \times U(1)$ gauge
symmetry. In turn, for finite $U/4t$ values only its interacting term is
invariant under $SU(2)\times SU(2) \times U(1)$ and therefore 
$SU(2)\times SU(2) \times U(1)$ becomes a group of permissible
unitary transformations. As discussed in that reference, 
the local $U(1)$ canonical transformation is not the
ordinary $U(1)$ gauge subgroup of electromagnetism, but instead
is a "nonlinear" transformation, i.e., a gauge transformation transforming a creation or
annihilation operator into a sum of nonlinear polynomials
of such operators.

Furthermore, the studies of Ref. \cite{bipartite} reveal
that for on-site repulsion $U>0$ the above local 
$SU(2)\times SU(2) \times U(1)$ gauge symmetry 
of the model for $U/4t\rightarrow\infty$ can 
be lifted to a global 
$[SU(2)\times SU(2) \times U(1)]/Z_2^2=SO(3)\times SO(3)\times U(1)$ 
symmetry in the presence of the kinetic-energy hopping
term of the Hamiltonian (\ref{H}) with $t>0$ and $U/4t>0$. The generator of the 
new found hidden global $U(1)$ symmetry 
is one half the rotated-electron number of singly-occupied sites 
operator. As mentioned above, addition of chemical-potential 
and magnetic-field operator 
terms to the Hamiltonian (\ref{H}) lowers its symmetry. 
However, such terms commute with it and therefore the  
rotated-electron occupancy configurations generate for 
all densities state representations of its global symmetry. 

Since the expression of the operator 
${\hat{S}}$ involves only the three kinetic operators 
$\hat{T}_0$, $\hat{T}_{+1}$, and $\hat{T}_{-1}$
given in Eq. (\ref{T-op}), the electron - rotated-electron 
unitary operator ${\hat{V}}={\tilde{V}}$ preserves the occurrence 
of nearest hopping only for rotated electrons alike for electrons and commutes 
with the momentum operator $\hat{{\vec{P}}}$ and, 
therefore, according to Eq. (\ref{OOr}) the latter
operator is such that $\hat{{\vec{P}}} ={\tilde{{\vec{P}}}}$ and thus has 
the same expression in terms of electron and rotated-electron 
creation and annihilation operators,
\begin{equation}
\hat{{\vec{P}}}  = \sum_{\sigma=\uparrow ,\,\downarrow }\sum_{\vec{k}}\,\vec{k}\,
c_{\vec{k},\,\sigma }^{\dag }\,c_{\vec{k},\,\sigma } =
\sum_{\sigma=\uparrow ,\,\downarrow }\sum_{\vec{k}}\,\vec{k}\,
{\tilde{c}}_{\vec{k},\,\sigma }^{\dag }\,{\tilde{c}}_{\vec{k},\,\sigma }
\, .
\label{P-invariant}
\end{equation}

In addition, according to the studies of Ref. \cite{bipartite} the unitary 
operator ${\hat{V}}$ commutes with the three generators
of the spin $SU(2)$ symmetry and three generators
of the $\eta$-spin $SU(2)$ symmetry. Hence, such generators 
also have the same expression in terms of electron and 
rotated-electron creation and annihilation operators and read,
\begin{eqnarray}
{\hat{S}}_{\eta}^z & = & -{1\over 2}[N_a^D-\hat{N}] \, ;
\hspace{0.25cm}
{\hat{S}}_s^z = -{1\over 2}[{\hat{N}}_{\uparrow}- {\hat{N}}_{\downarrow}] \, ,
\nonumber \\
{\hat{S }}_{\eta}^{\dag} & = & \sum_{j=1}^{N_a^D}e^{i\vec{\pi}\cdot\vec{r}_j}\,c_{\vec{r}_j,\downarrow}^{\dag}\,
c_{\vec{r}_j,\uparrow}^{\dag} =\sum_{j=1}^{N_a^D}e^{i\vec{\pi}\cdot\vec{r}_j}\,{\tilde{c}}_{\vec{r}_j,\downarrow}^{\dag}\,
{\tilde{c}}_{\vec{r}_j,\uparrow}^{\dag} \, ;
\hspace{0.25cm}
{\hat{S}}_{\eta} = \sum_{j=1}^{N_a^D}e^{-i\vec{\pi}\cdot\vec{r}_j}\,c_{\vec{r}_j,\uparrow}\,c_{\vec{r}_j,\downarrow} 
=\sum_{j=1}^{N_a^D}e^{-i\vec{\pi}\cdot\vec{r}_j}\,{\tilde{c}}_{\vec{r}_j,\uparrow}\,{\tilde{c}}_{\vec{r}_j,\downarrow} \, ,
\nonumber \\
{\hat{S}}_s^{\dag} & = &
\sum_{j=1}^{N_a^D}\,c_{\vec{r}_j,\downarrow}^{\dag}\,c_{\vec{r}_j,\uparrow} =
\sum_{j=1}^{N_a^D}\,{\tilde{c}}_{\vec{r}_j,\downarrow}^{\dag}\,{\tilde{c}}_{\vec{r}_j,\uparrow} 
\, ; \hspace{0.25cm}
{\hat{S}}_s = \sum_{j=1}^{N_a^D}c_{\vec{r}_j,\uparrow}^{\dag}\,
c_{\vec{r}_j,\downarrow}=\sum_{j=1}^{N_a^D}{\tilde{c}}_{\vec{r}_j,\uparrow}^{\dag}\,
{\tilde{c}}_{j,\,\downarrow} \, , 
\label{Scs}
\end{eqnarray}
where for the model on the square (and 1D) lattice the vector $\vec{\pi}$ has 
Cartesian components $\vec{\pi}=[\pi,\pi]$ (and component $\pi$).

In turn, alike the Hamiltonian of Eq. (\ref{H}), the generator ${\tilde{S}}_c$ 
of the charge independent $U(1)$ symmetry does not commute with
the unitary operator ${\hat{V}}$. On the contrary of the
Hamiltonian, that operator has a complicated expression in terms
of electron creation and annihilation operators and a simple expression
in terms of rotated-electron creation and annihilation operators \cite{bipartite},
which reads,
\begin{equation}
{\tilde{S}}_c = {1\over 2}\sum_{j=1}^{N_a^D}\sum_{\sigma =\uparrow
,\downarrow}\,{\tilde{n}}_{\vec{r}_j,\sigma}\,(1- {\tilde{n}}_{\vec{r}_j,-\sigma}) \, .
\label{Or-ope}
\end{equation}
It follows that ${\tilde{S}}_c= {\hat{V}}^{\dag}\,{\hat{S}}_c\,{\hat{V}}$
where ${\hat{S}}_c= {1\over 2}{\hat{Q}}$ and the operator
${\hat{Q}}$ is given in Eq. (\ref{H}). 
The six operators provided in Eq. (\ref{Scs}) plus that
given in Eq. (\ref{Or-ope}) are the seven generators
of the group $[SO(4)\times U(1)]/Z_2=SO(3)\times SO(3)\times U(1)$ associated with
the global symmetry of the Hamiltonian (\ref{H}). 

That for $U/4t>0$ the eigenvalue $S_c$ of the generator (\ref{Or-ope}) is a good quantum 
number implies that the rotated-electron doubly occupancy 
$D_c=[N/2-S_c]$ is also a good quantum number. Alike for the
more general case considered in Ref. \cite{bipartite}, following the unitary 
character of the operator $\hat{V}=\tilde{V}$ associated with the
rotated electrons as constructed in this paper one can either consider 
that ${\tilde{V}}\,{\tilde{H}}\,{\tilde{V}}^{\dag}= {\tilde{H}}+ [{\tilde{H}},\,{\tilde{S}}\,]+...$
is the Hubbard model written in terms of rotated-electron
operators or another Hamiltonian with an involved expression
and whose operators ${\tilde{c}}_{\vec{r}_j\sigma}^{\dag}$ and 
${\tilde{c}}_{\vec{r}_j\sigma}$ correspond to electrons. 
The studies of Ref. \cite{bipartite}
refer to any unitary operator $\hat{V}$ associated with
the general type of unitary transformations considered
in Ref. \cite{Stein}. In turn, the operator $\hat{V}$ used here corresponds to a 
suitable set of $4^{N_a^D}$ $U/4t\rightarrow\infty$ energy eigenstates
$\{\vert\Psi_{\infty}\rangle\}$ such that for $U/4t>0$ the $V$ tower states 
$\vert \Psi_{U/4t}\rangle={\hat{V}}^{\dag}\vert\Psi_{\infty}\rangle$
are energy and momentum eigenstates.
In the more general case considered in 
that reference the $4^{N_a^D}$ states 
${\hat{V}}^{\dag}\vert\Psi_{\infty}\rangle$ are 
eigenstates of the generator of
the charge $U(1)$ symmetry and of 
the $\eta$-spin and spin square operators of the
$SU(2)$ symmetries so that the values of $S_c$, $S_{\eta}$, 
and $S_s$ are well defined yet such states are not
in general energy and momentum eigenstates. 

Following the procedures of Ref. \cite{Stein}, the Hamiltonian
${\tilde{V}}\,{\tilde{H}}\,{\tilde{V}}^{\dag}= {\tilde{H}}+ [{\tilde{H}},\,{\tilde{S}}\,]+...$
is built up by use of the conservation of singly 
occupancy $2S_c=\langle {\tilde{Q}}\rangle$ 
by eliminating terms in the $t>0$
Hubbard Hamiltonian so that $S_c$ is an 
eigenvalue of the operator (\ref{Or-ope}).
According to the studies of that reference this can be done to all 
orders of $4t/U$ provided that $U/4t\neq 0$. The original lattice
remains invariant under the electron - rotated-electron unitary
transformation and the energy eigenstate description introduced in this paper has been constructed to 
inherently each site of that lattice being unoccupied, doubly occupied, 
singly occupied by a spin-up rotated
electron, or singly occupied by a spin-down rotated electron.

\section{Three elementary quantum objects and 
corresponding $c$, $\eta$-spin, and spin effective lattices}

\subsection{Elementary quantum objects and their operators}

The rotated electrons are not the ultimate objects whose occupancy configurations that generate the 
energy eigenstates of the model (\ref{H}) in the one- and two-electron subspace have simplest expressions. 
There emerge from the electron - rotated-electron unitary transformation
considered here three basic objects: the spin-less
and $\eta$-spin-less $c$ fermions, the spin-$1/2$
spinons, and the $\eta$-spin-$1/2$ $\eta$-spinons. 
Such objects are a generalization for $U/4t>0$ of those
introduced in Ref. \cite{Ostlund-06} for $U/4t\gg 1$,
as discussed below. In Appendix B it is confirmed
that for the 1D Hubbard model and all $U/4t$ values suitable occupancy 
configurations of such objects generate a complete set of energy eigenstates, 
which span the whole Hilbert space. In this paper and in Ref. \cite{companion} 
evidence is provided that suitable $c$ and $s1$
fermion momentum occupancy configurations generate the exact ground states
and the excited energy eigenstates that span the one- and two-electron subspace defined below in Section V. 
  
Here we profit from the result of Appendix A that full information about the quantum problem
can be achieved by defining it in the LWS-subspace. Within the 
corresponding LWS representation the 
$c$ fermion creation operator can be expressed in terms
of the rotated-electron operators and electron operators as follows,
\begin{eqnarray}
f_{\vec{r}_j,c}^{\dag} & = &
{\tilde{c}}_{\vec{r}_j,\uparrow}^{\dag}\,
(1-{\tilde{n}}_{\vec{r}_j,\downarrow})
+ e^{i\vec{\pi}\cdot\vec{r}_j}\,{\tilde{c}}_{\vec{r}_j,\uparrow}\,
{\tilde{n}}_{\vec{r}_j,\downarrow} 
\, ; \hspace{0.35cm}
f_{\vec{q}_j,c}^{\dag} =
{1\over {\sqrt{N_a^D}}}\sum_{j'=1}^{N_a^D}\,e^{+i\vec{q}_j\cdot\vec{r}_{j'}}\,
f_{\vec{r}_{j'},c}^{\dag} \, ,
\nonumber \\
{\tilde{n}}_{\vec{r}_j,\sigma} & = & {\tilde{c}}_{\vec{r}_j,\sigma}^{\dag}\,{\tilde{c}}_{\vec{r}_j,\sigma}
\, ; \hspace{0.35cm}
{\tilde{c}}_{\vec{r}_j,\sigma}^{\dag} =
{\hat{V}}^{\dag}\,c_{\vec{r}_j,\sigma}^{\dag}\,{\hat{V}}
\, ; \hspace{0.35cm}
{\tilde{c}}_{\vec{r}_j,\sigma} =
{\hat{V}}^{\dag}\,c_{\vec{r}_j,\sigma}\,{\hat{V}} \, ,
\label{fc+}
\end{eqnarray}
where we have introduced the corresponding $c$ fermion momentum-dependent
operators as well, $e^{i\vec{\pi}\cdot\vec{r}_j}$ is $\pm 1$ depending on which
sublattice site $\vec{r}_j$ is on, and ${\hat{V}}^{\dag}$ is the
uniquely defined electron - rotated-electron unitary operator introduced in this paper.
(For the 1D lattice that phase factor can be written as $(-1)^j$.) The expression in
terms of electron operators involves the electron - rotated-electron unitary
operator ${\hat{V}}$ as defined in this paper.
The $c$ momentum band is studied in Ref. \cite{companion} and has the same
shape and momentum area as the electronic first-Brillouin zone.

The three spinon local operators
$s^l_{\vec{r}_j}$ and three $\eta$-spinon local operators $p^l_{\vec{r}_j}$
such that $l=\pm,z$ are given by,
\begin{equation}
s^l_{\vec{r}_j} = n_{\vec{r}_j,c}\,q^l_{\vec{r}_j} \, ; \hspace{0.25cm}
p^l_{\vec{r}_j} = (1-n_{\vec{r}_j,c})\,q^l_{\vec{r}_j} \, , 
\hspace{0.15cm} l =\pm,z \, .
\label{sir-pir}
\end{equation}
Here $q^{\pm}_{\vec{r}_j}= q^{x}_{\vec{r}_j}\pm i\,q^{y}_{\vec{r}_j}$
are the rotated quasi-spin operators whose Cartesian coordinates 
$x,y,z$ are often denoted in this paper by $x_1,x_2,x_3$, 
respectively, and, 
\begin{equation}
n_{\vec{r}_j,c} = f_{\vec{r}_j,c}^{\dag}\,f_{\vec{r}_j,c} \, ,
\label{n-r-c}
\end{equation}
is the $c$ fermion local density operator.

In terms of rotated-electron creation and annihilation
operators the rotated quasi-spin operators read,
\begin{equation}
q^+_{\vec{r}_j} = ({\tilde{c}}_{\vec{r}_j,\uparrow}^{\dag}
- e^{i\vec{\pi}\cdot\vec{r}_j}\,{\tilde{c}}_{\vec{r}_j,\uparrow})\,
{\tilde{c}}_{\vec{r}_j,\downarrow} \, ; \hspace{0.25cm}
q^-_{\vec{r}_j} = (q^+_{\vec{r}_j})^{\dag} \, ;
\hspace{0.25cm}
q^z_{\vec{r}_j} = {1\over 2} - {\tilde{n}}_{\vec{r}_j,\downarrow} \, .
\label{rotated-quasi-spin}
\end{equation}

Since the electron - rotated-electron 
transformation generated by the operator $\hat{V}$ 
is unitary, the operators ${\tilde{c}}_{\vec{r}_j,\sigma}^{\dag}$ 
and ${\tilde{c}}_{\vec{r}_j,\sigma}$
have the same anticommutation relations as 
$c_{\vec{r}_j,\sigma}^{\dag}$ and $c_{\vec{r}_j,\sigma}$, respectively.
Straightforward manipulations based on Eqs.
(\ref{fc+})-(\ref{rotated-quasi-spin}) then lead
to the following algebra for the $c$ fermion operators,
\begin{equation}
\{f^{\dag}_{\vec{r}_j,c}\, ,f_{\vec{r}_{j'},c}\} = \delta_{j,j'} \, ;
\hspace{0.25cm}
\{f_{\vec{r}_j,c}^{\dag}\, ,f_{\vec{r}_{j'},c}^{\dag}\} =
\{f_{\vec{r}_j,c}\, ,f_{\vec{r}_{j'},c}\} = 0 \, ,
\label{albegra-cf}
\end{equation}
$c$ fermion operators and rotated quasi-spin operators,
\begin{equation}
[f_{\vec{r}_j,c}^{\dag}\, ,q^l_{\vec{r}_{j'}}] =
[ f_{\vec{r}_j,c}\, ,q^l_{\vec{r}_{j'}}] = 0 \, ,
\label{albegra-cf-s-h}
\end{equation}
and rotated quasi-spin operators,
\begin{equation}
[q^p_{\vec{r}_j}\, ,q^{p'}_{\vec{r}_{j'}}] =
i\,\delta_{j,j'}\sum_{p''} \epsilon_{pp'p''}\,q^{p''}_{\vec{r}_j} 
\, ; \hspace{0.15cm} p=x,y,z \, ,
\label{albegra-s-h}
\end{equation}
\begin{equation}
\{q^{+}_{\vec{r}_j},q^{-}_{\vec{r}_j}\} = 1 \, ,
\hspace{0.5cm}
\{q^{\pm}_{\vec{r}_j},q^{\pm}_{\vec{r}_j}\} = 0 \, ,
\label{albegra-qs-p-m}
\end{equation}
\begin{equation}
[q^{+}_{\vec{r}_j},q^{-}_{\vec{r}_{j'}}] =
[q^{\pm}_{\vec{r}_j},q^{\pm}_{\vec{r}_{j'}}]=0 \, .
\label{albegra-q-com}
\end{equation}
Hence the operators $q^{\pm}_{\vec{r}_j}$ anticommute 
on the same site and commute on different sites.

The relations provided in Eqs. (\ref{albegra-cf})-(\ref{albegra-q-com})
confirm that the $c$ fermions associated with the global $U(1)$ symmetry
are $\eta$-spinless and spinless fermionic objects and the spinons and $\eta$-spinons are 
spin-$1/2$ and $\eta$-spin-$1/2$ objects whose local operators
obey the usual corresponding spin and $\eta$-spin $SU(2)$ 
algebras, respectively. 

We can now fulfill task (i) suggested in Subsection II-A: Writing the rotated-electron creation 
and/or annihilation operators in the expression of any one- or two-electron operator ${\hat{O}}$ of Eq. (\ref{OOr}) in terms of the operators 
of the objects whose simple occupancy configurations generate the energy eigenstates 
that span the one- and two-electron subspace of the model on the square lattice and the
whole Hilbert space of the 1D model. For the LWS-subspace considered here this is simply achieved by 
inverting the relations given in Eqs. (\ref{fc+}) and (\ref{rotated-quasi-spin}) with the result,
\begin{equation}
{\tilde{c}}_{\vec{r}_j,\uparrow}^{\dag} =
f_{\vec{r}_j,c}^{\dag}\,\left({1\over 2} +
q^z_{\vec{r}_j}\right) + e^{i\vec{\pi}\cdot\vec{r}_j}\,
f_{\vec{r}_j,c}\,\left({1\over 2} - q^z_{\vec{r}_j}\right) \, ;
\hspace{0.25cm}
{\tilde{c}}_{\vec{r}_j,\downarrow}^{\dag} =
q^-_{\vec{r}_j}\,(f_{\vec{r}_j,c}^{\dag} -
e^{i\vec{\pi}\cdot\vec{r}_j}\,f_{\vec{r}_j,c}) \, .
\label{c-up-c-down}
\end{equation}

For $U/4t\rightarrow\infty$ the rotated electrons
become electrons and Eqs. (\ref{fc+})-(\ref{c-up-c-down}) 
are equivalent to Eqs. (1)-(3) of Ref. \cite{Ostlund-06} with
the rotated-electron operators replaced by the
corresponding electron operators and the $c$ fermion
creation operator $f_{\vec{r}_j,c}^{\dag}$ replaced
by the quasicharge annihilation operator $\hat{c}_r$.
Therefore, in that limit the $c$ fermions are the "holes"
of the quasicharge particles of that reference and
the spinons and $\eta$-spinons are associated with the
local spin and pseudospin operators, respectively. 

Since the transformation considered in Ref. \cite{Ostlund-06} 
does not introduce Hilbert-space constraints, suitable occupancy 
configurations of the objects associated with the local 
quasicharge, spin, and pseudospin operators generate a complete set 
of states. However, only in the $U/4t\rightarrow\infty$ limit simple occupancy configurations
of the corresponding quasicharge, spin, and pseudospin objects generate the energy eigenstates
that span the one- and two-electron subspace. In turn, for the
model on the square lattice the same occupancy configurations of the above 
corresponding finite-$U/4t$ objects generate the energy eigenstates belonging to the
same $V$ tower. 

Our next main goal is the fulfillment of the task (ii) 
also suggested in Subsection II-A: Finding the suitable occupancy 
configurations of the $c$ fermions, spinons, and
$\eta$-spinons that generate energy eigenstates of the one- and two-electron subspace. The study of the relationship of the transformation laws 
of such objects under the electron - rotated-electron
unitary transformation to the state representations
of the group $SO(3)\times SO(3)\times U(1)$ associated with
the model global symmetry provides valuable and 
useful yet partial information about that complex problem. 
Here we start a preliminary analysis of the degrees of freedom 
of the rotated-electron occupancy configurations that generate a complete set of
$S_{\eta}$, $S_{\eta}^z$, $S_s$, $S_s^z$, and $S_c$ eigenstates and
introduce the theory vacua. Such $4^{N_a^D}$ states correspond to representations of the model global 
$SO(3)\times SO(3)\times U(1)$ symmetry. As discussed in Section IV, they are momentum eigenstates
of the square-lattice model in the whole Hilbert space.
For the 1D Hubbard model the states of such a complete set are shown in Appendix B to be
both momentum and energy eigenstates.

\subsection{Interplay of the global symmetry
with the transformation laws under the 
operator $\hat{V}$: three basic effective lattices
and the theory vacua}

The rotated-electron occupancy configurations associated with the 
state representations of the global 
$SO(3)\times SO(3)\times U(1)=[SU(2)\times SU(2)\times U(1)]/Z_2^2$ 
symmetry are well defined for $U/4t>0$. It is found in
Subsection IV-D that the $\eta$-spin and spin 
state representations correspond to independent rotated-electron
occupancy configurations of $[N_a^D-2S_c]$ sites and $2S_c$ sites,
respectively, whereas the $U(1)$ symmetry refers to the
relative positions of the sites involved in each of these two types of
configurations. In the present $N_a\gg 1$ limit it is useful to introduce
the numbers $N_{a_{\eta}}$ and $N_{a_{s}}$ such that 
$N_{a_{\eta}}^D=[N_a^D-2S_c]$ and $N_{a_{s}}^D=2S_c$, 
\begin{equation}
N_{a_{\eta}} = (N_a^D-2S_c)^{1/D} 
\, ; \hspace{0.25cm}
N_{a_{s}} = (2S_c)^{1/D} \, ; \hspace{0.25cm}
N_a^D = N_{a_{\eta}}^D + N_{a_{s}}^D  \, .
\label{Na-eta-s}
\end{equation}
Here $N_{a_{\eta}}^D$ and $N_{a_s}^D$ are integer numbers
that below are identified with the number of sites of
a $\eta$-spin and spin effective lattice, respectively.
For the $D=2$ square lattice the number $N_a^D$ 
is chosen so that the number $N_a$ of sites in a row or column  
is an integer. However, the designations $N_{a_{\eta}}^D$ and
$N_{a_s}^D$ do not imply that the corresponding numbers 
$N_{a_{\eta}}$ and $N_{a_{s}}$ are integers. In general they
are not integers but for finite values of $x$ and $(1-x)$
the closest integer numbers to $N_{a_{\eta}}$ and $N_{a_{s}}$ 
give the mean value of the number of sites in a row or column
of the $\eta$-spin and spin effective lattices, respectively.

The degrees of freedom of the rotated-electron occupancy configurations of each 
of the $N_{a_{\eta}}^D=[N_a^D-2S_c]$ and  $N_{a_{s}}^D=2S_c$ sites of the original 
lattice that generate the $S_{\eta}$, $S_{\eta}^z$, $S_s$, $S_s^z$, and $S_c$ eigenstates
studied in Section IV, which correspond to state representations of the model global 
$SO(3)\times SO(3)\times U(1)$ symmetry, naturally separate as follows: 

i) The occupancy configurations of the $c$ fermions associated with the 
operators $f_{\vec{r}_j,c}^{\dag}$ of Eq. (\ref{fc+}) where $j=1,...,N_a^D$
correspond to the state representations
of the global $U(1)$ symmetry found in Ref. \cite{bipartite}. Such $c$
fermions live on the $c$ effective lattice, which is identical to the original lattice. Its occupancies are related to those of the 
rotated electrons, the number of $c$ fermion occupied and unoccupied sites 
being given by $N_c = N_{a_{s}}^D=2S_c$
and $N_c^h = N_{a_{\eta}}^D=[N_a^D-2S_c]$, respectively. Indeed,
the $c$ fermions occupy the sites singly occupied by the
rotated electrons whereas the rotated-electron doubly-occupied
and unoccupied sites are those unoccupied by the $c$ fermions. Hence the
$c$ fermion occupancy configurations describe the relative positions
in the original lattice of the $N_{a_{\eta}}^D=[N_a^D-2S_c]$ sites of the
{\it $\eta$-spin effective lattice} and $N_{a_{s}}^D=2S_c$ sites
of the {\it spin effective lattice}.

ii) In turn, the remaining degrees of freedom of rotated-electron
occupancies of the $N_{a_{\eta}}^D=[N_a^D-2S_c]$ and  
$N_{a_{s}}^D=2S_c$ original lattice sites correspond to the occupancy configurations 
associated with the $\eta$-spin $SU(2)$ symmetry
and spin $SU(2)$ symmetry representations,
respectively. The occupancy configurations of the 
$N_{a_{\eta}}^D=[N_a^D-2S_c]$ sites of the
$\eta$-spin effective lattice and $N_{a_{s}}^D=2S_c$ sites
of the spin effective lattice are independent. They 
refer to the operators $p^l_{\vec{r}_j}$ of Eq. (\ref{sir-pir}), which
act only onto the $N_{a_{\eta}}^D=[N_a^D-2S_c]$ sites of
the $\eta$-spin effective lattice, and to the operators 
$s^l_{\vec{r}_j}$ given in the same equation, which
act onto the $N_{a_{s}}^D=2S_c$ sites of
the spin effective lattice, respectively.
This is assured by the operators $(1-n_{\vec{r}_j,c})$
and $n_{\vec{r}_j,c}$ in their expressions provided
in that equation, which play the role of projectors
onto the $\eta$-spin and spin effective lattice, respectively.

If follows that for $U/4t>0$ the degrees of freedom of each rotated-electron singly 
occupied site separate into a spin-less $c$ fermion carrying the electronic
charge and a spin-down or spin-up spinon. Furthermore, the degrees of freedom of 
each rotated-electron doubly-occupied or unoccupied site separate 
into a $\eta$-spin-less {\it $c$ fermion hole} and a $\eta$-spin-down or 
$\eta$-spin-up $\eta$-spinon, respectively. The $\eta$-spin-down or 
$\eta$-spin-up $\eta$-spinon refers to the $\eta$-spin degrees of
freedom of a rotated-electron doubly-occupied
or unoccupied site, respectively, of the original lattice. Above and in the remaining of this
paper we often call $c$ fermion hole an unoccupied site of the $c$ effective
lattice.

A key point of our description is that for $U/4t>0$ its quantum objects correspond 
to rotated-electron configurations whose spin-down and
spin-up singly occupancy, doubly occupancy, and no occupancy
refer to good quantum numbers. This is in contrast
to descriptions in terms of electronic configurations, whose
singly occupancy refers to a good quantum number for
$U/4t\gg 1$ only \cite{2D-MIT,Feng,Fazekas,Xiao-Gang}.
 
The transformation laws of the $\eta$-spinons (and spinons) under the 
electron - rotated-electron unitary transformation associated with the 
operator $\hat{V}$ play a major role in the description of the $\eta$-spin 
(and spin) state representations in terms of occupancy configurations of the 
$N_{a_{\eta}}^D=[N_a^D-2S_c]$ sites of the $\eta$-spin
effective lattice (and $N_{a_{s}}^D=2S_c$ sites of the spin
effective lattice). Importantly, a well-defined number of
$\eta$-spinons (and spinons) remain invariant under the unitary 
transformation generated by $\hat{V}$. Those are called 
independent $\pm 1/2$ $\eta$-spinons (and independent 
$\pm 1/2$ spinons) and as further discussed below play the role of unoccupied 
sites of the $\eta$-spin (and spin) effective lattice. The 
values of the numbers $L_{\eta,\,\pm 1/2}$
of independent $\pm 1/2$ $\eta$-spinons and 
$L_{s,\,\pm 1/2}$ of independent $\pm 1/2$ spinons 
are fully controlled by those of the 
$\eta$-spin $S_{\eta}$ and $\eta$-spin projection $S_{\eta}^z=-x\,N_a^D$ 
and spin $S_{s}$ and spin projection $S_{s}^z=-m\,N_a^D$, respectively, and are
given by,
\begin{equation}
L_{\alpha} = [L_{\alpha,-1/2}+L_{\alpha,+1/2}]=2S_{\alpha} \, ;
\hspace{0.25cm}
L_{\alpha,\,\pm 1/2} = [S_{\alpha}\mp S_{\alpha}^z]
\, ; \hspace{0.25cm} \alpha = \eta \, , s \, .
\label{L-L}
\end{equation}
The invariance of such independent $\eta$-spinons (and spinons)
stems from the off diagonal generators of the $\eta$-spin
(and spin) algebra, which flip their $\eta$-spin (and spin), commuting with 
the unitary operator $\hat{V}$ and hence
having for $U/4t>0$ the same expressions in terms of electron and
rotated-electron operators, as given in Eq. (\ref{Scs}).

It follows that the number of sites of the $\eta$-spin ($\alpha =\eta$) and spin
($\alpha =s$) effective lattice can be written as,
\begin{equation}
N_{a_{\alpha}}^D = [2S_{\alpha} + 2C_{\alpha}] 
\, ; \hspace{0.25cm} \alpha = \eta \, , s \, .
\label{2S-2C}
\end{equation}
The occupancy configurations of $2S_{\alpha}$
sites out of $N_{a_{\alpha}}^D$ have for $U/4t>0$ the same form in terms
of electrons and rotated electrons. However, that refers to the
$\eta$-spin or spin effective lattices and hence only to the
$\eta$-spin ($\alpha=\eta$) or spin ($\alpha=s$) degrees of
freedom, respectively, of the rotated-electron occupancies of the original lattice. 
Indeed, for finite values of $U/4t$ and spin density $m<(1-x)$
the $c$ fermion occupancy configurations are not invariant 
under the electron - rotated-electron unitary transformation.

Nevertheless, the two sets of $2S_{\eta}$ and $2C_{\eta}$ (and
$2S_{s}$ and $2C_{s}$) sites of the $\eta$-spin (and spin) 
effective lattice correspond to two well-defined sets of
$2S_{\eta}$ and $2C_{\eta}$ (and $2S_{s}$ and $2C_{s}$) sites
of the original lattice. Out of the set of $2S_{\eta}=[L_{\eta,-1/2}+L_{\eta,+1/2}]$
(and $2S_{s}=[L_{s,-1/2}+L_{s,+1/2}]$) sites of the original
lattice, $L_{\eta,-1/2}$ are doubly occupied and $L_{\eta,+1/2}$
unoccupied both by electrons and rotated electrons
(and $L_{s,-1/2}$ and $L_{s,+1/2}$ are singly occupied 
both by spin-down and spin-up, respectively, electrons and 
rotated electrons.) Out of the $2C_{\eta}$ (and $2C_{s}$) sites of the
original lattice left over, the corresponding
$C_{\eta}$ (and $C_{s}$) sites are unoccupied
by rotated electrons (and singly occupied by spin-up  
rotated electrons) and the remaining $C_{\eta}$ (and $C_{s}$) 
sites are doubly occupied by rotated electrons (and 
singly occupied by spin-down rotated electrons).
In turn, in terms of electrons these $[2C_{\eta}+2C_{s}]$ 
sites of the original lattice have for finite $U/4t$ values
very involved occupancies such that singly and doubly 
occupancy are not good quantum numbers and thus are
not conserved.

The site numbers $C_{\eta}\geq 0$ and $C_{s}\geq 0$ are good quantum
numbers given by,
\begin{equation}
C_{\eta} = [N_{a_{\eta}}^D/2 - S_{\eta}] =
[N_a^D/2 - S_c - S_{\eta}] \, ; \hspace{0.25cm}  
C_{s} = [N_{a_{s}}^D/2 - S_{s}] = [S_c - S_{s}] \, .
\label{C-C}
\end{equation}
Hence their values are fully determined by those of the eigenvalue $S_c $ of the global $U(1)$ symmetry
generator and $\eta$-spin $S_{\eta}$ or spin $S_s$, respectively, so that $C_{\eta}$ and $C_{s}$
are not independent quantum numbers. 

The physics behind the $U(1)$ 
symmetry found in Ref. \cite{bipartite} includes that
brought about by the rotated-electron occupancy
configurations of the $[2C_{\eta}+2C_{s}]$ sites of Eq. (\ref{C-C}).   
The use of the corresponding model global 
$SO(3)\times SO(3)\times U(1)$ symmetry confirms that 
the integer numbers $2C_{\alpha}\geq 0$ are always even
and that application onto $S_{\alpha}=0$ states of the off-diagonal generators of the 
$\eta$ spin $(\alpha =\eta)$ or spin $(\alpha =s)$
algebra provided in Eq. (\ref{Scs}) gives zero. 
For such states $N_{a_{\alpha}}^D = 2C_{\alpha}$.

On the other hand, application of these generators onto $2S_{\alpha}>0$ states flips the
$\eta$-spin ($\alpha=\eta$) or spin ($\alpha=s$)
of an independent $\eta$-spinon ($\alpha=\eta$) or independent spinon ($\alpha=s$)
but leaves invariant the rotated-electron occupancy 
configurations of the above considered $2C_{\alpha}$ sites.  
It follows that such $2C_{\eta}$ (and $2C_{s}$) sites 
refer to $\eta$-spin-singlet (and spin-singlet) configurations involving
$C_{\eta}$ (and $C_{s}$) $-1/2$ $\eta$-spinons
(and $-1/2$ spinons) and an equal number of $+1/2$ 
$\eta$-spinons (and $+1/2$ spinons). 

One then concludes that concerning the rotated-electron  
occupancies of the original lattice,
besides the $[2S_{\eta}+2S_s]$ sites whose occupancies 
are the same in terms of electrons and rotated electrons,
there are $C_{\eta}\geq 0$ sites doubly occupied by
rotated electrons, $C_{\eta}\geq 0$ sites unoccupied by
rotated electrons, $C_{s}\geq 0$ sites singly occupied by
spin-down rotated electrons, and $C_{s}\geq 0$ sites singly 
occupied by spin-up rotated electrons.

Let $M_{\alpha} = M_{\alpha,-1/2}+M_{\alpha,+1/2}$ denote 
the total number of  $\eta$-spinons ($\alpha =\eta$) and spinons 
($\alpha =s$). It is given by $M_{\alpha} =N_{a_{\alpha}}^D$ and hence
$M_{\eta} = N_{a_{\eta}}^D=[N_a^D-2S_c]$ and $M_s=N_{a_{s}}^D=2S_c$.
Out of the $M_{\eta} = N_{a_{\eta}}^D=[N_a^D-2S_c]$ $\eta$-spinons
(and $M_s=N_{a_{s}}^D=2S_c$ spinons), $L_{\eta}=2S_{\eta}$ 
(and $L_{s}=2S_{s}$) are invariant under 
the unitary operator $\hat{V}$ and
$2C_{\eta}=[M_{\eta}-2S_{\eta}]$ (and $2C_{s}=[M_{s}-2S_{s}]$)
are not invariant under that unitary operator. Also the 
$c$ fermions are not invariant under $\hat{V}$. 
Indeed, the generators of the occupancy configurations of the latter objects
are for finite values of $U/4t$ and $m<(1-x)$ different when expressed in terms of 
rotated-electron and electron operators, respectively. 
The number of $\eta$-spinons ($\alpha =\eta$) 
and spinons ($\alpha =s$) can then be written as, 
\begin{equation}
M_{\alpha,\,\pm 1/2} = [L_{\alpha,\,\pm 1/2} + C_{\alpha}] \, ;
\hspace{0.25cm}
M_{\alpha} = N_{a_{\alpha}}^D = [2S_{\alpha} + 2C_{\alpha}]
\, ; \hspace{0.25cm} \alpha = \eta \, , s \, .
\label{M-L-C}
\end{equation}

The $\eta$-spinon and spinon operator
algebra refers to well-defined subspaces spanned by states
whose number of each of these basic objects is
constant and given by $M_{\eta}=N_{a_{\eta}}^D=[N_a^D-2S_c]$ 
and $M_{s}=N_{a_{s}}^D=2S_c$,
respectively. Hence in such subspaces the number $2S_c$ of 
rotated-electron singly occupied sites and the numbers $N_{a_{\eta}}^D$
and $N_{a_{s}}^D$ of sites of the $\eta$-spin and spin effective lattices,
respectively, are constant. For hole concentrations $0\leq x<1$ and
maximum spin density $m=(1-x)$ reached at a critical magnetic field
$H_c$ parallel to the square-lattice plane for $D=2$ and 
pointing along the chain for $D=1$ the $c$ fermion operators are 
invariant under the electron - rotated-electron unitary 
transformation and there is a fully polarized  
vacuum $\vert 0_{\eta s}\rangle$, which remains
invariant under such a transformation. It reads,
\begin{equation}
\vert 0_{\eta s}\rangle = \vert 0_{\eta};N_{a_{\eta}}^D\rangle\times\vert 0_{s};N_{a_{s}}^D\rangle
\times\vert GS_c;2S_c\rangle \, ,
\label{vacuum}
\end{equation}
where the $\eta$-spin $SU(2)$ vacuum $\vert 0_{\eta};N_{a_{\eta}}^D\rangle$ 
associated with $N_{a_{\eta}}^D$ independent $+1/2$
$\eta$-spinons, the spin $SU(2)$ vacuum $\vert 0_{s};N_{a_{s}}^D\rangle$ 
with $N_{a_{s}}^D$ independent $+1/2$ spinons, and the $c$ $U(1)$
vacuum $\vert GS_c;2S_c\rangle$ with $N_c=2S_c$ $c$ fermions
remain invariant under the electron - rotated-electron unitary transformation. 
The explicit expression of the state $\vert GS_c;2S_c\rangle$ used
below in Subsection IV-E is $\prod_{{\vec{q}}}f^{\dag}_{{\vec{q}},c}\vert GS_c;0\rangle$
where the vacuum $\vert GS_c;0\rangle$ corresponds to the electron and
rotated-electron vacuum of form (\ref{vacuum}) referring to
$N_{a_{\eta}}^D=N_a^D$ and $N_{a_{s}}^D=2S_c=0$. Only for 
a $m=(1-x)$ fully polarized state are the state $\vert GS_c;2S_c\rangle$
and the corresponding $N_c=2S_c$ fermions 
invariant under the electron - rotated-electron unitary transformation for $U/4t>0$. 
For the vacuum $\vert 0_{\eta};N_{a_{\eta}}^D\rangle$
(and $\vert 0_{s};N_{a_{s}}^D\rangle$) the $M_{\eta}=N_{a_{\eta}}^D$ independent 
$+1/2$ $\eta$-spinons refer to $N_{a_{\eta}}^D$ sites of
the original lattice unoccupied 
by rotated electrons (and the $M_{s}=N_{a_{s}}^D$ 
independent $+1/2$ spinons to the spins of $N_{a_{s}}^D$ 
spin-up rotated electrons that singly occupy sites of such
a lattice). At maximum spin density $m=(1-x)$ the $c$ fermions are the non-interacting spinless 
fermions that describe the charge degrees of freedom of the electrons of
the fully polarized ground state. At that spin density 
there are no electron doubly occupied sites and the quantum
problem is non-interacting for $U/4t>0$.

The studies of Ref. \cite{companion} reveal that 
the $x=0$ and $m=0$ absolute ground state such that
$N_{a_{\eta}}^D=0$, $N_{a_{s}}^D=N_a^D=2N_{s1}$,
and $2S_c=N_a^D$ where $N_{s1}$ is the number of $s1$ fermions considered below
is also invariant under the electron - rotated-electron transformation. Such an 
invariance is related to the Mott-Hubbard insulator physics. 
This is consistent with the $S_c=N_a^D/2$
vacuum referring to a spin-up fully polarized state, whose
numbers are also $N_{a_{\eta}}^D=0$, $N_{a_{s}}^D=N_a^D$,
and $2S_c=N_a^D$ and correspond to a particular
case of the general state (\ref{vacuum}), being invariant 
under that transformation.

Following the result of Appendix A that full information about the quantum
problem described by the Hamiltonian (\ref{H})
can be achieved by defining it in the LWS-subspace spanned
by states such that $L_{\alpha,\,-1/2}=0$ and $L_{\alpha,\,+1/2}=2S_{\alpha}$ 
for $\alpha = \eta \, , s$, in the
ensuing section we confirm that within the description
introduced in this paper, out of the 
$N_{a_{\alpha}}^D = [2S_{\alpha} + 2C_{\alpha}]$
sites of the $\eta$-spin ($\alpha =\eta$) 
and spin ($\alpha =s$) effective lattice, the $2S_{\alpha}$
lattices whose occupancies are invariant under
$\hat{V}$ play the role of unoccupied sites,
whereas the remaining $2C_{\alpha}$ sites play
the role of occupied sites. This is a natural consequence of
the $\eta$-spin $SU(2)$ vacuum $\vert 0_{\eta};N_{a_{\eta}}^D\rangle$ 
(and spin $SU(2)$ vacuum $\vert 0_{s};N_{a_{s}}^D\rangle$)
being for all $U/4t$ and $m$ values invariant under the electron - rotated-electron
unitary transformation and such that $N_{a_{\eta}}^D = 2S_{\eta}$ 
so that $2C_{\eta}=0$ (and $N_{a_{s}}^D = 2S_{s}$ 
so that $2C_{s}=0$).

\section{Composite $\alpha\nu$ fermions, $\alpha\nu$ bond particles, and 
corresponding $\alpha\nu$ effective lattices}

In this section we consider a uniquely chosen complete set of
$S_{\eta}$, $S_{\eta}^z$, $S_s$, $S_s^z$, $S_c$, and momentum eigenstates
that within the present description for the Hubbard model on the 1D or square lattices is 
generated by $c$ fermion, $\eta$-spinon, and spinon occupancy configurations.
As discussed in Appendix B, for the 1D Hubbard model with $U/4t>0$ such states are 
a complete set of energy eigenstates. For the model on the square lattice 
such states are in some particular cases energy eigenstates, including
ground states and the states that span the one- and two-electron subspace 
as defined in Section V. Here we access partial yet valuable information 
about the $\eta$-spinon and spinon occupancy configurations that
generate the $\eta$-spin and spin degrees of freedom of such
$S_{\eta}$, $S_{\eta}^z$, $S_s$, $S_s^z$, $S_c$, and momentum eigenstates
from the suitable use of the transformations 
laws of the above objects under the electron - rotated-electron unitary transformation. 

Specifically, within the general description introduced in this paper both for
the Hubbard model on the square and 1D lattice, the complete set of 
$S_{\eta}$, $S_{\eta}^z$, $S_s$, $S_s^z$, $S_c$, and momentum eigenstates 
can in the $N_a^D\rightarrow\infty$ limit be generated by occupancy configurations of
$c$ fermions, several branches of composite $\alpha\nu$ fermions
labeled by the indices $\alpha=\eta,s$ and $\nu=1,2,...$, and a well-defined
number of independent $\eta$-spinons and independent spinons. The latter objects are invariant
under the above unitary transformation and thus have a non-interacting character for
$U/4t>0$. The composite $\alpha\nu$ fermions and corresponding $\alpha\nu$ bond particles involve
$2\nu$ $\eta$-spinons ($\alpha =\eta$) or $2\nu$ spinons ($\alpha =s$) 
where $\nu=1,2,...$ is the number of $\eta$-spin-neutral $\eta$-spinon (and spin-neutral spinon) pairs. 

\subsection{The composite $\alpha\nu$ bond particles and $\alpha\nu$ fermions}

\subsubsection{The $\alpha\nu$ fermion operators and $\alpha\nu$ translation generators}

Alike for the $s1$ bond particles further studied in Refs. \cite{companion0,companion}, one 
can introduce within the $N_a^D\rightarrow\infty$ limit suitable creation 
operators $g^{\dag}_{{\vec{r}}_{j},\alpha\nu}$ for the $\alpha\nu$ bond particles.
Provided that $(1-x)>0$ for $\alpha\nu=s1$ and $S_{\alpha}/N_a^D>0$ for the
remaining $\alpha\nu$ branches, the $\alpha\nu$ bond-particle operators have been constructed to 
inherently upon acting onto their $\alpha\nu$ effective lattice introduced
below in Subsection IV-D anticommuting on 
the same site and commuting on different sites, so that they are hard-core like
and can be transformed onto fermionic operators whose general expression reads,
\begin{eqnarray}
f^{\dag}_{{\vec{r}}_{j},\alpha\nu} & = & e^{i\phi_{j,\alpha\nu}}\,
g^{\dag}_{{\vec{r}}_{j},\alpha\nu} \, ; \hspace{0.35cm}
\phi_{j,\alpha\nu} = \sum_{j'\neq j}f^{\dag}_{{\vec{r}}_{j'},\alpha\nu}
f_{{\vec{r}}_{j'},\alpha\nu}\,\phi_{j',j,\alpha\nu} \, ; \hspace{0.35cm}
\phi_{j',j,\alpha\nu} = \arctan \left({{x_{j'}}_2-{x_{j}}_2\over {x_{j'}}_1-{x_{j}}_1}\right) \, ,
\nonumber \\
f_{\vec{q}_j,\alpha\nu}^{\dag} & = & 
{1\over {\sqrt{N_{a_{\alpha\nu}}^D}}}\sum_{j'=1}^{N_{a_{\alpha\nu}}^D}\,e^{+i\vec{q}_j\cdot\vec{r}_{j'}}\,
f_{\vec{r}_{j'},\alpha\nu}^{\dag} \, ,
\hspace{0.15cm} (1-x)>0  \hspace{0.15cm}{\rm for}\hspace{0.15cm}
\alpha\nu = s1 \hspace{0.15cm}{\rm and}\hspace{0.15cm}
S_{\alpha}/N_a^D>0 \hspace{0.15cm}{\rm for}\hspace{0.15cm}
\alpha\nu\neq s1 \, .
\label{f-an-operators}
\end{eqnarray}
Here $\phi_{j,\alpha\nu}$ is the Jordan-Wigner phase \cite{Wang} operator,
the indices $j'$ and $j$ refer to sites of the $\alpha\nu$ effective lattice,
and $f_{\vec{q}_j,\alpha\nu}^{\dag}$ are the corresponding momentum-dependent 
$\alpha\nu$ fermion operators. The number $N_{a_{\alpha\nu}}^D$ of discrete momentum values of the
$\alpha\nu$ momentum band equals that of sites of the $\alpha\nu$ effective lattice derived below in Subsection IV-D.
In turn, if for $\alpha\nu\neq s1$ branches one has that $S_{\alpha}=0$ and $N_{\alpha\nu}/N_a^D\ll 1$,
provided that $N_{\alpha\nu'}=0$ for all remaining $\alpha\nu'$ branches with a number of $\eta$-spinon
or spinon pairs $\nu'>\nu$ one finds below that $N_{a_{\alpha\nu}}^D=N_{\alpha\nu}$, the phase operator $\phi_{j,\alpha\nu}$
may be replaced by the average phase $\phi_{j,\alpha\nu} \approx \phi_{j,\alpha\nu}^0 = 
\sum_{j'\neq j}\phi_{j',j,\alpha\nu}$, and the momenta of the operators $f_{\vec{q}_j,\alpha\nu}^{\dag}$
are given by $\vec{q}_j \approx 0$. In the latter case all sites of the $\alpha\nu$ effective lattice are 
occupied and the $\alpha\nu$ momentum band is full. If the value of $N_{\alpha\nu}$ is finite
one has $N_{a_{\alpha\nu}}^D= N_{\alpha\nu}$ discrete momentum values $\vec{q}_j\approx 0$
distributed around zero momentum, whose Cartesian components momentum 
spacing is $2\pi/L$. A case of interest is when $N_{\alpha\nu}=1$ so that the $\alpha\nu$ effective lattice
has a single site and the corresponding $\alpha\nu$ band a single discrete momentum
value, $\vec{q} = 0$. In that case $\phi_{\alpha\nu}=\phi_{j,\alpha\nu}=0$ so that $f^{\dag}_{{\vec{r}},\alpha\nu} 
= g^{\dag}_{{\vec{r}},\alpha\nu}$ and $f_{\vec{q},\alpha\nu}^{\dag}=f_{\vec{r},\alpha\nu}^{\dag}$
where $\vec{q} = 0$.

As further discussed in Subsection IV-E, the operators
$f_{\vec{q}_j,\alpha\nu}^{\dag}$ act onto subspaces with constant values for the set of
numbers $S_{\alpha}$, $N_{\alpha\nu}$, and $\{N_{\alpha\nu'}\}$ for $\nu'>\nu$ and equivalently 
of the numbers $S_c$, $N_{\alpha\nu}$, and $\{N_{\alpha\nu'}\}$ for all $\nu'\neq \nu$. 
Such subspaces are spanned by mutually neutral states, that is states with constant values for the numbers 
of $\alpha\nu$ fermions and $\alpha\nu$ fermion holes, respectively. Hence such states 
can be transformed into each other by $\alpha\nu$ band particle-hole processes. The phases 
$\phi_{j,\alpha\nu}$ given in Eq. (\ref{f-an-operators}) are associated with an effective
vector potential \cite{Wang,Giu-Vigna},
\begin{eqnarray}
{\vec{A}}_{\alpha\nu} ({\vec{r}}_j) & = & \Phi_0\sum_{j'\neq j}
n_{\vec{r}_{j'},\alpha\nu}\,{{\vec{e}}_{x_3}\times ({\vec{r}}_{j'}-{\vec{r}}_{j})
\over ({\vec{r}}_{j'}-{\vec{r}}_{j})^2} \, ; \hspace{0.35cm} 
n_{\vec{r}_j,\alpha\nu} = f_{\vec{r}_j,\alpha\nu}^{\dag}\,f_{\vec{r}_j,\alpha\nu} \, ,
\nonumber \\
{\vec{B}}_{\alpha\nu} ({\vec{r}}_j) & = & {\vec{\nabla}}_{\vec{r}_j}\times {\vec{A}}_{\alpha\nu} ({\vec{r}}_j)
=  \Phi_0\sum_{j'\neq j}
n_{\vec{r}_{j'},\alpha\nu}\,\delta ({\vec{r}}_{j'}-{\vec{r}}_{j})\,{\vec{e}}_{x_3}
\, ; \hspace{0.35cm} \Phi_0 = 1 \, ,
\label{A-j-s1-3D}
\end{eqnarray}   
where ${\vec{e}}_{x_3}$ is the unit vector perpendicular to the plane and we use units such 
that the fictitious magnetic flux quantum is given by $\Phi_0=1$. It follows from
the form of the effective vector potential ${\vec{A}}_{\alpha\nu} ({\vec{r}}_j)$ that
the present description leads to the intriguing situation where the $\alpha\nu$ 
fermions interact via long-range forces while all interactions in the original Hamiltonian are onsite. 

The components of the microscopic momenta of the $\alpha\nu$ fermions are 
eigenvalues of the two (and one for 1D) $\alpha\nu$ translation generators 
${\hat{q}}_{\alpha\nu\,x_1}$ and ${\hat{q}}_{\alpha\nu\,x_2}$
in the presence of the fictitious magnetic field ${\vec{B}}_{\alpha\nu} ({\vec{r}}_j)$. That
seems to imply that for the model on the square lattice the components 
${q}_{x1}$ and ${q}_{x2}$ of the microscopic momenta $\vec{q}=[{q}_{x1},{q}_{x2}]$ refer to operators that do not commute.
However, such operators commute
in the subspaces where the operators $f_{\vec{q}_j,\alpha\nu}^{\dag}$ act onto
because those are spanned by neutral states \cite{Giu-Vigna}. Since $[{\hat{q}}_{\alpha\nu\,x_1},{\hat{q}}_{\alpha\nu\,x_2}]=0$
in such subspaces, for the model on
the square lattice the $\alpha\nu$ fermions carry a microscopic momentum $\vec{q}=[{q}_{x1},{q}_{x2}]$
where the components ${q}_{x1}$ and ${q}_{x2}$ are well-defined simultaneously.
Importantly, it follows that the momentum operator $\hat{{\vec{P}}}$ of Eq. (\ref{P-invariant})
commutes with the $\alpha\nu$ generators ${\hat{{\vec{q}}}}_{\alpha\nu}$ whose
eigenvalues are the $\alpha\nu$ fermion microscopic momenta $\vec{q}$. Hence,
within our description the momentum operator (\ref{P-invariant}) can be written as,
\begin{equation}
\hat{{\vec{P}}} = {\hat{{\vec{q}}}}_c 
+ \sum_{\nu =1}^{C{s}}{\hat{{\vec{q}}}}_{s\nu} 
+ \sum_{\nu =1}^{C{\eta}}{\hat{{\vec{q}}}}_{\eta\nu} 
+ \vec{\pi}\,{\hat{M}}_{\eta\, ,-1/2} \, ,
\label{P-c-alphanu}
\end{equation}
where the $c$ and $\alpha\nu$ translation generators read,
\begin{equation}
{\hat{{\vec{q}}}}_c  = \sum_{{\vec{q}}}{\vec{q}}\, \hat{N}_c ({\vec{q}})
\, ; \hspace{0.35cm}
{\hat{{\vec{q}}}}_{s\nu} = \sum_{{\vec{q}}}{\vec{q}}\, \hat{N}_{s\nu} ({\vec{q}})
\, ; \hspace{0.35cm}
{\hat{{\vec{q}}}}_{\eta\nu} = \sum_{{\vec{q}}}[\vec{\pi} -{\vec{q}}]\,\hat{N}_{\eta\nu} ({\vec{q}}) \, .
\label{m-generators}
\end{equation}
Here $\vec{\pi}$ is the momentum carried by a $-1/2$ $\eta$-spinon, 
the operator ${\hat{M}}_{\eta\, ,-1/2}$ counts the number 
$M_{\eta\, ,-1/2}=[L_{\eta\, ,-1/2}+C_{\eta}]$ of such objects, that
the $\eta\nu$ generators involve $[\vec{\pi} -{\vec{q}}]$ instead of ${\vec{q}}$
follows from the anti-bounding character of the $\eta\nu$ fermions discussed
below, and $\hat{N}_{c}({\vec{q}})$ and $\hat{N}_{\alpha\nu}({\vec{q}})$ are the momentum 
distribution-function operators,
\begin{equation}
\hat{N}_{c}({\vec{q}}) = f^{\dag}_{{\vec{q}},c}\,f_{{\vec{q}},c} \, ;
\hspace{0.35cm}
\hat{N}_{\alpha\nu}({\vec{q}}) = f^{\dag}_{{\vec{q}},\alpha\nu}\,f_{{\vec{q}},\alpha\nu} \, ,
\label{Nc-s1op}
\end{equation}
respectively. 

For the Hubbard model both on the square and 1D lattices the Hamiltonian $\hat{H}$ of 
Eq. (\ref{H}) and the momentum operator $\hat{{\vec{P}}}$ of Eqs. (\ref{P-invariant}) and
(\ref{P-c-alphanu}) obey the commutation relations,
\begin{equation}
[\hat{H},\hat{{\vec{P}}}] = [\hat{H},{\hat{M}}_{\alpha\, ,\pm 1/2}]
= [\hat{H},{\hat{{\vec{q}}}}_c] = 0 \, ; \hspace{0.35cm} 
[\hat{{\vec{P}}},{\hat{M}}_{\alpha\, ,\pm 1/2}] = [\hat{{\vec{P}}},{\hat{{\vec{q}}}}_c]
= [\hat{{\vec{P}}},{\hat{{\vec{q}}}}_{\alpha\nu}] = 0 \, ,
\label{H-P-commutators}
\end{equation}
for $\nu =1,...,C_{\alpha}$ and $\alpha =\eta,s$. In turn, the set of commutators $[\hat{H},{\hat{{\vec{q}}}}_{\alpha\nu}]$
vanish for the 1D model whereas for the model on the square lattice one has in general
that $[\hat{H},{\hat{{\vec{q}}}}_{\alpha\nu}]\neq 0$. 
It follows from Eq. (\ref{P-c-alphanu}) that the momentum eigenvalues $\vec{P}$
can be expressed as a sum of the filled $c$ and $\alpha\nu$ fermion microscopic momenta. 
       
In the limit $N_a\rightarrow\infty$ one has that $\sum_{\alpha =\eta ,s}\,C_{\alpha}\rightarrow\infty$ for  
most subspaces so that the set of $\alpha\nu$ translation generators 
${\hat{{\vec{q}}}}_{\alpha\nu}$ of Eq. (\ref{m-generators}) in the presence of 
the fictitious magnetic fields ${\vec{B}}_{\alpha\nu}$ of Eq. (\ref{A-j-s1-3D})
is infinite. That for 1D all such operators commute both with
the Hamiltonian and momentum operator is behind the integrability of the 1D Hubbard
model in that limit. Indeed, such an infinite set of translation generators is equivalent to
the infinite set of conservation laws, which are known to be behind that model integrability
\cite{Martins,Prosen}. Such laws are also equivalent to the conservation of the set of $\alpha\nu$
fermion numbers $\{N_{\alpha\nu}\}$ where $\alpha =\eta,s$, $\nu=1,...,C_{\alpha}$
and the maximum $C_{\alpha}$ magnitude is $N_a^D/2$ and thus
$N_a/2$ for 1D \cite{Prosen}. In turn, the Hubbard model on the square lattice is not integrable 
and consistently the set of $\alpha\nu$ translation generators ${\hat{{\vec{q}}}}_{\alpha\nu}$ of Eq. (\ref{m-generators})
do not commute in general with the Hamiltonian so that the corresponding $\alpha\nu$ fermion discrete 
momentum values $\vec{q}_j=[{q}_{x1},{q}_{x2}]$ are not good quantum numbers. 
As further discussed in Subsection IV-E, that as given in Eq. (\ref{H-P-commutators}) for the square-lattice 
model such $\alpha\nu$ translation generators 
commute with momentum operator is consistent with it being invariant under the 
electron - rotated-electron unitary transformation and the $\alpha\nu$ fermion operators acting onto subspaces
spanned by mutually neutral states. In general such subspaces refer to constant values of the number
of $\alpha\nu$ fermions. 

It turns out that processes within the one- and two-electron subspace
as defined in this paper where two $s1$ fermions are annihilated and one $s2$ fermion created are also
neutral in the sense that for the model on the square lattice the $s1$ fermion microscopic momenta $\vec{q}=[{q}_{x1},{q}_{x2}]$ 
are for such excitations associated with $s1$ translation generators 
${\hat{q}}_{s1\,x_1}$ and ${\hat{q}}_{s1\,x_2}$ that commute. Moreover, in such a subspace
there are no $\alpha\nu$ fermions other than $s1$ fermions and none or one zero-momentum 
$s2$ fermion. The interest of our operator description lies in that for the Hubbard model on the square lattice 
in the neutral subspaces of the one- and two-electron subspace the $s1$ translation generators 
${\hat{{\vec{q}}}}_{s1}$ of Eq. (\ref{m-generators}) in the presence of the fictitious magnetic field ${\vec{B}}_{s1}$ 
commuting with both the Hamiltonian and momentum operator. As justified below, in spite of the lack of the 
square-lattice model integrability, for that model in such a subspace 
the $s1$ translation generators have been constructed to 
inherently these commutation relations holding. 
Therefore, for the square-lattice quantum liquid corresponding to the
Hubbard model on a square lattice in the one- and two-electron subspace
as defined in this paper the $s1$ fermion discrete momentum values 
$\vec{q}=[{q}_{x1},{q}_{x2}]$ are good quantum numbers and thus are conserved. 
We emphasize that according to Eq. (\ref{H-P-commutators}) the commutator
$[\hat{H},{\hat{{\vec{q}}}}_c]$ vanishes so that for the model on the square lattice 
the $c$ band momenta are good quantum numbers for the whole Hilbert space, alike for the 1D model.
Hence for the model on the square lattice in the one- and two-electron subspace
both the $c$ band and $s1$ band discrete momentum values are good quantum numbers.
This result follows from the way that our operator
description is constructed. However, the shape of the $s1$ momentum band and of its
boundary as well as the form of the $s1$ and $c$ fermion energy dispersions
remain unsolved problems, which are addressed in Ref. \cite{companion}.
                     
The advantage of using momentum-dependent $c$ fermion operators $f_{\vec{q}_j,c}^{\dag}$ 
and $\alpha\nu$ fermion operators $f_{\vec{q}_j,\alpha\nu}^{\dag}$
given in Eqs. (\ref{fc+}) and (\ref{f-an-operators}), respectively,
is then that for the 1D model in the whole Hilbert space and the model on the
square lattice in the one- and two-electron subspace as defined in this
paper such microscopic momenta are good quantum numbers.
That for the model on the square lattice the $\alpha\nu$ translation generators 
${\hat{{\vec{q}}}}_{\alpha\nu}$ of Eq. (\ref{m-generators}) in the presence 
of the fictitious magnetic field ${\vec{B}}_{\alpha\nu}$ do not commute in general
with the Hamiltonian is consistent with the set of $\alpha\nu$ fermion numbers $\{N_{\alpha\nu}\}$ which label the 
$S_{\eta}$, $S_{\eta}^z$, $S_s$, $S_s^z$, $S_c$, and momentum eigenstates considered
in Subsection IV-E not being in general good quantum numbers. 

The related studies of Refs. \cite{companion,cuprates0} focus mainly
on the square-lattice quantum liquid in the one- and two-electron subspace. In turn,
here we consider as well general ${\cal{N}}$-electron subspaces where  
${\cal{N}}=1,2,3,4,...$. The $\alpha\nu$ fermion operators $f^{\dag}_{{\vec{q}},\alpha\nu}$ 
and $f_{{\vec{q}},\alpha\nu}$ act onto subspaces spanned by neutral states. Nevertheless,
creation of one $\alpha\nu$ fermion is a well-defined process whose generator
is the product of an operator, which fulfills small changes in the $\alpha\nu$ effective lattice
and corresponding $\alpha\nu$ momentum band and the operator 
$f^{\dag}_{{\vec{q}},\alpha\nu}$ appropriate to the excited-state subspace.
It turns out that for the model on the 1D and square lattices
creation of one $\eta\nu$ fermion onto a $[N_a^2-N]= 2\nu$ ground state
generates for $U/4t>0$ a $2\nu$-electron charge excited energy eigenstate.
Moreover, creation of one $s\nu$ fermion onto a $[N_{\uparrow}-N_{\downarrow}]= 2\nu$ 
ground state generates a $2\nu$-electron spin excited energy eigenstate.
Furthermore and as discussed below in Subsection IV-E, the created $\eta\nu$ fermion or $s\nu$ fermion
is invariant under the electron - rotated-electron unitary transformation.
However, the detailed internal structure of the composite $\alpha\nu$ fermions is 
in general a unsolved complex problem.

Since the microscopic momenta of the $c$ fermions are good quantum numbers 
for the model on the square lattice, one can define an energy dispersion $\epsilon_c (\vec{q})$ \cite{companion}.
For the 1D model the $\alpha\nu$ energy dispersions $\epsilon_{\alpha\nu} (\vec{q})$ are
well defined as well. In turn, for the model on the square lattice the energies
$\epsilon_{\alpha\nu} (\vec{q})$ have for $\alpha\nu\neq s1$ no uncertainty for the momentum
values for which such objects are invariant under the electron - rotated-electron
unitary transformation. This is the case of the above $\eta\nu$ fermion created
onto a $[N_a^2-N]= 2\nu$ ground state and  $s\nu$ fermion onto a $[N_{\uparrow}-N_{\downarrow}]= 2\nu$ 
ground state whose excited-state $\alpha\nu$ momentum band corresponds
to the zero momentum value only, as confirmed below. For general $\alpha\nu$ fermion momentum values there is some
uncertainty in the magnitude of $\epsilon_{\alpha\nu} (\vec{q})$
yet fortunately it refers to a well-defined energy window. 
There is in general a uncertainty in the magnitude of $\epsilon_{s1} (\vec{q})$ as well
yet such an energy dispersion is well defined for the one- and two-electron
subspace \cite{companion}. 

Within our $N_a^D\gg 1$ limit on can access general useful information
on the energies $\epsilon_c$ and $\epsilon_{\alpha\nu}$ without knowing in detail the form of the
$\alpha\nu$ bond-particle operators $g^{\dag}_{{\vec{r}}_{j},\alpha\nu}$
and corresponding $\alpha\nu$ fermion operators $f^{\dag}_{{\vec{r}}_{j},\alpha\nu}$,
which for $\alpha=\eta$ (and $\alpha=s$) involve a set of $\eta$-spinon operators 
$p^l_{\vec{r}_j} = (1-n_{\vec{r}_j,c})\,q^l_{\vec{r}_j}$ (and spinon operators
$s^l_{\vec{r}_j} = n_{\vec{r}_j,c}\,q^l_{\vec{r}_j}$) given in Eqs. (\ref{sir-pir})-(\ref{rotated-quasi-spin})
associated with the $\eta$-spin (and spin) neutral superposition of the
corresponding $2\nu=2,4,6,...$ $\eta$-spinons (and spinons). 
(The $s1$ bond-particle operators $g^{\dag}_{{\vec{r}}_{j},s1}$ are studied in Ref. \cite{companion0}.)
Since as confirmed below, the ground-state occupancy configurations involve only the charge $c$ fermions 
and composite two-spinon $s1$ fermions, that problem is addressed for the one- and two-electron subspace in Refs. 
\cite{companion,cuprates0} concerning the $s1$ fermions. The studies of such references 
profit from the $c$ fermions and $s1$ fermions being the only objects playing an active role in that 
subspace. In turn, in the following we access general
information about such composite objects including for instance
the anti-bounding and bounding character of the $2\nu$-$\eta$-spinon 
composite $\eta\nu$ fermions and $2\nu$-spinon 
composite $s\nu$ fermions, respectively, and the
energy uncertainty of their energy spectrum, which is smaller than or equals the corresponding energy bandwidth. For the Hubbard
model on the square lattice in the one- and two-electron subspace such objects play the role of ``quasiparticles''. 
Due to integrability, for the 1D model they have zero-momentum forward-scattering
interactions only. In turn, for the model on the square lattice the unusual scattering
properties of the quantum problem are controlled and determined by the
inelastic $c$ - $s1$ fermion collisions \cite{cuprates}.

For the limit $N_a^D\gg 1$ considered in this
paper the problem of the internal degrees of freedom
of the composite $\alpha\nu$ fermions separates from
that of their positions in the corresponding $\alpha\nu$ effective
lattice defined below and corresponding discrete momentum values
in the $\alpha\nu$ band, whose number equals that of sites in
such a lattice. In this section it is confirmed
that for the subspaces spanned by states with constant values 
of $S_{\eta}$, $S_s$, and $S_c$ the momentum occupancy configurations 
of the composite $\alpha\nu$ fermions and $c$ fermions 
generate the correct number of state configurations
of the group $SO(3)\times SO(3)\times U(1)$. Addition of the dimensions
of all such subspaces leads to a total number $4^{N_a^2}$ of independent
state representations, which indeed is the dimension of the Hilbert space.
Furthermore, we initiate the study of the
$c$ and $s1$ effective lattices for the states that span the 
one- and two-electron subspace as defined in this paper.    
                                      
For the particular case of the 1D Hubbard model, a preliminary version 
of the general operator description introduced here for both the model on
the square and 1D lattices is presented in Ref. \cite{1D}. 
The studies of that reference profit from the exact
solution of the 1D model. No explicit relations to the rotated-electron operators 
as those given in Eqs. (\ref{fc+})-(\ref{c-up-c-down}) are derived
and no relation to the $U(1)$ symmetry contained in the model 
$SO(3)\times SO(3)\times U(1)$ global symmetry is established.  
Indeed, the eigenvalue of the generator (\ref{Or-ope}) of the 
$U(1)$ symmetry is one half the number of rotated-electron
singly occupied sites $2S_c$. Such an eigenvalue plays an
important role in the present description. It determines
the values of the number $N_c=2S_c$ of $c$ fermions, $M_s=2S_c$
of spinons, $N_c^h=[N_a^D-2S_c]$ of $c$ fermion holes,
and $M_{\eta} = [N_a^D-2S_c]$ of $\eta$-spinons. 
In reference \cite{1D} the $\eta$-spinons, $c$ fermions,
independent $\eta$-spinons, and independent spinons
are called holons, $c$ pseudoparticles, Yang holons, and HL 
spinons, respectively, where HL stands for Heilmann and Lieb.
Moreover, the $\alpha\nu$ pseudoparticles considered in 
that reference are the $\alpha\nu$ fermions.    
                         
\subsubsection{More about the states representation in terms of composite $\alpha\nu$ fermion occupancy configurations}

For the model (\ref{H}) on the square lattice the
sites of the $\eta$-spin (and spin) effective lattice are
distributed along well-defined $ox_1$ and $ox_2$ directions,
defined by the fundamental vectors ${\vec{a}}_{x_1} =a\,{\vec{e}}_{x_1}$
and ${\vec{a}}_{x_2} =a\,{\vec{e}}_{x_2}$ of the original lattice where ${\vec{e}}_{x_1}$ and ${\vec{e}}_{x_2}$ 
are the usual Cartesian unit vectors and $a$ is the lattice constant. 
The average distance between such sites when going 
through these directions is $a_{\alpha} = L/N_{a_{\alpha}} = [N_a/N_{a_{\alpha}}]\,a$
where $\alpha = \eta , s$. For 1D this is also the average distance between the sites of such effective
lattices. $a_{\eta}$ is the average distance along the $ox_1$ and/or
the $ox_2$ direction between the sites that are either doubly occupied 
or unoccupied by rotated electrons. $a_{s}$ is the average distance along
that and/or those directions between the 
sites that are singly occupied by rotated electrons. 

The concept of a $\eta$-spin (and spin) effective lattice 
is well defined for finite values of the hole concentration $x$ (and electronic density $n=(1-x)$)
and the present $N_a^D\gg 1$ limit.
The general representation introduced in this paper contains
full information about the relative positions of the sites of 
the $\eta$-spin and spin effective lattices in the 
original lattice. For each rotated-electron occupancy configuration that generates a given
momentum eigenstate, such an information
is stored in the corresponding occupancy 
configurations of the $c$ fermions in 
their $c$ effective lattice, which is identical to the original lattice. The latter configurations correspond to the
state representations of the $U(1)$ symmetry in the subspaces
spanned by states with constant 
values of $S_c$, $S_{\eta}$, and $S_s$. Indeed, the
sites of the $\eta$-spin (and spin) effective lattice have
in the original lattice the same real-space coordinates as
the sites of the $c$ effective lattice unoccupied (and
occupied) by $c$ fermions. 
That such an information is stored in the $c$ fermion
occupancy configurations is consistent with
the $\eta$-spinon and spinon occupancy 
configurations of the $\eta$-spin and spin effective lattices, 
respectively, being independent of each other. The latter occupancy configurations 
refer to the state representations of the $SU(2)$ $\eta$-spin and spin
symmetries, respectively, in the subspaces with constant 
values of $S_c$, $S_{\eta}$, and $S_s$.

For the model on the 1D lattice the $\eta$-spin (and spin) effective lattice
is straightforwardly and uniquely obtained by considering only the rotated-electron
doubly and unoccupied sites (and singly occupied sites) of the
original lattice and keeping the same site order as for that
lattice. In turn, for the model on the square lattice the
$\eta$-spin (and spin) effective lattice is in the limit $x\rightarrow 1$ 
(and $x\rightarrow 0$) exactly a square
lattice. Indeed, in that limit it becomes identical to the original lattice. 
Moreover, it is an excellent approximation to consider that for
small electron density $n=(1-x)$ (and small hole
concentration $x$) the $\eta$-spin (and spin) effective lattice
is a square lattice. 

Since the $\eta$-spinon, spinon, and $c$ fermion description 
contains full information about the relative positions of the sites of 
the $\eta$-spin and spin effective lattices in the original lattice, it turns out that  
within the $N_a^D\gg 1$ limit and for finite values of
the hole concentration $x$ (and electron density $n=(1-x)$) the 
$\eta$-spin (and spin) effective lattice can for 2D be represented by a
square lattice with lattice constant $a_{\eta}$ (and $a_s$) given by, 
\begin{equation}
a_{\alpha} = {L\over N_{a_{\alpha}}} = {N_a\over N_{a_{\alpha}}}\, a
\, ; \hspace{0.25cm} \alpha = \eta \, , s \, .
\label{a-alpha}
\end{equation}

The expression of the operator ${\hat{S}}$ 
such that ${\hat{V}}^{\dag} = e^{{\hat{S}}}$ and ${\hat{V}} = e^{-{\hat{S}}}$
involves only the three kinetic operators 
$\hat{T}_0$, $\hat{T}_{+1}$, and $\hat{T}_{-1}$
given in Eq. (\ref{T-op}). Therefore, when expressed in terms of
rotated-electron creation and annihilation operators
the expression of the electron - rotated-electron 
unitary operator ${\hat{V}}={\tilde{V}}= e^{-{\tilde{S}}}$ 
involves only the corresponding three rotated kinetic operators 
$\tilde{T}_0$, $\tilde{T}_{+1}$, and $\tilde{T}_{-1}$
and thus preserves the occurrence 
of only nearest hopping for rotated electrons, alike for electrons. 
Indeed, the effective rotated-electron hopping
between second and third nearest neighboring
sites is generated by products of such operators so
that the corresponding elementary hopping processes correspond
only to nearest hopping. This confirms that
in the square lattice the rotated electrons hop along the $ox_1$ 
or the $ox_2$ direction between the sites. Hence the same
holds for $\eta$-spinons (and spinons) moving in 
the corresponding $\eta$-spin and
spin effective lattices, consistently with those being
square lattices. 

The $\eta$-spin (and spin) effective lattice includes only the 
rotated-electron doubly and unoccupied sites (and singly occupied sites) of the original 
lattice. However, within periodic boundary conditions 
a pair of rotated electrons or rotated holes (and a rotated electron)
moving once around the chain in 1D and through the whole crystal along 
the $ox_1$ or $ox_2$ directions in the square lattice and doubly occupying 
(and singly occupying) a site in each step must pass an overall distance 
$L$. Therefore, consistently with expression 
$a_{\alpha} = L/N_{a_{\alpha}} = [N_a/N_{a_{\alpha}}]\,a$
introduced below where $\alpha = \eta , s$, the 
$\eta$-spin (and spin) effective lattice has both for 1D and 2D the 
same length and edge length $L$, respectively, as the original lattice. Furthermore, 
the requirement that for the 2D case when going 
through the whole crystal along the $ox_1$ or $ox_2$ directions 
a $\eta$ spinon (and spinon) passes an overall distance 
$L$ is met by a square effective $\eta$-spin (and spin) lattice.
However, since the number of sites sum-rule
$[N_{a_{\eta}}^D+N_{a_{s}}^D]=N_a^D$ holds,
the $\eta$-spin and spin effective lattices have in general a number of 
sites $N_{a_{\alpha}}^D$ smaller than that of the original lattice, $N_a^D$.
It follows that their lattice constants given in Eq. (\ref{a-alpha}) are larger 
than that of the original lattice, $a_{\alpha} \geq a$ where 
$\alpha =\eta ,s$.   

The average distance between the sites of the
$\eta$-spin and spin effective lattices provided in that equation 
then plays the role of lattice constant of such lattices.
The validity for $N_a^D\gg 1$ of the square spin effective lattice constructed in that way
is confirmed by the average value $\langle\Psi\vert \delta d \vert\Psi\rangle$ 
relative to any energy eigenstate $\vert\Psi\rangle$ belonging to a
subspace with constant number of rotated-electron singly occupied sites
of the distance $\delta d$ in real space of any of the $N_{a_{s}}^D$ 
sites of the spin effective lattice from the rotated-electron singly occupied site of the original
lattice closest to it vanishing in the limit $N_a^D\rightarrow\infty$. We
recall that the number of rotated-electron singly occupied sites of the
original lattice equals that of sites of the spin effective lattice. 

It is useful to consider the occupancy configurations of the $2S_{\eta}$ (and $2S_s$) "unoccupied sites" 
and the $2C_{\eta}$ (and $2C_{s}$) "occupied sites" of the $\eta$-spin 
(and spin) effective lattice that generate state representations of
the model global $SO(3)\times SO(3)\times U(1)$ symmetry. For the LWS-subspace considered
here the $2S_{\eta}$ (and $2S_s$) "unoccupied sites" of
such a lattice correspond to $2S_{\eta}$ 
independent $+1/2$ $\eta$-spinons 
(and $2S_s$ independent $+1/2$ spinons)
that remain invariant under the unitary 
transformation associated with the operator $\hat{V}$. 
In turn, the $C_{\eta}$ $+1/2$ $\eta$-spinons and $C_{\eta}$
$-1/2$ $\eta$-spinons (and $C_{s}$ $+1/2$ spinons and $C_{s}$
$-1/2$ spinons) that refer to the $2C_{\eta}$  (and $2C_{s}$) 
"occupied sites" of such a lattice do not remain invariant 
under that unitary transformation. 

From the relation of the transformation laws under 
the electron - rotated-electron unitary transformation to the 
number of $\eta$-spin (and spin) singlet configurations of the 
subspaces with constant values of $S_c$, $S_{\eta}$, and $S_s$
one finds that such configurations can be expressed in terms
of occupancies of $\eta$-spin-neutral 
$2\nu$-$\eta$-spinon composite {\it $\eta\nu$ bond particles} 
(and spin-neutral $2\nu$-spinon composite {\it $s\nu$ bond 
particles}) in suitable and independent {\it $\eta\nu$ effective
lattices} (and {\it $s\nu$ effective lattices}). 
We recall that here $\nu=1,2,...,C_{\eta}$ (and $\nu=1,2,...,C_{s}$)
is the number of $\eta$-spinon pairs (and spinon pairs)
of opposite $\eta$-spin projection (and spin projection)
in each composite $\eta\nu$ bond particle 
(and $s\nu$ bond particle). Among  such composite $\alpha\nu$ bond particles,
the two-spinon $s1$ bond particles play a major
role since they are found below to be the only of the composite objects with
finite occupancy in the ground states. Such
two-spinon objects are found in Ref. \cite{companion0}
to be related to the resonating-valence-bond pictures considered long ago 
\cite{Fazekas,Pauling,Fradkin,Auerbach}
for spin-singlet occupancy configurations of ground states.

The $\alpha\nu$ fermions have the same internal structure
as the corresponding $\alpha\nu$ bond particles and live on the $\alpha\nu$ effective lattice
as well. Hence a $\eta\nu$ fermion (and $s\nu$ fermion) is a $\eta$-spin-neutral 
$2\nu$-$\eta$-spinon composite (and spin-neutral $2\nu$-spinon composite) object.
Since within our description the momentum occupancy configurations of the $c$ and $\alpha\nu$ fermions 
generate momentum eingestates, in the following
we refer mostly to $\alpha\nu$ fermions, yet many of our considerations apply
as well to the corresponding $\alpha\nu$ bond particles. Indeed, as given in Eq.
(\ref{f-an-operators}) the creation operator $f^{\dag}_{{\vec{r}}_{j},\alpha\nu}$ of a local
$\alpha\nu$ fermion of real-space coordinate ${\vec{r}}_{j}$ differs from
the creation operator $g^{\dag}_{{\vec{r}}_{j},\alpha\nu}$ of the
corresponding $\alpha\nu$ bond particle with the same real-space
coordinate by a phase factor $e^{i\phi_{j,\alpha\nu}}$, $f^{\dag}_{{\vec{r}}_{j},\alpha\nu} = e^{i\phi_{j,\alpha\nu}}\,
g^{\dag}_{{\vec{r}}_{j},\alpha\nu}$. Otherwise, such objects have the same internal structure.

Our description is consistent with the Abelian
and non-Abelian character of the $U(1)$ symmetry
and two $SU(2)$ symmetries, respectively, of
the model global $[SU(2)\times SU(2)\times U(1)]/Z_2^2=SO(3)\times SO(3)\times U(1)$ symmetry.
Indeed, the $c$ fermions occupancy configurations refer to the 
state representations of a $U(1)$ symmetry associated with an
Abelian group and hence involve a single $c$ effective lattice. In contrast, 
the $\eta$-spinon (and spinon) occupancy configurations 
refer to the state representations
of a $SU(2)$ symmetry associated with a non-Abelian group 
so that several types of $\eta$-spin (and spin) singlet configurations emerge,
each corresponding to a $\eta\nu$ effective lattice 
(and $s\nu$ effective lattice)
occupied by $\eta$-spin-zero (and spin-zero) 
composite objects of $2\nu$ $\eta$-spinons
(and $2\nu$ spinons). As confirmed below, there 
exists an independent $\alpha\nu$ effective lattice for each
$\alpha\nu$ branch where $\alpha=\eta ,s$ and 
$\nu=1,...,C_{\alpha}$. Here $\alpha$ refers to
$\eta$-spin ($\alpha =\eta$) and spin ($\alpha =s$) and
$\nu$ to the number of $\eta$-spin-singlet $\eta$-spinon
pairs ($C_{\alpha}=C_{\eta}$) and spin-singlet spinon
pairs ($C_{\alpha}=C_{s}$). 

Within chromodynamics the quarks have color but all quark-composite physical 
particles are color-neutral \cite{Martinus}. Here the $\eta$-spinon (and spinons) that
are not invariant under the electron - rotated-electron unitary
transformation have $\eta$ spin $1/2$ (and spin $1/2$) but are
part of $\eta$-spin-neutral (and spin-neutral) $2\nu$-$\eta$-spinon (and $2\nu$-spinon) 
composite $\eta\nu$ fermions (and $s\nu$ fermions).
The occurrence of such $\eta$-spin neutral $\eta\nu$ fermions 
(and spin neutral $s\nu$ fermions) leads to
the following expression for the number $2C_{\alpha}$ 
of "occupied sites" of the $\eta$-spin ($\alpha=\eta$) and spin 
($\alpha=s$) effective lattice of Eq. (\ref{2S-2C}), which 
refers to the subspaces spanned by states with constant values 
of $S_c$, $S_{\eta}$, and $S_s$, 
\begin{equation}
2C_{\alpha} =
2\sum_{\nu =1}^{C_{\alpha}}\nu\,N_{\alpha\nu} \, ; \hspace{0.25cm}
N_{a_{\alpha}}^D = M_{\alpha} = 2S_{\alpha} + 2C_{\alpha}
\, ; \hspace{0.25cm} \alpha = \eta \, , s \, .
\label{M-L-Sum}
\end{equation}
Here $N_{\alpha\nu}$ denotes the number of composite $\alpha\nu$ fermions. 
Each subspace spanned by states with constant values of $S_c$, $S_{\eta}$, 
and $S_s$ and hence also with constant values of $C_{\eta}$ and $C_s$
can be divided into smaller subspaces spanned by states with constant values for the
set of numbers $\{N_{\eta\nu}\}$ and $\{N_{s\nu}\}$, which must
obey the sum rules of Eq. (\ref{M-L-Sum}). The numbers $\{N_{\eta\nu}\}$ and $\{N_{s\nu}\}$
correspond to operators that commute with the momentum operator and operators
associated with the numbers $S_{\eta}$, $S_{\eta}^z$, $S_s$, $S_s^z$, and $S_c$.
In turn, for the model on the square lattice such operators do not commute with the
Hamiltonian so that the numbers $\{N_{\eta\nu}\}$ and $\{N_{s\nu}\}$ 
are not in general good quantum numbers. The same applies to the set
of $\alpha\nu$ translation generators ${\hat{{\vec{q}}}}_{\alpha\nu}$ of Eq. (\ref{m-generators}) 
in the presence of the fictitious magnetic 
field ${\vec{B}}_{\alpha\nu}$ of Eq. (\ref{A-j-s1-3D}), which as given in Eq. (\ref{H-P-commutators}) 
commute with the momentum operator yet in general do not commute with the Hamiltonian of the 
model on the square lattice. For 1D both such $\alpha\nu$ translation generators
commute with the Hamiltonian and the set of numbers $\{N_{\alpha\nu}\}$ are
within our limit $N_a^D\rightarrow\infty$ associated with the set of conservation laws
behind the model integrability \cite{Prosen} and thus are conserved.
Fortunately, we confirm below that they are good quantum numbers as well 
for the model on the square lattice in the one- and two-electron subspace and read $N_{\eta\nu}=0$, 
$N_{s1}=[S_c -S_s -2N_{s2}]$, $N_{s2}=0,1$, and $N_{s\nu}=0$ for $\nu\geq 3$, consistently with
and the $s1$ translation generator ${\hat{{\vec{q}}}}_{s1}$ of Eq. (\ref{m-generators})
in the presence of the fictitious magnetic field ${\vec{B}}_{s1}$ of Eq. (\ref{A-j-s1-3D}) commuting 
with both the Hamiltonian and momentum operator.

In the following we consider the complete set of $S_{\eta}$, $S_{\eta}^z$, $S_s$, $S_s^z$, $S_c$,
and momentum eigenstates also labelled by the numbers $\{N_{\eta\nu}\}$ and $\{N_{s\nu}\}$.
The expressions of the corresponding $\alpha\nu$ fermion operators involve the spinon $(\alpha =s)$ and  $\eta$-spinon $(\alpha =\eta)$
operators of Eqs. (\ref{sir-pir}), (\ref{n-r-c}), and (\ref{rotated-quasi-spin}). Both
such spinon and  $\eta$-spinon operators and the $c$ fermion operators are
expressed in terms of the rotated-electron operators given in Eq. (\ref{fc+}), which are
generated from the corresponding electron operators 
by the electron - rotated-electron unitary operator as defined in this paper.
Hence and as discussed below in subsection IV-E, for the model on the
square lattice both the present $S_{\eta}$, $S_{\eta}^z$, $S_s$, $S_s^z$, $S_c$,
and momentum eigenstates and the unknown energy eigenstates are
generated by application onto the corresponding sets of $U/4t\rightarrow\infty$
states by the same electron - rotated-electron unitary operator ${\hat{V}}^{\dag}={\hat{V}}^{\dag} (U/4t)$.
Indeed, for that model the energy eigenstates are a superposition of a set
of the former $S_{\eta}$, $S_{\eta}^z$, $S_s$, $S_s^z$, $S_c$,
and momentum eigenstates with the same momentum and 
$S_{\eta}$, $S_{\eta}^z$, $S_s$, $S_s^z$, and $S_c$ values. For the 1D
model both sets of states are identical, due to the infinite conservation laws,
which for $N_a^D\rightarrow\infty$ are associated with the model integrability.

For $\alpha\nu$ fermion branches with $N_{\alpha\nu}>0$ finite occupancy in a given state 
the maximum $\nu$ value $\nu_{max}$ obeys the inequality $\nu_{max}\leq C_{\alpha}$. 
The absolute maximum value 
$\nu_{max}= C_{\alpha}$ is reached for a subspace with one
$\alpha\nu$ fermion and vanishing occupancies for the remaining  
$\alpha\nu'$ branches such that $\nu'\neq\nu$. For the states spanning such a
subspace the $2C_{\alpha}$ $\eta$-spinons ($\alpha =\eta$) or spinons 
($\alpha =s$) other than the independent $\eta$-spinons or 
independent spinons, respectively, are part of a single 
$\alpha\nu$ fermion, which moves around in a $\alpha\nu$
effective lattice with $2S_{\alpha}$ unoccupied sites. 
For $S_{\alpha}=0$ such a single $\alpha\nu$ bond
particle has a "Big-Bang" like character in that the
$\eta$-spin ($\alpha =\eta$) or spin ($\alpha =s$) degrees of freedom
of a half-filling $N=N_a^D$ state with an equal number $N_a^D/2=\nu_{max}$  
of doubly and unoccupied sites or $N_a^D=2\nu_{max}$
sites singly occupied by an equal number $N_a^D/2=\nu_{max}$
of spin-up and spin-down electrons, respectively,
are described by such an object. Such two Big-Bang states have
$\eta$-spin and spin lattice site numbers: (i) $N_{a_{\eta}}^D=2\nu_{max}=N_a^D$ 
and $N_{a_{s}}^D=0$ referring to a single $\eta$-spin-neutral $\nu =N_a^D/2$ and 
$N_a^D$-$\eta$-spinon composite $\eta\nu$ fermion in a $\eta\nu$ band
lattice with a single vanishing momentum value and an empty $c$ momentum band
with $N^h_c=N_a^D$ $c$ fermion holes;
(ii) $N_{a_{\eta}}^D=0$ and $N_{a_{s}}^D=2\nu_{max}=N_a^D$ 
corresponding to a
single spin-neutral $\nu =N_a^D/2$ and $N_a^D$-spinon
composite $s\nu$ fermion in a $s\nu$ band with a single 
vanishing momentum value and a full $c$ momentum band
filled by $N_c=N_a^D$ $c$ fermions. It turns out
that such Big Bang $\alpha\nu$ fermions are
invariant under the electron - rotated-electron unitary 
transformation. That invariance implies that they
are $N$-electron objects and hence the corresponding
Big-Bang states have vanishing overlap with 
one- and two-electron excitations: Their creation requires
application onto the ground state whose occupancy 
configurations we study below of a suitable
$N$-electron operator. As a result in part of their invariance under the
electron - rotated-electron unitary transformation,
the $N$-electron excited state generated by creation
of such objects is an exact energy eigenstate both for the
model on the 1D and square lattice. 

\subsection{Processes that conserve and do not
conserve the number of sites of the $\eta$-spin and spin effective lattices}

The vacuum (\ref{vacuum}) 
of the theory corresponds to $N_{a_{\eta}}^D$ independent $+1/2$ $\eta$-spinons 
and $N_{a_{s}}^D$ independent $+1/2$ spinons so that 
as mentioned above such 
objects play the role of "unoccupied sites" of the $\eta$-spin 
and spin effective lattices, respectively. Consistently, the latter objects 
have vanishing energy and momentum and relative to that vacuum their
$\eta$-spin and spin flip processes correspond 
to "creation" processes of independent $-1/2$ $\eta$-spinons
and independent $-1/2$ spinons, respectively.

Furthermore, creation of a local $\eta\nu$ fermion involves the 
replacement of $2\nu$ independent $+1/2$ $\eta$-spinons
corresponding to $2\nu$ sites of the 
$\eta$-spin effective lattice by $\nu$ $-1/2$ $\eta$-spinons and
$\nu$ $+1/2$ $\eta$-spinons in the $\eta$-spin-neutral
configuration associated with such a $\eta\nu$ fermion.
Also creation of a local $s\nu$ fermion involves the 
replacement of $2\nu$ independent $+1/2$ spinons
referring to $2\nu$ sites of the spin effective lattice by 
$\nu$ $-1/2$ spinons and $\nu$ $+1/2$ spinons in the 
spin-neutral configuration associated with such a $s\nu$ fermion.
Finally, creation (and annihilation) of a $c$ fermion involves annihilation
(and creation) of a $c$ fermion hole. The latter process involves
removal (and addition) of one site from (and to)
the $\eta$-spin effective lattice and addition 
(and removal) of one site to (and from) the spin
effective lattice.

The momentum eigenstates considered in this paper are generated by microscopic momentum
occupancy configurations of $c$ and $\alpha\nu$ fermions. Such configurations can be expressed
as a superposition of local rotated-electron occupancies in the original lattice.
The degrees of freedom of the rotated-electron occupancy of 
a given site of the original lattice are always of two
types. For the sites singly occupied by a rotated electron
of spin projection $-1/2$ or $+1/2$, those are a
spinless $c$ fermion of charge $-e$ associated with
the $U(1)$ symmetry and a spinon 
of spin projection $-1/2$ or $+1/2$ associated with the 
spin $SU(2)$ symmetry, respectively.
In turn, for the sites doubly occupied or unoccupied
by rotated electrons, those are a $\eta$-spinless $c$ 
fermion hole associated with the $U(1)$  
symmetry and a $\eta$-spinon of $\eta$-spin projection 
$-1/2$ or $+1/2$ associated with the    
$\eta$-spin $SU(2)$ symmetry, respectively. 

It follows that the $2\nu$ sites
of the spin effective lattice occupied by one
local $s\nu$ fermion correspond to the spin
degrees of freedom of $2\nu$ rotated-electron singly 
occupied sites whose charge degrees of freedom
are described by $2\nu$ occupied sites of the $c$
fermion lattice. Therefore, these $2\nu$ rotated-electron singly 
occupied sites are described both by the local 
$s\nu$ fermion and $2\nu$ local $c$ fermions.

Moreover, the $2\nu$ sites
of the $\eta$-spin effective lattice occupied by one
local $\eta\nu$ fermion correspond to the $\eta$-spin
degrees of freedom of $\nu$ rotated-electron doubly  
occupied sites and $\nu$ rotated-electron unoccupied 
sites whose $U(1)$ symmetry degrees of freedom
are described by $2\nu$ unoccupied sites of the $c$
fermion lattice. As a result, the $\nu$ rotated-electron doubly  
occupied sites and $\nu$ rotated-electron unoccupied 
sites are described both by the local 
$\eta\nu$ fermion and $2\nu$ local $c$ fermion holes
($c$ effective lattice unoccupied sites).

In summary, local $s\nu$ fermions share $2\nu$ sites 
of the original lattice with $2\nu$ local $c$ fermions and 
local $\eta\nu$ fermions share $2\nu$ sites of that lattice with 
$2\nu$ local $c$ fermion holes.
Any pair of local $\alpha\nu$ and $\alpha'\nu'$ fermions
always refer to two {\it different} sets of $2\nu$ and
$2\nu'$ sites, respectively, of the original lattice. 
Creation of local $\eta\nu$ and $s\nu$ fermions are
processes that conserve the total number of $\eta$-spinons
and spinons, respectively, and hence also conserve the number
of $c$ fermions and $c$ fermion holes. In contrast, creation
(and annihilation) of one local $c$ fermion is a process that
involves addition (and removal) of one site to 
(and from) the spin effective lattice and removal (and addition)
of one site from (and to) the $\eta$-spin effective lattice. Therefore,
the spin and $\eta$-spin effective lattices are exotic, since the
number of their sites $N_{a_s}^D=2S_c$ and
$N_{a_{\eta}}^D=[N_a^D-2S_c]$, respectively, changes by $\pm 1$
and $\mp 1$ upon creation/annihilation of one
$c$ fermion. Indeed, such processes change the eigenvalue
$S_c$ of the generator (\ref{Or-ope}) of the global $U(1)$
symmetry. A subspace with constant $S_c$ value
and hence constant $N_{a_{\eta}}^D=[N_a^D-2S_c]$
and $N_{a_s}^D=2S_c$ values as well is associated
with a well-defined vacuum $\vert 0_{\eta s}\rangle$ 
given in Eq. (\ref{vacuum}). An excitation involving a change of
such values drives the system into a new subspace referring
to a different vacuum $\vert 0_{\eta s}\rangle$. 
In turn, $\eta$-spinon and spinon creation and annihilation
processes refer to excitations within the same
quantum-liquid subspace associated with a well-defined
vacuum $\vert 0_{\eta s}\rangle$. 

It follows that from the point of view of the $\eta$-spin and spin
degrees of freedom, $c$ fermion creation and annihilation
processes correspond to a change of quantum system.
Indeed, the $\eta$-spin and spin lattices and corresponding
number of sites change along with the quantum-system
vacuum $\vert 0_{\eta s}\rangle$ of Eq. (\ref{vacuum}).
Therefore, within the $\eta$-spinon and spinon
representation there is a different quantum system for each
eigenvalue $S_c$ of the generator (\ref{Or-ope}) of the global $U(1)$
symmetry. However, from the
point of view of the degrees of freedom associated with
the latter symmetry, the model (\ref{H}) corresponds
to a single quantum system and the local $c$ fermions live on a
lattice whose number of sites $N_a^D=[N_{a_s}^D+N_{a_{\eta}}^D]$
is constant, alike that of the electrons and rotated electrons.
Consistently, the $c$ effective lattice is identical to the 
original lattice. This property follows from the invariance of the latter
lattice under the electron - rotated-electron unitary
transformation, which is related to the invariance of
the momentum operator of Eq. (\ref{P-invariant}). 

Creation of local $\eta\nu$ (and $s\nu$) fermions always involves virtual processes 
where $2\nu$ independent $+1/2$ $\eta$-spinons (and $2\nu$
independent $+1/2$ spinons) are replaced by the
$\eta$-spin-singlet (and spin-singlet)
$2\nu$-site occupancy configurations of the local 
$\eta\nu$ fermions (and $s\nu$ fermions) in the
$\eta$-spin (and spin) effective lattice. For instance,
a given process for which two local $s1$ fermions of the initial state are replaced by
one local $s2$ fermion in the final state is
divided into two virtual processes: first,
two local $s1$ fermions are annihilated, {\it i.e.}
the four sites of the spin effective lattice occupied
in the initial state by the local $s1$ fermions 
are upon two spin-flip processes 
occupied in an intermediate virtual state
by four independent $+1/2$ spinons (annihilation
of two $s1$ bond particles); second, one local $s2$
fermion is created on such four sites.
That involves two opposite
spin-flip processes and rearrangement of the
spinons associated with the creation of the 
local $s2$ fermion spin-neutral four-spinon occupancy
configurations in the spin effective lattice.

Concerning creation (and annihilation) of one local $c$ fermion,
the corresponding overall excitation always involves
a virtual process for which the site of the $\eta$-spin
effective lattice removed (and added) by such an
elementary process is occupied in the initial (and
final) state by an independent $+1/2$ $\eta$-spinon. 
Also the site of the spin effective lattice added 
(and removed) by such an elementary process is occupied in the final 
(and initial) state by an independent $+1/2$ spinon.

If as occurs for one-electron addition (and removal),
creation (and annihilation) of a local 
$c$ fermion involves creation (and annihilation)
of a local $s1$ fermion, the overall process 
is divided into two virtual processes. 
For instance, complementarily to creation of the local $c$ fermion in 
its effective lattice, the virtual
processes occurring in the spin effective
lattice are: First, a site occupied by an independent 
$+1/2$ spinon is added to that lattice;
Second, a local $s1$ fermion is created, one of
the two independent $+1/2$ spinons involved in
the final-state two-site $s1$ bond configuration being that
located on the site added to the spin effective lattice.
(A corresponding momentum eigenstate involves the superposition
of many local configurations for which that site has different
positions.) In turn, if annihilation of a local $c$ fermion at its
effective lattice involves annihilation of a local $s1$
fermion, one has the following virtual
processes in the spin effective lattice: First, 
a local $s1$ fermion is annihilated,
what involves a rearrangement process, which 
leads to the occupancy of its two sites
by two independent $+1/2$ spinons in the
intermediate virtual state; Second, the
site of the spin effective lattice occupied by  
one of these two independent spinons
is removed along with it.

Creation (and annihilation) of both one local $c$ 
fermion and one local $s1$ fermion corresponds
to creation (and annihilation) of a spin-down 
electron. In turn, the corresponding process of
creation (and annihilation) of a spin-up electron 
involves addition (and removal) of one 
local $c$ fermion to (and from) its effective lattice and addition
(and removal) of one
site occupied by an independent $+1/2$ spinon
to (and from) the spin effective lattice.  
Alike creation and annihilation of local $c$ fermions
leads to addition and removal (and removal 
and addition) of sites in the spin (and $\eta$-spin)
effective lattice, respectively, it is confirmed below
that creation of one local $\alpha\nu'$ fermion
gives rise to addition of $2\nu'$ sites to the $\alpha\nu$ effective
lattices of $\alpha\nu$ fermion branches such that $\nu<\nu'$. 

It is confirmed below in Section V that
for the states that span the one- and two-electron subspace
only $c$ fermions and $s1$ fermions play an active role. 
Concerning excitations belonging to such a subspace that conserve the
number $N_c=2S_c$ of $c$ fermions and thus the eigenvalues
$S_c$ of the generator (\ref{Or-ope}) of the global $U(1)$
symmetry and the number $N_{s1}$ of $s1$ fermions, the $c$ fermions
can move through their effective lattice independently of
the motion of the corresponding $s1$
fermions. In that case when a given $c$ fermion moves around in the lattice it
is not attached always to the same spinon and hence
to the same two-spinon $s1$ fermion. The investigations
of Ref. \cite{cuprates0} reveal that there are correlations between
a spin-neutral two-spinon $s1$ fermion and a well defined
set of $c$ fermion pairs whose centre of mass is located
at the $s1$ fermion position.

\subsection{Ranges of the $\alpha\nu$ fermion and
$c$ fermion energies, transformation laws of such objects,
and the ground-state occupancies}

The quantum-object occupancy configurations of the ground state are found below. Self-consistency of our description then confirms
that for $m\geq 0$ and $x\geq 0$ the elementary energies $\epsilon_{s,-1/2} = 2\mu_B\,H$ and  $\epsilon_{\eta,-1/2} = 2\mu$ of Eq.
(\ref{energies}) of Appendix A correspond to creation onto the ground state of an independent $-1/2$ spinon and an independent
$-1/2$ $\eta$-spinon, respectively. We recall that $\mu_B$ is the Bohr magneton and $H$
the magnitude of a magnetic field aligned parallel to the plane (2D) or chain (1D).
The energy $\epsilon_{s,-1/2} = 2\mu_B\,H$ (and  $\epsilon_{\eta,-1/2} = 2\mu$)
refers to an elementary spin-flip (and $\eta$-spin-flip) process,
which transforms an independent $+1/2$ spinon (and
$+1/2$ $\eta$-spinon) into an independent $-1/2$ spinon 
(and $-1/2$ $\eta$-spinon). Such elementary energies control the range of
several physically important energy scales. Since within our LWS
representation an independent $+1/2$ spinon (and
$+1/2$ $\eta$-spinon) has vanishing energy and an 
independent $-1/2$ spinon (and $-1/2$ $\eta$-spinon)
has an energy given by $\epsilon_{s,-1/2} = 2\mu_B\,H$ (and  $\epsilon_{\eta,-1/2} = 2\mu$),
the energy of a pair of independent spinons (and $\eta$-spinons) 
with opposite projections is $2\mu_B\,H$ (and $2\mu$).
Indeed, due to the invariance of such objects under the
electron - rotated-electron unitary transformation, 
they are not energy entangled and the total energy
is the sum of their individual energies.

The magnetization curve is such that the spin density $m$ vanishes at $H=0$ and is finite and positive for $H>0$. 
(We recall that within the LWS representation the convention that $\mu>0$ for $x>0$ and $H>0$ for $m>0$ is used.) 
Consistently with the properties of the $x=0$ and $m=0$ absolute
ground state discussed below and in Ref. \cite{companion}, for $x=0$ and $m=0$ the
chemical potential $\mu$ belongs to the range
$\mu \in (-\mu^0,\mu^0)$ where the energy
scale $\mu^0\equiv \lim_{x\rightarrow 0}\mu$  
equals one half the Mott-Hubbard gap and
is such that $\mu^0\rightarrow 0$
for $U/4t\rightarrow 0$ and $\mu^0>0$
for $U/4t>0$. For $0<x<1$ and 
$m=0$ the chemical potential is an increasing 
function of the hole concentration $x$ such that,
\begin{equation}
\mu^0\leq\mu (x)\leq\mu^1 \, ; \hspace{0.25cm}  0<x<1
\, , \hspace{0.15cm} m= 0 \, ,
\label{mu-x}
\end{equation}
where $\mu^1\equiv \lim_{x\rightarrow 1}\mu$.
$\mu^1$ reads,
\begin{equation}
\mu^1 =  U/2 + 2Dt \, ;
\hspace{0.25cm} D = 1,2 \, .
\label{mu-1}
\end{equation}
In turn, $\mu^0$ has the following limiting behaviors,
\begin{equation}
\mu^0 \approx {U\over 2\pi^2}\left({[8\pi]^2 t\over U}\right)^{D/2}e^{-2\pi \left({t\over U}\right)^{1/D}} 
\, , \hspace{0.25cm} U/4t\ll 1 \, ;
\hspace{0.5cm}
\mu^0 \approx  [U/2 - 2Dt] \, ,
\hspace{0.25cm}  U/4t\gg 1  \, ,
\hspace{0.25cm} D = 1,2\, .
\label{DMH}
\end{equation}
Hence indeed $\mu^0\rightarrow 0$ as $U/4t\rightarrow 0$ whereas $\mu^0\rightarrow\infty$ for
$U/4t\rightarrow\infty$ for both the model on the 1D and square lattices. 

For $D=1$ the expressions provided in Eq. (\ref{mu-x})-(\ref{DMH}) are obtained by use
of the exact Bethe-ansatz solution \cite{Lieb}. Expression (\ref{mu-1}) is exact both for
1D and the square lattice. It can be derived
explicitly for both lattices, since it refers to the 
non-interacting limit
of vanishing electronic density. In turn, the
expressions given in Eq. (\ref{DMH}) 
for the model on the square lattice are obtained by the
use of a result discussed in Subsection VI-A:
the energy below which the long-range 
antiferromagnetic order survives for $x=0$, 
$m=0$, and zero temperature is 
$\mu^0$ where $2\mu_0$ is the Mott-Hubbard gap. 
Indeed, according to the analysis of that subsection
the Mott-Hubbard gap $2\mu^0$, which refers to the charge degrees of freedom, 
affects the spin degrees of freedom as well.
Consistently, the $D=2$ limiting behaviors
$\mu^0 \approx 32\,t\,e^{-2\pi\sqrt{t/U}}$ and $\mu^0 \approx  U/2$
of Eq. (\ref{DMH}) for $U/4t\ll 1$ and $U/4t\gg 1$, respectively,
are those of the zero-temperature spin gap of Eq. (13) of Ref. \cite{Hubbard-T*-x=0}.

In the following we confirm that ground states 
have no $\eta\nu$ fermions and no $s\nu$ fermions
with $\nu\geq 2$ spinon pairs. Hence the corresponding
energies $\epsilon_{\eta\nu}$ and $\epsilon_{s\nu}$, respectively,
considered below refer to creation onto the ground state of one of such objects 
carrying a given momentum. We start by providing a set of useful properties.
We emphasize that some of these properties are not
valid for descriptions generated by rotated-electron
operators associated with the general unitary operators ${\hat{V}}$ considered
in Ref. \cite{bipartite}. Indeed, the following properties refer to
the operator description associated with the rotated-electron
operators ${\tilde{c}}_{\vec{r}_j,\sigma}^{\dag} =
{\hat{V}}^{\dag}\,c_{\vec{r}_j,\sigma}^{\dag}\,{\hat{V}}$ of Eq. (\ref{fc+}) as
defined in this paper, whose electron - rotated-electron unitary 
operator ${\hat{V}}={\hat{V}}(U/4t)$ is such that 
the set of $4^{N_a^D}$ states of the form
$\vert \Psi_{U/4t}\rangle ={\hat{V}}^{\dag}\vert\Psi_{\infty}\rangle$
are energy eigenstates for $U/4t>0$. Here $\vert\Psi_{\infty}\rangle$
corresponds to exactly one of the many set of 
$U/4t\rightarrow\infty$ energy eigenstates, chosen according
to the well-defined criterion discussed above in Subection II-A.
For the model on the square lattice our operator description refer to a related complete set of
$S_{\eta}$, $S_{\eta}^z$, $S_s$, $S_s^z$, $S_c$, and momentum eigenstates
$\vert \Phi_{U/4t}\rangle ={\hat{V}}^{\dag}\vert\Phi_{\infty}\rangle$
generated from the corresponding $U/4t\rightarrow\infty$ states $\vert\Phi_{\infty}\rangle$
by the same electron - rotated-electron unitary operator ${\hat{V}}={\hat{V}}(U/4t)$
as the energy eigenstates $\vert \Psi_{U/4t}\rangle ={\hat{V}}^{\dag}\vert\Psi_{\infty}\rangle$.
Some of the states $\vert \Phi_{U/4t}\rangle$ are identical to the
energy eigenstates $\vert \Psi_{U/4t}\rangle$, which are also 
$S_{\eta}$, $S_{\eta}^z$, $S_s$, $S_s^z$, $S_c$, and momentum eigenstates.
(For the 1D model both sets of states are identical for the whole Hilbert space.)

Let us profit from the interplay of the transformation
laws of the $\alpha\nu$ fermions under the 
electron - rotated-electron unitary transformation 
with the model global $SO(3)\times SO(3)\times U(1)$ symmetry
to reach valuable information about the range of the energy 
$\epsilon_{\alpha\nu}$ for addition onto the  
ground state of one $\alpha\nu$ fermion. For the model on the square lattice
there is in general a small uncertainty $\delta\epsilon$ in the energy $\epsilon_{\alpha\nu}$ 
of a $\alpha\nu$ fermion of a given momentum smaller than or equal to the corresponding
energy bandwidth and such that  
$\epsilon_{\alpha\nu}\pm \delta\epsilon$ belongs to that
energy range. For that model and $\alpha\nu$ fermions that are invariant under
the electron - rotated-electron unitary transformation one has $\delta\epsilon=0$. 
That holds for instance for the $\alpha\nu$ fermions whose
occupancy configurations generate above states $\vert \Phi_{U/4t}\rangle$ 
that are identical to energy eigenstates $\vert \Psi_{U/4t}\rangle$.
For both the model on the 1D and square
lattices the latter range corresponds to the energy bandwidth
of the energy dispersion $\epsilon_{\alpha\nu}$.
For the description associated with rotated-electrons, 
corresponding $c$ fermions, $\eta$-spinons, and spinons
whose operators are given in Eqs. (\ref{fc+})-(\ref{c-up-c-down}), 
and composite $\eta$-spinon and spinon fermions the
following properties hold. 

\subsubsection{Energy range of a $\eta\nu$ fermion}

A $\eta\nu$ fermion is a $\eta$-spin-neutral anti-bounding
configuration of $\nu$ $-1/2$ $\eta$-spinons and $\nu$ $+1/2$ $\eta$-spinons.
Symmetry implies that for $U/4t>0$ there is no energy overlap between the
ranges of bandwidth $W_{\eta\nu}$ of the energy $\epsilon_{\eta\nu}$
corresponding to different branches $\nu=1,...,C_{\eta}$  
associated with addition onto the ground state of one $\eta\nu$ fermion
of a given momentum. 
Since fermions belonging to neighboring branches 
$\eta\nu$ and $\eta\nu+1$ differ in the number of single $\eta$-spinon pairs by one,
the requirement for the above lack of energy overlap 
is that $W_{\eta\nu}\leq 2\mu$ where $2\mu$
equals the energy of a pair of independent $\eta$-spinons
of opposite $\eta$-spin projections.
Such properties imply that the energy $\epsilon_{\eta\nu}$ 
obeys the following inequality,
\begin{equation}
2\nu\mu\leq \epsilon_{\eta\nu} < 2(\nu + i_{\eta\nu})\mu 
\, ; \hspace{0.25cm} 0\leq i_{\eta\nu} \leq 1
\, ,
\label{energy-e-n}
\end{equation}
where $2\nu\mu$ is the energy for creation of 
$\nu$ $-1/2$ independent $\eta$-spinons and $\nu$ 
independent $+1/2$ $\eta$-spinons onto the ground
state. Since the latter objects are invariant under
the electron - rotated-electron unitary transformation associated with the 
operator $\hat{V}$, they are non interacting
and thus their energies are additive. For $m=0$
the number $i_{\eta\nu}$ decreases continuously 
for increasing values of $U/4t$, having the limiting
behaviors $i_{\eta\nu}\rightarrow 1$ for $U/4t\rightarrow 0$
and $i_{\eta\nu}\rightarrow 0$ for $U/4t\rightarrow\infty$
so that $W_{\eta\nu}\rightarrow 0$ as $U/4t\rightarrow\infty$. 
The latter behavior is associated with the full
degeneracy of the $\eta$-spin configurations reached
for $U/4t\rightarrow\infty$ when the spectrum of
the $2\nu$-$\eta$-spinon composite $\eta\nu$ fermion becomes dispersionless.

\subsubsection{Energy range of a $s\nu$ fermion}

A $s\nu$ fermion of a given momentum is a spin-neutral bounding
configuration of $\nu$ $-1/2$ spinons and $\nu$ $+1/2$ spinons.
Again, symmetry implies that for $U/4t>0$ there is no energy overlap
between the ranges of bandwidth $W_{s\nu}$ of the one-$s\nu$ fermion
energy $\epsilon_{s\nu}$ associated with different
branches $\nu=1,...,C_s$. For $s\nu$ branches with a
number of spinon pairs $\nu\geq 2$ the bandwidth $W_{s\nu}$ of 
such an energy range is for the present bounding
configurations and for the same reasoning as for the $\eta\nu$ fermion
such that $W_{s\nu}\leq 2\mu_B\,H$ where 
$2\mu_B\,H$ equals the energy of a pair of independent spinons
of opposite spin projections.
The energy $\epsilon_{s\nu}$ for addition onto the 
ground state of one $s\nu$ fermion with $\nu\geq 2$ spinon
pairs obeys the inequality,
\begin{equation}
2(\nu - i_{s\nu})\mu_B\,H \leq \epsilon_{s\nu} \leq 2\nu\mu_B\,H 
\, ; \hspace{0.25cm} \nu > 1
\, , \hspace{0.15cm} 0\leq i_{s\nu} \leq 1 \, ,
\label{energy-s-n}
\end{equation}
where $2\nu\mu_B\,H$ is the energy for creation of 
$\nu$ independent $-1/2$ spinons and $\nu$ 
independent $+1/2$ spinons onto the ground
state. Since the latter objects are invariant under
the unitary transformation associated with the operator 
$\hat{V}$, they are not
energy entangled and their energies are additive.
For $m=0$ the number $i_{s\nu}$ decreases continuously 
for increasing values of $U/4t$, having the limiting
behaviors $i_{s\nu}\rightarrow 1$ for $U/4t\rightarrow 0$
and $i_{s\nu}\rightarrow 0$ for $U/4t\rightarrow\infty$
so that $W_{s\nu}\rightarrow 0$ as $U/4t\rightarrow\infty$.
Such a behavior is associated with the full
degeneracy of the spin configurations reached
for $U/4t\rightarrow\infty$ when the spectrum of
the $2\nu$-spinon composite $s\nu$ fermion becomes dispersionless.

In turn, for a $m=0$ and $x\geq 0$ ground state all sites of the 
$s1$ effective lattice are occupied so that the corresponding $s1$
momentum band is full and the energy $-\epsilon_{s1}$ 
for removal from that state of one $s1$ fermion carrying
momentum obeys the inequality,
\begin{equation}
0 \leq -\epsilon_{s1} \leq {\rm max}\,\{W_{s1},\vert\Delta\vert\} \, ,
\label{energy-s-in}
\end{equation}
where the energy scale $\vert\Delta\vert$ is introduced below in Subsection VI-B and further studied in Ref. \cite{companion}.
For small hole concentrations $0<x\ll 1$ it vanishes both in the limits $U/4t\rightarrow 0$ 
and $U/4t\rightarrow \infty$ and goes through a maximum value at $U/4t=u_0\approx 1.302$
and at constant $U/4t$ decreases for increasing $x$ and vanishes for $x>x_*$ where
the critical hole concentration $x_*$ is studied in Ref. \cite{companion} for approximately $U/4t\geq u_0$. 
As discussed in that subsection, due to a sharp quantum phase transition it has a singular behavior at $x=0$, having for $U/4t>0$
different magnitudes at $x=0$ and  $x\rightarrow 0$, respectively. In turn,
for $m=0$ the energy bandwidth $W_{s1}$ refers to the auxiliary $s1$ fermion dispersion 
defined in Ref. \cite{companion} and has it maximum 
magnitude at $U/4t=0$. For $U/4t>0$ it decreases monotonously for increasing values of $U/4t$, 
vanishing for $U/4t\rightarrow\infty$. That both $W_{s1}\rightarrow 0$ and $\vert\Delta\vert\rightarrow 0$ 
for $U/4t\rightarrow\infty$ is associated with the full 
degeneracy of the spin configurations reached in that limit for which the spectrum of the two-spinon composite
$s1$ fermions becomes dispersionless. For instance, we 
could access the explicit limiting behaviors of the $m=0$ energy bandwidth
$W_{s1}^0\equiv \lim_{x\rightarrow 0}\,W_{s1}=W_{s1}\vert_{x=0}$,
\begin{equation}
W_{s1}^0 = 2Dt  \, ,
\hspace{0.25cm}  U/4t = 0 
\, ; \hspace{0.50cm} 
W_{s1}^0 \approx 2D\pi\,{t^2\over U }\, ,
\hspace{0.25cm} U/4t\gg 19^{D-1} \, ,
\hspace{0.15cm} D = 1, 2 \, ,
\label{W-s-0}
\end{equation}
where that for the $D=2$ square lattice the large-$U/4t$ expression is
valid for $U/4t\gg 19$ is justified below in Section VI.  

In the $U/4t\rightarrow\infty$ limit 
the $s1$ fermion occupancy configurations that generate the
spin degrees of freedom of spin-density
$m=0$ ground states considered below become for the 1D model
those of the spins of the spin-charge factorized wave function introduced both
by Woynarovich \cite{Woy} and Ogata and Shiba \cite{Ogata}. In turn, for the model
on the square lattice such configurations become in that limit and
within a mean-field approximation for the fictitious magnetic field 
${\vec{B}}_{s1}$ of Eq. (\ref{A-j-s1-3D}) those of a full lowest Landau level with 
$N_{s1}=N_{a_{s1}}^2=N/2$ one-$s1$-fermion degenerate states of the $2D$ quantum  
Hall effect \cite{companion}. Here $N_{a_{s1}}^2$ is the number of 
both sites of the square $s1$ effective lattice and $s1$ band discrete
momentum values. For finite $U/4t$ values and $x>0$ the degeneracy of the $N_{a_{s1}}^2=N/2$ 
one-$s1$ fermion states of the square-lattice quantum liquid is removed by the emergence
of a finite-energy-bandwidth $s1$ fermion dispersion yet the number of $s1$
band discrete momentum values remains being given by $N_{a_{s1}}^2=B_{s1}\,L^2/\Phi_0$
and the $s1$ effective lattice spacing by $a_{s1}=l_{s1}/\sqrt{2\pi}$
where $l_{s1}$ is the fictitious-magnetic-field length and we recall that in our units the fictitious-magnetic-field 
flux quantum reads $\Phi_0=1$ \cite{companion}.

\subsubsection{The energy range of the $c$ fermions}
 
The energy $\epsilon_{c}$ for addition onto the ground state of 
one $c$ fermion of a given momentum and the energy $-\epsilon_{c}$ 
for removal from that state such a $c$ fermion obey the inequalities,
\begin{equation}
0 \leq \epsilon_{c}\leq W_c^h=[4Dt - W_c^p]   \, ; \hspace{0.5cm}
0 \leq -\epsilon_{c}\leq W_c^p  
\, , \hspace{0.25cm} D = 1,2 \, ,
\label{energy-c}
\end{equation}
respectively. Here $W_c^h=[4Dt-W_c^p]\in (0,4Dt)$ 
increases monotonously for increasing 
values of hole concentration $x\in (0,1)$.
The energy bandwidth $W_c^p$ depends little 
on $U/4t$ and for $U/4t>0$ has the following limiting behaviors,
\begin{equation}
W_c^p = 4Dt  \, ,
\hspace{0.25cm}  x = 0 \, ; \hspace{0.50cm}
W_c^p = 0  \, ;
\hspace{0.25cm}  x = 1 \, .
\label{W-c}
\end{equation}

\subsubsection{Transformation laws of $\alpha\nu$ fermions and $c$ fermions under the electron - rotated-electron 
unitary transformation}

An useful property is that $\eta\nu$ fermions and $s\nu$ fermions with
$\nu\geq 2$ spinon pairs that remain invariant under the electron - rotated-electron
unitary transformation have energy given by,
\begin{equation} 
\epsilon_{\eta\nu} = 2\nu\mu \, , \hspace{0.25cm} \nu =1,...,C_{\eta}
\, ; \hspace{0.50cm}
\epsilon_{s\nu} = 2\nu\mu_B\,H 
\, , \hspace{0.25cm} \nu =2,...,C_{s} \, .
\label{invariant-V}
\end{equation}
Those are non-interacting objects such that
their energy is additive in the individual energies
of the corresponding $2\nu$ $\eta$-spinons
and spinons, respectively. Therefore, for $U/4t>0$ 
they refer to the same occupancy configurations in terms
of both rotated electrons and electrons. However, note that
in contrast to the independent $\eta$-spinons or independent spinons here the
objects that are invariant under the electron - rotated-electron
unitary transformation are the composite $2\nu$-$\eta$-spinon $\eta\nu$ fermions or 
$2\nu$-spinon $s\nu$ fermions and not the corresponding individual 
$2\nu$ $\eta$-spinons or $2\nu$-spinons, respectively.
In turn, $\eta\nu$ fermions and $s\nu$ fermions with
$\nu\geq 2$ spinon pairs whose energies obey the inequalities
$\epsilon_{\eta\nu} > 2\nu\mu$ and $\epsilon_{s\nu}< 2\nu\mu_B\,H$
are not invariant under the electron - rotated-electron 
transformation. Furthermore, for finite $U/4t$ values     
$c$ fermions and $s1$ fermions are not invariant under that
transformation. 

Both for the model on the 1D and square lattices the initial ground state and the excited states
generated by creation of one $\eta\nu$ fermion or $s\nu$ fermion whose energy is 
given by (\ref{invariant-V}) are energy eingenstates. 

\subsubsection{Ground state occupancies}

Fulfillment of the requirement of self-consistency concerning the set of 
properties given here, reveals that both for the model on the 1D and 
square lattices the subspace spanned by the $x>0$ and $m>0$ LWS 
ground states and their excited energy eigenstates of 
energy $\omega<{\rm min}\,\{2\mu,2\mu_B\,H\}$ there are 
no $-1/2$ $\eta$-spinons, $\eta\nu$ fermions, independent $-1/2$ spinons, and 
$s\nu'$ fermions with $\nu'\geq 2$ spinon
pairs so that $N_{\eta\nu}=0$ and $N_{s\nu'}=0$ for $\nu'\geq 2$.
Hence the number of $c$ fermions is $N_c =2S_c = N=(1-x)\,N_a^D$,
independent $+1/2$ $\eta$-spinons $L_{\eta ,\,+1/2} = 2S_{\eta}=[N_a^D-N]=x\,N_a^D$,
independent $+1/2$ spinons 
$L_{s ,\,+1/2} = 2S_{s}=[N_{\uparrow}-N_{\downarrow}]=m\,N_a^D$, and
$s1$ fermions $N_{s1}=N_{\downarrow}$.

From analysis and comparison of the occupancies of the spin LWS
ground states ($m>0$) and spin highest-weight state (HWS) ground states ($m<0$),
one finds that a $m=0$ ground state for which $N$ is even and $x\geq 0$ has
no $-1/2$ $\eta$-spinons, $\eta\nu$ fermions, independent $\pm 1/2$ spinons, and 
$s\nu'$ fermions with $\nu'>1$ spinon pairs so that 
$N_{\eta\nu}=0$ and $N_{s\nu'}=0$ for $\nu'>1$.
Hence the number of $c$ fermions is $N_c = 2S_c = N=(1-x)\,N_a^D$,
independent $+1/2$ $\eta$-spinons $L_{\eta ,\,+1/2} = 2S_{\eta}=[N_a^D-N]=x\,N_a^D$,
and $s1$ fermions $N_{s1}=N/2=(1-x)\,N_a^D/2$. 

\subsection{The number of sites of the $\alpha\nu$ effective lattices and
discrete momentum values of the $\alpha\nu$ bands}

A local $\eta\nu$ (and $s\nu$) fermion refers to a superposition of 
well-defined $\eta$-spinon (and spinon) 
occupancy configurations involving $2\nu$ sites of the 
$\eta$-spin (and spin) effective lattice. 
The number of sites of the $s1$ effective lattice plays an
important role in the study of the model in the one- and two-electron
subspace considered below in Subsection V-A.
In order to evaluate its expression one needs to solve the same problem for the remaining
$\alpha\nu$ effective lattices as well. Indeed, the values of the
set of numbers $\{N_{a_{\alpha\nu}}^D\}$ are dependent of
each other. Fortunately, in the limit $N_a^D\gg 1$ one does not 
need detailed information about 
the occupancy configurations of the $\eta$-spin (and spin) 
effective lattice whose superposition defines
the internal structure of a $\eta\nu$ (and $s\nu$) fermion to achieve such a goal.
The only needed information is that  
the $2\nu$ sites of the $\eta$-spin (or spin) effective
lattice involved in such configurations are centered
at a point of real-space coordinate $\vec{r}_j$, which defines
the real-space coordinate of the local $\alpha\nu$ fermion.
Here $j=1,...,N_{a_{\alpha\nu}}^D$ of a given subspace where
$N_{a_{\alpha\nu}}^D$ is the number of sites of the $\alpha\nu$
effective lattice. Our goal is the derivation of the
expression for that number in terms of the set of
numbers $N_c$ and $\{N_{\alpha'\nu'}\}$ where
$\alpha'=\eta ,s$ and $\nu'=1,2,3,...,C_{\alpha'\nu'}$. The real-space coordinate $\vec{r}_j$ plays the role
of "centre of mass" of the local $\alpha\nu$ fermion and 
has $N_{a_{\alpha\nu}}^D$ well-defined values
associated with the sites of the $\alpha\nu$ effective
lattice. 

For $N_a^D\gg 1$ (i) the internal structure of a local $\alpha\nu$ fermion 
and (ii) its real-space position $\vec{r}_j$ are separated problems. 
The present analysis refers to the problem (ii) only. The 
problem (i) is addressed for the $s1$ bond
particle associated with the local $s1$ fermion in Ref. \cite{companion0}. 
Concerning the internal structure of a local $\alpha\nu$ fermion, the only issue that matters for the present
analysis is that the $2\nu$ sites of the $\eta$-spin 
(and spin) effective lattice occupied by a given local 
$\eta\nu$ (and $s\nu$) fermion correspond to
$2\nu$ sites of the original lattice that are not 
simultaneously occupied by any other such fermions.

The number of sites $N_{a_{\alpha\nu}}^D$ of the $\alpha\nu$ effective 
lattice is an integer number. We emphasize, however, that for the model on the
square lattice for which $D=2$ the related number $N_{a_{\alpha\nu}}$ is 
not in general integer so that the $\alpha\nu$ effective lattice is
not a perfect square. However, for $N_{a_{\alpha\nu}}/N_a$
finite and $N_a^2\gg 1$ it is nearly a square
lattice so that in that limit we use the notation $N_{a_{\alpha\nu}}^2$ 
for its number of sites. 

The $\alpha\nu$ effective lattice and its $N_{a_{\alpha\nu}}^D$ sites
are well-defined concepts in a subspace for which the values of the set 
of numbers $N_c=2S_c$ and $\{N_{\alpha\nu'}\}$ where $\nu'=1,2,3,...,C_{\alpha}$
remain constant. As justified below, this is equivalent to the $\eta$-spin $S_{\eta}$ ($\alpha=\eta$) or spin $S_{s}$ ($\alpha=s$)
and values of the set of numbers $\{N_{\alpha\nu'}\}$ 
where $\nu'=\nu,\nu +1,...,C_{\alpha}$ remaining constant.
For such subspaces the motion of the $\alpha\nu$ 
fermion through its effective lattice corresponds to a set of elementary 
steps where it hops from a given site of well-defined real-space coordinate 
to another. For the model on the square lattice such elementary steps
involve horizontal and vertical virtual
steps \cite{companion0}. 

For a local $\alpha\nu$ fermion the 
$2\nu\,N_{\alpha\nu}$ sites of the $\eta$-spin ($\alpha =\eta$)
or spin ($\alpha =s$) effective lattice occupied by
the $N_{\alpha\nu}$ local fermions belonging to the same
$\alpha\nu$ branch play the role of the $N_{\alpha\nu}$
occupied sites of the $\alpha\nu$ effective lattice.
In turn, the $2S_{\eta}$ (and $2S_s$) sites of the 
$\eta$-spin (and spin) effective lattice occupied by 
independent $+1/2$ $\eta$-spinons (and spinons) and 
$2(\nu'-\nu)$ sites out of the $2\nu'$ sites of
that lattice occupied by each local $\eta\nu'$ (and $s\nu'$)
fermion such that $\nu'>\nu$ play  
the role of unoccupied sites of the 
$\eta\nu$ (and $s\nu$) effective lattice.

The number of sites of the 
$\alpha\nu$ effective lattice is then given by,
\begin{equation}
N_{a_{\alpha\nu}}^D = [N_{\alpha\nu} + N^h_{\alpha\nu}] \, ,
\label{N*}
\end{equation}
where the number of unoccupied sites reads,
\begin{equation}
N^h_{\alpha\nu} = 
[2S_{\alpha}+2\sum_{\nu'=\nu+1}^{C_{\alpha}}(\nu'-\nu)N_{\alpha\nu'}] =
[N_{a_{\alpha}}^{D} - 
\sum_{\nu' =1}^{C_{\alpha}}(\nu +\nu' - \vert \nu-\nu'\vert)N_{\alpha\nu'}] \, .
\label{N-h-an}
\end{equation}
The equivalence of the two expressions given here confirms that the numbers $N_c=2S_c$ and $\{N_{\alpha\nu'}\}$ 
where $\nu'=1,2,3,...,C_{\alpha}$ remaining constant is equivalent to the $\eta$-spin $S_{\eta}$ ($\alpha=\eta$) or 
spin $S_{s}$ ($\alpha=s$) and values of the set of numbers $\{N_{\alpha\nu'}\}$ where $\nu'=\nu,\nu +1,...,C_{\alpha}$ 
remaining constant as well. In both cases that implies that $N_{a_{\alpha\nu}}^D$ remains constant. 

In the following and in Appendix C it is confirmed that the expressions given in Eqs.
(\ref{N*}) and (\ref{N-h-an}), which equal the $\alpha\nu$ band corresponding numbers of discrete momentum
values, are compatible with the number
of representations of the group $SO(3)\times SO(3)\times U(1)$ in each
subspace with constant values of $S_c$, $S_{\eta}$, and
$S_s$. From the use of these equations one finds that the number 
of unoccupied sites of the $\alpha 1$ effective lattices reads,
\begin{equation}
N^h_{\alpha 1} = [N_{a_{\alpha}}^{D} - 2B_{\alpha}]
\, ; \hspace{0.35cm} B_{\alpha} = \sum_{\nu =1}^{C_{\alpha}}N_{\alpha\nu}
\, ; \hspace{0.25cm}  \alpha = \eta \, , s \, . 
\label{Nh+Nh}
\end{equation}
This number equals that of $\alpha 1$ fermion holes in the $\alpha 1$ band. 

Straightforward manipulations of the above equations lead to
the following general expressions for $S_{\eta}$, $S_s$, and $S_c$,
\begin{equation}
S_{\alpha} = [{1\over 2} N^h_{\alpha 1} -
C_{\alpha} + B_{\alpha}]
\, , \hspace{0.25cm} \alpha = \eta \, , s 
\, ; \hspace{0.50cm}
S_c = [{N_a^D\over 2} - {1\over 2} N^h_{\eta 1}
- B_{\eta} =
{1\over 2} N^h_{s1} + B_{s}] \, .
\label{S-S-S}
\end{equation}
It follows from the equality of the two $S_c$ expressions 
given Eq. (\ref{S-S-S}) that,
\begin{equation}
\sum_{\alpha =\eta,s} {1\over 2} N^h_{\alpha 1} =
[{N_a^D\over 2} - \sum_{\alpha =\eta,s}B_{\alpha}] \, .
\label{sum-Nh+Nh}
\end{equation}

Equation (\ref{M-L-Sum}), the first expression of Eq. (\ref{S-S-S}),
and Eq. (\ref{sum-Nh+Nh}) are equivalent to the following sum-rules
for the numbers of $\alpha\nu$ fermions,
\begin{equation}
C_{\alpha} = \sum_{\nu =1}^{C_{\alpha}}\nu\,N_{\alpha\nu} = 
{1\over 2}\,[N_{a_{\alpha}}^{D} - 2S_{\alpha}] = 
\, ; \hspace{0.35cm}
B_{\alpha} = \sum_{\nu =1}^{C_{\alpha}}N_{\alpha\nu} = 
{1\over 2}\,[N_{a_{\alpha}}^{D} - N^h_{\alpha 1}] \, , \hspace{0.15cm} \alpha = \eta \, , s \, ,
\label{sum-rules}
\end{equation}
respectively, consistently with the expressions $C_{s}=[S_c-S_s]$ and $C_{\eta}=[N_a^D-S_c-S_{\eta}]$
given in Eq. (\ref{C-C}).  
  
Each subspace with constant values of $S_c$ and hence also with constant values 
of $N_{a_{\eta}}^D=[N_a^D-2S_c]$ and $N_{a_{s}}^D=2S_c$ that the 
vacuum of the theory given in Eq. (\ref{vacuum}) refers to can be divided into smaller 
subspaces with constant values of $S_c$, $S_{\eta}$, and $S_s$ and hence also 
with constant values of $C_{\eta}$ and $C_s$. Furthermore, the latter subspaces 
can be further divided into even smaller subspaces with constant values for the 
set of numbers $\{N_{\eta\nu}\}$ and $\{N_{s\nu}\}$, which must obey the 
sum-rules of Eq. (\ref{M-L-Sum}). 

Let us confirm that expressions (\ref{N*}) and (\ref{N-h-an}) for the number of discrete
momentum values of the $\alpha\nu$ band and $\alpha\nu$ fermion holes in that
band, repectively, are compatible with the
number of state representations of the $U(1)$,
$\eta$-spin $SU (2)$  ($\alpha =\eta$), and spin $SU (2)$ ($\alpha =s$) 
symmetries of the model global 
$SO(3)\times SO(3)\times U(1)=[SU(2)\times SU(2)\times U(1)]/Z_2^2$
symmetry in the subspaces with constant values of $S_c$, $S_{\eta}$,
and $S_s$. Those are subspaces of the larger subspace that the 
$S_c>0$ vacuum of Eq. (\ref{vacuum}) refers to. Alike in Ref. \cite{bipartite}, 
let us divide the Hilbert space of the model (\ref{H})
in a set of subspaces spanned by the states 
with constant values of $S_c$, $S_{\eta}$, and $S_s$ and
hence also of $N_c=2S_c$, $M_{\eta}=N_{a_{\eta}}^D=[N_a^D-2S_c]$,
and $M_{s}=N_{a_{s}}^D=2S_c$. We recall that 
for the subspace with constant values of $S_c$, $S_{\eta}$,
and $S_s$ under consideration $M_{\eta}=N_{a_{\eta}}^D$
(and $M_{s}=N_{a_{s}}^D$) is both the total number of
$\eta$-spinons (and spinons) and the number of sites
of the $\eta$-spin (and spin) effective lattice.

According to the studies of Ref. \cite{bipartite} the dimension of each 
such a subspace is,
\begin{equation}
d_r\cdot\prod_{\alpha=\eta,s}{\cal{N}}(S_{\alpha} ,M_{\alpha}) \, ,
\label{dimension}
\end{equation}
where $d_r$ and ${\cal{N}} (S_{\alpha},M_{\alpha})$ are given by,
\begin{equation}
d_r = {N_a^D\choose 2S_c} \, ; \hspace{0.35cm}
{\cal{N}} (S_{\alpha},M_{\alpha}) = (2S_{\alpha} +1)\left\{
{M_{\alpha}\choose M_{\alpha}/2-S_{\alpha}} - {M_{\alpha}\choose
M_{\alpha}/2-S_{\alpha}-1}\right\} \, ,
\label{N-Sa-Ma}
\end{equation}
and are the number of $U(1)$ symmetry state representations
and that of $\eta$-spin $SU(2)$ ($\alpha=\eta$)
or spin $SU(2)$ ($\alpha =s$) symmetry state representations, respectively.

The dimension $d_r$ given in Eq. (\ref{N-Sa-Ma}) is here straightforwardly recovered 
as $d_r = {N_a^D\choose N_c}$ and equals the number 
of occupancy configurations of the $N_c=2S_c$ $c$ fermions 
in their $c$ band with $N_a^D$ discrete momentum values. 
On the other hand, the values of the numbers $N_{a_{\alpha\nu}}^D$ of discrete
momentum values of the $\alpha\nu$ band must obey exactly the following 
equality {\it for all} subspaces,
\begin{equation}
{{\cal{N}} (S_{\alpha},M_{\alpha})\over (2S_{\alpha} +1)} = \sum_{\{N_{\alpha\nu}\}}\, \prod_{\nu
=1}^{C_{\alpha}}\,{N_{a_{\alpha\nu}}^D\choose N_{\alpha\nu}} \, , 
\hspace{0.25cm} \alpha=\eta,s \, .
\label{Ncs-cpb}
\end{equation}
Here ${N_{a_{\alpha\nu}}^D\choose N_{\alpha\nu}}$ 
is the number of occupancy configurations of the $N_{\alpha\nu}$ 
$\alpha\nu$ fermions in their $\alpha\nu$ band with $N_{a_{\alpha\nu}}^D$ 
discrete momentum values and the $\{N_{\alpha\nu}\}$ summation runs over 
{\it all} sets of $N_{\alpha\nu}$ numbers for $\nu =1,2,...,C_{\alpha}$ that owing to
the conservation of $C_{\alpha}$ exactly obey the subspace sum-rule,
\begin{equation} 
2C_{\alpha} =\sum_{\nu =1}^{C{\alpha}}\,2\nu\,N_{\alpha\nu} = [M_{\alpha} - 2S_{\alpha}] 
\, , \hspace{0.25cm} \alpha=\eta,s \, .
\label{sum-M-2S}
\end{equation}

The general expression of the number $N_{a_{\alpha\nu}}^D$ of $\alpha\nu$ band 
discrete momentum values that obeys Eq. (\ref{Ncs-cpb}) for {\it all} subspaces 
corresponds indeed to that given in Eq. (\ref{N*}).
Consistently, the occupancies of the independent $\eta$-spinons 
and independent spinons give rise
to the usual factors $(2S_{\eta} +1)$ and $(2S_s +1)$,
respectively, appearing in the expressions provided in Eqs. (\ref{N-Sa-Ma}) 
and (\ref{Ncs-cpb}) for the dimensions 
${\cal{N}} (S_{\eta},M_{\eta})$ and ${\cal{N}} (S_s,M_s)$.
In Appendix C it is shown that the subspace-dimension summation,
\begin{equation}
{\cal{N}}_{tot} =  
\sum_{S_c=0}^{[N_a^D/2]}\,\sum_{S_{\eta}=0}^{[N_a^D/2-S_c]}\,
\sum_{S_s=0}^{S_c}{N_a^D\choose
2S_c} \prod_{\alpha =\eta,s}{[1+(-1)^{[2S_{\alpha}+2S_c]}]\over 2}
\,{\cal{N}}(S_{\alpha},M_{\alpha}) = 4^{N_a^D} \, , 
\label{Ntot}
\end{equation}
leads indeed to the dimension $4^{N_a^D}$ of the full Hilbert space.
It follows that the present description in terms of momentum eigenstates is complete. 

In the $N_a^D\gg 1$ limit considered
here and alike for the $\eta$-spin and spin effective lattices, provided
that $N_{a_{\alpha\nu}}^D/N_a^D$ is finite the related
$\alpha\nu$ effective lattices can for 1D and 2D be represented by 
1D and square lattices, respectively, of lattice constant,
\begin{equation}
a_{\alpha\nu} = {L\over N_{a_{\alpha\nu}}} = 
{N_a\over N_{a_{\alpha\nu}}}\, a
= {N_{a_{\alpha}}\over N_{a_{\alpha\nu}}}\, a_{\alpha}
\, ; \hspace{0.25cm} 
N_{a_{\alpha\nu}} \geq 1 \, ,
\label{a-a-nu}
\end{equation}
where $\nu = 1,...,C_{\alpha}$ and $\alpha =\eta ,s$. 
In turn, the corresponding $\alpha\nu$ bands whose number of
discrete momentum values is also given by $N_{a_{\alpha\nu}}^D$
are well defined even when $N_{a_{\alpha\nu}}^D$ is
given by a finite small number, $N_{a_{\alpha\nu}}^D=1,2,3,...$

For the model on the square lattice and states with numbers 
$N_{\alpha\nu}=N_{a_{\alpha\nu}}^2$ for $\alpha\nu\neq s1$ and
$N_{s1}\approx N_{a_{s1}}^2$ for $\alpha\nu= s1$ where $[N_{a_{s1}}^2-N_{s1}]$ vanishes
or is of order $1/N_a^2$ for $N_a^2\rightarrow\infty$, the lattice constant
$a_{\alpha\nu}$ is directly related to the fictitious magnetic-field
length $l_{\alpha\nu}$ associated with the field of Eq. (\ref{A-j-s1-3D}).
Indeed, in that case one has that
$\langle n_{\vec{r}_j,\alpha\nu}\rangle\approx 1$ and such a fictitious magnetic field reads
${\vec{B}}_{\alpha\nu} ({\vec{r}}_j) \approx \Phi_0\sum_{j'\neq j}\delta ({\vec{r}}_{j'}-{\vec{r}}_{j})\,{\vec{e}}_{x_3}$.
It acting on one $\alpha\nu$ fermion differs from zero only at the positions
of other $\alpha\nu$ fermions. In the mean field approximation one replaces it
by the average field created by all $\alpha\nu$ fermions at position 
$\vec{r}_j$. This gives,
${\vec{B}}_{\alpha\nu} ({\vec{r}}_j) \approx \Phi_0\,n_{\alpha\nu} (\vec{r}_j)\,{\vec{e}}_{x_3}
\approx \Phi_0\,[N_{a_{\alpha\nu}}^2/L^2]\,{\vec{e}}_{x_3}=[\Phi_0/a_{\alpha\nu}^2]\,{\vec{e}}_{x_3}$. 
One then finds that the number $N_{a_{\alpha\nu}}^2$
of the $\alpha\nu$ band discrete momentum values equals $[B_{\alpha\nu}\,L^2]/\Phi_0$ and the
$\alpha\nu$ effective lattice spacing $a_{\alpha\nu}$ is expressed in terms to the fictitious 
magnetic-field length $l_{\alpha\nu}$ as $a_{\alpha\nu}^2=2\pi\,l_{\alpha\nu}^2$.
This is consistent with for such states each $\alpha\nu$ fermion having a flux
tube of one flux quantum on average attached to it. For the states under consideration
the $\alpha\nu$ fermion problem is related to the Chern-Simons theory  
\cite{Giu-Vigna}, the number of flux quanta being one being consistent with the
$\alpha\nu$ fermion and $\alpha\nu$ bond-particle wave functions obeying Fermi and Bose statistics, respectively. 
Hence the composite $\alpha\nu$ fermion consists of $2\nu$ $\eta$-spinons ($\alpha =\eta$)
or spinons ($\alpha =s$) plus an infinitely thin flux tube attached to it. 

The states that span the one- and two-electron subspace belong to such a class of states 
so that for the model on the square lattice in that subspace the composite $s1$ fermion 
consists of two spinons in a spin-singlet configuration plus an infinitely thin flux tube attached 
to it \cite{companion}. Thus, each $s1$ fermion appears to carry a fictitious magnetic solenoid
with it as it moves around in the $s1$ effective lattice.
Finally, since for such a quantum liquid the $s1$ fermions play a major role 
we find from straightforward manipulations of Eqs. (\ref{N*}) and (\ref{N-h-an}) for $\alpha\nu =s1$
that the number $N_{a_{s1}}^D = [N_{s1} + N^h_{s1}]$ of $s1$ effective lattice sites and thus of
$s1$ band discrete momentum values is given by,
\begin{equation}
N_{a_{s1}}^D = [S_c + S_{s}+\sum_{\nu=3}^{C_{s}}(\nu-2)N_{s\nu}] \, .
\label{N-a-s1-D}
\end{equation}
Hence for a subspace where $N_{s\nu}=0$ for $\nu\geq 3$ one has that
$N_{a_{s1}}^D = [S_c + S_{s}]$ is a good quantum number. 

\subsection{The momentum eigenstates of our description and
the good and quasi-good quantum numbers}

\subsubsection{The momentum eigenstates of our description}

The low-energy physics of a 3D perturbative and isotropic many-electron system for which
each energy eigenstate $\vert \Psi (0)\rangle$ of the corresponding non-interacting system
evolves upon adiabatically switching on the interaction $U$ on an
energy eigenstate $\vert \Psi (U)\rangle$ is in general successfully described by Fermi liquid theory \cite{Pines}.
(In contrast to the Hubbard model, here the interaction $U$ is not
necessarily onsite.) Since the two sets of orthogonal and normalized states $\{\vert \Psi (0)\rangle\}$
and $\{\vert \Psi (U)\rangle\}$ are complete and refer to the same Hilbert space,
there is a well-defined unitary transformation such that for each $U=0$ energy 
eigenstate $\vert \Psi (0)\rangle$ there is exactly one $U>0$ energy
eigenstate $\vert \Psi (U)\rangle={\hat{\cal{V}}}^{\dag}\vert \Psi (0)\rangle$ where
${\hat{\cal{V}}}^{\dag}={\hat{\cal{V}}}^{\dag}(U)$ is the corresponding unitary operator.
For the exact ground state such a statement is equivalent to the Gell-Mann
and Low theorem \cite{Fetter-Walecka}. By perturbative system we mean above that 
$\langle \Psi (0)\vert\Psi (U)\rangle\neq 0$ for all such pairs of states of each of the 
two complete sets.

If the $U=0$ energy eigenstates can be written as Slatter determinants of the form, 
\begin{equation}
\vert \Psi (0)\rangle =
\prod_{\sigma}\prod_{{\vec{k}}}c^{\dag}_{{\vec{k}},\sigma}\vert 0\rangle \, ,
\label{FL-0}
\end{equation}
where $c^{\dag}_{{\vec{k}},\sigma}$ creates an electron of momentum ${\vec{k}}$ and spin
projection $\sigma$ and $\vert 0\rangle$ is the electronic vacuum, then the corresponding 
states $\vert \Psi (U)\rangle={\hat{\cal{V}}}^{\dag}\vert \Psi (0)\rangle$ read,
\begin{equation}
\vert \Psi (U)\rangle =
\prod_{\sigma}\prod_{{\vec{k}}}{\hat{\cal{V}}}^{\dag}c^{\dag}_{{\vec{k}},\sigma}\vert 0\rangle 
= \prod_{\sigma}\prod_{{\vec{k}}}[{\hat{\cal{V}}}^{\dag}c^{\dag}_{{\vec{k}},\sigma}{\hat{\cal{V}}}]{\hat{\cal{V}}}^{\dag}\vert 0\rangle 
= \prod_{\sigma}\prod_{{\vec{k}}}{\tilde{c}}^{\dag}_{{\vec{k}},\sigma}\vert 0\rangle 
\, ; \hspace{0.25cm} {\tilde{c}}_{\vec{k},\sigma}^{\dag} =
{\hat{\cal{V}}}^{\dag}\,c_{\vec{k},\sigma}^{\dag}\,{\hat{\cal{V}}}
\, ; \hspace{0.25cm} {\hat{\cal{V}}}^{\dag}\vert 0\rangle = \vert 0\rangle
\, .
\label{FL-(U)}
\end{equation}
Here ${\tilde{c}}_{\vec{k},\sigma}^{\dag} ={\hat{\cal{V}}}^{\dag}\,c_{\vec{k},\sigma}^{\dag}\,{\hat{V}}$ is
the creation operator of a quasiparticle of momentum ${\vec{k}}$ and spin
projection $\sigma$.

Note that since for $U/4t\rightarrow 0$ the unitary operator ${\hat{\cal{V}}}^{\dag}$ becomes the
unit operator, a state $\vert \Psi (0)\rangle$ of Eq. (\ref{FL-0}) equals the corresponding state
$\vert \Psi (U)\rangle$ of Eq. (\ref{FL-(U)}) for $U/4t\rightarrow 0$.
Let us consider the one-electron Green function,
\begin{equation}
G ({\vec{k}},\sigma,t-t') = -i \langle\Psi_{GS} (U)\vert T\,c_{{\vec{k}},\sigma}(t)c^{\dag}_{{\vec{k}},\sigma}(t')
\vert \Psi_{GS} (U)\rangle \, ,
\end{equation}
where for $U>0$ (and  $U=0$) the $N+1$ states $c^{\dag}_{{\vec{k}},\sigma}(t)\vert \Psi_{GS} (U)\rangle$ and
$c^{\dag}_{{\vec{k}},\sigma}(t')\vert \Psi_{GS} (U)\rangle$ 
and $N-1$ states $c_{{\vec{k}},\sigma}(t)\vert \Psi_{GS} (U)\rangle$ and
$c_{{\vec{k}},\sigma}(t')\vert \Psi_{GS} (U)\rangle$  
are not (and are) energy eigenstates. In turn, for $U>0$ the 
``$N+1$ quasiparticle states'' ${\tilde{c}}^{\dag}_{{\vec{k}},\sigma}(t)\vert \Psi_{GS} (U)\rangle$ and
${\tilde{c}}^{\dag}_{{\vec{k}},\sigma}(t')\vert \Psi_{GS} (U)\rangle$ and
``$N-1$ quasiparticle states'' ${\tilde{c}}_{{\vec{k}},\sigma}(t)\vert \Psi_{GS} (U)\rangle$ and
${\tilde{c}}_{{\vec{k}},\sigma}(t')\vert \Psi_{GS} (U)\rangle$ are energy eigenstates,
yet they cannot be proved in real experiments because only the electrons are physical
particles. Indeed, quasiparticles exist only inside the many-electron system. Therefore, 
quasiparticles can only be added to or removed from the system 
upon addition to and removal from it of electrons. That for $U>0$ the above one-electron
states are not energy eigenstates implies that the renormalization factor
$Z ({\vec{k}})$ is smaller than one and for momenta ${\vec{k}}$ in the vicinity of the isotropic Fermi surface
and small excitation energy $\omega$ such states have an inverse lifetime proportional
to $\omega^2$ so that the ``quasiparticle lifetime'' is finite for $\omega>0$ and becomes infinite
as $\omega\rightarrow 0$ \cite{Pines}. 

We emphasize that in spite of such an one-electron lifetime being finite, the quasiparticle occupancy 
configurations (\ref{FL-(U)}) generate for $U>0$ exact energy eigenstates, whose continuous changes 
with time occur according to the many-electron Schr\"odinger equation.
Concerning one- and two-electron excitations, the concept of a quasiparticle is well-defined
for low-energy eigenstates whose quasiparticle occupancy configurations differ from those of the initial
ground state only near the Fermi surface. In turn, in the absence of one- and two-electron quantum measurement processes 
the quasiparticle occupancy configurations (\ref{FL-(U)}) as defined here generate all $U>0$ energy eigenstates.
Hence solution of the quantum problem is equivalent to derivation of the unitary operator 
${\hat{\cal{V}}}^{\dag}={\hat{\cal{V}}}^{\dag}(U)$ for $U>0$. Unfortunately, this is in general a
very involved and unsolved problem. Landau's Fermi liquid theory is a very useful scheme that allows solution of it
for the subspace spanned by excited energy eigenstates generated from the initial $U>0$ ground state by a 
finite number of quasiparticle processes near the Fermi surface.

In turn, for the Hubbard model on the square or 1D lattices our choice of $4^{N_a^D}$ energy eigenstates 
$\{\vert\Psi_{\infty}\rangle\}$ for $U/4t\rightarrow\infty$ reported is, as in Subsection II-A, 
such that the states $\vert \Psi_{U/4t}\rangle={\hat{V}}^{\dag}\vert\Psi_{\infty}\rangle$ are energy eigenstates 
for finite values $U/4t>0$. Following the result of Appendix A that the general problem of that model in the
whole Hilbert space can be fully described by the same model in the LWS subspace,
for simplicity to start with our discussion refers to that subspace. That just means that for states with
independent $\eta$-spinon and/or independent spinon occupancy the $\eta$-spin and spin
projection, respectively, of such objects is given by $+1/2$. Within our operator description for the model on the square lattice each of the
$U/4t\rightarrow\infty$ LWS energy eigenstates of the corresponding set $\{\vert \Psi_{LWS;\infty}\rangle\}$ can 
be expressed as a suitable superposition of $S_{\eta}$, $S_{\eta}^z$, $S_s$, $S_s^z$, $S_c$, and momentum
eigenstates $\{\vert \Phi_{LWS;\infty}\rangle\}$ with the same values for such physical quantities and also for
the numbers $C_{\alpha}=\sum_{\nu=1}^{C_{\alpha}}\nu\,N_{\alpha\nu}$ of Eqs. (\ref{2S-2C}), (\ref{C-C}), 
(\ref{M-L-Sum}), and (\ref{sum-rules}) where $\alpha =\eta,s$. The latter states can be generated by occupancy 
configurations of the microscopic momenta ${\vec{q}}$ carried by the $c$ and $\alpha\nu$ fermions, which
as justified below are of the general form,
\begin{equation}
\vert \Phi_{LWS;\infty}\rangle =
[\prod_{\alpha}\prod_{\nu}\prod_{{\vec{q}}\,'}{\mathcal{F}}^{\dag}_{{\vec{q}}\,',\alpha\nu}\vert 0_{\alpha};N_{a_{\alpha}^D}\rangle]
[\prod_{{\vec{q}}}{\mathcal{F}}^{\dag}_{{\vec{q}},c}\vert GS_c;0\rangle] \, .
\label{LWS-full-el-infty}
\end{equation}
Here ${\mathcal{F}}^{\dag}_{{\vec{q}}\,',\alpha\nu}$ and ${\mathcal{F}}^{\dag}_{{\vec{q}},c}$ are creation operators
of a $U/4t\rightarrow\infty$ $\alpha\nu$ fermion of momentum ${\vec{q}}\,'$ and $c$ fermion of
momentum ${\vec{q}}$, respectively, and the $\eta$-spin $SU(2)$ vacuum $\vert 0_{\eta};N_{a_{\eta}}^D\rangle$ 
associated with $N_{a_{\eta}}^D$ independent $+1/2$ $\eta$-spinons, the spin $SU(2)$ vacuum 
$\vert 0_{s};N_{a_{s}}^D\rangle$ with $N_{a_{s}}^D$ independent $+1/2$ spinons, and the $c$ $U(1)$
vacuum $\vert GS_c;0\rangle$ such that $\prod_{{\vec{q}}}{\mathcal{F}}^{\dag}_{{\vec{q}},c}\vert GS_c;0\rangle=\vert GS_c;2S_c\rangle$ 
has $N_c=2S_c$ $c$ fermions are those of the vacuum given in Eq. (\ref{vacuum}). 
We recall that for $U/4t\rightarrow\infty$ Eqs. (\ref{fc+})-(\ref{c-up-c-down}) are equivalent to Eqs. (1)-(3) of Ref. \cite{Ostlund-06} with
the $c$ fermion creation operator $f_{\vec{r}_j,c}^{\dag}$ replaced by the quasicharge annihilation operator $\hat{c}_r$.
Therefore, the operator ${\mathcal{F}}^{\dag}_{{\vec{q}},c}$ is given by,
\begin{equation}
{\mathcal{F}}^{\dag}_{{\vec{q}},c}=
{1\over {\sqrt{N_a^D}}}\sum_{r}\,e^{+i\vec{q}\cdot r}\,\hat{c}_r \, ,
\label{F+-cr}
\end{equation}
where within the notation of Ref. \cite{Ostlund-06}, $r \equiv\vec{r}_j$ are the real-space coordinates
of the quasicharge particles of that reference. The holes of such quasicharge particles are for
$U/4t\rightarrow\infty$ the spinless fermions that describe the charge degrees of freedom
of the electrons of the singly occupied sites. Moreover, the spinons and $\eta$-spinons of the
$2\nu$-spinon operators ${\mathcal{F}}^{\dag}_{{\vec{q}}\,',s\nu}$ and
$2\nu$-$\eta$-spinon operators ${\mathcal{F}}^{\dag}_{{\vec{q}}\,',\eta\nu}$, respectively, are associated with the
local spin and pseudospin operators, respectively, defined in that reference.
As discussed in Appendix B, for the 1D model all states (\ref{LWS-full-el-infty}) are energy eigenstates \cite{companion},
so that $\vert \Psi_{LWS;\infty}\rangle=\vert \Phi_{LWS;\infty}\rangle$.
For the model on the square lattice that holds only for some of these states. For instance, it
holds for the energy eigenstates that span the one- and two-electron subspace as defined below 
so that $\vert \Psi_{LWS;\infty}\rangle= \vert \Phi_{LWS;\infty}\rangle$ for that subspace.

Symmetry implies that for the Hubbard model on a square lattice at $U/4t>0$ the $N_{a_{\alpha\nu}}^2$ $\alpha\nu$ band
discrete momentum values are $U/4t$ independent for states belonging to the same $V$ tower.
Importantly, the vacua of the $U/4t\rightarrow\infty$ states (\ref{LWS-full-el-infty}) are invariant 
under the electron - rotated-electron unitary transformation so that for finite
values of $U/4t$ the corresponding $S_{\eta}$, $S_{\eta}^z$, $S_s$, $S_s^z$, $S_c$, and momentum eigenstates read,
\begin{equation}
\vert \Phi_{LWS;U/4t}\rangle =
[\prod_{\alpha}\prod_{\nu}\prod_{{\vec{q}}\,'}f^{\dag}_{{\vec{q}}\,',\alpha\nu}\vert 0_{\alpha};N_{a_{\alpha}^D}\rangle]
[\prod_{{\vec{q}}}f^{\dag}_{{\vec{q}},c}\vert GS_c;0\rangle] \, ; \hspace{0.25cm} f^{\dag}_{{\vec{q}}\,',\alpha\nu} =
{\hat{V}}^{\dag}\,{\mathcal{F}}^{\dag}_{{\vec{q}}\,',\alpha\nu}\,{\hat{V}} \, ; \hspace{0.25cm} f^{\dag}_{{\vec{q}},c} =
{\hat{V}}^{\dag}\,{\mathcal{F}}^{\dag}_{{\vec{q}},c}\,{\hat{V}} \, .
\label{LWS-full-el}
\end{equation}

Alike in the $U/4t\rightarrow\infty$ limit, for the one- and two-electron
subspace the energy eigenstates $\vert \Psi_{LWS;U/4t}\rangle$ of the model on the square lattice 
are such that $\vert \Psi_{LWS;U/4t}\rangle= \vert \Phi_{LWS;U/4t}\rangle$, whereas for the 1D
model the equality $\vert \Psi_{LWS;U/4t}\rangle= \vert \Phi_{LWS;U/4t}\rangle$ holds for all 
$U/4t>0$ energy eigenstates \cite{companion,1D}. In Appendix B it is confirmed that the discrete momentum
values of the $c$ and $\alpha\nu$ fermion operators appearing on the right-hand side of Eq. (\ref{LWS-full-el})
are indeed the good quantum numbers of the exact solution whose occupancy configurations generate
the energy eigenstates. Furthermore, in that Appendix the relation of the $c$
and $\alpha\nu$ fermion operators whose suitable products refer to the operators whose application
onto the theory vacua generates the states  (\ref{LWS-full-el}) to the
creation and annihilation fields of the charge ABCDF algebra  
\cite{Martins} and more traditional spin ABCD Faddeev-Zamolodchikov algebra \cite{ISM} of the 
algebraic formulation of the 1D exact solution of Ref. \cite{Martins} is discussed and the consistency 
between the two corresponding operational representations confirmed.     

We recall that addition of chemical-potential and magnetic-field operator 
terms to the Hamiltonian (\ref{H}) lowers its symmetry. However, such 
operator terms commute with that Hamiltonian and the momentum operator so that its energy 
and momentum eigenstates correspond to state representations of the 
$SO(3)\times SO(3)\times U(1)$ group for all values of the densities $n=(1-x)$ and $m$.
The same holds for the $4^{N_a^D}$ momentum eigenstates considered here, whose $c$ fermion and 
$\alpha\nu$ fermion occupancy configurations and independent $\eta$-spinon and spinon occupancies 
are of the form,
\begin{equation}
\vert \Phi_{U/4t}\rangle = \prod_{\alpha
=\eta,\,s}\frac{({\hat{S}}^{\dag}_{\alpha})^{L_{\alpha,\,-1/2}}}{
\sqrt{{\cal{C}}_{\alpha}}}\vert \Phi_{LWS;U/4t}\rangle =
[\prod_{\alpha}\frac{({\hat{S}}^{\dag}_{\alpha})^{L_{\alpha,\,-1/2}}}{
\sqrt{{\cal{C}}_{\alpha}}}\prod_{\nu}\prod_{{\vec{q}}\,'}f^{\dag}_{{\vec{q}}\,',\alpha\nu}\vert 0_{\alpha};N_{a_{\alpha}^D}\rangle]
[\prod_{{\vec{q}}}f^{\dag}_{{\vec{q}},c}\vert GS_c;0\rangle] \, .
\label{non-LWS}
\end{equation}
Such states correspond indeed to the state representations of the $SO(3)\times SO(3)\times U(1)$ group.
In this expression $\vert \Phi_{LWS;U/4t}\rangle$ are the states of Eq. (\ref{LWS-full-el})
and the normalization constant ${\cal{C}}_{\alpha}$ is given in Eq. 
(\ref{Calpha}) of Appendix A. The set of $4^{N_a^D}$ momentum eigenstates of form
(\ref{non-LWS}) is complete and corresponds exactly to the $c$ fermion and $\alpha\nu$
momentum occupancy configurations associated with the subspace dimensions of Eq.
(\ref{dimension}) and the corresponding subspace dimensions,
\begin{equation}
d_r\cdot\prod_{\alpha=\eta,s}(2S_{\alpha} +1)\cdot{\cal{N}}(S_{\alpha} ,M_{\alpha}) \, ,
\label{dimension-non-LWS}
\end{equation}
where $d_r$ and ${\cal{N}} (S_{\alpha},M_{\alpha})$ are given in Eq.
(\ref{N-Sa-Ma}) and here we have accounted for the occupancies of the independent $\eta$-spinons 
and independent spinons associated with the usual factors $(2S_{\eta} +1)$ and $(2S_s +1)$,
respectively. Use of the summations of the subspace dimensions of Eq.
(\ref{dimension-non-LWS}) performed in Appendix C confirms that the correct Hilbert
space dimension $4^{N_a^D}$ is indeed obtained. 
(For the 1D model one has that such $4^{N_a^D}$ states
$\{\vert \Phi_{U/4t}\rangle\}$ are both momentum and energy eigenstates so that
$\vert \Psi_{U/4t}\rangle =\vert \Phi_{U/4t}\rangle$.)

For the the operators $f_{\vec{q}_j,\alpha\nu}^{\dag}$ there is an independent problem for
the model in each subspace with constant values of $S_c$, $S_s$, and
number of sites of the $\alpha\nu$ effective lattice
and discrete momentum values of the $\alpha\nu$ band
$N_{a_{\alpha\nu}}^D = [N_{\alpha\nu} + N^h_{\alpha\nu}]$ given in Eq. (\ref{N*}). 
Hence the operators $f_{\vec{q}_j,\alpha\nu}^{\dag}$ and $f_{\vec{q}_j,\alpha\nu}$ act onto 
such subspaces, where they are the building blocks of the generators that change the $\alpha\nu$ fermion occupancy 
configurations. Hence and as mentioned above, such operators act onto subspaces
spanned by mutually neutral states and
that assures that for the model on the square lattice the components 
$q_{x1}$ and $q_{x2}$ of the microscopic momenta $\vec{q}=[q_{x1},q_{x2}]$ refer to
commuting $s1$ translation generators $\hat{q}_{s1\,x_1}$ and  $\hat{q}_{s1\,x_2}$ \cite{Giu-Vigna}. 
In turn, for the operators $f_{\vec{q}_j,c}^{\dag}$ there are in 1D two types of quantum
problems depending on the even or odd character of the number $[B_{\eta}+B_s]=\sum_{\alpha,\nu}N_{\alpha\nu}$.
Indeed, in spite of the $c$ effective lattice being identical to the original lattice
for the whole Hilbert space, as discussed in Appendix B the $c$ fermions feel the Jordan-Wigner 
phases of the $\alpha\nu$ fermions created or annihilated under subspace transitions
that change the number $[B_{\eta}+B_s]$. Hence for 1D the $c$ band discrete momenta have
different values for $[B_{\eta}+B_s]$ odd and even, respectively. Also for the model on
the square lattice the $c$ fermions feel the Jordan-Wigner 
phases of the $\alpha\nu$ fermions created or annihilated under subspace transitions
that change the number $[B_{\eta}+B_s]$ \cite{companion}.

The generators that transform into each other the states with constant values of $S_c$, $S_s$, and
$N_{a_{\alpha\nu}}^D = [N_{\alpha\nu} + N^h_{\alpha\nu}]$, which span such a subspace, 
can be expressed in terms of creation and annihilation $\alpha\nu$ fermion operators. These states
can be formally written as given in Eqs. (\ref{LWS-full-el}) and (\ref{non-LWS})
provided that the $\alpha\nu$ momentum bands are those of the
state under consideration. The $\alpha\nu$ fermion operators $f^{\dag}_{{\vec{q}},\alpha\nu}$ 
and $f_{{\vec{q}},\alpha\nu}$ act onto subspaces spanned by neutral states. However,
creation of one $\alpha\nu$ fermion is a well-defined process whose generator
is the product of an operator that fulfills small changes in the $\alpha\nu$ effective lattice
and corresponding $\alpha\nu$ momentum band and the operator 
$f^{\dag}_{{\vec{q}},\alpha\nu}$ appropriate to the excited-state subspace.
Since in the case of Eqs. (\ref{LWS-full-el}) and (\ref{non-LWS}) it is assumed
the $\alpha\nu$ momentum bands are those of the state under consideration
the corresponding generators on the vacua are simple products of
$f^{\dag}_{{\vec{q}},\alpha\nu}$ operators, as given in these equations.
  
As discussed in Appendix B, the use of the exact solution confirms that for 1D the states
(\ref{LWS-full-el}) and (\ref{non-LWS}) are momentum and energy eigenstates. Following the way 
that our operator description is constructed they are momentum eigenstates of the model on the square lattice
as well. This is consistent with general properties that play a key role in
the model physics: Both for the 1D and square lattices and in contrast to the Hamiltonian (\ref{H}), the momentum 
operator commutes with the set of $\alpha\nu$ translation generators ${\hat{{\vec{q}}}}_{\alpha\nu}$ of Eq. 
(\ref{m-generators}) and unitary operator $\hat{V}$ associated with the electron - rotated-electron unitary 
transformation. Since the Hamiltonian does not commute with the $\alpha\nu$ translation generators ${\hat{{\vec{q}}}}_{\alpha\nu}$,
the $\alpha\nu$ fermion operators labelled by the 
$\alpha\nu$ band discrete momentum values ${\vec{q}}_j$ 
where $j=1,...,N_{a_{\alpha\nu}}^2$ act onto and are
defined in subspaces with constant values of $N_{a_{\alpha\nu}}^2$,
and the values of their discrete momenta are subspace dependent,
the microscopic momenta ${\vec{q}}_j$ of such objects are not in general conserved.
Consistently, the states of Eqs. (\ref{LWS-full-el}) and (\ref{non-LWS})
generated by momentum occupancy configurations of such 
$\alpha\nu$ fermions are not in general energy eigenstates.

The unitary operator $\hat{V}^{\dag}$ considered in Subsection II-A has been constructed to 
inherently generating exact $U/4t>0$ energy and momentum eigenstates 
$\vert \Psi_{LWS;U/4t}\rangle={\hat{V}}^{\dag}\vert \Psi_{LWS;\infty}\rangle$
and $\vert \Psi_{U/4t}\rangle = {\hat{V}}^{\dag}\vert \Psi_{\infty}\rangle
=\prod_{\alpha}[({\hat{S}}^{\dag}_{\alpha})^{L_{\alpha,\,-1/2}}/
\sqrt{{\cal{C}}_{\alpha}}]\vert \Psi_{LWS;U/4t}\rangle$. For the
model on the square lattice the energy and momentum eigenstates 
$\vert \Psi_{U/4t}\rangle$ are within our description a well-defined superposition of momentum
eigenstates $\vert \Phi_{U/4t}\rangle$ of Eq. (\ref{non-LWS}) of equal
$S_{\eta}$, $S_{\eta}^z$, $S_s$, $S_s^z$, and $S_c$ values and momentum eigenvalue. 
The energy eigenvalues of the former states
are $U/4t$ dependent. In turn, the invariance of the momentum operator under $\hat{V}$ imposes
that the momentum eigenvalues of both the states  
$\vert \Psi_{U/4t}\rangle$ and $\vert \Phi_{U/4t}\rangle$ are independent
of $U/4t$. Therefore, provided that the 
corresponding set $\{\vert \Psi_{\infty}\rangle\}$
of $U/4t\gg 1$ energy and momentum eigenstates and thus that 
$\{\vert \Phi_{\infty}\rangle\}$ of momentum eigenstates associated with the same unitary operator
${\hat{V}}^{\dag}$ are suitably chosen according to the recipe of Subsection II-A, the momentum 
eigenvalues of such states are the same as those of the corresponding $U/4t>0$ momentum and 
energy eigenstates $\vert \Psi_{U/4t}\rangle={\hat{V}}^{\dag}\vert \Psi_{\infty}\rangle$ and
momentum eigenstates $\vert \Phi_{U/4t}\rangle={\hat{V}}^{\dag}\vert \Phi_{\infty}\rangle$,
respectively.

Hence the rotated electrons used as starting building blocks of the description of the quantum problem 
in terms of $c$ fermions, $\alpha\nu$ fermions, and independent $\eta$-spinons and spinons
have been constructed to inherently the momentum eigenvalues being for $U/4t>0$ fully determined by those of the 
corresponding $U/4t\rightarrow\infty$ system. For both the model on the square and 1D lattices 
it follows from Eqs. (\ref{P-c-alphanu}) and (\ref{m-generators}) that the momentum eigenvalues are indeed
independent of $U/4t$ and given by,
\begin{equation}
\vec{P} =\sum_{{\vec{q}}}{\vec{q}}_j\, N_c ({\vec{q}})
+ \sum_{\nu =1}^{C{s}}\sum_{{\vec{q}}}{\vec{q}}\, N_{s\nu} ({\vec{q}}) 
+ \sum_{\nu =1}^{C{\eta}}\sum_{{\vec{q}}}[\vec{\pi} -{\vec{q}}]\,N_{\eta\nu} ({\vec{q}})
+ \vec{\pi}\,M_{\eta\, ,-1/2} \, .
\label{P-1-2-el-ss}
\end{equation}
Here $N_{c}({\vec{q}})$ and $N_{\alpha\nu}({\vec{q}})$ are the expectation values of the momentum 
distribution-function operators (\ref{Nc-s1op}).
For the model on the 1D lattice the $c$ and $\alpha\nu$ fermion discrete momentum values are good quantum numbers
so that the momentum distribution functions $N_c (q)$ and $N_{\alpha\nu} (q)$ are 
eigenvalues of the operators (\ref{Nc-s1op}) and have values $1$ and $0$ for occupied 
and unoccupied momentum values, respectively \cite{companion,1D}. As justified below, for the model on the square lattice 
that remains true for $N_{c}({\vec{q}})$ and for such a model in the one- and two-electron subspace defined in Section V  
this is true for $N_{s1}({\vec{q}})$ as well.

For $D=1$ the validity of expression (\ref{P-1-2-el-ss}) is confirmed by the exact solution 
\cite{companion,1D}. Let us provide further information on why  
for the Hubbard model on the square lattice the momentum operator can for $U/4t>0$ be 
expressed as given in Eqs. (\ref{P-c-alphanu}) and (\ref{m-generators}) so that the momentum 
eigenvalues are of the general form (\ref{P-1-2-el-ss}), alike for the 1D model. 
For $U/t\rightarrow\infty$ electron single and doubly occupancy 
become good quantum numbers so that the electrons that singly occupy 
sites do not feel the on-site repulsion. That implies that the discrete momentum values 
${\vec{q}}$ associated with their occupancy configurations separate into (i) those associated with electron hopping, 
which are good quantum numbers and 
involve the charge degrees of freedom and finite kinetic energy and are carried by the 
$c$ fermions and (ii) those associated with the $\eta$-spinon and spinons, including 
the composite $\eta\nu$ and $s\nu$ fermion occupancy configurations, respectively, whose 
momentum bands are for $U/t\rightarrow\infty$ dispersionless. 

That for the model on the square lattice the momentum eigenvalues
have for $U/4t>0$ the general form,
\begin{equation}
\vec{P} =\sum_{{\vec{q}}}{\vec{q}}\, N_c ({\vec{q}})
+ {\vec{P}}_{\eta-spin} + {\vec{P}}_{spin}  \, ,
\label{P-1-2-el-ss-2D}
\end{equation}
where ${\vec{P}}_{\eta-spin}$ and ${\vec{P}}_{spin}$ are the momentum contributions
associated with the $\eta$-spin and spin configurations, respectively, which for $U/t\rightarrow\infty$ 
do not contribute to the kinetic energy, is an exact result. Indeed, in that limit the 
rotated electrons are the electrons, which for the finite-energy physics can only singly occupy 
sites and due to the exclusion principle behave as spinless and $\eta$-spinless fermions. Since the
only effect of the interactions on their charge degrees of freedom 
is to impose such a single occupancy constrain, those lead to a momentum
contribution $\sum_{\sigma}\sum_{{\vec{q}}}{\vec{q}}\,N_{\sigma} ({\vec{q}})
=\sum_{{\vec{q}}}{\vec{q}}\,N ({\vec{q}})$
where $N ({\vec{q}})=\sum_{\sigma}N_{\sigma} ({\vec{q}})$
is the $U/t\rightarrow\infty$ electron momentum
distribution and $N_{\sigma} ({\vec{q}})=\langle \Psi_{\infty}\vert c_{\vec{q},\sigma}^{\dag}
c_{\vec{q},\sigma}\vert \Psi_{\infty}\rangle$. In that limit one then has  
$N_c ({\vec{q}}) =N ({\vec{q}})$.
For $U/4t$ finite the $U/t\rightarrow\infty$ distribution $N ({\vec{q}})$ is 
replaced by the rotated-electron momentum distribution,
\begin{equation}
N_{rot} ({\vec{q}}) = \sum_{\sigma}\langle \Psi_{U/4t}\vert{\tilde{c}}_{\vec{q},\sigma}^{\dag}
{\tilde{c}}_{\vec{q},\sigma}\vert \Psi_{U/4t}\rangle =
\sum_{\sigma}\langle \Psi_{\infty}\vert{\hat{V}}[{\hat{V}}^{\dag} c_{\vec{q},\sigma}^{\dag}
{\hat{V}}{\hat{V}}^{\dag}c_{\vec{q},\sigma}{\hat{V}}]{\hat{V}}^{\dag}\vert \Psi_{\infty}\rangle
= \sum_{\sigma}\langle \Psi_{\infty}\vert c_{\vec{q},\sigma}^{\dag}
c_{\vec{q},\sigma}\vert \Psi_{\infty}\rangle \, ,
\label{N-rot}
\end{equation}
and the $c$ fermion
momentum distribution reads $N_c ({\vec{q}}) =N_{rot} ({\vec{q}})$.
Indeed, for rotated electrons singly occupancy remains a
good quantum number for $U/4t>0$ and the $c$ effective lattice
occupied sites have exactly the same spatial variables
as the sites singly occupied by rotated electrons in the
original lattice. 

For the model on the square lattice an energy eigenstate $\vert \Phi_{U/4t}\rangle$ is a superposition of
$S_{\eta}$, $S_{\eta}^z$, $S_s$, $S_s^z$, $S_c$, and momentum
eigenstates $\vert \Phi_{U/4t}\rangle$ of Eq. (\ref{non-LWS}) with
the same momentum eigenvalue, $S_{\eta}$, $S_{\eta}^z$, $S_s$, $S_s^z$, $S_c$
values, and $c$ fermion momentum distribution function $N_c ({\vec{q}}) =N_{rot} ({\vec{q}})$.
Hence all states $\vert \Phi_{U/4t}\rangle$ of such a set have
the same distribution function $N_c ({\vec{q}}) = \sum_{\sigma}\langle \Phi_{U/4t}\vert{\tilde{c}}_{\vec{q},\sigma}^{\dag}
{\tilde{c}}_{\vec{q},\sigma}\vert \Phi_{U/4t}\rangle$. As discussed below, 
consistently with the commutator of the Hamiltonian with the $\alpha\nu$ translation generators 
${\hat{{\vec{q}}}}_{\alpha\nu}$ of Eq. (\ref{m-generators}) not vanishing,
such states differ in general in the $\alpha\nu$ fermion momentum occupancies.
Since for momentum eigenstates $\vert \Phi_{U/4t}\rangle ={\hat{V}}^{\dag}\vert \Phi_{\infty}\rangle$
of Eq. (\ref{non-LWS}) (one for each value of $U/4t>0$) belonging to the same $V$ tower 
the momentum eigenvalues are independent
of $U/4t$, the contribution to the momentum of the $c$
fermions is $\sum_{{\vec{q}}}{\vec{q}}\,N_c ({\vec{q}})$
where alike in Eq. (\ref{N-rot}) the distribution function 
$N_c ({\vec{q}})=N_{rot} ({\vec{q}}) =  \sum_{\sigma}\langle \Phi_{U/4t}\vert{\tilde{c}}_{\vec{q},\sigma}^{\dag}
{\tilde{c}}_{\vec{q},\sigma}\vert \Phi_{U/4t}\rangle$ has for $U/4t>0$ the
same magnitude $N ({\vec{q}})=\sum_{\sigma}\langle \Phi_{\infty}\vert c_{\vec{q},\sigma}^{\dag}
c_{\vec{q},\sigma}\vert \Phi_{\infty}\rangle$ as for the corresponding $U/4t\rightarrow\infty$
initial momentum eigenstate $\vert \Phi_{\infty}\rangle$.

This holds both for the model on the 1D and square lattices.
For the model on the former lattice the electron momentum distribution $N (q)$ is 
for the $x=0$ and $m=0$ ground state
plotted for $U/t\rightarrow\infty$ in Fig. 3 (a) of Ref. \cite{88}.
Note that the energy spectrum depends on $U/4t$
so that for the model on the square lattice the $x>0$ and $m=0$ ground states 
may belong to different $V$ towers for different values of $U/4t$. 
The exception is the $x=0$ and $m=0$ ground state, which belongs to the same $V$ tower
for all $U/4t>0$ values, as confirmed in Ref. \cite{companion}.

The above analysis confirms that consistently with as given in Eq. (\ref{H-P-commutators})
the Hamiltoninan and momentum operator commute with the $c$ translation generator
${\hat{{\vec{q}}}}_{c}$ of Eq. (\ref{m-generators}), for $U/4t>0$ the distribution
$N_c ({\vec{q}})$ is an eigenvalue of the operator 
${\hat{N}}_c ({\vec{q}})$ of Eq. (\ref{Nc-s1op}) and has 
values $1$ and $0$ for filled and unfilled 
momentum values, respectively. That result combined with the 
validity of the momentum spectrum (\ref{P-1-2-el-ss-2D})
for all occupancy configurations of the $c$ fermions that
generate the momentum eigenstates of form (\ref{non-LWS}) confirms
that the discrete momentum values ${\vec{q}}$ of the $c$
fermion band are good quantum numbers for the Hubbard
model on the square lattice. 

In turn, for $U/4t\rightarrow\infty$ all $\eta$-spin and spin configurations
are degenerate and do not contribute to the kinetic energy. Nevertheless
they lead to overall momentum contributions ${\vec{P}}_{\eta-spin}$ and 
${\vec{P}}_{spin}$, respectively. The main point is that the 
$\alpha\nu$ fermion operators are defined in and act onto subspaces
spanned by mutually neutral states for which for the square-lattice model
the $\alpha\nu$ translation generators $\hat{q}_{\alpha\nu\,x_1}$ and
$\hat{q}_{\alpha\nu\,x_2}$ commute. That implies that alike in 1D the $\alpha\nu$ translation 
generators ${\hat{{\vec{q}}}}_{\alpha\nu}$ of Eq. (\ref{m-generators})
in the presence of the fictitious magnetic field ${\vec{B}}_{\alpha\nu}$ 
of Eq. (\ref{A-j-s1-3D}) commute with the momentum operator both in the 
$U/4t\rightarrow\infty$ limit and for $U/4t$ finite, as given in Eq. (\ref{H-P-commutators}).
Consistently with the momentum operator expression provided in Eqs. (\ref{P-c-alphanu}) and (\ref{m-generators})
and as given in Eq. (\ref{P-1-2-el-ss}), that then implies that both in the 
$U/4t\rightarrow\infty$ limit and for $U/4t$ finite the
momentum contributions ${\vec{P}}_{\eta-spin}$ and 
${\vec{P}}_{spin}$ of Eq. (\ref{P-1-2-el-ss-2D}) have the form
${\vec{P}}_{\eta-spin}=\sum_{\nu}\sum_{{\vec{q}}}[\vec{\pi}-{\vec{q}}]\,N_{\eta\nu} ({\vec{q}})+\vec{\pi}\,M_{\eta\, ,-1/2}$
and ${\vec{P}}_{spin}=\sum_{\nu}\sum_{{\vec{q}}}{\vec{q}}\, N_{s\nu} ({\vec{q}})$,
respectively. 

For the Hubbard model on the square lattice the momentum spectrum of Eq. (\ref{P-1-2-el-ss}) is
found in \cite{companion} to lead to the correct spin spectrum in some limiting cases. 
Note however that the momentum area $N_{a_{\alpha\nu}}^2\,[2\pi/L]^2$ and 
discrete momentum values number $N_{a_{\alpha\nu}}^2$ of the $\alpha\nu$ bands are known. 
In contrast, the shape of their boundary remains an open issue. The shape of the $c$ band is
the same as that of the first Brillouin zone, yet the discrete momentum values may
have small overall shifts under transitions between subspaces with different
$[B_{\eta}+B_s]=\sum_{\alpha\nu}N_{\alpha\nu}$ values. The studies of Ref. \cite{companion}
use several approximations to obtain information on the $c$ and $s1$ bands momentum values and
the corresponding $c$ and $s1$ fermion energy dispersions of the model on the square
lattice in the one- and two-electron subspace, for which as justified below the $s1$ fermion discrete momentum
values are good quantum numbers.

The square-lattice quantum liquid studied in this paper corresponds to
the Hubbard model on the square lattice in such a subspace.
For that quantum problem the $c$ and $s1$ fermions play the role of ``quasiparticles''. 
There are three main differences relative to an isotropic Fermi liquid \cite{Pines}:
i) Concerning the charge degrees of freedom, the non-interacting limit of the theory refers to $4t^2/U\rightarrow 0$
rather than to the limit of zero interaction $U\rightarrow 0$; ii)
In the $4t^2/U\rightarrow 0$ limit the $c$ fermions
and $s1$ fermions become the holes of the ``quasicharges'' of Ref.
\cite{Ostlund-06} and spin-singlet two-spin configurations of the
spins of such a reference rather than electrons and only the
charge dynamical factor becomes that of non-interacting spinless
fermions, the one-electron and spin spectral distributions remaining non-trivial; iii) For
$U/4t>0$ the $s1$ band is full and 
displays a single hole for initial $m=0$ ground states and their one-electron excited states,
respectively, and as found in Ref. \cite{companion} its boundary line is anisotropic,
what is behind anomalous one-electron scattering properties \cite{cuprates}. 

\subsubsection{The good and quasi-good quantum numbers}

That $S_{\eta}$, $S_{\eta}^z$, $S_s$, and $S_s^z$ are good quantum numbers for both the Hubbard model
on the 1D and square lattices implies that the numbers of independent $\eta$-spinons ($\alpha=\eta$) and 
independent spinons ($\alpha=s$) $L_{\alpha,\,\pm 1/2} = [S_{\alpha}\mp S_{\alpha}^z]$ of Eq. (\ref{L-L})
are good quantum numbers as well. That the eigenvalue $S_c$ of the generator of the new 
$U(1)$ symmetry of the model global $SO(3)\times SO(3)\times U(1)=[SU(2)\times SU(2)\times U(1)]/Z_2^2$
symmetry found in Ref. \cite{bipartite} is a good quantum number implies that the numbers
$N_{a_{s}}^D=N_c=M_s=2S_c$ and $N_{a_{\eta}}^D=N_c^h=M_{\eta}=[N_a^D-2S_c]$ are also good quantum
numbers. It then follows that the numbers $C_{\alpha}$ of Eqs. (\ref{2S-2C}), (\ref{C-C}), 
(\ref{M-L-Sum}), and (\ref{sum-rules}), which can be expressed as $C_{\alpha} =[N_{a_{\alpha}}^D/2-S_{\alpha}]$,
are good quantum numbers as well. The same then applies to the total numbers $M_{\alpha,\,\pm 1/2} = L_{\alpha,\,\pm 1/2} + C_{\alpha}$
of $\pm 1/2$ $\eta$-spinons ($\alpha=\eta$) and $\pm 1/2$ spinons ($\alpha=s$).

The integrability of the 1D Hubbard model is for $N_a\rightarrow\infty$ associated with an infinite number of
conservation laws such that the Hamiltonian commutes with the infinite $\alpha\nu$ translation generators
${\hat{{\vec{q}}}}_{\alpha\nu}$ of Eq. (\ref{m-generators})
in the presence of the fictitious magnetic fields ${\vec{B}}_{\alpha\nu}$. According to the results 
of Ref. \cite{Prosen} such laws are equivalent to the independent conservation of the corresponding
set of numbers $\{N_{\alpha\nu}\}$ of $\alpha\nu$ fermions, which are good quantum
numbers. The lack of integrability of the Hubbard model on the square lattice is behind 
the $\alpha\nu$ translation generators not commuting with the Hamiltonian and the
set of numbers $\{N_{\alpha\nu}\}$ not being in general good quantum numbers. It follows that for the 1D model
the numbers $B_{\alpha}$ of Eqs. (\ref{Nh+Nh}) and (\ref{sum-rules}) and related numbers
$N^h_{\alpha 1}$ and $N_{s1}$ are good quantum numbers as well, whereas for the model on the
square lattice are {\it quasi-good quantum numbers}. By that we mean that in general they are
not good quantum numbers for that model, yet they are good quantum numbers for it defined 
in the following subspaces:
\vspace{0.25cm}

A) $B_{\eta}$ and $N^h_{\eta 1}$ (and $B_{s}$, $N^h_{s 1}$, and $N_{s1}$) are good quantum numbers for the model on the square
lattice in subspaces spanned by $S_{\eta}=0$ (and $S_s=0$) momentum eigenstates provided that $N_{\eta\nu}>0$ 
is finite for a single $\eta\nu$ branch (and $N_{s\nu}>0$ is finite for the $s1$ branch and a single 
$s\nu$ branch other than it) and $N_{\eta\nu'}=0$ (and $N_{s\nu'}=0$) for the remaining
$\eta\nu'$ (and $s\nu'$) branches and $B_{\eta}=N_{\eta\nu}$ 
(and $[B_{s}-N_{s1}]=N_{s\nu}$) is such that $2\nu N_{\eta\nu}/N_a^2\rightarrow 0$
(and $2\nu N_{s\nu}/N_a^2/N_a^2\rightarrow 0$ and $N_{a_{s1}}^2/N_a^2$ is finite) for $N_a^2\rightarrow\infty$.
In such subspaces the Hamiltonian commutes with the $\eta\nu$ translation generators
$\hat{q}_{\eta\nu\,x_1}$ and $\hat{q}_{\eta\nu\,x_2}$ 
in the presence of the corresponding fictitious magnetic field ${\vec{B}}_{\eta\nu}$ (and
$s1$ translation generators $\hat{q}_{s1\,x_1}$ and $\hat{q}_{s1\,x_2}$  
plus $s\nu$ translation generators $\hat{q}_{s\nu\,x_1}$ and $\hat{q}_{s\nu\,x_2}$
in the presence of the corresponding fictitious magnetic fields ${\vec{B}}_{s1}$ and
${\vec{B}}_{s\nu}$, respectively). In these subspaces the $N_{\alpha\nu}$ $\alpha\nu\neq s1$ fermions correspond to a zero-momentum and
vanishing-energy occupancy and are invariant under the electron - rotated-electron unitary
transformation. Indeed, they obey the criterion of Eq. (\ref{invariant-V}). When all 
$\alpha\nu\neq s1$ fermions of a given branch have that invariance their number
$N_{\alpha\nu}$ is conserved so that in the present case the numbers $B_{\alpha}$ and $N^h_{\alpha 1}$
where $\alpha =\eta, s$ and $N_{s1}$ if $\alpha=s$ can be expressed solely in
terms of good quantum numbers so that they are good quantum numbers as well.
\vspace{0.25cm}

B) In subspaces where $N_{a_{\eta\nu}}^2=2S_{\eta}$ (and $N_{a_{s\nu}}^2=2S_{s}$) for all branches with $\nu\geq 1$
(and $\nu\geq 2$) then $N_{\eta\nu}=0$ and $N_{\eta\nu}^h=2S_{\eta}$
(and $N_{s\nu}=0$ and $N_{s\nu}^h=2S_{s}$) are good quantum numbers for all such branches. 
In addition, $B_{\eta}=0$ (and $B_{s}=[S_c-S_s]$, $N_{a_{s1}}^2=[S_c+S_{s}]$,
$N_{s1}=[S_c-S_s]$, and $N_{s1}^h=2S_{s}$) are good quantum numbers as well.
This follows straightforwardly from $S_{\eta}$, $S_s$, and $S_c$ being conserved.
In such subspaces the Hamiltonian commutes with the $s1$ translation generators
$\hat{q}_{s1\,x_1}$ and $\hat{q}_{s1\,x_2}$  
in the presence of the corresponding fictitious magnetic field ${\vec{B}}_{s1}$.
\vspace{0.25cm}

The Hubbard model on a square-lattice in a subspace (A) and or in a subspace (B) provided that $S_s=0$ 
has in the limit $U/4t\rightarrow\infty$ an interesting physics. For such subspaces the $\alpha\nu$ fermion occupancy configurations 
are in that limit and within a mean-field approximation for the fictitious magnetic field ${\vec{B}}_{\alpha\nu}$ of Eq. (\ref{A-j-s1-3D})  
closely related to the physics of a full lowest Landau level of the 2D quantum Hall effect (QHE). Consider for instance 
a subspace (A) with $S_s=0$, $N_{s1}$ and $N_{s\nu}$ finite where $\nu\neq 1$, $N_{s\nu'}=0$ for all remaining $s\nu'$ branches, 
and $2\nu N_{s\nu}/N_a^2/N_a^2\rightarrow 0$ and $N_{a_{s1}}^2/N_a^2$ finite for $N_a^2\rightarrow\infty$. 
From the use of Eqs. (\ref{N*}) and (\ref{N-h-an}) one then finds that $N_{a_{s1}}^2=N_{s1}[1+{\cal{O}} (1/N_a^D)]=N_{s1}$
and $N_{a_{s\nu}}^2=N_{s\nu}$ for $N_a^2\rightarrow\infty$. That corresponds
to the case when $\langle n_{\vec{r}_j,s1}\rangle\approx 1$ and $\langle n_{\vec{r}_j,s\nu}\rangle\approx 1$ 
and if in the mean field approximation one replaces the corresponding fictitious magnetic fields by the 
average fields created by all $s1$ and $s\nu$ fermions, respectively, at a given position one finds as above
${\vec{B}}_{s1} ({\vec{r}}_j) \approx [\Phi_0/a_{s1}^2]\,{\vec{e}}_{x_3}$ and
${\vec{B}}_{s\nu} ({\vec{r}}_j) \approx [\Phi_0/a_{s\nu}^2]\,{\vec{e}}_{x_3}$. The numbers of $s1$ and $s\nu$ 
band discrete momentum values can then be written as,
\begin{equation}
N_{a_{s1}}^2 = {B_{s1}\,L^2\over \Phi_0} 
\, ; \hspace{0.35cm} \nu_{s1} = {N_{s1}\over N_{a_{s1}}^2} = 1
\, ; \hspace{0.35cm}
N_{a_{s\nu}}^2 = {B_{s\nu}\,L^2\over \Phi_0} 
\, ; \hspace{0.35cm} \nu_{s\nu} = {N_{s\nu}\over N_{a_{s\nu}}^2} = 1 \, .
\label{N-B-Phi}
\end{equation}
A similar relation $N_{a_{\eta\nu}}^2=B_{\eta\nu}\,L^2/\Phi_0$ is found for subspaces (A)
when there is an occupied $\eta\nu$ band and for the $s1$ fermions alone for subspaces (B) and
$S_s=0$. It is found in Ref. \cite{companion} that the $\alpha\nu$ fermions have a momentum dependent energy dispersion
and only in the limit $U/4t\rightarrow\infty$ their energy bandwidth vanishes. Hence only in that
limit are the $N_{a_{s1}}^2$ one-$s1$-fermion states and $N_{a_{s\nu}}^2$ one-$s\nu$-fermion states 
of Eq. (\ref{N-B-Phi}) degenerate. It follows that in such a limit $N_{a_{s1}}^2=B_{s1}\,L^2/\Phi_0$ and
$N_{a_{s\nu}}^2=B_{s\nu}\,L^2/\Phi_0$ play the role of the number of degenerate states in each 
Landau level of the 2D QHE. In the subspaces considered here the $\alpha\nu$ fermion occupancies 
correspond to a full lowest Landau level with filling factor one, as given in Eq. (\ref{N-B-Phi}). 
Only for the $U/4t\rightarrow\infty$ limit there is
fully equivalence between the $\alpha\nu$ fermion occupancy configurations
and, within a mean-field approximation for the fictitious magnetic field 
${\vec{B}}_{\alpha\nu}$ of Eq. (\ref{A-j-s1-3D}), the 2D QHE with a $\nu_{\alpha\nu}=1$ full lowest Landau level. 
In spite of the lack of state degeneracy emerging upon decreasing the value of $U/4t$,
there remains for finite $U/4t$ values some relation to the 2D QHE. The occurrence of QHE-type behavior 
in the square-lattice quantum liquid shows that a magnetic field is not 
essential to the 2D QHE physics. Indeed, here the fictitious magnetic fields
arise from expressing the effects of the electronic correlations in terms of the 
$\alpha\nu$ fermion interactions. 

The mean-field analysis associated with Eq. (\ref{N-B-Phi}) is consistent with in the
subspaces (A) and (B) the square-lattice model Hamiltonian commuting with the $\alpha\nu$ translation
generators ${\hat{{\vec{q}}}}_{\alpha\nu}$ in the $U/4t\rightarrow\infty$ limit so that their eigenvalues are good 
quantum numbers. Indeed, in that limit such an analysis refers to an effective
description where the Hamiltonian is the sum of a $c$ fermion kinetic-energy
term and a QHE like Hamiltonian for each $\alpha\nu$ branch. The $\alpha\nu$ translation
generators ${\hat{{\vec{q}}}}_{\alpha\nu}$ then commute with all such Hamiltonian terms so that  in the $U/4t\rightarrow\infty$ limit
their eigenvalues are indeed good quantum numbers for the model on the square 
lattice in a subspace (A) or (B). Since the electron - rotated-electron
transformation is unitary such commutation relations also hold for
$U/4t>0$ so that in such subspaces the $\alpha\nu$ fermion discrete momentum
values are indeed good quantum numbers for $U/4t$ finite as well.  

In summary, for the Hubbard model on the square lattice each $U/4t>0$ energy eigenstate
$\vert \Psi_{U/4t}\rangle={\hat{V}}^{\dag}\vert \Psi_{\infty}\rangle$ where ${\hat{V}}$ is
the uniquely defined electron - rotated-electron unitary operator considered in this paper can 
be expressed as a superposition of a set of $S_{\eta}$, $S_{\eta}^z$, $S_s$, $S_s^z$, $S_c$, and momentum
eigenstates $\vert \Phi_{U/4t}\rangle={\hat{V}}^{\dag}\vert \Phi_{\infty}\rangle$ of general form
given in Eq. (\ref{non-LWS}) with equal values for such physical quantities, $c$ fermion momentum distribution
function $N_c ({\vec{q}}) =N_{rot} ({\vec{q}})$ of Eq. (\ref{N-rot}), and
numbers $C_{\alpha}=\sum_{\nu=1}^{C_{\alpha}}\nu\,N_{\alpha\nu}$ of Eqs. (\ref{2S-2C}), (\ref{C-C}), 
(\ref{M-L-Sum}), and (\ref{sum-rules}) where $\alpha =\eta,s$ but different values for
the $\alpha\nu$ fermion numbers $\{N_{\alpha\nu}\}$. However, in some limiting
cases such as for the energy eigenstates that span the one- and two-electron subspace
that set reduces to a single state $\vert \Phi_{U/4t}\rangle$ so that
$\vert \Psi_{U/4t}\rangle=\vert \Phi_{U/4t}\rangle$. In turn, that for the 1D model the whole
set of $\alpha\nu$ fermion numbers $\{N_{\alpha\nu}\}$ are good quantum numbers 
and the corresponding $\alpha\nu$ translation generators ${\hat{{\vec{q}}}}_{\alpha\nu}$ commute with the Hamiltonian
is behind the equality $\vert \Psi_{U/4t}\rangle=\vert \Phi_{U/4t}\rangle$ always holding.

\subsection{Number of $s1$ fermion holes of the elementary excitations
of $x>0$ and $m=0$ ground states}

The number $N^h_{s1}$ of unoccupied sites of the $s1$ effective lattice equals that of
$s1$ fermion holes in the $s1$ momentum band and plays an
important role in the general energy spectrum of the square-lattice 
quantum liquid introduced in this paper and further studied
in Refs. \cite{companion,cuprates0}. Indeed, for the one- and two-electron subspace
it is a good quantum number. Here we address the problem of the 
$N^h_{s1}$ expression in terms of the number $N$ of electrons,
which contains important physical information. For the index choice $\alpha =s$, Eq. (\ref{Nh+Nh}) reads 
$N^h_{s1} = [N_{a_{s}}^{D} - 2B_s]$ where $N_{a_{s}}^{D}=2S_c$ is the number of sites
of the spin effective lattice and $B_s$ the number defined in Eqs. (\ref{Nh+Nh}) and (\ref{sum-rules}).
From the use of expression $a_{\alpha} = L/N_{a_{\alpha}} = [N_a/N_{a_{\alpha}}]\,a$
given in Eq. (\ref{a-alpha}) where $\alpha = \eta, s$, the number of unoccupied
sites of the $s1$ effective lattice can be rewritten as
$N^h_{s1} = [N_a^D-N_{a_{\eta}}^{D} - 2B_s]$
where $N_{a_{\eta}}^{D}=[N_a^D-2S_c]$ is the number of sites
of the $\eta$-spin effective lattice.
From combination of the expressions given in Eqs. (\ref{L-L})
and (\ref{M-L-C}) one finds
$N_{a_{\eta}}^{D}=[N_a^{D}-N+2C_{\eta}+2L_{\eta,-1/2}]$
where $C_{\eta}$ is given in Eq. (\ref{M-L-Sum}) and $L_{\eta,-1/2}$
is the number of independent $-1/2$ $\eta$-spinons.
The use of that result in the above general $N^h_{s1}$ expression leads to, 
\begin{equation}
N^h_{s1} = [N - 2B_s-2C_{\eta}-2L_{\eta,-1/2}] \, .
\label{N-h-s1-N}
\end{equation}
For the model on the square lattice this number labels the momentum eigenstates of
Eq. (\ref{non-LWS}) but in general is not a good quantum number because the number  
$B_s$ on the right-hand side of Eq. (\ref{N-h-s1-N}) given in Eqs. (\ref{Nh+Nh}) and (\ref{sum-rules})
is not in general conserved. (All remaining numbers on its right-hand side are.)

Since all sites of the $s1$ effective lattice of $x>0$ and $m=0$ ground states 
are occupied and hence there are no unoccupied sites, from the use of such a $N^h_{s1}$ expression  
one finds that for the corresponding excited states the number 
of unoccupied sites of the $s1$ effective lattice and thus of
$s1$ fermion holes in the $s1$ momentum band reads,
\begin{eqnarray}
N^h_{s1} & = & [\delta N - 2\delta N_{s1} - 2\sum_{\nu =2}^{C_{s}}N_{s\nu}
-2C_{\eta}-2L_{\eta,-1/2}] \hspace{0.25cm} ({\rm general})
\, , \nonumber \\
N^h_{s1} & = & [\delta N - 2\delta N_{s1} - 2N_{s2}] \hspace{0.25cm}
({\rm one-}\hspace{0.10cm}{\rm and}\hspace{0.10cm}{\rm two-electron}\hspace{0.10cm}{\rm subspace}) \, .
\label{Ns1h-general}
\end{eqnarray}
Here $\delta N$ and $\delta N_{s1}$ are the electron and $s1$ fermion number 
deviations, respectively, generated in the transitions from the above ground
states to the excited states. Since $x>0$ and $m=0$ ground states
have no $2\nu$-spinon composite $s\nu$ fermions with $\nu\geq 2$ 
spinon pairs, no $2\nu$-$\eta$-spinon composite $\eta\nu$ fermions,
and no independent $-1/2$ $\eta$-spinons, the deviations in the
corresponding numbers $N_{s\nu}$, $C_{\eta} = 
\sum_{\nu =1}^{C_{\eta}}\nu\,N_{\eta\nu}$, and $L_{\eta,-1/2}$
equal the corresponding excited-state numbers. The expressions
given in Eq. (\ref{Ns1h-general}) refer to the latter numbers.

The first expression is valid for the whole Hilbert space of the Hubbard model 
and hence takes into account all available excited states. It reveals that
the number of holes in the $s1$ momentum band can
be written as $N^h_{s1} = \delta [N- N^{\delta N}_{s1}]$ where the absolute value of 
the number $N^{\delta N}_{s1}\equiv [\delta N-N^h_{s1}]$ 
given in Eq. (\ref{Ns1h-general}) is always an even number. 
Therefore, for $\delta N=\pm 1$ one-electron excited states $N^h_{s1}$ 
must be always an odd integer. In turn, for $\delta N=0$ and $\delta N=\pm 2$
two-electron excited states $N^h_{s1}$ must be
always an even integer. Hence one-electron (and two-electron)
excitations do not couple to excited states with two
holes (and a single hole) in the $s1$ momentum band.

The second expression provided in Eq. (\ref{Ns1h-general}) refers to the one- and 
two-electron subspace defined in the ensuing section. The states that span such a subspace 
are of type (A) or (B) so that for them all numbers 
involved in that expression are good quantum numbers for the model on the square lattice,
as confirmed below.

\section{The square-lattice quantum liquid: A two-component fluid
of charge $c$ fermions and spin-neutral two-spinon $s1$ fermions}

In this section we introduce and define the one- and two-electron subspace and
corresponding spin and $s1$ effective lattices, whereas the $c$ and $s1$ momentum
bands associated with the $c$ and $s1$ effective lattices, respectively,
are studied in Ref. \cite{companion}. Moreover, here we
address the problem of the generation of the one- and 
two-electron spectral weight in terms of processes of $c$ fermions and spinons.
The picture which emerges is that of a two-component quantum liquid
of charge $c$ fermions and spin neutral two-spinon $s1$ fermions.
For the model on the square lattice it refers to the
square-lattice quantum liquid further studied in
Refs. \cite{companion,cuprates0}. 

\subsection{The one- and two-electron subspace as defined in this paper}

Let $\vert \Psi_{GS}\rangle$ be
the exact ground state for $x\geq 0$ and $m=0$ and
${\hat{O}}$ denote an one- or two-electron operator.
Then the state,
\begin{eqnarray}
{\hat{O}}\vert \Psi_{GS}\rangle & = & \sum_j C_j \vert \Psi_{j}\rangle 
\, ; \hspace{0.5cm} C_j = \langle\Psi_{j}\vert{\hat{O}}\vert \Psi_{GS}\rangle \, ,
\label{1-2-subspace}
\end{eqnarray}
generated by application of ${\hat{O}}$ onto that ground
state is contained in the general one- and two-electron subspace.
This is the subspace spanned by the set of energy eigenstates
$\vert \Psi_{j}\rangle$ such that the corresponding coefficients $C_j $ are not vanishing.

Our goal is finding a set of excited states that have 
nearly the whole spectral weight $\sum_j \vert C_j\vert^2\approx 1$ of the above 
one- and two-electron excitations. For a $x> 0$ (and $x=0$) and $m=0$ ground state 
$\vert \Psi_{GS}\rangle$ it is found here that such states have 
excitation energy $\omega<2\mu$ (and $\omega<\mu^0$). 
The inequality $\omega<2\mu$ applies to hole concentrations
$0<x<1$ and $m=0$. In turn, for the $\mu=0$ and $m=0$ absolute ground state with the
chemical-potential zero level at the middle of the Mott-Hubbard gap
the smallest energy required for creation
of both one rotated-electron and one rotated-hole doubly occupied site is
instead given by the energy scale $\mu^0$. This justifies why
for such initial ground state the above inequality
$\omega<2\mu$ is replaced by $\omega<\mu^0$.

From the use of the invariance under the electron - rotated-electron
unitary transformation of the independent $\pm 1/2$ spinons,
one finds that the number of such objects 
generated by application of $\cal{N}$-electron operators 
onto a ground state is exactly restricted to the following range,
\begin{equation}
L_s = [L_{s,\,-1/2} + L_{s,\,+1/2}] = 2S_s = 0,1,2,...,{\cal{N}} 
\, . \label{srcsL}
\end{equation}
It follows that for the one- and two-electron
subspace only the values  
$L_s =[L_{s,\,-1/2} + L_{s,\,+1/2}] = 2S_s = 0,1,2$ are
allowed. Such a restriction is exact 
for both the model on the square and 1D lattices.

An important property is that for a number $\nu\geq 2$
of spinon pairs the $s\nu$ fermions created onto 
a $x\geq 0$ and $m=0$ ground state have vanishing energy and momentum.
Since at $m=0$ one has that $H=0$, such objects obey 
the criterion $\epsilon_{s\nu} = 2\nu\mu_B\,H = 0$ of
Eq. (\ref{invariant-V}). Hence they are invariant under 
the electron - rotated-electron unitary transformation. Therefore,
for $U/4t>0$ they correspond to the same occupancy 
configurations in terms of both rotated electrons and electrons.
That reveals that such $s\nu$ fermions describe
the spin degrees of freedom of a number $2\nu$ of electrons.
It follows that nearly the whole spectral weight 
generated by application onto the ground state of 
$\cal{N}$-electron operators refers to a subspace 
spanned by energy eigenstates with numbers in
the following range,
\begin{equation}
[L_{s,\,-1/2} + L_{s,\,+1/2} +2C_s - 2B_s] =0,1,2,...,
{\cal{N}} \, . 
\label{srs0}
\end{equation}
Note that owing to the above invariance of the $s\nu$ fermions with $\nu\geq 2$ spinon
pairs under consideration provided that ${\cal{N}}/N_a^D\rightarrow 0$ and $B_s/N_a^D\rightarrow 0$ for
$N_a^D\rightarrow\infty$ the number $B_s=\sum_{\nu}N_{s\nu}$ is a good quantum number.
(This is a generalization of the subspace (A) considered in Subsection IV-D.) 
Consistently, the $x> 0$ (and $x=0$ and $\mu=0$) and $m=0$ ground state 
and the set of excited states of energy $\omega<2\mu$
(and $\omega<\mu^0$) that span the one- and two-electron subspace introduced in this
paper have no $-1/2$ $\eta$-spinons, $\eta\nu$ fermions, and $s\nu'$ fermions with $\nu'\geq 3$ spinon pairs so that 
$N_{\eta\nu}=0$ and $N_{s\nu'}=0$ for $\nu'\geq 3$. As further discussed below, summation over 
the states that span that subspace as defined here gives
$\sum_j \vert C_j\vert^2\approx 1$ for the coefficients of the one- or two-electron excitation
${\hat{O}}\vert \Psi_{GS}\rangle$ of Eq. (\ref{1-2-subspace}) and there is both for the
model on the 1D and square lattices an extremely
small weight corresponding mostly to states with $N_{s3}=1$, which is neglected within
our definition of the one- and two-electron subspace.

Thus, according to Eq. (\ref{srs0}) the numbers of independent $\pm 1/2$ spinons and 
that of $s2$ fermions of the excited states that span such a subspace are restricted to the 
following ranges,
\begin{eqnarray}
L_{s,\,\pm 1/2} & = & 0,1  \, ;
\hspace{0.35cm} N_{s2} = 0 \, ,
\hspace{0.25cm}{\rm for}\hspace{0.25cm} {\cal{N}} = 1 \, ,
\nonumber \\
2S_s + 2N_{s2} & = & [L_{s,\,-1/2} + L_{s,\,+1/2} + 2N_{s2}] =
0,1,2   \, ,
\hspace{0.25cm}{\rm for}\hspace{0.25cm} {\cal{N}} = 2 \, .
\label{srs-0-ss}
\end{eqnarray}
Here ${\cal{N}}=1,2$ refers to the ${\cal{N}}$-electron
operators ${\hat{O}}$ whose application onto the ground
state $\vert \Psi_{GS}\rangle$ generates the above excited states,
as given in Eq. (\ref{1-2-subspace}). Furthermore, $N_c =N=(1-x)\,N_a^D$ and
$N_{s1} = [N/2-2N_{s2}-S_s]=(1-x)\,N_a^D/2 -[2N_{s2}+S_s]$.
If in addition we restrict our
considerations to the LWS-subspace of the one- and
two-electron subspace, then $L_{s,\,-1/2}=0$ 
in Eq. (\ref{srs-0-ss}), whereas
the values $L_{s,\,+1/2} = 0,1$ for $N_{s2} = 0$ and
${\cal{N}} = 1$ remain valid and in $[2S_s + 2N_{s2}] =0,1,2$
one has that $2S_s=L_{s,\,+1/2}$ for ${\cal{N}} = 2$.
\begin{table}
\begin{tabular}{|c|c|c|c|c|c|c|c|c|c|c|c|c|c|c|} 
\hline
numbers & charge & +1$\uparrow$el. & -1$\downarrow$el. & +1$\downarrow$el. & -1$\uparrow$el. & singl.spin & 
tripl.spin & tripl.spin & tripl.spin & $\pm$2$\uparrow\downarrow$el. & +2$\uparrow$el. & -2$\downarrow$el. & +2$\downarrow$el. & -2$\uparrow$el. \\
\hline
$\delta N_c^h$ & $0$ & $-1$ & $1$ & $-1$ & $1$ & $0$ & $0$ & $0$ & $0$ & $\mp 2$ & $-2$ & $2$ & $-2$ & $2$ \\
\hline
$N_{s1}^h$ & $0$ & $1$ & $1$ & $1$ & $1$ & $2$ & $2$ & $2$ & $2$ & $0$ & $2$ & $2$ & $2$ & $2$ \\
\hline
$\delta N_{\uparrow}$ & $0$ & $1$ & $0$ & $0$ & $-1$ & $0$ & $1$ & $-1$ & $0$ & $\pm 1$ & $2$ & $0$ & $0$ & $-2$ \\
\hline
$\delta N_{\downarrow}$ & $0$ & $0$ & $-1$ & $1$ & $0$ & $0$ & $-1$ & $1$ & $0$ & $\pm 1$ & $0$ & $-2$ & $2$ & $0$ \\
\hline
$L_{s,\,+1/2}$ & $0$ & $1$ & $1$ & $0$ & $0$ & $0$ & $2$ & $0$ & $1$ & $0$ & $2$ & $2$ & $0$ & $0$ \\
\hline
$L_{s,\,-1/2}$ & $0$ & $0$ & $0$ & $1$ & $1$ & $0$ & $0$ & $2$ & $1$ & $0$ & $0$ & $0$ & $2$ & $2$ \\
\hline
$N_{s2}$ & $0$ & $0$ & $0$ & $0$ & $0$ & $1$ & $0$ & $0$ & $0$ & $0$ & $0$ & $0$ & $0$ & $0$ \\
\hline
$S_s$ & $0$ & $1/2$ & $1/2$ & $1/2$ & $1/2$ & $0$ & $1$ & $1$ & $1$ & $0$ & $1$ & $1$ & $1$ & $1$ \\
\hline
$\delta S_c$ & $0$ & $1/2$ & $-1/2$ & $1/2$ & $-1/2$ & $0$ & $0$ & $0$ & $0$ & $\pm 1$ & $1$ & $-1$ & $1$ & $-1$ \\
\hline
$\delta N_{s1}$ & $0$ & $0$ & $-1$ & $0$ & $-1$ & $-2$ & $-1$ & $-1$ & $-1$ & $\pm 1$ & $0$ & $-2$ & $0$ & $-2$ \\
\hline
$\delta N_{a_{s1}}$ & $0$ & $1$ & $0$ & $1$ & $0$ & $0$ & $1$ & $1$ & $1$ & $\pm 1$ & $2$ & $0$ & $2$ & $0$ \\
\hline
\end{tabular}
\caption{The deviations $\delta N_c^h=-2\delta S_c$ and numbers $N_{s1}^h=[2S_s+2N_{s2}]$ of Eq. (\ref{deltaNcs1}) for the 
fourteen classes of one- and two-electron excited states of the $x>0$ and $m=0$ ground state that span the one- and two-electron 
subspace as defined in this paper and Ref. \cite{companion}, corresponding electron number deviations 
$\delta N_{\uparrow}$ and $\delta N_{\downarrow}$, and independent-spinon numbers $L_{s,\,+1/2}$ and 
$L_{s,\,-1/2}$ and $s2$ fermion numbers $N_{s2}$ restricted to the values provided in Eq. (\ref{srs-0-ss}).
The spin $S_s$ and deviations $\delta S_c$, $\delta N_{s1}=[\delta S_c-S_s-2N_{s2}]$, and $\delta N_{a_{s1}}=[\delta S_c +S_s]$ 
of each excitation are also provided.}
\label{table1}
\end{table} 

We recall that the numbers of independent $\eta$-spinons ($\alpha=\eta$) and 
independent spinons ($\alpha=s$) $L_{\alpha,\,\pm 1/2}$, total numbers of $\eta$-spinons 
($\alpha=\eta$) and spinons ($\alpha=s$) $M_{\alpha,\,\pm 1/2} = [L_{\alpha,\,\pm 1/2} + C_{\alpha}]$,
number of sites of the spin effective lattice $N_{a_{s}}^D=2S_c$, number of sites of the $\eta$-spin
effective lattice $N_{a_{\eta}}^D=[N_a^D-2S_c]$, number of $c$ fermions $N_c=2S_c$,
and number of $c$ fermion holes $N_c^h=[N_a^D-2S_c]$ are good quantum numbers of
both the Hubbard model on the 1D and square lattices. The good news is that in the one- and
two-electron subspace as defined here the numbers $N_{a_{s1}}^2$, $N_{s1}$, $N_{s1}^h$, 
and $N_{s2}$ are good quantum numbers of the Hubbard model on the square lattice
as well. Indeed, for spin $S_s=0$ that subspace is a subspace (A) as defined in Subsection
IV-D so that $N_{s1}=[S_c-2N_{s2}]$, $N_{s1}^h=2N_{s2}$, and $N_{s2}$ and hence $N_{a_{s1}}^2=[N_{s1}+N_{s1}^h]=S_c$
are good quantum numbers. Furthermore, for the remaining spin values $S_s=1/2$ and $S_s=1$ such a subspace is
a subspace (B) as defined in that subsection so that $N_{s1}=[S_c-S_s]$, $N^h_{s 1} =2S_s$, and $N_{a_{s1}}^2=
[S_c+S_s]$ are good quantum numbers. That implies that the microscopic momenta ${\vec{q}}$ carried by 
the $s1$ fermions are good quantum numbers. 

The number of sites, unoccupied sites, and occupied sites of the $c$ and $s1$ effective
lattices equal those of discrete momentum values, unfilled momentum values, and
filled momentum values of the $c$ and $s1$ bands, respectively.
Use of the general expressions of $s1$ band discrete momentum values and
number $N_{s1}^h$ of $s1$ band unfilled momentum values given in Eq. (\ref{Nh+Nh}) for $\alpha =s$ together with 
the restrictions in the values of the numbers of Eqs. (\ref{srcsL}), (\ref{srs0}), and (\ref{srs-0-ss})
and the result proved above in Subection IV-E that one-electron (and two-electron) excitations have
no overlap with excited states with none and two (and one) $s1$ band holes 
(and hole) reveals that nearly the whole one- and two-electron spectral weight is contained in the subspace 
spanned by states whose deviation $\delta N_c^h$ in the number of $c$ band holes
and number $N_{s1}^h$ of $s1$ band holes are given by, 
\begin{eqnarray}
\delta N_c^h & = & -2\delta S_c = - \delta N = 0, \mp 1, \mp 2 \, ,
\nonumber \\
N_{s1}^h & = & 2S_s+2N_{s2} = \pm (\delta N_{\uparrow} - \delta N_{\downarrow})
+ 2L_{s,\,\mp 1/2} + 2N_{s2} = L_{s,\,-1/2} + L_{s,\,+1/2} + 2N_{s2} = 0, 1, 2 \, .
\label{deltaNcs1}
\end{eqnarray}
Here $\delta N$ is the deviation in the number of electrons, 
$\delta N_{\uparrow}$ and $\delta N_{\downarrow}$ those in
the number of spin-projection $\uparrow$ and $\downarrow$ 
electrons, respectively, $N_{s2}$ the number of the excited-state $s2$
fermions, and $L_{s,\,\pm1/2}$ that of independent
spinons of spin projection $\pm1/2$. The deviations $\delta N_c^h$ and numbers $N_{s1}^h$ 
of Eq. (\ref{deltaNcs1}) for the fourteen classes of one- and two-electron excited states of the $x>0$ 
and $m=0$ ground state that span the one- and two-electron subspace, 
corresponding electron number deviations $\delta N_{\uparrow}$ 
and $\delta N_{\downarrow}$, and independent-spinon numbers $L_{s,\,+1/2}$ and 
$L_{s,\,-1/2}$ and $s2$ fermion numbers 
$N_{s2}$ restricted to the values provided in Eq. (\ref{srs-0-ss})
are given in Table \ref{table1}. For excited states with $N_{s2}=1$ 
the $s2$ effective lattice has a single site and the corresponding $s2$ band a single vanishing 
discrete momentum value, $\vec{q} = 0$, occupied by the $s2$ fermion. Such a $s2$ fermion
is invariant under the electron - rotated-electron unitary transformation and has zero
energy, consistently with the invariance condition of Eq. (\ref{invariant-V}) for
$\alpha\nu=s2$ and zero magnetic field $H=0$.

The initial $x>0$ and $m=0$ ground states of the one- and two-electron subspace 
have zero holes in the $s1$ band so that $\delta N_{s1}^h=N_{s1}^h$ for the
excited states. The one- and two-electron subspace as defined here 
is spanned by the states of Table \ref{table1} generated by creation or annihilation 
of $\vert\delta N_c^h\vert=0,1,2$ holes in the
$c$ momentum band and $N_{s1}^h=0,1,2$ holes in the $s1$ band 
plus small momentum and low energy particle-hole processes in the $c$ band. 
The charge excitations of $x>0$ and $m=0$ initial ground states
consist of a single particle-hole process in the $c$ band of arbitrary momentum and 
energy compatible with its momentum and energy bandwidths, 
plus small-momentum and low-energy $c$ fermion particle-hole processes.
Such charge excitations correspond to state
representations of the global $U(1)$ symmetry and
refer to the type of states denoted by "charge" in the table.
The one-electron spin-doublet excitations correspond 
to the four types of states denoted by "$\pm1\sigma$el." in
Table \ref{table1} where $+1$ and $-1$ denotes creation and
annihilation, respectively, and $\sigma =\uparrow ,\downarrow$.
The spin-singlet and spin-triplet excitations refer to the four types of states 
denoted by "singl.spin" and "tripl.spin" in the table.
The two-electron excitations whose electrons are in a spin-singlet
configuration and those whose two created or annihilated electrons are 
in a spin-triplet configuration correspond to the five types of 
states "$\pm 2\uparrow\downarrow$el." and "$\pm 2\sigma$el." of that table  
where $+2$ and $-2$ denotes creation and
annihilation, respectively, of two electrons.

Alike for the 1D model, for the model on the square lattice such fourteen types of
states are energy eigenstates  
and contain nearly the whole one- and two-electron spectral
weight. Excited states of classes other than those of the table contain
nearly no one- and two-electron spectral weight. Such a weight analysis
applies to the 1D Hubbard model as well.
Moreover, for the quantum liquid corresponding to the 
Hamiltonian (\ref{H}) in the one- and two-electron 
subspace as defined in this paper, the numbers $2S_c$,
$2S_{\eta}$, $2S_s$, and $-2S_s^z$ associated with the global
$SO(3)\times SO(3)\times U(1)$ symmetry are given by,
\begin{equation}
2S_c = (1-x)\,N_a^D \, ; \hspace{0.25cm}
2S_{\eta} = -2S_{\eta}^z= x\,N_a^D \, ; \hspace{0.25cm}
2S_s = (1-x)\,N_a^D - 2[N_{s1}+2N_{s2}] \, ; \hspace{0.25cm}
-2S_s^z = m\,N_a^D = 2S_s - 2L_{s,-1/2} \, .
\label{NN-SSS}
\end{equation}
For such a quantum liquid the number of sites and
lattice spacing of the spin effective lattice read,
\begin{equation}
N_{a_{s}}^D = (1-x)\,N_a^D 
\, ; \hspace{0.35cm}
a_{s} = {a\over (1-x)^{1/D}} \, ,
\hspace{0.25cm} (1-x)\geq 1/N_a^D \, .
\label{NNCC}
\end{equation}

We emphasize that for the model in the one- and two-electron subspace
the concept of a $\eta$-spin lattice has no physical significance because the $\eta$-spin degrees of freedom
refer to the $C_{\eta}=(N_a^D/2-S_c-S_{\eta})=0$ vacuum $\vert 0_{\eta};N_{a_{\eta}}^D\rangle$ of Eq. (\ref{vacuum}),
which for $S_c>0$ corresponds to a single occupancy configuration of the $N_{a_{\eta}}^D$ $+1/2$ $\eta$-spinons.
Only when both $N_{a_{\eta}}^D/N_a^D=[1-2S_c/N_a^D]>0$ and $C_{\eta}=(N_a^D/2-S_c-S_{\eta})>0$ is the 
concept of a $\eta$-spin effective lattice well defined and useful.   
Indeed, for that subspace the numbers $C_{\eta}$ and $C_{s}$ of sites of the original lattice
whose electron occupancy configurations that generate the energy and momentum eigenstates are not invariant under
the electron - rotated-electron unitary  transformation are given by $C_{\eta} = 0$ and 
$C_s=[N_{s1}+2N_{s2}]$, respectively.
For $x=0$ (and $x=1$) one finds $N_{a_{\eta}}^D=0$
and $N_{a_{s}}^D = N_a^D$ (and 
$N_{a_{\eta}}^D= N_a^D$
and $N_{a_{s}}^D = 0$), so that there is no
$\eta$-spin (and no spin) effective lattice and the
spin (and $\eta$-spin) effective lattice equals
the original lattice.

As further confirmed below in Subsection V-D,
for the one- and two-electron subspace defined here only the $c$ and $s1$
fermions have an active role. Straightforward manipulations of the above general
expressions given in Eqs. (\ref{N*})-(\ref{a-a-nu})
reveal that for that subspace the number $N_{a_{s1}}^D$
of $s1$ band discrete momentum values, 
$N_{s1}$ of $s1$ fermions, and $N^h_{s 1}$ of $s1$ fermion holes are given by,
\begin{equation}
N_{a_{s1}}^D = [N_{s1} + N^h_{s1}] = 
[S_c + S_s] \, ;
\hspace{0.25cm}
N_{s1} = [S_c - S_s -2N_{s2}]  \, ;
\hspace{0.25cm}
N^h_{s 1} = [2S_s +2N_{s2}] =0,1,2 \, ,
\label{Nas1-Nhs1}
\end{equation}
respectively. In turn,
the corresponding $c$ effective lattice, $c$ momentum band,
and $c$ fermion numbers read,
\begin{equation}
N_{a_{c}}^D = [N_{c} + N^h_{c}] = 
N_{a}^D  \, ;
\hspace{0.25cm}
N_{c} = 2S_c = (1-x)\,N_{a}^D  \, ;
\hspace{0.25cm}
N^h_{c} = x\,N_{a}^D \, .
\label{Nac-Nhc}
\end{equation}
 
As mentioned above, for states belonging to the one- and two-electron subspace
the $\eta$-spin degrees of freedom
correspond to a single occupancy configuration
referring to the $\eta$-spin vacuum 
$\vert 0_{\eta};N_{a_{\eta}}^D\rangle$ of Eq. 
(\ref{vacuum}) (and to no configuration since $N_{a_{\eta}}^D=0$
in that vacuum). Therefore, when defined in 
that subspace the Hamiltonian (\ref{H}) has an effective
global $U(2)/Z_2 \equiv SO(3)\times U(1)$ symmetry.
This is anyway the global symmetry relevant to
more general models belonging to the same universality
class as that considered here. Indeed,
the $\eta$-spin $SU(2)$ symmetry disappears if one adds 
to the Hamiltonian (\ref{H}) kinetic-energy terms 
involving electron hopping beyond nearest neighbors. In contrast, the global  
$SO(3)\times U(1)$ symmetry is robust under the addition
of such terms. 

\subsection{Generation of the one- and two-electron spectral weight
in terms of processes of $c$ fermions and spinons}

Expression by use of the operator relations given in Eq. (\ref{c-up-c-down}) of
the operator ${\tilde{O}}$ corresponding to the one- and two-electron 
operators of the excitations ${\hat{O}}\vert \Psi_{GS}\rangle$ 
of Eq. (\ref{1-2-subspace}) in terms of the of $c$ fermion and rotated quasi-spin 
operators reveals that the elementary processes associated with
the number value ranges of Eqs. (\ref{srs0}), (\ref{srs-0-ss}), and (\ref{deltaNcs1}) are
for the one- and two-electron subspace fully generated by 
the operator ${\tilde{O}}$.  

In turn, expressing by means of the operator relations provided in Eq. (\ref{c-up-c-down}) 
the operator terms of Eq. (\ref{OOr}) containing
commutators involving the operator ${\tilde{S}} = (t/U)\,[\tilde{T}_{+1} -\tilde{T}_{-1}] 
+ {\cal{O}} (t^2/U^2)$ in terms of the of $c$ fermion and rotated quasi-spin operators
and taking into account that independently of their form, the additional operator terms 
${\cal{O}} (t^2/U^2)$ of higher order are products of the kinetic operators 
$\tilde{T}_0$, $\tilde{T}_{+1}$, and $\tilde{T}_{-1}$ of Eq. (\ref{T-op}), 
one finds that the processes generated by such
operator terms refer only to excitations whose number value ranges are different
from those of Eqs. (\ref{srs0}), (\ref{srs-0-ss}), and (\ref{deltaNcs1}) 
for the one- and two-electron subspace. That confirms that
such processes generate very little one- and two-electron
spectral weight, consistently with the discussions and analysis of Subsections II-A and V-A. 

For the model on the 1D lattice the spectral-weight
distributions can be calculated explicitly by the pseudofermion
dynamical theory associated with the model exact solution, 
exact diagonalization of small chains, and other methods. The relative 
one-electron spectral weight generated by different types of 
microscopic processes is studied by means of such methods in Ref. \cite{1EL-1D}.
The results of that reference confirm the dominance of
the processes associated with the number value ranges provided in
Eqs. (\ref{srs-0-ss}) and (\ref{deltaNcs1}). In this case
the general operators ${\hat{O}}$ and ${\tilde{O}}$ are
electron and rotated-electron, respectively, creation or
annihilation operators. The operator ${\tilde{O}}$
generates {\it all} processes associated with the number value ranges
of Eqs. (\ref{srs-0-ss}) and (\ref{deltaNcs1}) and number values of
Table \ref{table1}. In addition, it also generates some of the non-dominant processes. That
is confirmed by the weights given in Table 1 of Ref.
\cite{1EL-1D}, which correspond to the dominant processes
associated with only these ranges. The small missing
weight refers to excitations whose number value ranges
are not those of Eqs. (\ref{srs-0-ss}) and (\ref{deltaNcs1}) but whose
weight is also generated by the operator ${\tilde{O}}$.
Indeed, that table refers to $U/4t\rightarrow\infty$
so that ${\hat{O}}={\tilde{O}}$ and the operator terms of 
Eq. (\ref{OOr}) containing commutators involving the 
operator ${\tilde{S}}$ vanish because all such 
commutators vanish in that limit.

On the other hand, for finite values of $U/4t$ all dominant processes associated 
with the number value ranges of Eqs. (\ref{srs-0-ss}) and (\ref{deltaNcs1}) 
and number values provided in Table \ref{table1} 
are generated by the operator ${\tilde{O}}$ whereas
the small spectral weight associated with excitations whose
number value ranges are different from those are generated both
by that operator and the operator terms of 
Eq. (\ref{OOr}) containing commutators involving the 
operator ${\tilde{S}}$. For the model on the 1D lattice the small 
one-electron spectral weight generated by the non-dominant 
processes is largest at half filling and $U/4t\approx 1$. 
For the range of hole concentrations $x\geq 0$ considered in
this paper the one- and two-electron subspace is spanned
by states with vanishing rotated-electron double
occupancy. Its generalization for the range $x\leq 0$ reveals
that then it is spanned by states with vanishing rotated-hole double
occupancy. That property combined with the particle-hole
symmetry explicit at $x=0$ and $\mu=0$, implies that
the relative spectral-weight contributions from different types of excitations   
given in Fig. 2 of Ref. \cite{1EL-1D} for half-filling one-electron addition 
leads to similar corresponding relative weights for   
half-filling one-electron removal. Analysis of that figure 
confirms that for the corresponding one-electron removal
spectrum the dominant processes associated 
with the number value ranges of Eqs. (\ref{srs-0-ss}) and (\ref{deltaNcs1}) and 
number values given in Table \ref{table1}
refer to the states called in the figure 1 holon - 1 $s1$ hole states, whose
minimum relative weight of about $0.95$ is reached
at $U/4t\approx 1$. For other hole concentrations $x>0$ 
and values of $U/4t$ the relative weight of such states is always 
larger than $0.95$, as confirmed from analysis of 
Figs. 1 and 2 and Table 1 of Ref. \cite{1EL-1D}.

For the Hubbard model on the square lattice the
explicit derivation of one- and two-electron spectral
weights is a more involved problem. The number value ranges
of Eqs. (\ref{srs0}), (\ref{srs-0-ss}), and (\ref{deltaNcs1}) 
and number values provided in Table \ref{table1} 
for the one- and two-electron subspace also apply,
implying similar results for the relative spectral weights
of one- and two-electron excitations. Furthermore and
as discussed above, analysis of expression (\ref{Ns1h-general}) reveals that
for $\delta N=\pm 1$ (and $\delta N=\pm 0$ and $\delta N=\pm 2$) 
excited states the number $N^h_{s1}$ of holes in the $s1$ 
band must be always an odd 
(and even) integer so that one-electron (and two-electron) excitations do not 
couple to excited states with two holes 
(and one hole) in the $s1$ band. 
Both such selection rules hold also for the one- and two-electron
subspace for which $N^h_{s1} = [\delta N - 2\delta N_{s1} - 2N_{s2}]$,
as given in Eq. (\ref{Ns1h-general}).

\subsection{The spin and $s1$ effective lattices
for the one- and two-electron subspace}

Within the operator description
introduced in this paper, the subspace with relevance 
for the one- and two-electron physics is the one- and
two-electron subspace as defined above, which 
is spanned by states with no $\eta\nu$ fermions
and no $s\nu$ fermions with a number $\nu\geq 3$ of spinon
pairs. According to the restrictions and numbers values of
Eqs. (\ref{srs-0-ss}) and (\ref{deltaNcs1}) and Table \ref{table1}, such states may involve none 
or one $s2$ fermion. As confirmed in Refs. \cite{companion,cuprates0,cuprates},
it is convenient to express the one- and two-electron excitation spectrum relative
to initial $x>0$ and $m=0$ ground states in terms of the deviations in the numbers
of $c$ effective lattice and $s1$ effective lattice unoccupied sites given 
explicitly in Eq. (\ref{deltaNcs1}) and Table \ref{table1}. Note that the number of $s1$ fermions 
provided in Eq. (\ref{Nas1-Nhs1}) can be written as 
$N_{s1}=[(1-x)N_a^D/2-S_s -2N_{s2}]$ 
where $S_s=0$ for $N_{s2}=1$ and $S_s=0,1/2,1$ for $N_{s2}=0$.
The square-lattice quantum liquid associated with the
Hubbard model on the square lattice in the
one- and two-electron subspace is further studied in Refs.
\cite{companion0,companion}. The studies of Ref. \cite{cuprates0} refer to
the same quantum liquid perturbed by small 3D anisotropy associated with
weak plane coupling. 

Since for $N_{s2}=1$ states the $s2$ fermion has vanishing energy 
and momentum and in addition consistently with Eq. (\ref{invariant-V})
for vanishing magnetic field $H=0$ is invariant under the unitary transformation 
associated with the operator $\hat{V}$, the only effect of its creation
and annihilation is on the numbers
of occupied and unoccupied sites of the $s1$ effective lattice and
thus in the occupancies of the discrete momentum values of the $s1$ band, as discussed below.
Therefore, for the study of the Hamiltonian (\ref{H}) in 
the one- and two-electron subspace the only composite object whose
internal occupancy configurations in the spin effective lattice are important
for the physics is the two-spinon $s1$ fermion and associated $s1$ bond particle
studied in Ref. \cite{companion0}.

It is found below in Subsection V-D that for the
model in the one- and two-electron subspace as defined in Subsection V-A
the presence of independent $+1/2$ spinons or a composite $s2$ 
fermion is felt through the numbers of occupied and unoccupied sites of
the $s1$ effective lattice, whereas the number of independent
$+1/2$ $\eta$-spinons equals that of the unoccupied 
sites of the $c$ effective lattice. Therefore, when acting onto that
subspace, the Hubbard model refers to a two-component quantum liquid described
in terms of $c$ fermions and $s1$ fermions. 

For $N_a^D\gg 1$ the description introduced 
in Ref. \cite{companion0} of both the spinon occupancy
configurations of a $s1$ bond-particle associated with a local
$s1$ fermion and of the underlying
$s1$ effective lattice is a good approximation 
provided that the ratio $N_{a_s}^D/N_a$ is finite. 
Since the number of sites of the spin effective
lattice is given by $N_{a_s}^D=(1-x)N_a^D$, that requirement is met provided 
that the electronic density $n=(1-x)$
is finite. Such a description is exact for hole concentrations 
$0\leq x\ll 1$ when the spin-effective lattice becomes the original lattice. In that limit
all sites of the spin effective lattice become sites
of the original square or 1D lattice. In turn, for the 
model on the square lattice it does not apply
when $x\approx 1$ and there is a finite number of sites
in the spin effective lattice only. So it is a good approximation 
for the range $(1-x)\geq 1/N_a^D$ and the limit $N_a^D\gg 1$
considered in our studies. Particularly, it is a good approximation for hole concentrations
$x\in (0,x_*)$ considered in Refs. \cite{companion,cuprates0}, where $x_*\in (0.23,0.28)$ for the
range $U/4t\in (u_0,u_1)$ of interest for the studies of these references where
$u_0\approx 1.3$ and $u_1\approx 1.6$, respectively.

For $x<1$ the spin effective
lattice has a number of sites $N_{a_{s}}^D=(1-x)\,N_a^D$ 
smaller than that of the original lattice. Nevertheless for $N_a^D\gg 1$
a site of the spin effective lattice with real-space coordinate
$\vec{r}_j$ in that lattice has a well-defined position $\vec{r}=\vec{r}_j$
in the original real space contained within the system area 
or length $L^D$ where $D=2$ and $D=1$, respectively. 
This is consistent with the lattice
constant $a_s$ defined in Eq. (\ref{a-alpha}) for $\alpha = s$,
which for the one- and two-electron subspace reads,
\begin{equation}
a_{s} = {a\over (1-x)^{1/D}} \, ; \hspace{0.25cm}
x \ll  1 \, , \hspace{0.15cm} D=1,2 \, ,
\label{a-s-1-2-el}
\end{equation}
being such that the area $(D=2)$ or length $(D=1)$ $L^D$ of the system 
is preserved. Obviously, any real-space point within the spin effective lattice corresponds to 
the same real-space point in the system original lattice. So the real-space coordinates 
$\vec{r}_{j'}$ of the $j'=1,...,N_{a_{s}}^D$ sites of the spin effective 
lattice are for $x>0$ different from the real-space coordinates
$\vec{r}_{j}$ of the $j=1,...,N_{a}^D$ sites of the original lattice.
The vectors $\vec{r}_{j}$ and $\vec{r}_{j'}$ refer though to well-defined
positions in the same 1D or 2D real space contained within such a lattice.

The suitability of the description introduced in this paper in terms
of occupancies of effective lattices is confirmed for 1D by
the results of Appendix B: The discrete 
momentum values of the $c$ fermions and $\alpha\nu$ fermions 
correspond to quantum numbers of the
model exact solution. In turn, no exact solution exists for
the model on the square lattice. However, the discrete momentum values
obtained for such objects refer to good 
quantum numbers for that model in the one- and two-electron
subspace. 

For such a $N_a^D\gg 1$ subspace there is commensurability 
between the real-space distributions of the $N_{a_{s1}}^D\approx N_{s1}$
sites of the $s1$ effective lattice and the $N_{a_{s}}^D\approx 2N_{s1}$ 
sites of the spin effective lattice. For $(1-x)\geq 1/N_a^D$ and $N_a^D\gg 1$
the spin effective lattice has $N_{a_s}^D=(1-x)\,N_a^D$ sites and from the use of 
the expression given in Eq. (\ref{Nas1-Nhs1}) for $N_{a_{s1}}^D$
and Eq. (\ref{a-a-nu}) for $a_{s1}$ we find,
\begin{equation}
a_{s1} = 2^{1/D}\,a_s\,{1\over \left(1+{2S_s\over (1-x)N_a^D}\right)^{1/D}}
\approx 2^{1/D}\,a_s\, \left(1-{2S_s\over D(1-x)}{1\over N_a^D}\right)
\approx 2^{1/D}\,a_s \, , \hspace{0.25cm} 2S_s =0, 1, 2 \, .
\label{a-a-s1-sube}
\end{equation}

For $N^h_{s1}=0$ states
the bipartite 1D and square spin effective lattices have two
well-defined sub-lattices. As discussed in Ref. \cite{companion0}, for $N^h_{s1}=1,2$ 
states the spin effective lattice has two bipartite lattices as well,
with one or two extra sites corresponding to suitable boundary conditions. 
For the model on the square lattice the two spin 
effective sub-lattices have lattice constant $a_{s1} =\sqrt{2}\,a_s$. 
In turn, for 1D the sites of each spin effective sub-lattice are 
distributed alternately along the chain, the corresponding 
nearest-neighboring sites being separated by $a_{s1} =2a_{s}$.
The fundamental translation vectors of such sub-lattices read,
\begin{equation}
{\vec{a}}_{s1} = a_{s1}\,{\vec{e}}_{x_1}
\hspace{0.15cm} {\rm for}  \hspace{0.10cm} 1D \, ;
\hspace{0.50cm}
{\vec{a}}_{s1} = {a_{s1}\over\sqrt{2}}({\vec{e}}_{x_1}+{\vec{e}}_{x_2})
\, , \hspace{0.25cm} 
{\vec{b}}_{s1} = -{a_{s1}\over\sqrt{2}}({\vec{e}}_{x_1}-{\vec{e}}_{x_2}) 
\hspace{0.15cm} {\rm for} \hspace{0.10cm} 2D \, ,
\label{a-b-s1}
\end{equation}
where ${\vec{e}}_{x_1}$ and ${\vec{e}}_{x_2}$ are the unit vectors
and $x_1$ and $x_2$ Cartesian coordinates. As confirmed in Ref. \cite{companion0}, 
the vectors given in Eq. (\ref{a-b-s1}) are the fundamental translation vectors 
of the $s1$ effective lattice.
     
\subsection{The two-component quantum liquid
of charge $c$ fermions and spin-neutral two-spinon $s1$ fermions}

In the case of $x\geq 0$, $m=0$, and $N^h_{s1}=0$ ground states of the model on the 
square lattice whose $s1$ momentum band is full and all $N_{a_{s1}}^2$ sites of the $s1$ effective lattice 
are occupied we consider that the square root $N_{a_s}$ of the number $N_{a_s}^2$ 
of sites of the spin effective lattice is an integer so that the spin effective lattice is a
square lattice with $N_{a_s}\times N_{a_s}$ sites. It follows that the square root $N_{a_{s1}}$ of
the number $N_{a_{s1}}^2$ of sites of the $s1$ effective lattice cannot in general
be an integer number yet $N_{a_{s1}}^2$ is. However, within the $N_a^D\gg 1$ limit considered here
and as mention above for general $\alpha\nu$ branches, we use the notation $N_{a_{s1}}^D$ for the number
of sites of the $s1$ effective lattice where $D=1$ and $D=2$ for the model on the 1D and square
lattice, respectively. 

For the one- and two-electron subspace as defined in this paper the 
number $N_{a_{s1}}^D$ of sites of the $s1$ effective lattice and $s1$ band discrete momentum values, 
$N_{s1}$ of $s1$ fermions, and $N^h_{s 1}$ of $s1$ fermion holes have expressions
given in Eq. (\ref{Nas1-Nhs1}). The corresponding numbers for the $c$ effective lattice and $c$ band
are provided in Eq. (\ref{Nac-Nhc}). For that subspace the $s1$ band is either full
by $N_{s1}=N_{a_{s1}}^D =2S_c =N_{a_s}^D/2$ $s1$
fermions or has one or two holes.  
Furthermore, one-electron and two-electron excitations have
no overlap with excited states with two holes and one hole 
in the $s1$ band, respectively. Hence as given in Table \ref{table1}
excited states with a single hole
in the $s1$ band correspond to one-electron
excitations and those with $N^h_{s 1} =0,2$ 
holes in that band refer to
two-electron excitations. Excited states with 
$N^h_{s1} =3$ (and $N^h_{s 1} =4$) holes
in the $s1$ momentum band correspond to very little 
one-electron (and two-electron) spectral weight
and are ignored within our definition of one- and
two-electron subspace.

According to the number value ranges of Eqs. (\ref{srs-0-ss}) and (\ref{deltaNcs1}) and 
number values provided in Table \ref{table1}, the
one- and two-electron subspace is spanned by excited
states having either none $N_{s2}=0$ or one $N_{s2}=1$
spin-neutral four-spinon $s2$ fermion. When $N_{s2}=1$
there are no independent spinons
and one finds for the $N_{s2}=1$ states spanning that subspace that $N_{a_{s2}}^D=1$ for 
$N_{s2}=1$ so that the $s2$ fermion occupies a $s2$ 
band with a single vanishing momentum value and hence $N^h_{s2}=0$
holes. Indeed, when $N_{s2}=1$ the $s2$ fermion has vanishing energy and momentum and 
consistently with Eq. (\ref{invariant-V}) is for $H=0$ invariant
under the electron - rotated-electron unitary transformation. 
Therefore, the only effect of its creation and annihilation 
is on the numbers of occupied and unoccupied sites of 
the $s1$ effective lattice and corresponding numbers of 
$s1$ band $s1$ fermions and $s1$ fermion holes. Specifically, according to the
expressions provided in Eq. (\ref{Nas1-Nhs1}) and number values of Table \ref{table1} the
deviations $\delta S_c =\delta S_s =0$ and $\delta N_{s2}=1$ generated by a state
transition involving creation of one $s2$ fermion lead to deviations in the number
of $s1$ fermions and $s1$ fermion holes given by 
$\delta N_{s1} = -2\delta N_{s2}=-2$ and
$\delta N^h_{s1} = 2\delta N_{s2}=2$, respectively.
 
It follows that for $N_{s2}=1$ excited states one has from the
ranges of Eqs. (\ref{srs-0-ss}) and (\ref{deltaNcs1}) and 
number values of Table \ref{table1} that $S_s=0$ so that 
according to Eq. (\ref{Nas1-Nhs1}) the number of holes of the 
$s1$ band is given by $N^h_{s 1} = 2N_{s2} =2$, 
in contrast to $N^h_{s 1}=0$ for the initial ground state.
This is consistent with what was mentioned previously: 
Upon creation onto the ground state of a $s2$ fermion 
the number $N_{a_{s 1}}^D$ of sites of the $s1$ effective lattice
remains unaltered and following the annihilation of two $s1$ fermions and
creation of one $s2$ fermion two unoccupied sites appear in it and 
hence two holes emerge in the $s1$ band. The emergence of these 
unoccupied sites and holes 
involves two virtual processes where 
(i) two $s1$ fermions are annihilated and
four independent spinons are created and (ii)
the latter independent spinons are annihilated
and the $s2$ fermion is created. 

The only net effect of creation of the vanishing-energy and zero-momentum
$s2$ fermion is the annihilation of two $s1$ fermions and corresponding emergence
of two holes in the $s1$ band and two unoccupied sites in the $s1$ effective lattice. Therefore, 
in the case of the one- and two-electron subspace one can
ignore that object in the theory provided that the corresponding changes in the
$s1$ band and $s1$ effective lattice occupancies are accounted for.
For excited states with a single $s2$ fermion two of the four 
spinons of such an object are used within neutral $s1$ fermion particle-hole
excitations in the motion of $s1$ fermions around in the $s1$ effective lattice as unoccupied sites of 
that lattice \cite{companion0}, consistently with the expression
$N^h_{s 1} = 2N_{s2}$ given in Eq. (\ref{Nas1-Nhs1}).

A similar situation occurs for the $L_s =2S_s$ independent 
spinons, whose number $L_s =[L_{s\,,-1/2}+L_{s\,,+1/2}]=2S_s$ belongs in the present subspace
to the ranges given in Eqs. (\ref{srs-0-ss}) and (\ref{deltaNcs1}) and Table \ref{table1} such that
$2S_s=0,1,2$. For $2S_s=1,2$ one has that $N_{s2} =0$ so that 
following Eq. (\ref{Nas1-Nhs1}) the number of unoccupied 
sites of the $s1$ effective lattice and of $s1$ fermion holes is for the
excited states under consideration given by $N^h_{s 1} = 2S_s=1,2$.
However, in contrast to the $s2$ fermion,
a deviation $\delta 2S_s=1,2$ generated by a state
transition may lead to deviations in the numbers
of occupied and unoccupied sites of the $s1$
effective lattice and corresponding $s1$ fermion and $s1$
fermion holes that do not obey the
usual equality $\delta N_{s1}= -\delta N^h_{s1}$.
Indeed, now $2\delta S_c=\pm 1$ for 
$\delta N^h_{s 1} = 2\delta S_s=1$ and $2\delta S_c=0,\pm 2$
for $\delta N^h_{s 1} = 2\delta S_s=2$ and 
according to the expressions provided in Eq. (\ref{Nas1-Nhs1}) 
such deviations lead to deviations in the numbers
of occupied and unoccupied sites of the $s1$
effective lattice and corresponding numbers of $s1$ fermions and
$s1$ fermion holes given by 
$\delta N_{s1} = [\delta S_c-\delta S_s]$ and
$\delta N^h_{s1} = \delta 2S_s$, respectively.
It follows that the total number of sites and thus of
discrete momentum values of the $s1$ band may also 
change. That leads to an additional deviation
$\delta N_{a_{s1}}^D = [\delta S_c +\delta S_s]$.
As given in Table \ref{table1}, for one-electron excited states one has that
$\delta N^h_{s 1} = 2\delta S_s=1$
and $2\delta S_c=\pm 1$ so that
$\delta N_{s1} = \pm 1/2-1/2=-1,0$
and $\delta N_{a_{s1}}^D = \pm 1/2 +1/2=0,-1$.
For $N_{s2}=0$ two-electron excited states one has 
$\delta N^h_{s 1} = 2\delta S_s=2$
and $2\delta S_c=0,\pm 2$ so that
$\delta N_{s1} = -1, \pm 1-1=-2,-1,0$
and $\delta N_{a_{s1}}^D = 1,\pm 1+1=0,1,2$. 

Excitations that involve changes $\delta N_{a_{s1}}^D = [\delta S_c +\delta S_s]$
in the number of sites and discrete momentum values of the $s1$ effective
lattice and $s1$ band, respectively, correspond for the $s1$ fermion operators 
$f^{\dag}_{{\vec{q}},s1}$ and $f_{{\vec{q}},s1}$ to transitions between different quantum problems. Indeed,
such operators act onto subspaces spanned by neutral states, which conserve 
$S_c$, $S_s$, and $N_{a_{s1}}^D$. In turn, the generator of non-neutral excitations
is the product of an operator that makes small changes associated
with the small variations $\delta N_{a_{s1}}^D = [\delta S_c +\delta S_s]$ both
in the $s1$ effective lattice and corresponding $s1$ momentum band with
a $s1$ fermion operator or a product of such operators appropriate to
the excited-state subspace.

Also the vanishing momentum and energy $L_{\eta,+1/2}=x\,N_a^D$ independent $+1/2$
$\eta$-spinons are invariant under the electron - rotated-electron unitary 
transformation and have no direct effect on the physics. For 
$x>0$ and the one- and two-electron subspace 
considered here they correspond to a single
occupancy configuration associated with
the $\eta$-spin vacuum $\vert 0_{\eta};N_{a_{\eta}}^D\rangle$
of Eq. (\ref{vacuum}). In turn, the degrees of freedom
of the rotated-electron occupancies of such $x\,N_a^D$ sites of the original 
lattice associated with the $U(1)$ symmetry
refer to the $N^h_c=x\,N_a^D$ unoccupied
sites of the $c$ effective lattice of Eq. (\ref{Nac-Nhc}) and
corresponding $c$ band holes.
Therefore, alike for the one- and two-electron subspace
the effects of creation and annihilation of 
a $s2$ fermion or independent 
spinons are taken into account through the
corresponding changes in the 
numbers of occupied and unoccupied sites of the $s1$ effective lattice
and corresponding $s1$ fermion and $s1$ fermion hole numbers, 
the creation and annihilation of 
independent $+1/2$ $\eta$-spinons has no effects
on the physics except that their number equals that of 
unoccupied sites of the $c$ effective lattice and $c$ band holes. 

In summary, when acting onto the one- and two-electron
subspace as defined in Subsection V-A, the Hubbard 
model refers to a two-component quantum liquid described
in terms of two types of objects on the corresponding effective lattices
and momentum bands: The charge $c$ fermions and spin-neutral two-spinon $s1$ fermions. For the model
on the square lattice that is the square-lattice quantum
liquid further studied in Refs. \cite{companion,cuprates0}.
The presence of independent $+1/2$ spinons or a composite $s2$ 
fermion is taken into account through the numbers of occupied and unoccupied sites of
the $s1$ effective lattice and corresponding $s1$ fermion and $s1$ fermion
holes, whereas the number of independent
$+1/2$ $\eta$-spinons equals that of the unoccupied 
sites of the $c$ effective lattice and $c$ band holes. Otherwise, the
presence of vanishing momentum and energy independent $+1/2$ spinons,
spin-neutral four-spinon $s2$ fermion, and independent $+1/2$ $\eta$-spinons 
has no effects on the physics. This property follows from all such objects being
invariant under the electron - rotated-electron unitary transformation.

Spin-singlet excitations generated by application onto a $m=0$ and $x\geq 0$
initial ground state of the operator $f^{\dag}_{0,s2}\,f_{{\vec{q}},s1}\,f_{{\vec{q}}\,',s1}$ where ${\vec{q}}$ and
${\vec{q}}\,'$ are the momenta of the two emerging $s1$ fermion holes are
neutral states which conserve $S_c$, $S_s$, and $N_{a_{s1}}^D$. (See Table \ref{table1}.) For
the model on the square lattice the role of the $s2$ fermion creation operator
$f^{\dag}_{0,s2}$ is exactly canceling the contributions of the
annihilation of the two $s1$ fermions of momenta ${\vec{q}}$ and ${\vec{q}}\,'$   
to the commutator $[\hat{q}_{s1\,x_1},\hat{q}_{s1\,x_2}]$ of the $s1$ translation generators
in the presence of the fictitious magnetic field ${\vec{B}}_{s1}$ of Eq. (\ref{A-j-s1-3D}),
so that the overall excitation is neutral. Since the $s2$ fermion has vanishing
energy and momentum and the $s1$ band and its number $N_{a_{s1}}^2$ of
discrete momentum values remain unaltered, one can effectively consider that the generator
of such an excitation is $f_{{\vec{q}},s1}\,f_{{\vec{q}}\,',s1}$ and omit the $s2$
fermion creation, whose only role is assuring that the overall excitation is neutral
and the two components of the $s1$ fermion microscopic momenta can be specified. It follows
that for the one- and two-electron subspace the operators $f_{{\vec{q}},s1}\,f_{{\vec{q}}\,',s1}$,
$f^{\dag}_{{\vec{q}}\,',s1}\,f^{\dag}_{{\vec{q}},s1}$, $f^{\dag}_{{\vec{q}},s1}\,f_{{\vec{q}}\,',s1}$,
and $f_{{\vec{q}},s1}\,f^{\dag}_{{\vec{q}}\,',s1}$ generate neutral excitations.
   
The quantum-liquid $c$ fermions are $\eta$-spinless and spinless
objects without internal degrees of freedom and 
structure whose effective lattice 
is identical to the original lattice. For the complete set of energy eigenstates that
span the Hilbert space the occupied sites (and unoccupied
sites) of the $c$ effective lattice correspond to those
singly occupied (and doubly occupied and unoccupied)
by the rotated electrons. The corresponding $c$ band
has the same shape and momentum area as the
first Brillouin zone.

In contrast, the quantum-liquid composite spin-neutral two-spinon $s1$ fermions 
have internal structure and the definition of the $s1$ effective lattice in terms of
both the original lattice and the spin effective lattice as well
as the spinon occupancy configurations that describe
such objects is for the one- and two-electron subspace 
a more complex problem, which deserves and requires
further studies \cite{companion0}. It is simplified by the
property that the $s1$ fermion occupancy configurations
of the states that span the one- and two-electron subspace 
refer to a $s1$ effective lattice with none, one, or two unoccupied sites. 
Also the shape of the corresponding $s1$ momentum band and
corresponding boundary and the form of the $c$ and $s1$
energy dispersions are complex unsolved problems investigated in Ref. \cite{companion}.

\section{Long-range antiferromagnetic order and short-range spin order of 
the model on the square lattice for $x=0$ and $0<x<x_*$, respectively}

Here we profit from the rotated-electron description introduced in this
paper to study the occurrence at zero temperature of a long-range antiferromagnetic 
order and a short-range spin order in the Hubbard model on the square
lattice at $x=0$ and for a well defined range of finite hole concentrations, 
respectively. For small hole concentrations $0<x\ll1$ and intermediate and
large $U/4t$ values the latter order has a spiral-incommensurate character 
\cite{Mura}.     

\subsection{Extension of the Mermin and Wagner Theorem to the
half-filling Hubbard model for $U/4t>0$}

We recall that for $U/4t\gg 1$
the spin degrees of freedom of the half-filling Hubbard model 
on a square lattice are described by a spin-$1/2$ isotropic Heinsenberg
model on a square lattice. It follows that the Mermin and 
Wagner Theorem \cite{MWT} is valid for the former
model at half filling and $U/4t\gg 1$. The theorem states 
that then there is no long-range antiferromagnetic order 
for finite temperatures $T>0$. 

Let us provide evidence that the Mermin and Wagner Theorem 
applies to the half-filling Hubbard model on a square lattice
for all values $U/4t>0$. The possibility of such an extension 
to $U/4t>0$ is strongly suggested by evidence involving
the transformation laws of the spin configurations under the 
electron - rotated-electron unitary transformation.
We recall that in terms of rotated electrons, the occupancy
configurations that generate the energy eigenstates
are the same for all finite values $U/4t>0$. Moreover,
such rotated-electron occupancy configurations equal
those that generate such states in terms of electrons 
in the $U/4t\rightarrow\infty$ limit. 

However, the electronic occupancy configurations that
in the $U/4t\rightarrow\infty$ limit generate the energy eigenstates
that span the one- and two-electron subspace
associated with our description correspond to an involved problem. Indeed,
the suitable objects whose occupancy configurations that generate
such states have a simple form, rather than the electrons, are the ``quasicharges'', spins, and pseudospins
of Ref. \cite{Ostlund-06}, which are our $c$ fermion holes,
spinons, and $\eta$-spinons, respectively, in that limit. 
For the model on the square lattice in the one- and two-electron subspace
the energy bandwidths of the $\alpha\nu$ fermion dispersions vanish for $x>0$
in the $U/4t\rightarrow\infty$ limit and the $c$ fermion dispersion has for $0<x<1$ the simple form
$\epsilon_c ({\vec{q}}) = -2t \sum_{i=1}^2\,[\cos (q_{x_i})-\cos (q_{Fcx_i})]$ in terms of the $c$ band momentum
components \cite{companion}. 
 
The use of the following two properties provides below strong evidence that 
the Mermin and Wagner Theorem holds for the half-filling Hubbard model on the
square lattice for $U/4t>0$:
I) The $x=0$ and $m=0$ absolute ground state is invariant under the electron - rotated-electron
unitary transformation \cite{companion} so that the occurrence of long-range antiferromagnetic
order for $U/4t\rightarrow\infty$, associated with that of the spin-$1/2$ isotropic Heisenberg
model, implies the occurrence of that long-range order for $U/4t>0$ as well;
II) Since in terms of rotated electrons for $U/4t>0$ single and double occupancy are good quantum
numbers, the rotated-electron occupancy configurations that generate
the energy eigenstates are more ordered than the corresponding electron occupancy
configurations. It follows that the lack of long-range antiferromagnetic
order in terms of the spins of the rotated electrons implies as well 
a lack of such an order in terms of the spins of the electrons whose
occupancy configurations generate the same states. 

We recall that the rotated-electron occupancy configurations that for $U/4t>0$ generate
the energy eigenstates are identical to the electron occupancy configurations that 
for $U/4t\rightarrow\infty$ generate the energy eigenstates belonging to the same $V$ tower.
Hence concerning the original-lattice 
rotated-electron occupancies, the Mermin and Wagner Theorem applies 
to all finite values of $U/4t>0$: for the occupancy configurations
of the rotated-electron spins there is no long-range antiferromagnetic
order for temperatures $T>0$. 
That the lack of long-range antiferromagnetic order for $U/4t>0$ 
and $T>0$ of the rotated-electron spins implies a similar
lack of such an order for the spins of the original electrons
follows from the above property II, being
consistent with for finite values of $U/4t$ the
spin occupancy configurations being more ordered 
for the rotated electrons than for the electrons. 

This is also consistent with the existence of long-range antiferromagnetic
order for the Hubbard model on a square lattice at hole concentration $x=0$, 
temperature $T=0$, and $U/4t>0$ only, associated with that of the $x=0$ and $m=0$ 
absolute ground state, which the above property I refers to,
in agreement with the numerical results of Refs. 
\cite{2D-A2,2D-NM}. As discussed below, there is strong evidence that 
for $x=0$, $T>0$, and $U/4t>0$ such an order is replaced by a short-range spin order. 

Since the electron - rotated-electrom unitary operator $\hat{V}$ does not
commute with the Hamiltonian, the rotated-electron
occupancy configurations that generate the energy
eingenstates are for $U/4t>0$ the same as for
$U/4t\rightarrow\infty$ in terms of electrons. However the corresponding
energies depend on the value of $U/4t$. It turns out that
the rotated-electron occupancy configurations associated with
the long-range antiferromagnetic order refer to a subspace
that for $U/4t$ finite exists below an energy given by
the energy scale $\mu^0$, Eq. (\ref{DMH}). Such a subspace 
is spanned by energy eigenstates with $N_{a_{\eta}}^2=[N_a^2-2S_c]=0$ 
and $N_{a_{s}}^2=2S_c=N_a^2$ sites so 
that there is no $\eta$-spin effective lattice and
the spin effective lattice has as many sites as the original lattice and is identical
to it. In such a subspace there are no rotated-electron 
doubly occupied sites and no rotated-electron unoccupied sites. 

The point is that the number of sites of the spin effective 
lattice being given by $N_{a_{s}}^2=N_a^2$ is a necessary condition 
for the occurrence of long-range antiferromagnetic order:
Concerning the spins of the rotated-electron occupancy
configurations such an order occurs when the $\eta$-spin effective lattice
does not exist and the quantum-system vacuum
(\ref{vacuum}) becomes
$\vert 0_{s}\rangle = \vert 0_{s};N_{a}^D\rangle
\times\vert GS_c;N_{a}^D\rangle$ and the system
global symmetry is $U(2)/Z_2=SO(3)\times U(1)$.
For $U/4t\rightarrow\infty$ when the spin degrees of freedom
of the half-filling Hubbard model are described by the Heisenberg model,
the subspace for which $N_{a_{s}}^2=N_a^2$ describes 
the whole finite-energy physics. This follows from according to Eq. (\ref{DMH}) 
the energy scale $\mu^0$ being given by $\mu^0 = U/2\rightarrow\infty$.  
However, for finite $U/4t>0$ one has that the spin occupancy 
configurations in a spin effective lattice with $N_{a_{s}}^2=N_a^2$
sites, which are expected to be behind the long-range antiferromagnetic order, 
exist only for energies below $\mu^0$. The Mott-Hubbard
gap $2\mu^0$, which refers to the charge degrees of freedom, 
affects the spin degrees of freedom as well: $\mu^0$ is the energy above
which the $\eta$-spin effective lattice emerges and hence the 
number of sites of the spin effective lattice decreases so that  
$N_{a_{s}}^2<N_a^2$. It follows that for the Hubbard model 
on a square lattice $\mu^0$ is the energy below which the long-range 
antiferromagnetic order of the rotated-electron
spins survives for $x=0$, $m=0$, and temperature $T=0$. 

Finally, it follows from the above property II that
since the ground-state rotated-electron occupancy configurations
are for finite values of $U/4t$ more ordered than those of
the electrons for the same state, a lack of long-range antiferromagnetic
order of the rotated-electron spins for
$N_{a_{s}}^2< N_a^2$ implies a similar lack of 
long-range antiferromagnetic order for the spins of the
original electrons. In contrast to the $x=0$ and $m=0$ absolute
ground state of property I, $0<x<1$ and $m=0$ 
ground states are not invariant under the electron - rotated-electron
unitary transformation.

\subsection{Short-range spin order of the model on the square lattice
for $m=0$ and $0<x\ll 1$}

The requirement for the occurrence of long-range antiferromagnetic 
order at zero temperature $T=0$ in the Hubbard model
on the square lattice that the number of sites of the $\eta$-spin and spin effective 
lattices are given by $N_{a_{\eta}}^2=0$ and $N_{a_{s}}^2=N_a^2$,
respectively, implies that there is no such a long-range
order for $m=0$, $x>0$, and $T=0$. Indeed, then 
$2S_c<N_a^2$ so that $N_{a_{\eta}}^2=[N_a^2-2S_c]>0$ and 
$N_{a_{s}}^2=2S_c<N_a^2$. For the one- and two-electron subspace 
such expressions read $N_{a_{\eta}}^2=x\,N_a^2>0$ and 
$N_{a_{s}}^2=(1-x)\,N_a^2<N_a^2$, respectively. 

One may use as zero-temperature order parameter of the $m=0$ and $x=0$ quantum phase 
the excitation energy $\mu^0$ below which the long-range antiferromagnetic order survives, whose
limiting behaviors for $U/4t\ll 1$ and $U/4t\gg 1$ are given in Eq. (\ref{DMH}) 
for $D=2$. In addition to those, here we provide an approximate formula valid for 
intermediate $U/4t$ values $u_0\leq U/4t\leq u_1$,
\begin{eqnarray}
\mu^0 & \approx & 32\,t\,e^{-\pi\sqrt{4t\over U}} \, ,
\hspace{0.25cm} U/4t\ll 1 \, ,
\nonumber \\
& \approx & {2e^1\,t\over \pi}\sqrt{1+(U/4t -u_0)}  \, ,
\hspace{0.25cm} u_0 \leq U/4t\leq u_1 \, ,
\nonumber \\
& \approx  & [U/2 - 4t] \, ;
\hspace{0.25cm}  U/4t\gg 1 \, ,
\label{Delta-0-s}
\end{eqnarray}
where the behavior $\mu_0 (U/4t)\approx \mu_0 (u_0)\sqrt{1+(U/4t -u_0)}$ is expected to be a good approximation for the 
small range $u_0 \leq U/4t\leq u_1$, $u_0\approx 1.302$, and $u_1\approx 1.600$. The approximate magnitude 
$\mu_0 (u_0)\approx [2e^1/\pi]\,t$ is that consistent with the relation $\mu_0 (u_0)\approx \mu_0 (u_*)/\sqrt{1+(u_* -u_0)}$
where $u_0=1.302$ is the $U/4t$ value at which the energy parameter $\Delta_0$ is found below to reach its maximum 
magnitude and $u_* = 1.525$ is that for which the studies of Ref. \cite{companion} lead by a completely different method to 
$\mu_0 (u_*)\approx 566$\,meV for $t=295$\,meV and $U/4t\approx u_* =1.525$. The use of $\mu_0 (u_0)\approx [2e^1/\pi]\,t$ 
in the formula $\mu_0 (U/4t)\approx \mu_0 (u_0)\sqrt{1+(U/4t -u_0)}$ leads for $t=295$\,meV, $U/4t\approx u_* =1.525$,
and $u_0=1.302$ to nearly the same magnitude, $\mu_0 (u_*)\approx 565$\,meV. 

There is strong evidence of the occurrence in the half-filling Hubbard model on the square lattice
of strong short-range antiferromagnetic correlations for $T>0$ and below a crossover temperature called 
$T_x$ in Ref. \cite{Hubbard-T*-x=0}, which here we denote by $T_0^*$. This is consistent with then the system 
being driven into a phase with short-range spin order. Furthermore, that the occurrence of long-range 
antiferromagnetic order requires $T=0$, $N_{a_{\eta}}^2=0$, and $N_{a_{s}}^2=N_a^2$ is
consistent with the short-range spin order occurring for $m=0$, $0<x\ll 1$, and $0\leq T< T_0^*$ 
being similar to that occurring for $m=0$, $x=0$, and $0<T< T_0^*$ and was studied previously
in Ref. \cite{Hubbard-T*-x=0} for $0<T\ll T_0^*$. 

One may choose the excitation energy $2\Delta_0$ below which the short-range spin 
order with strong antiferromagnetic correlations survives to play the role of zero-temperature
order parameter of the quantum phase displaying such an order
occurring for $m=0$ and $0<x\ll 1$. That energy parameter has a $U/4t$ dependence qualitatively similar to that of the 
energy scale $2k_B\,T_0^*$, with the equality $2\Delta_0\approx 2k_B\,T_0^*$ holding approximately.
For both vanishing and finite temperatures a phase displaying a short-range spiral-incommensurate spin 
order occurs for (i) $m=0$, $0<x\ll 1$, and $0\leq T\ll T_0^*$
and (ii) $m=0$, $x=0$, and $0<T\ll T_0^*$. At $m=0$ and temperatures below $T_0^*$ the system is driven both for 
(i) $0<x\ll 1$ and $0\leq T<T_0^*$ and (ii) $x=0$ and $0<T<T_0^*$ into a renormalized classical regime where the 
$x=0$ and $T=0$ long-range antiferromagnetic order is replaced by such a short-range spin order, which
is a quasi-long-range spin order as that studied in Ref. \cite{Principles} for simpler spin systems. 

Let us study the $U/4t$ dependence of $\Delta_0\approx k_B\,T_0^*$ by combining 
the results obtained from the use of the general description introduced in this paper with
those of the low-temperature approach to the half-filled model 
of Ref. \cite{Hubbard-T*-x=0}, whose investigations focus on temperatures $0<T\ll T_0^*$. 
We use often in the following the magnitudes of the energy bandwidth $W_{s1}^0$ and energy parameter $\mu_0$ derived in 
Ref. \cite{companion} For the value $U/4t\approx 1.525$ found in Ref. \cite{cuprates} to be appropriate to the description of the
properties of several hole-doped cuprates on combining the description introduced in this paper
with results on the spin-wave dispersion of the half-filled Hubbard model in a 
spin-density-wave-broken symmetry ground state obtained by summing up an infinite set of ladder 
diagrams \cite{LCO-Hubbard-NuMi}.

The picture that emerges is that for the Hubbard model on a square 
lattice at $T=0$ there is an overall zero-temperature energy order parameter
$2\vert\Delta\vert$, which at $x=0$ has a singular behavior due to
a sharp quantum phase transition occurring at vanishing hole concentration, such that
there is long-range antiferromagnetic order at $x=0$ and, as justified below by
combining our description with the results of Refs. \cite{companion,duality,Mura}, a short-range 
spiral-incommensurate spin order for $0<x\ll 1$. Consistently, the energy scale 
$2\vert\Delta\vert$ reads,
\begin{equation}
2\vert\Delta\vert  = \mu^0
\, , \hspace{0.25cm} x = 0 \, ; \hspace{0.5cm}
2\vert\Delta\vert  \approx 2\Delta_0\left(1-{x\over x_*^0}\right)
\, , \hspace{0.25cm} 0<x\ll 1 \, , \hspace{0.25cm} U/4t\geq u_0 \approx 1.302 \, .
\label{Delta}
\end{equation}
The linear dependence of $2\vert\Delta\vert \approx 2\Delta_0 (1-x/x_*^0)$ on $x$ 
where $x_*^0\approx 2r_s/\pi$ and $r_s =\Delta_0/4W_{s1}^0$ is justified in Ref. \cite{companion}.  
Hence $2\vert\Delta\vert \vert_{x=0}= \mu^0$ is for $U/4t>0$ different from 
$2\Delta_0=\lim_{x\rightarrow 0}2\vert\Delta\vert$. Symmetry arguments imply that 
$\lim_{U/4t\rightarrow 0}2\Delta_0=2\vert\Delta\vert \vert_{x=0}= \mu^0$ so that
the two energy scales become the same in that limit.

The energy scale $\Delta_0$ is found in Refs. \cite{companion,cuprates0}
to play an important role in the square-lattice quantum-liquid physics.
Upon increasing $U/4t$, it interpolates between $\Delta_0\approx \mu^0/2\approx 16t\,e^{-\pi\sqrt{4t/U}}$ for
$U/4t\ll 1$ and $\Delta_0\approx 4W^0_{s1}\approx \pi\,[4t]^2/U$ for $U/4t\gg 19$,
going through a maximum magnitude at a $U/4t$ value found below to be approximately given by $U/4t=u_0\approx 1.302$.
The energy parameter $\mu_0/2$ is an increasing function of $U/4t$.
As given in Eq. (\ref{Delta-0-s}), it behaves as $\mu_0/2\approx 16\,t\,e^{-\pi\sqrt{4t\over U}}$
for $U/4t\ll 1$ and as $\mu_0/2\approx  [U/2 - 4t]$ for $U/4t\gg 1$.
In turn, the energy scale $4W_{s1}^0$ is a decreasing function $U/4t$, which according to 
Eq. (\ref{W-s-0}) is for $D=2$ given by $4W_{s1}^0 = 16t$ at $U/4t=0$ and decreases approximately as 
$4W_{s1}^0 \approx \pi\,[4t]^2/U$ for very large $U/4t$. The energy scale $\Delta_0$ interpolates between
these two behaviors, vanishing both for $U/4t\rightarrow 0$ and $U/4t\rightarrow\infty$
as $\Delta_0\approx 16t\,e^{-\pi\sqrt{4t/U}}$ and $\Delta_0 \approx \pi (4t)^2/U$ where
it becomes $\mu_0/2$ and an energy scale $\pi (4t)^2/U$ 
associated with the strong $0<x\ll 1$ antiferromagnetic correlations, respectively.
This is consistent with its maximum magnitude being reached at the intermediate $U/4t$ value
$U/4t=u_0$ for which $\mu_0/2\approx 4W_{s1}^0$ and we consider that such a $U/4t$ value is that
at which the equality $\mu_0 (u_0)/2= 4W_{s1}^0 (u_0)$ holds.
 
The ratio $r_s =\Delta_0/4W_{s1}^0$ is parametrized in the following as $r_s=e^{-\lambda_s}$ 
where $\lambda_s = \vert\ln (\Delta_0/4W_{s1}^0)\vert$ has an important role in such an interpolation 
behavior. The energy scale $\Delta_0$ can then be expressed as,
\begin{equation}
\Delta_0 = r_s\,4W_{s1}^0 = 4W_{s1}^0\,e^{-\lambda_s} 
\, ; \hspace{0.35cm} 
\lambda_s = \vert\ln (\Delta_0/4W_{s1}^0)\vert \, ,
\label{Delta-0-gen}
\end{equation}
where $\lambda_s$ has the limiting behaviors,
\begin{eqnarray}
\lambda_s & = & \pi\sqrt{4t/U} \, , \hspace{0.25cm} U/4t \ll 1 
\, ; \hspace{0.50cm}
\lambda_s \approx 4t\,u_0/U \, , \hspace{0.25cm} 
u_{00} \leq U/4t \leq u_1 \, ; \hspace{0.50cm}
\lambda_s = 0 \, , \hspace{0.25cm} U/4t \rightarrow\infty \, ,
\nonumber \\
u_{00} & \approx & (u_0/\pi )^2 \approx 0.171  \, ; \hspace{0.35cm} 
u_0 \approx 1.302 \, ; \hspace{0.35cm} u_1 \approx 1.600 \, .
\label{lambda-s}
\end{eqnarray}
Consistently, the ratio $r_s$ is an increasing function
of $U/4t$, which changes continuously from $r_s=0$ for $U/4t\rightarrow 0$
to $r_s=1$ for $U/4t\rightarrow\infty$. That for $u_{00} \leq U/4t \leq u_1$ it is approximately given by
$r_s\approx e^{-4t\,u_0/U}$ instead of $r_s\approx e^{-\pi\sqrt{4t/U}}$ for $U/4t\ll 1$ 
is consistent with for large $U/4t$ it being such that $(1-r_s)\propto 4t/U$ rather than
$(1-r_s)\propto \sqrt{4t/U}$.

The temperature $T_x$ which plays the role of $T_0^*\approx \Delta_0/k_B$ is plotted in Fig. 3 of
Ref. \cite{Hubbard-T*-x=0}. Its $U/4t$ dependence 
is qualitatively correct, $T_x$ vanishing both
in the limits $U/4t\rightarrow 0$ and $U/4t\rightarrow\infty$
and going through a maximum magnitude at an intermediate
value $5/4<U/4t<3/2$. Nevertheless, the interpolation function used to produce
it, provided in Ref. 74 of such a paper, is poor for intermediate
values of $U/4t$. However, that does not affect the validity of the results of
Ref. \cite{Hubbard-T*-x=0}. Indeed, the studies of that reference
refer to the temperature range $0<T\ll T_x$ for which the accurate dependence
of $T_x$ on $U/4t$ is not needed and the goal of its Fig. 3 is just illustrating
qualitatively the $T_x$ behavior over the entire coupling range
\cite{Nicolas}. 

Our goal is reaching a more quantitatively accurate $U/4t$ dependence of the
energy parameter $\Delta_0\approx k_B\,T_0^*$ for intermediate $U/4t$
values, consistent with the qualitative physical picture of Ref. \cite{Hubbard-T*-x=0}. 
That according to Eqs. (\ref{Delta-0-gen}) and (\ref{lambda-s}) one has that
$\Delta_0=4W_{s1}^0\,e^{-\lambda_s}$ for $U/4t>0$ where the parameter $\lambda_s$ 
is a continuous decreasing function of $U/4t$ given by
$\lambda_s =\infty$ for $U/4t\rightarrow 0$ and $\lambda_s =0$ for $U/4t\rightarrow\infty$
reveals that the magnitude $\lambda_s=1$ separates two physical regimes. Consistently,
$\lambda_s=1$ refers to the $U/4t$ value $U/4t=u_0$ at which
$\Delta_0$ reaches its maximum magnitude ${\rm max}\{\Delta_0\}= 
[\mu_0 (u_0)/2]\,e^{-1} =4W_{s1}^0(u_0)\,e^{-1}$

That as found in Ref. \cite{companion}, both the parameter $x_*^0$ is for
approximately $u_0\leq U/4t\leq u_1$ 
given by $x_*^0=2r_s/\pi=2e^{-4t\,u_0/U}/\pi$ and at $U/4t\approx u_*= 1.525$
reads $x_*^0\approx 0.27$ implies that $u_0\approx 1.3$. Moreover,
in that reference it is found that $4W_{s1}^0 (u_*)=[198.4/295]\,t\approx 0.673\,t$
at $U/4t\approx u_*= 1.525$. Since $W_{s1}^0 =\lim_{x\rightarrow}W_{s1}$ 
has the same magnitude as the $x=0$ energy bandwidth $W_{s1}$, such a result is obtained
in that reference from comparison of the $x=0$ and $m=0$ spin excitation spectra for the 
high symmetry directions found by use of the description introduced in this paper with those
estimated by the controlled approximation of Ref. \cite{LCO-Hubbard-NuMi}.
On combining the equality $\mu_0 (u_0)/2=4W_{s1}^0(u_0)$
with the magnitude $\mu_0 (u_0)\approx [2e^1/\pi]\,t$ of Eq. (\ref{Delta-0-s}) 
we find $4W_{s1}^0(u_0)\approx [e^1/\pi]\,t\approx 1.731\,t$ and
${\rm max}\{\Delta_0\}= [\mu_0 (u_0)/2]\,e^{-1} =4W_{s1}^0(u_0)\,e^{-1}\approx t/\pi$.

Note that the magnitude $W_{s1}^0(u_*)=[49.6/295]\,t\approx 0.168\,t$ 
obtained in Ref. \cite{companion} is about $12.25$ times smaller than that
found by use of the limiting expression $4\pi\,t^2/U\approx 2.060\,t$ at
$U/4t=1.525$. This reveals that for the Hubbard model on the square lattice
such a limiting expression is valid for a smaller range
of very large $U/4t$ values than for 1D. Specifically,
we find that it is valid for $U/4t\gg 19$. The
value $U/4t\approx 18.70\approx 19$ is that at which $4\pi\,t^2/U$
is given by $0.168\,t$, alike $W^0_{s1}$ at $U/4t\approx 1.525$.
Therefore, the relation $W^0_{s1} = J \approx 4\pi\,t^2/U$
is indeed valid for $U/4t\gg 19$ and for the model on the square lattice the energy scale 
$J \approx 4\pi\,t^2/U$ controls the physics for a smaller $U/4t$ range than
in 1D, which corresponds to very large $U/4t\gg 19$ values. 
Hence the intermediate-$U/4t$ range plays a major role in the
physics of that model, as confirmed by the related studies of 
Refs. \cite{companion,cuprates0,cuprates}.

For the model on the square lattice and approximately the
intermediate range $u_0\leq U/4t\leq u_1$ 
the $U/4t$ dependence of the energy parameter $4W_{s1}^0$ is of the
form $4W_{s1}^0\approx [e^1/\pi]\,t\,W(U/4t)$ where $W(u)$ is an unknown 
function of $u=U/4t$ such that $W (u_0)=1$. Since on combining Eqs. 
(\ref{Delta-0-gen}) and (\ref{lambda-s}) one finds $\Delta_0 = 4W_{s1}^0\,e^{-4t\,u_0/U}$, 
fulfillment of the maximum condition $\partial \Delta_0 (u)/\partial u =0$ at $u=u/4t=u_0$ requires
the function $W( u)$ be given by $W (u)\approx (2-u/u_0)$ for 
$0\leq [(u-u_0)/(u_1-u_0)]\ll 1$. In turn, the use of the above results 
$4W_{s1}^0(u_*)=[198.4/295]\,t\approx 0.673\,t$ and 
$4W_{s1}^0(u_0)\approx [e^1/\pi]\,t\approx 1.731\,t$ reveals that the
ratio $4W_{s1}^0(u_*)/4W_{s1}^0(u_0)$ may be expressed as
$4W_{s1}^0(u_*)/4W_{s1}^0(u_0)\approx [1-(u_*-u_0)]$ so that
$W (u)\approx [1-(u-u_0)]$ for $u\approx u_* =1.525$. We then
use a suitable interpolation function for $W (u)$, which has these two
limiting behaviors, so that the energy parameter $4W_{s1}^0$ is
for $U/4t$ intermediate values $U/4t\in (u_0,u_1)$ given approximately by, 
\begin{eqnarray}
4W_{s1}^0 & \approx & {e^1\,t\over \pi}\,W(U/4t) \, ;
\hspace{0.35cm} W (u) \approx 1 - (u-u_0)\,e^{-{u_*-u\over u_*-u_0}\ln (u_0)} \, ,
\hspace{0.25cm} u_0\leq U/4t\leq u_1 \, ,
\nonumber \\
u_0 & \approx & u_* -1 + e^{-1} \pi {198.4\over 295} \approx 1.302 \, ;
\hspace{0.35cm} u_* \approx 1.525 \, .
\label{W-s1-0-u0}
\end{eqnarray}
The value $u_0\approx 1.302$ is that obtained from the equation
$4W_{s1}^0 (u_*) = [e^1\,t/\pi]\,[1-(u_*-u_0)]= [198.4/295]\,t \approx 0.673\,t$.

On combining the above results we find the following magnitudes for the energy parameter $\Delta_0$,
\begin{eqnarray}
\Delta_0 & \approx & 16t\,e^{-\pi\sqrt{4t/U}} \, ,
\hspace{0.25cm}  U/4t\ll 1 \, ,
\nonumber \\
& = & {\rm max}\,\{\Delta_0\} \approx t/\pi \, ,
\hspace{0.25cm}  U/4t = u_0 \, ,
\nonumber \\
& \approx & e^{(1-4t\,u_0/U)}[t/\pi]\,W(U/4t) \, ,
\hspace{0.25cm} u_0\leq U/4t \leq u_1 \, ,
\nonumber \\
& = & 4W^0_{s1} \approx \pi\,[4t]^2/U \, ,
\hspace{0.25cm} U/4t\gg 19 \, ,
\label{Delta-0}
\end{eqnarray}
where $W(U/4t)$ is the interpolation function given in Eq. (\ref{W-s1-0-u0}). Note that
$\partial \Delta_0 (u)/\partial u =0$ at $u=u/4t=u_0$, consistently with
$\Delta_0$ reaching its maximum magnitude $t/\pi$ at that $U/4t$ value.
The overall $U/t$ dependence of $T_0^*\approx \Delta_0/k_B$ is
similar to that plotted in Fig. 3 of Ref. \cite{Hubbard-T*-x=0} for $T_x$
with the $U/t$ value at which the maximum magnitude is reached shifted from $U/t\approx 5.60$
to $U/t\approx 5.21$ and that magnitude lessened from ${\rm max}\,\{k_B\,T_x\} \approx 0.625\,t$
to ${\rm max}\,\{k_B\,T_0^*\}\approx {\rm max}\,\{\Delta_0\}\approx t/\pi\approx 0.318\,t$.
Otherwise, such a dependence has qualitatively a shape similar to that illustrated in the figure.
 
Next let us provide strong evidence that for intermediate and large values of $U/4t$ 
and small hole concentrations $0<x\ll1$ the short-range spin order of the $m=0$ 
ground state corresponds to that of a spin-singlet incommensurate 
spiral state. In terms of the rotated-electron spins occupancy
configurations the ground state is then a spin-singlet incommensurate 
spiral state for $U/4t>0$, $m=0$, and $0<x\ll1$. That evidence is
found on combining the above result that such a ground state
is a spin-singlet state with the recent results of Ref. \cite{Mura}
concerning the spin degrees of freedom of a related quantum
problem. 

As discussed in Section II, for intermediate and large values of $U/4t$ 
the Hubbard model on the square lattice given in Eqs. (\ref{H}) and (\ref{HHr}) in terms of
electron and rotated-electron operators, respectively, can for $D=2$ be mapped 
onto an effective $t-J$ model on a square lattice with $t$, $t'=t'(U/4t)$, 
and $t''=t''(U/4t)$ transfer integrals
where the role of the processes associated with 
$t'=t'(U/4t)$ and $t''=t''(U/4t)$ becomes increasingly important
upon decreasing the $U/4t$ value.
Rigorous results on the spin degrees of freedom of the 
$t-J$ model on a square lattice with $t$, $t'$, and $t''$ transfer 
integrals were recently achieved in Ref. \cite{Mura}. The investigations
of that paper refer to small values of the hole concentration $0<x\ll1$ and spin 
density $m=0$ and have as starting point a suitable action first introduced in Ref. \cite{Wieg}. 
The use in Ref. \cite{Mura} of a staggered CP$^1$ representation for the spin degrees of freedom
allows to resolve exactly the constraint against double occupancy,
which is equivalent to performing the electron - rotated-electron
unitary transformation. In order to achieve the rigorous result 
that for small hole concentrations there occurs a incommensurate-spiral 
spin order, the effective action for the spin degrees of freedom
is reached after integrating out the charge fermionic degrees of
freedom and the magnetic fast CP$^1$ modes. Importantly, the
dependence on the hole-concentration of the coupling constants of the effective
field theory is obtained explicitly for small $x$.  

From the mapping between the above effective $t-J$ model
and the Hubbard model of Eqs. (\ref{H}) and (\ref{HHr})
in the subspace with vanishing rotated-electron double-occupancy, the studies of Ref. \cite{Mura} 
reveal that for intermediate and large values of $U/4t$, spin-density $m=0$, and small
hole concentrations $0<x\ll1$ the ground state is an incommensurate spiral 
state. This is a rigorous result, yet the studies of Ref. \cite{Mura} are not 
conclusive on whether for $0<x\ll1$
the $m=0$ incommensurate spiral ground state has short-range
or long-range spin order. However, symmetry requires that the
occurrence of long-range and short-range incommensurate-spiral spin 
order is associated with a spin-singlet ground state and 
a broken-symmetry ground state without well defined spin, 
respectively. Therefore, combination of our above result that 
the $m=0$ and $0<x\ll1$ ground state of the Hubbard model 
(\ref{H}) is a spin-singlet state with the 
result of Ref. \cite{Mura} that such a state has a short-range
incommensurate-spiral spin order is consistent with for intermediate and large values of $U/4t$ the ground state
of the model being a spin-singlet state with short-range incommensurate-spiral spin 
order and thus strong antiferromagnetic correlations. 

Moreover, in terms of rotated electrons the short-range incommensurate-spiral spin order 
prevails for $U/4t>0$, $m=0$, and $0<x\ll1$ and implies
that in terms of electrons the system has a short-range spin order.
Indeed and as discussed above, the ground-state rotated-electron occupancy configurations
are for finite values of $U/4t$ more ordered than those of
the electrons for the same state so that  a lack of long-range spin
order of the rotated-electron spins for $m=0$ and $0<x\ll1$ implies a similar lack of 
long-range spin order for the spins of the original electrons.
The occurrence of a short-range incommensurate-spiral spin order for
$0<x\ll1$ is fully consistent with the occurrence for $U/4t>0$ of a 
long-range antiferromagnetic order and a short-range spin order for $x=0$ and 
$0<x\ll1$, respectively. Above the occurrence of such long-range and short-range
spin orders is associated with the number of sites of the
spin effective lattice obeying the relations $N_{a_{s}}^2= N_a^2$ and 
$N_{a_{s}}^2< N_a^2$, respectively. In turn, here the occurrence for $0<x\ll1$ of a short-range 
spiral-incommensurate spin order is reached by use of completely different arguments.

The energy order parameter $2\vert\Delta\vert$ of Eq. (\ref{Delta})
is both for $x=0$ and $x>0$ identified in the studies of Refs. \cite{companion,cuprates0} 
with the maximum pairing energy of the $-1/2$ and $+1/2$ spinons
of a composite spin-neutral two-spinon $s1$ fermion. 
Its magnitude has a singular behavior at $x=0$ due to the sharp quantum phase transition
from a ground state with long-range 
antiferromagnetic order at $x=0$ to a state with short-range spiral-incommensurate spin order 
and strong antiferromagnetic correlations for $0<x\ll 1$. 
The results of Refs. \cite{companion,cuprates0} extend the $m=0$ and $T=0$
short-range spin order to a well-defined range of hole 
concentrations $0<x<x_*$. According to these results, for $U/4t\geq u_0$ the $x$ dependence 
$2\vert\Delta\vert \approx 2\Delta_0 (1-x/x_*^0)$ given in Eq. (\ref{Delta}) for $0<x\ll1$ is valid for
$x\in (0,x_*)$ whereas for $x>x_*$ the energy scale $2\vert\Delta\vert$ vanishes. Here for approximately
$U/4t\in (u_0,u_{\pi})$ where $u_{\pi}>u_1$ is the $U/4t$ value at which $r_s=1/2$ the critical hole concentration
$x_*$ equals the $U/4t$-dependent parameter $x_* ^0$. For $U/4t\geq u_0$ the studies 
of the above references identify it with a critical 
hole concentration above which there is no short-range spin
order so that $2\vert\Delta\vert \rightarrow 0$ as $x\rightarrow x_*$. Indeed, 
the energy order parameter $2\vert\Delta\vert$ vanishes in the absence of that order. 
Consistently, the spinon pairing energy vanishes for the 
Hubbard model on the 1D lattice for which $2\vert\Delta\vert =0$ for the
whole range of hole concentrations. The short-range incommensurate-spiral spin order discussed here
for $0<x\ll1$ corresponds then to a limiting case of the general short-range 
spin order that according to the investigations of Refs. \cite{companion,cuprates0}
occurs for $0<x<x_*$. 

The form of the $s1$ effective lattice spacing (\ref{a-a-s1-sube})
is for the square-lattice model directly related to the above spin orders. Indeed,
that at $x=0$ and $m=0$ the square $s1$ effective lattice has 
no unoccupied sites and its spacing is given by $a_{s1}=\sqrt{2}\,a$
reveals that then its periodicity has increased relative to that of
both the original lattice and $c$ effective lattice. 
This is because at $x=0$ the $s1$ effective lattice refers to a $\sqrt{2}\times\sqrt{2}$
reconstruction in which the periodicity of the spin-sub-system
real-space structure is increased. Such an effect
is consistent with the occurrence of the long-range antiferromagnetic
order at $x=0$ and $m=0$. 
In turn, that for $x>0$ and $m=0$ the square $s1$ effective lattice remains having 
no unoccupied sites but its spacing reads instead $a_{s1}=\sqrt{2/(1-x)}\,a$ 
is consistent with the emergence of the short-range incommensurate-spiral spin
order discussed here. Furthermore, that then the $s1$ effective-lattice spacing 
is given by $a_{s1}=\sqrt{2/(1-x)}\,a$ 
implies that the $s1$ fermion occupancy-configuration states break the 
translational symmetry. 

The studies of Refs. \cite{companion,cuprates0}
reveal that for the square-lattice quantum liquid the $s1$ bond particles associated 
with the $s1$ fermions are 
closely related to the spin-singlet bonds of Ref. \cite{duality}. 
The investigations of Ref. \cite{cuprates0} combine the description introduced
in this paper with some of the results of Ref. \cite{duality} on
a related problem to investigate the square-lattice quantum-liquid 
fluctuations and competing orders in the presence of a small 
3D anisotropy perturbation associated with weak plane coupling. 
They confirm the validity of the long-range and short-range spin orders found here at $x=0$
and $x>0$, respectively, by studying the fluctuations of the phases
that control such orders.

At $x=0$ and $m=0$ the spin effective lattice is identical to the
original lattice so that the two-spinon $s1$ bonds of the $s1$ fermions refer to the same lattice
as in Ref. \cite{duality} and the $s1$ effective lattice is one
of its two sub-lattices. Also for $0<x\ll1$ the spin effective
lattice is very similar to the original lattice and the $s1$ effective
lattice spacing $a_{s1}\approx [\sqrt{2}/(1+x/2)]\,a\approx \sqrt{2}\,a$ 
is close to that of the original sub-lattices. As further discussed in
Ref. \cite{cuprates0}, at zero temperature, $x=0$, and $m=0$ 
there are in the the square-lattice quantum liquid strong phase
fluctuations whose action has a local compact gauge symmetry.
That corresponds to the long-range antiferromagnetic phase studied
above for which monopole-antimonopole pairs of the type considered
in Ref. \cite{duality} are unbound and proliferate. 
For small values $0<x\ll1$ the motion of the $c$ fermions and the associated kinetic energy
play the role of symmetry-breaking Higgs terms. The presence
of such terms supresses free monopoles and is behind
the replacement of the long-range antiferromagnetic order
at $x=0$ by the short-range incommensurate-spiral spin
order for $0<x\ll1$.

\section{Concluding remarks}

In this paper we have profited from the interplay of the global 
$SO(3)\times SO(3)\times U(1)$ symmetry found in Ref.
\cite{bipartite} for the Hubbard model 
on any bipartite lattice with the transformation laws of
a well-defined set of operators and quantum objects
whose occupancy configurations generate a complete set
of $S_{\eta}$, $S_{\eta}^z$, $S_s$, $S_s^z$, $S_c$, and momentum
eigenstates under a suitably chosen electron -
rotated-electron unitary transformation to introduce a 
general description for the model on a square lattice
with $N_a^2\gg 1$ sites. Often the problem is also addressed for the model
on the 1D lattice.

Within the description introduced in this paper the above complete set of states 
is generated by occupancy configurations of $c$ fermions associated with the state representations 
of the $U(1)$ symmetry, $\eta$-spin-neutral $2\nu$-$\eta$-spinon 
composite $\eta\nu$ bond particles and independent $\eta$-spinons associated
with the state representations of the $\eta$-spin $SU(2)$ 
symmetry, and spin-neutral $2\nu$-spinon composite $s\nu$ 
bond particles and independent spinons associated
with the state representations of the spin $SU(2)$ 
symmetry. The index $\nu=1,2,...$ refers to the number of
$\eta$-spinon or spinon pairs in each composite object.
Evidence is provided in this paper that in addition to being $S_{\eta}$, $S_{\eta}^z$, $S_s$, $S_s^z$, $S_c$, and momentum
eigenstates, for the Hubbard model on the square lattice in the 
one- and two-electron subspace defined in this paper such states
are energy eigenstates.
(For the 1D model they are energy eigenstates for the whole
Hilbert space \cite{companion}.) 

The relation of the composite $\eta\nu$ bond 
particles (and $s\nu$ bond particles) to the $\eta$-spinons 
(and spinons) has similarities with that of the composite 
physical particles to the quarks in chromodynamics \cite{Martinus}. Within 
the latter theory the quarks have color but all quark-composite physical 
particles are color-neutral. Here the $\eta$-spinon (and spinons) that
are not invariant under the electron - rotated-electron unitary
transformation have $\eta$-spin $1/2$ (and spin $1/2$) but 
the $2\nu$-$\eta$-spinon (and $2\nu$-spinon) 
composite $\eta\nu$ bond particles (and $s\nu$ bond particles) are 
$\eta$-spin-neutral (and spin-neutral) objects. 
The components of the microscopic momentum values of the $\alpha\nu$ fermions
are eigenvalues of $\alpha\nu$ translation generators ${\hat{{\vec{q}}}}_{\alpha\nu}$ of Eq. (\ref{m-generators})
in the presence of the fictitious magnetic fields ${\vec{B}}_{\alpha\nu} ({\vec{r}}_j)$ of Eq. (\ref{A-j-s1-3D}). Those are associated with
extended Jordan-Wigner transformations from which the $\alpha\nu$ fermions
emerge from the corresponding $\alpha\nu$ bond particles. In the $U/4t\rightarrow\infty$ limit
the $\alpha\nu$ band momentum occupancy configurations that generate the energy
eigenstates of Hubbard model on a square lattice in subspaces (A) and (B)
defined in this paper are described by the 2D QHE physics. The structure of the spin 
configurations associated with a $s1$ fermion simplifies in the one- and 
two-electron subspace introduced in this paper
where only the charge $c$ fermions and spin-neutral two-spinon 
$s1$ fermions play an active role. The physical
picture that emerges is that of a two-component
quantum liquid of charge $c$ fermions and spin-neutral two-spinon $s1$ fermions
whose momentum values are good quantum numbers. 
For the Hubbard model on a square lattice in the one- and two-electron subspace 
the composite $s1$ fermion consists of two spinons in a spin-singlet configuration 
plus an infinitely thin flux tube attached to it. Thus, each $s1$ fermion appears to 
carry a fictitious magnetic solenoid with it as it moves around in the $s1$ effective lattice.
The square-lattice quantum liquid is further
investigated in Refs. \cite{companion,cuprates0}.
 
Concerning previous studies on the large-$U$ Hubbard
model and $t-J$ model on a square lattice involving for instance
the slave particle formalism \cite{2D-MIT,Fazekas,Xiao-Gang} or Jordan-Wigner 
transformations \cite{Feng}, the crucial requirement is to impose
the single occupancy constraint. Here that constraint is 
naturally implemented for all values of $U/4t>0$, since the
spins associated with the spin $SU(2)$ state representations
refer to the rotated electrons of the singly occupied sites.
Moreover, for the above schemes the spinless fermions arise from 
individual spins or spinons. In contrast, within the extended
Jordan-Wigner transformation considered in this paper and Ref. \cite{companion}
the $s1$ fermions emerge from spin-neutral two-spinon composite 
$s1$ bond particles. Elsewhere the two-spinon occupancy configurations in the
spin effective lattice of the $s1$ bond particles 
are studied in detail for the model (\ref{H}) in the one- 
and two-electron subspace \cite{companion0}. 
This is a needed step for the investigations of 
Refs. \cite{companion,cuprates0} of the square-lattice quantum liquid 
of $c$ and $s1$ fermions with residual interactions.
Following the results of these references, it is expected that
such a quantum liquid captures the universal properties of general
many-electron models with short-range interactions on the square lattice. 

The general rotated-electron description introduced here 
for the Hubbard model on the square lattice is consistent with a 
ground state with long-range antiferromagnetic order for half filling,
short-range incommensurate-spiral spin order for $0<x\ll 1$,
and short-range spin order for finite hole concentrations $x$ below 
a $U/4t$-dependent critical value $x_*$ above which the system is
driven into a disordered state without short-range spin order. 
Strong evidence is given in Refs. \cite{cuprates0,cuprates} 
that for a well-defined hole-concentration range $x\in (x_c,x_*)$ where $x_c\approx 10^{-2}$
and $x_* \in (0.23,0.31)$ for approximately $u_0\leq U/4t\leq u_{\pi}$
a long-range superconducting order coexists with the
short-range spin order in the square-lattice quantum liquid 
perturbed by small 3D anisotropy associated with weak plane coupling. 

Finally, the results of Refs. \cite{companion,cuprates0,cuprates}
suggest that the description introduced in this paper is useful for
the further understanding of the role plaid by the electronic
correlations in the unusual properties of the hole-doped cuprates.

\begin{acknowledgments}
I thank Daniel Arovas, Nicolas Dupuis, Alejandro Muramatsu, 
Stellan \"Ostlund, Karlo Penc, Nuno M. R. Peres, and Pedro D. Sacramento
for discussions and the support of the ESF Science Program INSTANS and grant PTDC/FIS/64926/2006.
\end{acknowledgments}
\appendix

\section{Full information about the quantum problem when
defined in the LWS-subspace}

In this Appendix we confirm that full information about the quantum problem
can be achieved by defining it in the LWS-subspace spanned
by the energy eigenstates that are both LWSs of the $\eta$-spin
and spin algebras. By that we mean that all ${\cal{N}}$-electron 
operator matrix elements between energy eigenstates such
that at least one of them is a non-LWS can be evaluated exactly 
in terms of a  corresponding quantum problem involving 
another well-defined ${\cal{N}}$-electron operator acting 
onto the LWS-subspace. Here ${\cal{N}}=1,2,...$ refers
to one, two, or any other finite number of electrons so that the expression
of the general ${\cal{N}}$-electron operator under consideration  
reduces to an elementary creation or annihilation electronic
operator (${\cal{N}}=1$) or involves the product of two or more
such elementary operators (${\cal{N}}\geq 2$). 

In addition, we show here that use of the model
global symmetry provides a simple relation between
the energy of any non-LWS and the corresponding
LWS so that all contributions to the physical quantities
from non-LWSs can be evaluated by considering 
a related problem defined in the LWS-subspace. 
The use of such a symmetry 
reveals that for hole concentration
$x>0$ and spin density $m>0$ the ground state
is always a LWS of both the $\eta$-spin and
spin algebras. For simplicity here we consider matrix elements
between the ground state and a non-LWS. Those
appear in Lehmann representations of  
zero-temperature spectral functions
and correlation functions, but similar results can be
obtained for matrix elements between any excited 
energy eigenstates. 
 
Expressions (\ref{fc+})-(\ref{rotated-quasi-spin})
for the $c$ fermion, $\eta$-spinon, and spinon operators refer to the
LWS-subspace. A similar representation can be used for instance for
the HWS subspace, which is spanned by all energy eigenstates 
with $S_{\alpha}=S^z_{\alpha}$ where $\alpha =\eta ,s$. 
(There are also two mixed subspaces such that $S_{\eta}=\pm S^z_{\eta}$
and $S_{s}=\mp S^z_{s}$.) The HWS representation is suitable 
for canonical ensembles referring to electronic densities larger than 
one and negative spin densities since then the ground states are 
HWSs of both the $\eta$-spin and spin algebras.
The studies of this paper refer to the LWS-subspace.
In this Appendix we provide the expressions for the  $c$ fermion, $\eta$-spinon, 
and spinon operators that are suitable for the HWS subspace.
  
\subsection{How to map the problem onto the LWS-subspace}

We start by confirming that there is a well-defined
${\cal{N}}$-electron operator
${\hat{\Theta}}_{{\cal{N}}}$ such that any matrix element 
$\langle f\vert\,{\hat{O}}_{{\cal{N}}}\vert\psi_{GS} \rangle$ of
a ${\cal{N}}$-electron operator
${\hat{O}}_{{\cal{N}}}$ where $\vert f\rangle$ is
a non-LWS of the $\eta$-spin algebra and/or spin algebra
and $\vert\psi_{GS} \rangle$ the ground state of the Hubbard
model on the 1D or square lattice can be written as
$\langle f\vert\,{\hat{O}}_{{\cal{N}}}\vert\psi_{GS} \rangle =
\langle f.LHS\vert{\hat{\Theta}}_{{\cal{N}}} \vert\psi_{GS} \rangle$.
Here $\vert f.LWS\rangle$ is the LWS that corresponds to the state 
$\vert f\rangle$. Within the two $SU(2)$ algebras the
latter state can be expressed as,
\begin{equation}
\vert f\rangle = \prod_{\alpha
=\eta,\,s}\frac{({\hat{S}}^{\dag}_{\alpha})^{L_{\alpha,\,-1/2}}}{
\sqrt{{\cal{C}}_{\alpha}}}\vert f.LWH\rangle \, , 
\label{IRREG}
\end{equation}
where
\begin{equation}
{\cal{C}}_{\alpha} = \delta_{L_{\alpha,\,-1/2},\,0} +
\prod_{l=1}^{L_{\alpha,\,-1/2}}l\,[\,L_{\alpha}+1-l\,] \, , 
\label{Calpha}
\end{equation}
is a normalization constant, $L_{\alpha,\,-1/2}\leq L_{\alpha}=2S_{\alpha}$,
and the $\eta$-spin flip ($\alpha =\eta$) and spin flip ($\alpha =s$)
operators ${\hat{S}}^{\dag}_{\alpha}$ are the off-diagonal generators of the
corresponding $SU(2)$ algebras provided in Eq. (\ref{Scs}).
These operators remain invariant under the electron - rotated-electron unitary
transformation and thus as given in that equation have the same expression 
in terms of electron and rotated-electron creation and annihilation operators. 

As confirmed in Section IV,
for a hole concentration $x\geq 0$ and spin density $m\geq 0$
the ground state is a LWS of both the $\eta$-spin and spin $SU(2)$ algebras 
and thus has the following property,
\begin{equation}
{\hat{S}}_{\alpha}\,\vert\psi_{GS} \rangle = 0 \, ; \hspace{0.5cm} \alpha = \eta,\,s \, .
\label{GS=0}
\end{equation}
The operator ${\hat{\Theta}}_{{\cal{N}}}$ is then such that,
\begin{eqnarray}
\langle f\vert\,{\hat{O}}_{{\cal{N}}}\vert\psi_{GS} \rangle & = & \langle f.LWS\vert 
\prod_{\alpha =\eta,\,s}{1\over
\sqrt{{\cal{C}}_{\alpha}}}({\hat{S}}_{\alpha})^{L_{\alpha,\,-1/2}}\,
{\hat{O}}_{{\cal{N}}} \vert\psi_{GS} \rangle 
\nonumber \\
& = & \langle f.LHS\vert
{\hat{\Theta}}_{{\cal{N}}} \vert\psi_{GS} \rangle  \, . \label{SYMME}
\end{eqnarray}

By the suitable use of Eq. (\ref{GS=0}), it is straightforward to show that 
for $L_{\eta,\,-1/2}>0$ and/or $L_{s,\,-1/2}>0$ the operator
${\hat{\Theta}}_{{\cal{N}}}$ is given by the following commutator,
\begin{equation}
{\hat{\Theta}}_{{\cal{N}}} = \Bigl[\prod_{\alpha =\eta,\,s}{1\over
\sqrt{{\cal{C}}_{\alpha}}}\Bigr]\,\Bigl[\prod_{\alpha
=\eta,\,s}({\hat{S}}_{\alpha})^{L_{\alpha,\,-1/2}},\,
{\hat{O}}_{{\cal{N}}}\Bigr] \, , 
\label{Th}
\end{equation}
and by,
\begin{equation}
{\hat{\Theta}}_{{\cal{N}}} = {\hat{O}}_{{\cal{N}}} \, ,  
\label{OL-0}
\end{equation}
for $L_{\eta,\,-1/2}=L_{s,\,-1/2}= 0$. If the commutator on
the right-hand side of Eq. (\ref{Th}) vanishes then the matrix element
$\langle f\vert\,{\hat{O}}_{{\cal{N}}}\vert\psi_{GS} \rangle$ under
consideration also vanishes.

Let $E_f$ denote the energy eigenvalue of the non-LWS $\vert f\rangle$ and 
$E_{f.LWS}$ that of the corresponding LWS $\vert f.LWS\rangle$. To find the 
relation between $E_f$ and $E_{f.LWS}$ one adds chemical-potential and 
magnetic-field operator terms to the Hamiltonian (\ref{H}), what lowers its 
symmetry. Since our description refers to the LWS-subspace, we
use units such that the sign of 
the chemical potential $\mu\geq 0$ is that of the hole concentration 
$x=[N_a^D-N]/N_a^D\geq 0$ and the sign of the magnetic field $H\geq 0$ 
that of the spin density $m=[N_{\uparrow}-N_{\downarrow}]/N_a^D\geq 0$.
Since such operator terms commute with 
the Hamiltonian (\ref{H}), the  rotated-electron occupancy configurations 
of all energy eigenstates correspond to state representations of its 
global symmetry for all densities. Moreover, the use of such commutation relations
reveals that the energy eigenvalues $E_f$ and $E_{f.LWS}$ are related as,
\begin{equation}
E_f = E_{f.LWS} + \sum_{\alpha =\eta,s}\mu_{\alpha}\,L_{\alpha,\,-1/2} \, ,
\label{energies}
\end{equation}
where $\mu_{\eta} =2\mu$ is twice the chemical potential
whose range is $2\mu^0\leq 2\mu\leq U+4Dt$ for $D=1,2$,
$2\mu^0$ is the half-filling Mott-Hubbard gap considered in
Subsection IV-C, $\mu_{s} =2\mu_B\,H \geq 0$, and $\mu_B$ is the Bohr 
magneton. 

It follows from Eq. (\ref{energies}) that $E_f\geq E_{f.LWS}$. 
For hole concentration $x>0$ and spin density $m>0$ 
such an inequality can be shown to be consistent with 
the ground state being always a LWS of both the $\eta$-spin and spin algebras.
On the other hand, the model global symmetry requires that $E_f = E_{f.LWS}$ for the
half-filing and zero-magnetization absolute ground state. Such a requirement
is fulfilled: for half filling one has that $\mu\in (-\mu^0,\mu^0)$. 
The value $\mu=0$ corresponds to the middle of the Mott-Hubbard gap,
whereas the magnetic field $H$ vanishes for zero magnetization. The
absolute ground state corresponds to $S_{\eta}=S_s=0$ and $S_c =N_a^D/2$ and 
hence is both a LWS and a HWS of the $\eta$-spin 
and spin algebras.

Within a Lehmann representation the ${\cal{N}}$-electron spectral functions
are expressed as a sum of terms, one for each excited energy eigenstate. In
spectral-function terms associated with non-LWSs one can then replace the 
matrix element $\langle f\vert\,{\hat{O}}_{{\cal{N}}}\vert\psi_{GS} \rangle$ by 
$\langle f.LHS\vert{\hat{\Theta}}_{{\cal{N}}} \vert\psi_{GS} \rangle$
provided that the excited-state energy $E_f$ is expressed as in Eq. (\ref{energies}).
Then the original ${\cal{N}}$-electron spectral function can be written as a 
sum of spectral functions of well-defined ${\cal{N}}$-electron operators 
of the form given in Eqs. (\ref{Th}) or (\ref{OL-0}), all acting onto the LWS-subspace.
 
These results confirm that full information about the present quantum problem
can be obtained by defining it in the LWS-subspace. 
 
\subsection{Operators of the three elementary objects within the HWS representation} 
 
Within the HWS representation the $c$ fermion operators read,
\begin{equation}
f_{\vec{r}_j,c}^{\dag} =
{\tilde{c}}_{\vec{r}_j,\downarrow}^{\dag}\,
(1-{\tilde{n}}_{\vec{r}_j,\uparrow})
+ e^{i\vec{\pi}\cdot\vec{r}_j}\,{\tilde{c}}_{\vec{r}_j,\downarrow}\,
{\tilde{n}}_{\vec{r}_j,\uparrow} \, .
\label{fc+HWS}
\end{equation}
The $\eta$-spinon and spinon operators are given by
Eq. (\ref{sir-pir}) where now the operator 
$n_{\vec{r}_j,c} = f_{\vec{r}_j,c}^{\dag}\,f_{\vec{r}_j,c}$
of Eq. (\ref{n-r-c})
involves the $c$ fermion operator (\ref{fc+HWS}) and
the rotated quasi-spin operators
$q^{\pm}_{\vec{r}_j}=q^{x}_{\vec{r}_j}\pm i\,q^{y}_{\vec{r}_j}$
and $q^z_{\vec{r}_j}$ read,
\begin{eqnarray}
q^+_{\vec{r}_j} & = & {\tilde{c}}^{\dag} _{\vec{r}_j,\uparrow}\,
({\tilde{c}}_{\vec{r}_j,\downarrow}
- e^{i\vec{\pi}\cdot\vec{r}_j}\,{\tilde{c}}^{\dag}_{\vec{r}_j,\downarrow}) \, ,
\nonumber \\
q^-_{\vec{r}_j} & = & (q^+_{\vec{r}_j})^{\dag} \, ;
\hspace{0.35cm}
q^z_{\vec{r}_j} = {\tilde{n}}_{\vec{r}_j,\downarrow} - {1\over 2} \, ,
\label{rotated quasi-spin-HWS}
\end{eqnarray}
respectively. After inverting such relations one finds that
within the HWS representation the local
rotated-electron operators can be expressed in terms of the 
local $c$ fermion and rotated quasi-spin operators as follows,
\begin{eqnarray}
{\tilde{c}}_{\vec{r}_j,\downarrow}^{\dag} & = &
f_{\vec{r}_j,c}^{\dag}\,\left({1\over 2} -
q^z_{\vec{r}_j}\right) + e^{i\vec{\pi}\cdot\vec{r}_j}\,
f_{\vec{r}_j,c}\,\left({1\over 2} + q^z_{\vec{r}_j}\right) \, ,
\nonumber \\
{\tilde{c}}_{\vec{r}_j,\uparrow}^{\dag} & = &
q^+_{\vec{r}_j}\,(f_{\vec{r}_j,c}^{\dag} -
e^{i\vec{\pi}\cdot\vec{r}_j}\,f_{\vec{r}_j,c}) \, .
\label{c-up-c-down-HWS}
\end{eqnarray}

The commutation and anti-commutation relations (\ref{albegra-cf})-(\ref{albegra-q-com})
are valid for all representations and hence are the same for both the LWS and 
HWS representations. 

\section{Equivalence of the quantum numbers of the present 
description and those of the exact solution for 1D}

\subsection{Relation to the quantum numbers of the exact solution
for $N_a\gg1$}

The studies of this paper establish that the spinons 
(and $\eta$-spinons) that are not invariant
under the electron rotated-electron transformation are
part of spin (and $\eta$-spin) neutral $2\nu$-spinon 
(and $2\nu$-$\eta$-spinon) composite $s\nu$ (and
$\eta\nu$) bond particles, which have a bounding (and an
anti-bounding) character. Moreover, for the 1D model some neutral $2\nu$-spinon 
(and $2\nu$-$\eta$-spinon) configurations exist 
that define suitable $s\nu$ (and $\eta\nu$) bond particles
whose occupancies generate the energy eigenstates for $U/4t>0$. 
The detailed structure of such
configurations remains an open problem. It
is studied in Ref. \cite{companion0} both
for the 1D and square lattices for the case of the one- and
two-electron subspace as defined in this paper, where only the $c$ fermions and
$s1$ bond particles have an active role.

By generalizing the procedures used in Ref. \cite{companion0} for
the $s1$ bond particles to the remaining $\alpha\nu$
bond-particle branches introduced in this paper, 
one finds that for the model on the 1D lattice
such objects can be associated with $\alpha\nu$ 
bond-particle operators $g^{\dag}_{x_{j},\alpha\nu}$ of the general form,
\begin{equation}
g_{x_{j},\alpha\nu} = \sum_{g} h_{\alpha\nu,g}\, a_{x_{j},\alpha\nu,g} 
\, ; \hspace{0.50cm} g_{x_{j},\alpha\nu}^{\dag} = \left(g_{x_{j},\alpha\nu}\right)^{\dag}  \, .
\label{g-an+general}
\end{equation}
Here $a_{x_{j},\alpha\nu,g}$ is a superposition of operators belonging
to a number of families, which increases for increasing values of
the number $\nu=1,2,...$ of spinon ($\alpha=s$) or $\eta$-spinon ($\alpha=\eta$)
pairs, the spatial-coordinate $x_j$ refers to the $\alpha\nu$ effective 
lattice, the index $j=1,...,N_{a_{\alpha\nu}}$ 
to the sites of that lattice, and the index $g$ to different types 
of $2\nu$-site links belonging to 
the same family as defined in that paper for the
$s1$ bond particles. The $c$ effective lattice equals 
the original lattice and thus has length $L=N_a\, a$ and lattice constant 
$a$ so that $x_j =j\,a$ for $N_a\gg1$. The $\alpha\nu$ effective lattices have also
length $L$ yet the lattice constants read 
$a_{\alpha\nu}=[N_a/N_{a_{\alpha\nu}}]\,a\geq a$ and are such
that $L=N_{a_{\alpha\nu}}\,a_{\alpha\nu}$ and
$x_j =j\,a_{\alpha\nu}$ where $j=1,...,N_{a_{\alpha\nu}}$. 
Furthermore, according to the same results the number of sites of the 
$\alpha\nu$ effective lattice $N_{a_{\alpha\nu}}$ is for 1D given by
expression (\ref{N*}) with $D=1$ so that
$N_{a_{\alpha\nu}} = N_{\alpha\nu} + N^h_{\alpha\nu}$
where $N_{a_{\alpha}}\equiv N_{a_{\alpha}}^1$
and $N_{\alpha\nu}$ is the number of composite $\alpha\nu$ bond 
particles and hence of occupied sites of the $\alpha\nu$ effective
lattice. The corresponding number of unoccupied sites 
$N^h_{\alpha\nu}$ of that lattice is provided in Eq. (\ref{N-h-an})
with $N_{a_{\alpha}}^D\equiv N_{a_{\alpha\nu}}$. 
Here $C_{\alpha}$ is the number of "occupied sites" of the $\eta$-spin 
($\alpha=\eta$) and spin ($\alpha=s$) effective lattice, which 
refers to the subspaces with constant values of $S_c$, $S_{\eta}$, 
and $S_s$ and is given in Eq. (\ref{M-L-Sum}) where
$N_{a_{\alpha}}^D\equiv N_{a_{\alpha\nu}}$ for 1D
is the total number of sites of these effective lattices.
Each subspace with constant values of $S_c$ and hence
also of $N_{a_{\eta}}=[N_a-2S_c]$ and $N_{a_{s}}=2S_c$
can be divided into smaller subspaces with constant values of $S_c$, $S_{\eta}$, 
and $S_s$ and hence also of $C_{\eta}$ and $C_s$.
Furthermore, the latter subspaces can be further divided into even smaller 
subspaces with constant values for the set of numbers $\{N_{\eta\nu}\}$ and 
$\{N_{s\nu}\}$, which must obey the sum-rule of Eq. (\ref{M-L-Sum}).   

Concerning the operators $a_{x_{j},\alpha\nu,g}$ of the
general expressions given in Eq. (\ref{g-an+general}),
the simplest case refers to the $s1$ bond particles
whose number of families is two and for instance for the
$N_{s1}^h=0$ configuration state the operator $a_{x_{j},s1,g}$ 
is given by \cite{companion0},
\begin{equation}
a_{x_{j},s1,g} = \sum_{l=\pm1}
\, b_{x_{j}+{\vec{r}_{1,l}}^{\,0},s1,l,g} 
\, ; \hspace{0.50cm}
b_{x,s1,l,g}^{\dag} = 
{1\over\sqrt{2}}\left(\left[{1\over 2}+s^z_{x-x_{l}}\right]
s^-_{x+x_{l}} - \left[{1\over 2}+s^z_{x+x_{l}}\right]
s^-_{x-x_{l}}\right) \, ,
\label{g-s1+general-1D}
\end{equation}
and $b_{x,s1,l,g} = \left(b_{x,s1,l,g}^{\dag}\right)^{\dag}$.
Here the spinon operators are given in Eq. (\ref{sir-pir})
and for simplicity we have omitted the family
index $d$, which reads $d=1$ for 1D so that 
$x_{l}$ corresponds to the general two-site
link vector $\vec{r}_{d,l}$ introduced in Ref. \cite{companion0}.

The studies of Ref. \cite{companion0} lead to $s1$ bond-particle
operators that are shown in that paper to obey a hard-core
algebra. For 1D it follows straightforwardly that
the two-$\eta$-spinons $\eta1$ bond-particle operators also
obey such an algebra. One then considers the four-spinon $s2$
bond-particle and four-$\eta$-spinon $\eta 2$ bond-particle
operator problems and finds that operators of the general form
(\ref{g-an+general}) can be constructed by procedures similar to
those used for the $s1$ and $\eta1$ bond-particle operators
so that the hard-core algebra is obeyed. 

We skip here the details of such derivations since 
the basic procedures are the same as those already 
used for the two-spinon $s1$ bond particles in 
Ref. \cite{companion0} except that the algebraic manipulations are for $\nu>1$ much 
more cumbersome. Moreover, it turns out that for interesting
applications within the dynamical theory of Ref. \cite{V} 
referring to the model on the 1D lattice,
such as the evaluation for finite excitation
energy $\omega$ of one- and two-electron
matrix elements between the ground state and excited states 
and corresponding spectral-weight distributions,
only excited states involving $c$ and 
$s1$ fermions and a finite number of $s\nu$
fermions with $\nu>1$ spinon pairs 
lead to finite spectral weight. The latter $s\nu$
fermions have vanishing energy and
momentum and remain invariant under the electron - rotated-electron
unitary transformation so that the only effect of
their creation and annihilation is in the
numbers of $s1$ fermions and $s1$ band
discrete momentum values. Therefore, for
the excited states contributing to the
dynamical theory of Ref. \cite{V} only
the $c$ fermions and $s1$ fermions 
play an active role.

The only general result this is needed for the goals 
of this Appendix is that alike for the two-site one-bond operators,
composite $\alpha\nu$ bond-particle operators of general form
given in Eq. (\ref{g-an+general}) can for 1D be constructed so that when acting 
onto the LWS Hilbert space they anticommute 
on the same $\alpha\nu$ effective-lattice site,
\begin{equation}
\{g^{\dag}_{x_{j},\alpha\nu},g_{x_{j},\alpha\nu}\} = 1 \, ;
\hspace{0.35cm}
\{g^{\dag}_{x_{j},\alpha\nu},g^{\dag}_{x_{j},\alpha\nu}\} =
\{g_{x_{j},\alpha\nu},g_{x_{j},\alpha\nu}\}=0 \, ,
\label{g-local-an}
\end{equation}
and commute on different sites,
\begin{equation}
[g^{\dag}_{x_{j},\alpha\nu},g_{x_{j'},\alpha\nu}] =
[g^{\dag}_{x_{j},\alpha\nu},g^{\dag}_{x_{j'},\alpha\nu}]
= [g_{x_{j},\alpha\nu},g_{x_{j'},\alpha\nu}] = 0 
\, ; \hspace{0.5cm} j\neq j' \, .
\label{g-non-local-an}
\end{equation}
Furthermore, such operators commute with the
$c$ fermion operators and operators corresponding to different 
$\alpha\nu$ branches also commute with each other.

In this Appendix we confirm that the general hard-core algebra
of Eqs. (\ref{g-local-an}) and (\ref{g-non-local-an}) combined with
the universal number expressions given in Section IV 
leads to discrete momentum values for the $c$ and $\alpha\nu$ fermions
that coincide with the quantum numbers of the exact 
solution for the whole LWS Hilbert subspace it refers to.

It follows from such an algebra that one can perform an 
extended Jordan-Wigner transformation 
that transforms the $\alpha\nu$ bond particles into $\alpha\nu$ fermions with 
operators $f^{\dag}_{x_{j},\alpha\nu}$. Alike in the general expressions
provided in Eq. (\ref{f-an-operators}), such operators are related to the corresponding 
bond-particle operators as,
\begin{equation}
f^{\dag}_{x_{j},\alpha\nu} = e^{i\phi_{j,\alpha\nu}}\,
g^{\dag}_{x_{j},\alpha\nu} 
\, ; \hspace{0.50cm} 
f_{x_{j},\alpha\nu} = e^{-i\phi_{j,\alpha\nu}}\,
g_{x_{j},\alpha\nu} \, ,
\label{JW-f+-an}
\end{equation}
where 
\begin{equation}
\phi_{j,\alpha\nu} = \sum_{j'\neq j}f^{\dag}_{x_{j'},\alpha\nu}
f_{x_{j'},\alpha\nu}\,\phi_{j',j,\alpha\nu} 
\, ; \hspace{0.50cm}
\phi_{j',j,\alpha\nu} = \arctan \left({y_{j'}-y_{j}\over x_{j'}-x_{j}}\right) 
\, ; \hspace{0.35cm}
0\leq \phi_{j',j,\alpha\nu}\leq 2\pi \, .
\label{JW-phi-an}
\end{equation}

However, for 1D the coordinate $z_{j}=x_{j}+i\,y_{j}$ is 
such that $y_j=0$ and hence reduces to the
real-space coordinate $x_j$ of the $\alpha\nu$ bond particle 
in its $\alpha\nu$ effective lattice. Therefore, 
for 1D the phase $\phi_{j',j,\alpha\nu}$ 
can for all $\alpha\nu$ branches have the values
$\phi_{j',j,\alpha\nu}=0$ and
$\phi_{j',j,\alpha\nu}=\pi$ only. Indeed, the
relative angle between two sites of the
$\alpha\nu$ effective lattice in a 1D chain
can only be one of the two values. Then
the $\alpha\nu$ phase factor of Eq. (\ref{JW-phi-an}) 
is such that,
\begin{equation}
e^{ia_{\alpha\nu}{\partial\over\partial x}\phi_{\alpha\nu} (x)\vert_{x=x_j}}
= e^{i(\phi_{j+1,\alpha\nu}-\phi_{j,\alpha\nu})}  
= e^{i\pi\,f^{\dag}_{x_{j},\alpha\nu}\,f_{x_{j},\alpha\nu}} \, ,
\label{1D-rel}
\end{equation}
where $\phi_{\alpha\nu} (x_j) \equiv \phi_{j,\alpha\nu}$.

The $c$ fermion operators have the anticommuting relations given in Eq. (\ref{albegra-cf}),
which for 1D read,
\begin{equation}
\{f^{\dag}_{x_j,c}\, ,f_{x_{j'},c}\} = \delta_{j,j'} 
\, ; \hspace{0.50cm}
\{f_{x_j,c}^{\dag}\, ,f_{x_{j'},c}^{\dag}\} =
\{f_{x_j,c}\, ,f_{x_{j'},c}\} = 0 \, .
\label{albegra-cf-1D}
\end{equation}
Moreover, the $\alpha\nu$ bond-particle operators that emerge from
the Jordan-Wigner transformation associated with Eqs.
(\ref{JW-f+-an}) and (\ref{JW-phi-an}) have similar anticommuting 
relations given by,
\begin{equation}
\{f^{\dag}_{x_{j},\alpha\nu}\, ,f_{x_{j'},\alpha\nu}\} =
\delta_{j,j'} 
\, ; \hspace{0.50cm}
\{f^{\dag}_{x_{j},\alpha\nu}\, ,f^{\dag}_{x_{j'},\alpha\nu}\} =
\{f_{x_{j},\alpha\nu}\, ,f_{x_{j'},\alpha\nu}\}  = 0 \, ,
\label{1D-anti-com-1D}
\end{equation}
and the $c$ fermion operators commute with 
the $\alpha\nu$ fermion operators and $\alpha\nu$ 
and $\alpha'\nu'$ fermion operators such that
$\alpha\nu\neq \alpha'\nu'$ also commute.

One can introduce $c$ fermion operators given
in Eq. (\ref{fc+}) associated with discrete momentum 
values, which for 1D read,
\begin{equation}
f_{q_j,c}^{\dag} = 
{1\over{\sqrt{N_a}}}\sum_{j'=1}^{N_a}\,e^{+iq_j x_{j'}}\,
f_{x_{j'},c}^{\dag} 
\, ; \hspace{0.50cm}
f_{q_j,c} = {1\over{\sqrt{N_a}}}\sum_{j'=1}^{N_a}\,e^{-iq_j x_{j'}}\,
f_{x_{j'},c} 
\, ; \hspace{0.35cm}
j = 1,...,N_a \, ; \hspace{0.35cm} x_j = j a
\, ; \hspace{0.35cm} L = a N_{a} \, .
\label{fc-q-x-1D}
\end{equation}
For the 1D Hubbard model the $\alpha\nu$ fermion operators $f^{\dag}_{q_{j},\alpha\nu}$
given in Eq. (\ref{f-an-operators})
labeled by the discrete momentum values $q_j$
such that $j=1,...,N_{a_{\alpha\nu}}$, which are the conjugate 
variables of the $\alpha\nu$ effective lattice spatial
coordinates $x_j$, can be introduced 
provided that the ratio $N_{a_{\alpha\nu}}/N_a$ involving the number
$N_{a_{\alpha\nu}}$ of sites of the $\alpha\nu$ effective lattice
given in Eqs. (\ref{N*}) and (\ref{N-h-an}) is finite for $N_a\rightarrow\infty$,
\begin{equation}
f_{q_j,\alpha\nu}^{\dag} = 
{1\over{\sqrt{N_{a_{\alpha\nu}}}}}\sum_{j'=1}^{N_{a_{\alpha\nu}}}\,e^{+iq_j x_{j'}}\,
f_{x_{j'},\alpha\nu}^{\dag} 
\, ; \hspace{0.5cm}
f_{q_j,\alpha\nu} = {1\over{\sqrt{N_{a_{\alpha\nu}}}}}
\sum_{j'=1}^{N_{a_{\alpha\nu}}}\,e^{-iq_j x_{j'}}\,
f_{x_{j'},\alpha\nu} 
\, ; \hspace{0.35cm}
j = 1,...,N_{a_{\alpha\nu}} \, ; \hspace{0.1cm} x_j = j a_{\alpha\nu} ,
\label{fan-q-x-1D}
\end{equation}
where $L = a_{\alpha\nu} N_{a_{\alpha\nu}}$.

In 1D the phase factor $e^{i\phi_{j,\alpha\nu}}$ does not have any effect 
when operating before $f^{\dag}_{x_{j},\alpha\nu}$. It follows that in 1D
the expression of the Hamiltonian does not
involve the phase $\phi_{j,\alpha\nu}$. Moreover, expression
of the 1D normal-ordered Hamiltonian in terms of the $c$
and $\alpha\nu$ fermion operators reveals 
that such objects have zero-momentum forward-scattering only. 
This is consistent with the integrability of the model in 1D and the 
existence of an infinite number of conservations laws for the
limit $N_a\rightarrow\infty$ that our description refers
to \cite{Martins}. For the 1D model the occurrence of such conservations laws is behind
the set of $\alpha\nu$ fermion numbers 
$\{N_{\alpha\nu}\}$ being good quantum numbers \cite{Prosen}.
This is in contrast to the model on the square lattice, for which
such numbers are not in general conserved, yet the
quantity $C_{\alpha}=\sum_{\nu}\nu\,N_{\alpha\nu}$ is.

The Jordan-Wigner transformations phases $\phi_{j,\alpha\nu}$
have direct effects on the boundary conditions, which
determine the discrete momentum values $q_j$ of both the
$c$ and $\alpha\nu$ fermion operators of Eqs. (\ref{fc-q-x-1D}) and
(\ref{fan-q-x-1D}), respectively. In 1D the periodic 
boundary conditions of the original electron problem are 
ensured provided that one takes into account the effects 
of the Jordan-Wigner transformation on the boundary conditions  
of the $c$ fermions and $\alpha\nu$ fermions
upon moving one of such objects around the chain of length
$L$ once. 

As discussed in Subsection IV-B, for both the model on 
the 1D and square lattices the rotated-electron 
occupancies of the sites of the original lattice separate into
two degrees of freedom only. (In the remaining of this Appendix
we limit our considerations to the 1D problem.) Those of the $2S_c$ sites
of the original lattice singly occupied by rotated electrons separate into
(i) $2S_c$ sites of the $c$ effective lattice occupied
by $c$ fermions and (ii) $2S_c$ sites of the spin effective
lattice occupied by spinons.
Those of the $[N_a-2S_c]$ sites
of the original lattice doubly occupied and unoccupied
by rotated electrons separate into
(i) $[N_a-2S_c]$ sites of the $c$ effective lattice unoccupied
by $c$ fermions and (ii) $[N_a-2S_c]$ sites of the $\eta$-spin effective
lattice occupied by $\eta$-spinons.

For instance, the $2\nu$ sites of the spin (and $\eta$-spin) 
effective lattice referring to the occupancy configuration
of one local $2\nu$-spinon composite $s\nu$ fermion
(and  $2\nu$-$\eta$-spinon composite $\eta\nu$ fermion)
correspond to the spin (and $\eta$-spin) degrees of
freedom of $2\nu$ sites of the original lattice whose
degrees of freedom associated with the 
global $U(1)$ symmetry found
in Ref. \cite{bipartite} are described  
by $2\nu$ sites of the $c$ effective lattice occupied
(and unoccupied) by $c$ fermions. 

An important point is that the independent spinons and 
independent $\eta$-spinons are not part of
the Jordan-Wigner transformations that transform
the $\alpha\nu$ bond particles onto $\alpha\nu$
fermions. It follows that when one $c$ fermion 
moves around its effective lattice of length
$L$ it feels the effects of the Jordan-Wigner transformations 
through the sites of the spin and $\eta$-spin lattices
associated with those of the $s\nu$ and $\eta\nu$
effective lattices occupied by $s\nu$ and $\eta\nu$ fermions,
respectively. Indeed, we recall that the sites of 
the $c$ effective lattice on the one hand and those 
of the spin and $\eta$-spin effective lattices on the
other hand correspond to the different degrees of
freedom of rotated-electron occupancies of the same sites of the original lattice.

Since the $c$ fermions do not emerge
from a Jordan-Wigner transformation and each $\alpha\nu$
fermion corresponds to a set of $2\nu$ sites of the original
lattice different from and independent of those of any other $\alpha'\nu'$
fermion, when a $\alpha\nu$ fermion moves
around its $\alpha\nu$ effective lattice of length $L$ it only feels the
Jordan-Wigner-transformation phases of its own 
lattice associated with both the $\alpha\nu$ fermions and 
$\alpha\nu$ fermion holes so that its discrete momentum 
values obey the following periodic or anti-periodioc boundary conditions,
\begin{equation}
e^{iq_j\,L} = \prod_{j=1}^{N_{a_{\alpha\nu}}}\left\{
\left[e^{i(\phi_{j+1,\alpha\nu}-\phi_{j,\alpha\nu})}\right]^{\dagger}
e^{i(\phi_{j+1,\alpha\nu}-\phi_{j,\alpha\nu})}\right\} =
e^{i\pi [N_{a_{\alpha\nu}}-1]} = -e^{i\pi N_{a_{\alpha\nu}}} \, .
\label{pbcanp}
\end{equation}
Here the phase factor reads $1$ and $-1$ for $[N_{a_{\alpha\nu}}-1]$
even and odd, respectively. The term $-1$ in $[N_{a_{\alpha\nu}}-1]$ can be
understood as referring to the site occupied by the $\alpha\nu$ fermion moving
around its effective lattice and must be excluded. For the $\alpha\nu$
fermions the unoccupied sites of their $\alpha\nu$ effective lattice
exist in their own right. Indeed, note that according to Eq. (\ref{JW-f+-an}) 
both the creation and annihilation operators of such objects involve 
the Jordan-Wigner-transformation phase $\phi_{j,\alpha\nu}$. 
As a result such a phase affects both the $\alpha\nu$ fermions and 
$\alpha\nu$ fermion holes. That justifies why the phase factor
$e^{i\pi [N_{a_{\alpha\nu}}-1]}$ of
Eq. (\ref{pbcanp}) involves all the $N_{a_{\alpha\nu}}=[N_{\alpha\nu}+
N_{\alpha\nu}^h]$ sites of the $\alpha\nu$ effective lattice except
that occupied by the moving $\alpha\nu$ fermion. Hence it involves both
the $[N_{\alpha\nu}-1]$ sites occupied by the remaining fermions of
the same $\alpha\nu$ branch and the corresponding $N_{\alpha\nu}^h$ 
$\alpha\nu$ fermion holes.

In contrast, the $c$ fermions are affected by the
sites occupied by $\alpha\nu$ fermions only. Indeed, only the 
sets of $2\nu$ sites of the spin (and $\eta$-spin) effective lattice
associated with each occupied site of the $s\nu$ (and $\eta\nu$)  
$\nu=1,2,3,...$ effective lattices and the sites of the spin (and $\eta$-spin) effective 
lattice occupied by independent spinons (and independent 
$\eta$-spinons) correspond to sites of the original lattice whose 
degrees of freedom associated with the $c$ fermion $U(1)$ 
symmetry are described by the occupancy configurations
of the $c$ effective lattice. However, the  
independent spinons (and independent 
$\eta$-spinons) do not undergo any 
Jordan-Wigner transformation
so that due to the Jordan-Wigner-transformation
phase $\phi_{j',\alpha\nu}$ of each of the $N_{\alpha\nu}$ 
$\alpha\nu$ fermions at sites $j'=1,...,N_{a_{\alpha\nu}}$ 
of their $\alpha\nu$ effective lattice the 
$c$ fermion discrete momentum values are determined by the 
following periodic or anti-periodioc boundary condition,
\begin{equation}
e^{iq_j\,L} = 
\prod_{\alpha\nu}\prod_{j'=1}^{N_{\alpha\nu}}e^{i(\phi_{j'+1,\alpha\nu}-\phi_{j',\alpha\nu})} 
= e^{i\pi\sum_{\alpha\nu}N_{\alpha\nu}} \, .
\label{pbccp}
\end{equation}
Again the phase factor on the right-hand side
of Eq. (\ref{pbccp}) reads $1$ and $-1$ for $\sum_{\alpha\nu}N_{\alpha\nu}$
even and odd, respectively. 

The above results imply that the discrete momentum values $q_j$ of both
$c$ and $\alpha\nu$ fermions have the usual 
momentum spacing $q_{j+1}-q_j=2\pi/L$ and read,
\begin{equation}
q_j = {2\pi\over L}\,I^{\alpha\nu}_j \, ; \hspace{0.35cm} j=1,...,N_{a_{\alpha\nu}} 
\, ; \hspace{0.5cm}
q_j = {2\pi\over L}\,I^c_j \, ; \hspace{0.35cm} j=1,...,N_a \, .
\label{q-j-f-repr}
\end{equation}
However, following the boundary conditions (\ref{pbcanp}) [and (\ref{pbccp})] the numbers 
$I^{\alpha\nu}_j$ (and $I^{c}_j$) where $j=1,2,...,N_{a_{\alpha\nu}}$ (and $j=1,2,...,N_a$)
appearing in this equation are not always integers. They are 
integers and half-odd integers for $[N_{a_{\alpha\nu}}-1]$
(and $\sum_{\alpha\nu}N_{\alpha\nu}$ ) even and odd, respectively. Furthermore,
as a result of the periodic or anti-periodic character of such boundary conditions
these numbers obey the inequality $\vert I^{\alpha\nu}_j\vert\leq [N_{a_{\alpha\nu}}-1]/2$ for both 
$[N_{a_{\alpha\nu}}-1]$ odd and even (and the inequality $\vert I^{c}_j\vert\leq [N_a-1]/2$ 
for $\sum_{\alpha\nu}N_{\alpha\nu}$ even
and  $-[N_a-2]/2\leq I^{c}_j \leq N_a/2$ for $\sum_{\alpha\nu}N_{\alpha\nu}$ odd).

For the one- and two-electron subspace one can separate the numbers $I^{s1}_j$ 
and $I^{c}_j$ of Eq. (\ref{q-j-f-repr}) in two terms corresponding to an integer
number and a small deviation as $I^{s1}_j\equiv [{\cal{N}}^{s1}_j + {q_{s1}^0\over 2\pi}{L\over N_{s1}}]$
and $I^{c}_j\equiv [{\cal{N}}^c_j + {q_{c}^0\over 2\pi}{L\over N_c}]$, respectively. The corresponding
$c$ and $s1$ fermion discrete momentum values then read,
\begin{eqnarray}
q_j & = & {2\pi\over L}\,{\cal{N}}^{c}_j + q_{c}^0/N_{c} 
\, ; \hspace{0.50cm}
{\cal{N}}^{c}_j= j - {N_{a}\over 2} = 0,\pm 1, \pm 2, ... 
\, ; \hspace{0.35cm} j=1,...,N_{a} \, .
\nonumber \\
q_j & = & {2\pi\over L}\,{\cal{N}}^{s1}_j + q_{s1}^0/N_{s1} 
\, ; \hspace{0.50cm}
{\cal{N}}^{s1}_j= j - {N_{a_{s1}}\over 2} = 0,\pm 1, \pm 2, ... 
\, ; \hspace{0.35cm} j=1,...,N_{a_{s1}} \, ,
\label{q-j-f-Q-c-0}
\end{eqnarray}
Here $q_{c}^0$ (and $q_{s1}^0$) is given either by $q_{c}^0=0$ or $q_{c}^0=\pi [N_c/L]$ 
(and $q_{s1}^0=0$ or $q_{s1}^0= \pi [N_{s1}/L]$) for all $j=1,...,N_a$ (and $j=1,...,N_{a_{s1}}$)
discrete momentum values of the $c$ (and $s1$) band whose momentum occupancy describes a given state
and the $s1$ effective lattice length $L=N_{a_{s1}}\,a_{s1}$ where
$a_{s1}=L/N_{a_{s1}}=[N_a/N_{a_{s1}}]\,a$ is the $s1$ effective lattice 
constant.

Importantly, the $c$ fermion and $\alpha\nu$ fermion
discrete momentum values obtained from our $N_a\gg1$ operational 
description of the quantum problem correspond to the Bethe-ansatz quantum 
numbers of the exact solution. Indeed, the discrete momentum 
values of Eq. (\ref{q-j-f-repr}) can be expressed as, 
\begin{equation}
q_j = {2\pi\over L}\,I_j \, ; \hspace{0.35cm} j=1,...,N_a 
\, ; \hspace{0.5cm}
q_j = {2\pi\over L}\,J^{'\nu}_j \, ; \hspace{0.35cm} j=1,...,N_{a_{\eta\nu}} 
\, ; \hspace{0.5cm}
q_j = {2\pi\over L}\,J^{\nu}_j \, ; \hspace{0.35cm} j=1,...,N_{a_{s\nu}} \, ,
\label{q-j}
\end{equation}
where $I_j\equiv I^c_j$, $J^{'\nu}_j\equiv I^{\eta\nu}_j$, and 
$J^{\nu}_j\equiv I^{s\nu}_j$ are the exact-solution
integers or half integers quantum numbers involved 
in Eqs. (2.12a)-(2.12c) of Ref. \cite{Takahashi} and defined 
in the unnumbered equations provided below these equations
(in the notation of that reference $\nu =n$ 
and $j=\alpha$ in $J^{'\nu}_j$ and $J^{\nu}_j$). 
Moreover, the numbers on the right-hand side of the two 
inequalities given just above
Eq. (2.13a) of that reference correspond to 
$N_{a_{\eta\nu}}/2$ and $N_{a_{s\nu}}/2$, respectively, so
that these inequalities read $\vert J^{'\nu}_j\vert<N_{a_{\eta\nu}}/2$
and $\vert J^{\nu}_j\vert<N_{a_{s\nu}}/2$. That
is fully consistent with the above inequality
$\vert I^{\alpha\nu}_j\vert\leq [N_{a_{\alpha\nu}}-1]/2$
where $\alpha=\eta,s$ and $N_{a_{\alpha\nu}}$ is 
given in Eq. (\ref{N*}). A careful comparison of the notations and definitions
used in Ref. \cite{Takahashi} and here confirms that
there is also full consistency between the even or odd character
of the integer numbers $[N_{a_{\eta\nu}}-1]$, $[N_{a_{s\nu}}-1]$,
and $\sum_{\alpha\nu}N_{\alpha\nu}$ considered here and those
that determine the integer of half-integer character
of the quantum numbers $J^{'\nu}_j\equiv I^{\eta\nu}_j$, 
$J^{\nu}_j\equiv I^{s\nu}_j$, and $I_j\equiv I^c_j$,
respectively, in that reference.

We emphasize that for the $\alpha\nu$ fermions the discrete momentum values
$q_j$ of Eq. (\ref{q-j}) are the eigenvalues of the translation generator in the
presence of the fictitious magnetic field of Eq. (\ref{A-j-s1-3D}), which for
1D reads ${\vec{B}}_{\alpha\nu} (x_j) = 
\sum_{j'\neq j} n_{x_{j'},\alpha\nu}\,\delta (x_{j'}-x_{j})\,{\vec{e}}_{x_3}$.
Hence the corresponding exact-solution quantum numbers are
the eigenvalues of such a translator operator in units of $2\pi/L$. 
 
Since for 1D the numbers $\{N_{\alpha\nu}\}$ of $\alpha\nu$ fermions are conserved,
that the discrete momentum values $q_j$ of the $c$ and $\alpha\nu$ fermions are
good quantum numbers is consistent with the momentum operator commuting with the
unitary operator $\hat{V}^{\dag}$ as defined in this paper. That operator generates exact
$U/4t>0$ energy and momentum eigenstates
$\vert \Psi_{LWS;U/4t}\rangle ={\hat{V}}^{\dag}\,\vert \Psi_{LWS;\infty}\rangle$ 
of general form given in Eq. (\ref{LWS-full-el}) where in 1D 
$\vert \Psi_{LWS;U/4t}\rangle=\vert \Phi_{LWS;U/4t}\rangle$ with $U/4t$-dependent energy
eigenvalues but $U/4t$-independent momentum eigenvalues from the 
$U/4t\rightarrow\infty$ energy and momentum eigenstates
$\vert \Psi_{LWS;\infty}\rangle$ of Eq. (\ref{LWS-full-el-infty}). Hence the momentum
eigenvalues are fully determined by the $U/4t\gg 1$ physics. 

The use of the exact solution of the 1D problem confirms that the momentum eigenvalues 
have the general form given in Eq. (\ref{P-1-2-el-ss}). For 1D they may be written as,
\begin{equation}
P =\sum_{j=1}^{N_a} q_j\, N_c (q_j)
+ \sum_{\nu}\sum_{j'=1}^{N_{a_{s\nu}}}
q_{j'}\, N_{s\nu} (q_{j'}) 
+ \sum_{\eta\nu}\sum_{j'=1}^{N_{a_{\eta\nu}}}
[\pi -q_{j'}]\, N_{\eta\nu} (q_{j'}) 
+\pi\,M_{\eta,\,-1/2} \, .
\label{P}
\end{equation}
Here $M_{\eta,\,-1/2}$ is the total number of $\eta$-spin-projection $-1/2$ $\eta$-spinons
and the distributions $N_c (q_j)$ and 
$N_{\alpha\nu} (q_{j'})$ are the eigenvalues of
the operators $\hat{N}_{c}(q_{j}) = f^{\dag}_{q_{j},c}\,f_{q_{j},c}$
and $\hat{N}_{\alpha\nu}(q_{j}) = 
f^{\dag}_{q_{j},\alpha\nu}\,f_{q_{j},\alpha\nu}$, respectively,
which have values $1$ and $0$ for occupied and unoccupied 
momentum values, respectively. One may obtain expression
(\ref{P}) from analysis of the 1D problem for $U/4t\gg 1$ without the use
of Bethe ansatz, whose starting point is that for $U/t\rightarrow\infty$ the 
electrons that singly occupy sites do not feel the 
on-site repulsion. Consistently, expression (\ref{P})
is that also provided by the exact solution after the
rapidities are replaced by the quantum numbers
$I_j\equiv I^c_j$, $J^{'\nu}_j\equiv I^{\eta\nu}_j$, and 
$J^{\nu}_j\equiv I^{s\nu}_j$ related to the discrete
momentum values $q_j$ by Eq. (\ref{q-j}).

That the physical momentum (\ref{P}) is for $U/4t>0$ additive 
in the $c$ and $\alpha\nu$ fermion discrete momentum values 
follows from the latter being good quantum numbers whose occupancy 
configurations generate the energy eigenstates. 
It follows from the direct relation to the thermodynamic Bethe
ansatz equations of Ref. \cite{Takahashi} that the $c$
fermions obtained here from the rotated electrons
through Eq. (\ref{fc+}) and the $\alpha\nu$ fermions 
that emerge from the Jordan-Wigner transformations
of Eq. (\ref{JW-f+-an}) are for the 1D lattice the
$c$ pseudoparticles and $\alpha\nu$ pseudoparticles,
respectively, associated in Ref. \cite{1D} with the 
Bethe-ansatz quantum numbers. The momentum quantum numbers
of Eq. (\ref{q-j}) are precisely those given in Eq. (A.1)
of Ref. \cite{1D} and the numbers $N_{a_{\alpha\nu}}$
of Eq. (\ref{N*}) equal the numbers $N_{\alpha\nu}^*$ 
defined by its Eqs. (B.6), (B.7), and (B.11) with the index $c$ replaced by
$\eta$. Furthermore, the $\eta$-spinons
and spinons considered here are for 1D the holons
and spinons of that reference, respectively. Also the
independent $\eta$-spinons and independent spinons
are for 1D the Yang holons
and HL spinons, respectively, of Ref. \cite{1D}.
(In the notation of that reference HL stands for Heilmann and
Lieb.) 

\subsection{Relation to the algebraic formulation of the exact solution}

We just confirmed that for 1D the discrete momentum values of
the $c$ fermion operators $f^{\dag}_{q_{j},c}$, which 
within our description emerge from 
the electron - rotated-electron unitary transformation, 
and those of the $\alpha\nu$ fermion operators $f^{\dag}_{q_{j},\alpha\nu}$,
which emerge from that transformation and an
extended Jordan-Wigner 
transformation, equal the quantum
numbers of the exact Bethe-ansatz solution. Such a result
was obtained in the limit $N_a\gg1$ that the description 
considered in this paper and in Refs. \cite{companion,companion0} 
refers to. However, such a connection corresponds to the quantum
numbers only and the relation of the $c$ and $\alpha\nu$
fermion operators to the exact solution of the 1D model
remains an open problem. 

The relation of the building blocks of our description to the
original electrons is uniquely defined yet corresponds to
a complex problem. Such building blocks are the
$c$ fermions, $\eta$-spinons, and spinons. For 
$U/4t\gg 1$ the rotated electrons
become electrons and the $c$ fermion creation operator 
$f_{\vec{r}_j,c}^{\dag}$ becomes the quasicharge annihilation 
operator $\hat{c}_r$ of Ref. \cite{Ostlund-06}.
Therefore, in that limit the $c$ fermions are the "holes"
of the quasicharge particles of that reference whereas
the spinons and $\eta$-spinons are associated with the
local spin and pseudospin operators, respectively, of
the same reference. The transformation considered in 
Ref. \cite{Ostlund-06} does not introduce Hilbert-space 
constraints. It follows that suitable occupancy 
configurations of the objects associated with the local 
quasicharge, spin, and pseudospin operators considered
in that reference exist that generate a complete set 
of states. However, only in the limit $U/4t\gg 1$ 
suitable occupancy configurations of such basic objects 
generate exact energy eigenstates. 

The point is that rotated electrons as defined in 
this paper are related to electrons by a unitary 
transformation. And such a transformation is such 
that for $U/4t>0$ rotated-electron occupancy configurations 
of the same form as those that generate energy eigenstates for
$U/4t\gg 1$ in terms of electron operators
do generate energy eigenstates for finite values of $U/4t$.
The $c$ fermion, $\eta$-spinon, and spinon operators
are related to the rotated-electron operators as the
quasicharge, spin, and pseudospin operators of
Ref. \cite{Ostlund-06} are related to electron operators.

Importantly, note that the validity of the $c$ fermion,
spinon, and $\eta$-spinon operational description
constructed in this paper and in Refs. \cite{companion,companion0}
for the Hubbard model on a square and 1D lattices
is for the 1D problem independent of its relation to the
exact solution. For the LWS subspace that such a solution refers
to the validity of our operational description
follows from the transformations behind it
not introducing Hilbert-space constraints.
Such a transformations correspond to explicit 
operator expressions in terms of rotated-electron 
operators: For the $c$ fermion operators such
an expression is given in Eq. (\ref{fc+}) 
and for the spinon and $\eta$-spinon operators 
in Eqs. (\ref{sir-pir})-(\ref{rotated-quasi-spin}). And
the rotated-electron operators are related to
the original electron operators by the unitary
transformation ${\tilde{c}}_{\vec{r}_j,\sigma}^{\dag} =
{\hat{V}}^{\dag}\,c_{\vec{r}_j,\sigma}^{\dag}\,{\hat{V}}$
whose unitary operator ${\hat{V}}^{\dag}$ is within our description uniquely defined. For $U/4t>0$ it 
has been constructed to inherently generating a complete set of energy
eigenstates of the general form 
$\vert \Psi_{LWS;U/4t}\rangle ={\hat{V}}^{\dag}\,\vert \Psi_{LWS;\infty}\rangle$ with 
$\{\vert \Psi_{LWS;\infty}\rangle\}$ being a complete
set of suitably chosen $U/4t\gg 1$ energy eigenstates.
For 1D such states are such that $\vert \Psi_{LWS;U/4t}\rangle =\vert \Phi_{LWS;U/4t}\rangle$
where $\vert \Phi_{LWS;U/4t}\rangle ={\hat{V}}^{\dag}\,\vert \Phi_{LWS;\infty}\rangle$ 
are the states of general form given in Eq. (\ref{LWS-full-el}).

In order to clarify the relation of the $c$ and $\alpha\nu$
fermion operators to the exact solution of the 1D model
rather than the so called coordinate 
Bethe ansatz \cite{Lieb,Takahashi} it is convenient to consider 
the solution of the problem by an algebraic operator formulation 
where the HWSs or LWSs of the $\eta$-spin and spin algebras are built up in 
terms of linear combination of products of several types of
creation fields acting onto the hole or electronic vacuum, 
respectively \cite{Martins,ISM}. The model energy eigenstates 
that are HWSs or LWSs of these algebras are often 
called {\it Bethe states}. Here we briefly discuss how in 1D
and for $N_a\gg1$ the $c$ and $\alpha\nu$ fermion
operators emerge from the creation fields of the 
algebraic formulation of the Bethe states. 

That the general description introduced in this paper for the Hubbard
model on the square and 1D lattices is consistent 
with the exact solution of the 1D problem at the operator level as
well confirms its validity for 1D and this is the only 
motivation and aim of this Appendix. However, 
since that refers to a side problem of the square-lattice
quantum liquid of $c$ and $s1$ fermions studied in
this paper, in the following discussion we skip 
most technical details that are unnecessary for its general 
goals. Nevertheless provided that our analysis
of the problem is complemented with the detailed information
provided in Refs. \cite{Takahashi,Martins} the resulting
message clarifies the main issues under consideration.
 
The algebraic formulation of the Bethe states refers to the 
transfer matrix of the classical coupled spin model, which is the "covering" 
1D Hubbard model \cite{CM}. Indeed, within the inverse scattering method 
\cite{Martins} the central object to be 
diagonalized is the quantum transfer matrix rather than the underlying 
1D Hubbard model. The transfer-matrix eigenvalues provide the spectrum 
of a set of $[N_a-1]$ conserved charges. 
The creation and annihilation fields are labeled by the Bethe-ansatz rapidities 
$\lambda$, which may be generally complex and are not the ultimate quantum 
numbers of the model. Many quantities are functions of 
such rapidities. For instance, the weights $a(\lambda)$ and $b(\lambda)$ 
considered in the derivation of Ref. \cite{Martins}
satisfy the free-fermion condition $a(\lambda)^2 +b(\lambda)^2 =1$. 
(A possible and often used parametrization is
$a(\lambda)=\cos (\lambda)$ and $b(\lambda) =\sin (\lambda)$.)
The reparametrization $\tilde{\lambda}=[a(\lambda)/b(\lambda)]\,e^{2h(\lambda)}-
[b(\lambda)/a(\lambda)]\,e^{-2h(\lambda)}-U/2$
where the constraint $h(\lambda)$ is defined by the relation
$\sinh [2h(\lambda)]=[U/2]\,a(\lambda)b(\lambda)$,
plays an important role in the derivation in the context of the quantum
inverse scattering method of the non-trivial Boltzmann
weights of the isotropic six-vertex model given in Eq. (33) of Ref.
\cite{Martins}.

The diagonalization of the charge degrees of freedom involves a transfer 
matrix of the form provided in Eq. (21) of that reference, whose off-diagonal 
entries are some of the above mentioned creation and annihilation fields. 
The commutation relations of such important operators play a major role 
in the theory and are given in Eqs. (25), (40)-(42), (B.1)-(B.3), (B.7)-(B.11), 
and (B.19)-(B.22) of the same reference. 
The solution of the spin degrees of freedom involves the 
diagonalization of the auxiliary transfer matrix associated with the monodromy 
matrix provided in Eq. (95) of Ref. \cite{Martins}. Again, the off-diagonal entries 
of that matrix play the role of creation and annihilation operators, whose 
commutation relations are given in Eq. (98) of that reference. 
The latter commutation relations correspond to the usual
Faddeev-Zamolodchikov algebra associated with the traditional
ABCD form of the elements of the monodromy matrix. In turn, the above 
relations associated with the charge monodromy matrix refer to a 
different algebra. The corresponding form of that matrix was
called ABCDF by the authors of Ref. \cite{Martins}.

The main reason why the solution of the problem by the algebraic
inverse scattering method \cite{Martins} was achieved only thirty years
after that of the coordinate Bethe ansatz \cite{Lieb,Takahashi} is
that it was expected that the charge and spin monodromy 
matrices had the same traditional
ABCD form \cite{ISM}, consistently with the occurrence
of a spin $SU(2)$ symmetry and a charge (and
$\eta$-spin) $SU(2)$ symmetry known long ago \cite{HL},
associated with a global $SO(4)=[SU(2)\times SU(2)]/Z_2$ 
symmetry \cite{Zhang}.
Fortunately, the studies of Ref. \cite{Martins} used an
appropriate representation of the charge and spin monodromy 
matrices whose structure is able to distinguish creation and
annihilation fields as well as possible {\it hidden symmetries},
as discussed by the authors of that reference.

A hidden symmetry beyond $SO(4)$ was indeed identified  
recently: It is the charge global $U(1)$ symmetry found 
in Ref. \cite{bipartite}. The studies of that reference reveal 
that for $U/4t>0$ the model charge and spin degrees of 
freedom are associated with $U(2)=SU(2)\times U(1)$ and
$SU(2)$ symmetries, rather than with two $SU(2)$ symmetries,
respectively. The occurrence of such charge $U(2)=SU(2)\times U(1)$
symmetry and spin $SU(2)$ symmetry is fully consistent
with the different ABCDF and ABCD forms of the charge and spin monodromy 
matrices of Eqs. (21) and (95) of Ref. \cite{Martins}, respectively.
Indeed, the former matrix is larger than the latter and involves 
more fields than expected from the global
$SO(4)=[SU(2)\times SU(2)]/Z_2$ symmetry alone,
consistently with the global $SO(3)\times SO(3)\times U(1)=[SO(4)\times U(1)]/Z_2$
symmetry of the model on the 1D and any other bipartite lattice 
\cite{bipartite}. 

Our general description of the model on a square and 1D lattices 
takes such an extended global symmetry into
account. Furthermore, for 1D it was also implicitly taken into
account by the appropriate representation of the charge
and spin monodromy matrices used in Ref. \cite{Martins}. Such
a consistency is a necessary condition for the $c$ and
$\alpha\nu$ fermion operators emerging from the
fields of the monodromy matrices upon diagonalization
of the 1D problem for $N_a\gg1$.

Initially the expressions obtained by the algebraic
inverse scattering method for the Bethe states include both
wanted terms and several types of unwanted terms. The latter terms are 
eliminated by imposing suitable restrictions to the rapidities. Such constrains 
lead to the Bethe-ansatz equations and ultimately to the real integer
and half-integer quantum numbers, which in units of
$2\pi/L$ are the discrete momentum values that label the $c$ and
$\alpha\nu$ fermion operators. 

After solving the Bethe ansatz 
for both the charge and spin degrees of freedom, one reaches the charge
rapidities $\lambda_j$ with $[N_a -2S_{\eta}]$ values 
such that  $j=1,...,[N_a -2S_{\eta}]$
and the spin rapidities ${\tilde{\lambda}}_j$ with $[N_a/2-S_{\eta}-S_s]$ values 
such that  $j=1,...,[N_a/2-S_{\eta}-S_s]$. Here $S_{\eta}=S_{\eta}^z$ and
$S_{s}=S_{s}^z$ for a HWS and $S_{\eta}=-S_{\eta}^z$ and
$S_{s}=-S_{s}^z$ for a LWS of both the $\eta$-spin and spin algebras
yet the rapidities can be extended to non-Bethe tower states provided
that the number of their values are expressed in terms of
$S_{\eta}$ and $S_{s}$ rather than of $S_{\eta}^z$ and
$S_{s}^z$, respectively. Equations of the same form as
those obtained by the coordinate Bethe anstaz are reached by
the algebraic operator formulation provided that one 
introduces the charge momentum rapidities $k_j$ and spin rapidities 
${\bar{\lambda}}_j$ given by \cite{Martins}
$z_{-}(\lambda_j)=[a(\lambda_j)/b(\lambda_j)]\,e^{2h(\lambda_j)}=
e^{ik_j a}$ where $j=1,...,[N_a -2S_{\eta}]$ and  
${\bar{\lambda}}_j = - i {\tilde{\lambda}}_j/2$ where
$j=1,...,[N_a/2-S_{\eta}-S_s]/2$. 

Before discussing the structure of the rapidities for $N_a\gg1$, 
let us confirm that the numbers $[N_a -2S_{\eta}]$
and $[N_a/2-S_{\eta}-S_s]$ of discrete charge momentum and spin rapidity 
values, respectively, provided by the Bethe-ansatz solution are closely
related to the occupancy configurations of the rotated electrons that
are not invariant under the electron - rotated-electron unitary
transformation. Indeed, such numbers can be rewritten as,
\begin{equation}
[N_a -2S_{\eta}] = [2S_c + 2C_{\eta}] \, ;
\hspace{0.50cm} [N_a/2-S_{\eta}-S_s] = [C_{\eta} + C_{s}] \, ,
\label{NN-CC}
\end{equation}
where the numbers $C_{\eta}$ and $C_{s}$ are those of Eqs. (\ref{M-L-Sum}), (\ref{sum-rules}), and (\ref{sum-M-2S}).

Here $2S_c$ is the number of elementary rotated-electron
charges associated with the singly
occupied sites and $2C_{\eta}=\sum_{\nu}2\nu\,N_{\eta\nu}$
the number of such charges associated with the rotated-electron
doubly occupied sites whose occupancy configurations are
not invariant under the above unitary transformation.
Moreover, $[C_{\eta} + C_{s}]=\sum_{\alpha\nu}\nu\,N_{\alpha\nu}$ 
is the number $C_{\eta}=\sum_{\nu}\nu\,N_{\eta\nu}$ of down spins
(and up spins) of such rotated-electron doubly occupied sites  
plus the number $C_{s}=\sum_{\nu}\nu\,N_{s\nu}$ of down spins 
(and up spins) of the rotated-electron singly occupied sites
whose occupancy configurations are
not invariant under that transformation.
 
The structure of the charge momentum and spin rapidities simplifies in the
limit $N_a\gg1$ \cite{Takahashi}. Then the $[N_a -2S_{\eta}] = 
[2S_c + 2C_{\eta}]$ charge momentum rapidity values separate 
into two classes corresponding to $2S_c$ and $2C_{\eta}$ of 
these values, respectively. The set of charge momentum 
rapidity values $k_j$ such that $j=1,...,2S_c$ are real and are
related to the integer or half-integer quantum numbers
$I^c_j$ of Eq. (\ref{q-j-f-repr}) such that $q_j = [2\pi/L]\,I^c_j$ are
the $2S_c$ discrete momentum values occupied by the $c$
fermions, out of a total number $N_a$ of
such values. That corresponds to the $2S_c$ elementary 
charges of the rotated electrons that singly occupy sites.

Moreover, for $N_a\gg1$ the $[N_a/2-S_{\eta}-S_s] = [C_{\eta} + C_{s}]$
spin rapidity values separate into two classes corresponding to 
$C_{\eta}$ and $C_{s}$ of these values, respectively. The point
is that the $C_{\eta}$ down spins and $C_{\eta}$ up spins
of the rotated electrons that doubly occupy sites and
whose occupancy
configurations are not invariant under the electron -
rotated-electron unitary transformation combine with 
the $2C_{\eta}$ elementary charges left over by the
above separation of the $[N_a -2S_{\eta}] = 
[2S_c + 2C_{\eta}]$ charge momentum rapidity values.
Therefore, $C_{\eta}$ spin rapidity values out of 
$[C_{\eta} + C_{s}]$ combine with $2C_{\eta}$ 
charge momentum rapidity values out of
$[2S_c + 2C_{\eta}]$, leading to $C_{\eta}$ new
rapidity values associated with the $\eta$-spin
singlet configurations. Those describe the
$\eta$-spin degrees of freedom of the 
rotated electrons that doubly occupy sites and
whose occupancy configurations are not invariant under the electron -
rotated-electron unitary transformation.

In turn, the $C_{s}$ spin rapidity values left over by
the separation of the $[N_a/2-S_{\eta}-S_s] = 
[C_{\eta} + C_{s}]$ spin rapidity values describe the 
spin-singlet configurations associated with the 
rotated electrons that singly occupy sites and
whose occupancy
configurations are not invariant under the electron -
rotated-electron unitary transformation. The charge
degrees of freedom of these rotated electrons are
described by the above set of charge momentum 
rapidity values $k_j$ such that $j=1,...,2S_c$.

According to the $N_a\gg1$ results of
Ref. \cite{Takahashi}, the $C_{\eta}$ $\eta$-spin rapidity values
(and $C_{s}$ spin rapidity values) further separate into
$\eta\nu$ rapidities of length $\nu=1,2,...$
with $N_{\eta\nu}$ values (and $s\nu$ rapidities of length $\nu=1,2,...$ with 
$N_{s\nu}$ values). Obviously, the sum-rules
$C_{\eta}=\sum_{\nu}\nu\,N_{\eta\nu}$ and 
$C_{s}=\sum_{\nu}\nu\,N_{s\nu}$ are obeyed. The
results of that reference reveal
that $s\nu$ rapidities of length $\nu=1$ are real and
the imaginary part of the remaining branches of
$\alpha\nu$ rapidities where $\alpha =\eta,s$ has a simple
form. 

Therefore, for $N_a\gg1$ the original Bethe-ansatz equations
lead to a system of $[1+C_{\eta}+C_s]$ (infinite in the limit
$N_a\rightarrow\infty$) coupled thermodynamic equations
whose solution gives the charge momentum 
rapidity values $k_j$ such that $j=1,...,2S_c$ and 
the values $\Lambda_{j,\alpha\nu}$ of the real part
of the $\alpha\nu$ rapidities such that $j=1,...,N_{\alpha\nu}$
as a function of the occupancies of the real integer
or half-integer quantum numbers $I^{c}_j$ 
and $I^{\alpha\nu}_j$ of Eq. (\ref{q-j-f-repr}).
Alike for the $c$ fermions, the quantum
numbers $I^{\alpha\nu}_j$ such that $q_j = [2\pi/L]\,I^{\alpha\nu}_j$ are
the $N_{\alpha\nu}$ discrete momentum values occupied by $\alpha\nu$
fermions, out of a total number $N_{a_{\alpha\nu}}$ of
such values. (Note that here we used a notation suitable
to our rotated-electron description of the problem and
that in Ref. \cite{Takahashi} such quantum numbers are 
denoted as in Eq. (\ref{q-j}) with $\nu$ replaced by $n$
and the values $\Lambda_{j,\eta\nu}$ and $\Lambda_{j,s\nu}$ 
of the real part of the rapidities by 
$\Lambda^{'n}_{\alpha}$ and $\Lambda^{n}_{\alpha}$,
respectively, where $\alpha$ plays the role of
$j$.)

Summarizing the above discussion, in order to reach the $c$ 
and $\alpha\nu$ fermion operators 
the algebraic operator formulation of the 
diagonalization of the quantum problem
starts by building up the Bethe states in terms of linear 
combination of products of the above mentioned
several types of creation fields 
acting onto a suitable vacuum. The diagonalization of the charge 
and spin degrees of freedom involves the transfer 
matrices given in Eqs. (21) and (95) of Ref. \cite{Martins}, 
respectively, whose off-diagonal entries are some of these
creation and annihilation fields. The different form of such
matrices is consistent with the model global $SO(3)\times SO(3)\times U(1)$
symmetry, which is taken into account by our general description 
both for the model on the 1D and
square lattices used in our studies of the square-lattice 
quantum liquid of $c$ and $s1$ fermions. 
The creation and annihilation fields 
obey the very involved commutation relations given in 
Eqs. (25), (40)-(42), (98), (B.1)-(B.3), (B.7)-(B.11), and
(B.19)-(B.22) of Ref. \cite{Martins} and are labeled by rapidities, 
which may be generally complex and are not the ultimate quantum numbers.

However, such creation and annihilation fields and their involved
algebra generate expressions for the Bethe states that include both
wanted terms and several types of unwanted and unphysical terms.
Indeed, they act onto an extended and partially unphysical Hilbert
space, larger than that of the model.
The unwanted and unphysical terms are eliminated by imposing suitable restrictions to 
the rapidities that change the nature of the fields and for
$N_a\gg1$ replace them by the $c$ and $\alpha\nu$ fermion operators
labeled by the real integer and half-integer quantum numbers
of the diagonalized model. Hence the Bethe-ansatz equations 
obtained by imposing suitable restrictions to the rapidities
describe the relation between the rapidities and the
ultimate quantum numbers associated with the
$c$ and $\alpha\nu$ fermion operators.

In addition to emerging from the elimination of the unwanted and 
unphysical terms of the Bethe states generated by the initial 
creation and annihilation fields, the $c$ and $\alpha\nu$ fermion operator
algebra refers to well-defined subspaces, which are spanned by energy eigenstates
whose number of $\eta$-spinons, spinons, and $c$ fermions is
constant and given by $N_{a_{\eta}}=[N_a-2S_c]$, $N_{a_{s}}=2S_c$,
and $N_c=2S_c$, respectively. Hence the number $2S_c$ of 
rotated-electron singly occupied sites and the numbers $N_{a_{\eta}}$
and $N_{a_{s}}$ of sites of the $\eta$-spin and spin lattices,
respectively, are constant. 

As discussed in this paper, for both the model on 1D lattice considered in this Appendix and
the model on the square lattice the $S_c>0$ vacuum of the
LWS subspace $\vert 0_{\eta s}\rangle$ 
given in Eq. (\ref{vacuum}) is 
invariant under the electron - rotated-electron unitary transformation.
For $D=1$ the $N_{a_{s}}$ independent $+1/2$ spinons 
of such a vacuum are the spins of $N_{a_{s}}$ 
spin-up electrons, the $N_{a_{\eta}}$ independent $+1/2$ $\eta$-spinons refer 
to the $N_{a_{\eta}}$ sites unoccupied by electrons, and
the $N_c$ $c$ fermions describe the charge degrees of freedom
of such electrons of the fully polarized state.
 
The corresponding LWSs that span the above subspace
refer to rotated electrons rather than to electrons and
have the general form given in Eq. (\ref{LWS-full-el})
where the set of numbers $\{N_{\alpha\nu}\}$ obey the
sum-rule associated with the expressions given in Eq.
(\ref{M-L-Sum}), the $\eta$-spin ($\alpha =\eta$) and
spin ($\alpha =s$) can have values $S_{\alpha}=0,..., N_{a_{\alpha}}/2$
such that $S_{\alpha} =[N_{a_{\alpha}}/2 -C_{\alpha}]$,
and the set of numbers $\{N_{a_{\alpha\nu}}\}$ of discrete 
momentum values of each $\alpha\nu$ band are well
defined and given by Eq. (\ref{N*}) and (\ref{N-h-an}).

As mentioned previously, each subspace with constant values of $S_c$ and hence
also with constant values of $N_{a_{\eta}}=[N_a-2S_c]$ and $N_{a_{s}}=2S_c$,
which the vacuum of Eq. (\ref{vacuum}) refers to,
can be divided into smaller subspaces with constant values of $S_c$, $S_{\eta}$, 
and $S_s$ and hence also with constant values of $C_{\eta}$ and $C_s$.
Furthermore, the latter subspaces can be further divided into even smaller 
subspaces with constant values for the set of numbers $\{N_{\eta\nu}\}$ and 
$\{N_{s\nu}\}$, which obey the sum-rule of Eq. (\ref{M-L-Sum}).   

Last but not least, since in contrast to the initial creation and annihilation fields
the $c$ and $\alpha\nu$ fermion operators generate the Bethe
states free from unwanted and unphysical terms, are labeled by the quantum 
numbers of the diagonalized model, and their algebra does
refer to the full LWS or HWS subspace but instead
to subspaces spanned by well-defined types of
Bethe states, there is no contradiction whatsoever between the 
charge ABCDF algebra \cite{Martins} and spin ABCD traditional Faddeev-Zamolodchikov 
algebra \cite{ISM} associated with involved commutation relations of the initial fields
and the anticommutation relations of the $c$ and $\alpha\nu$ fermion operators
provided in Eqs. (\ref{albegra-cf-1D}) and (\ref{1D-anti-com-1D}).
Indeed, the initial creation and annihilation fields act onto an
extended and partially unphysical Hilbert space whereas the
$c$ and $\alpha\nu$ fermion operators act onto well-defined 
subspaces of the model physical Hilbert space.

Upon acting onto such subspaces, the operators $f^{\dag}_{q_{j},c}$
of the $c$ fermions and $f^{\dag}_{q_{j},\alpha\nu}$ of the
$\alpha\nu$ fermions have the expected simple anticommutation
relations associated with those provided in Eqs. (\ref{albegra-cf-1D}) 
and (\ref{1D-anti-com-1D}) for the corresponding operators 
$f^{\dag}_{x_{j},c}$ and $f^{\dag}_{x_{j},\alpha\nu}$, respectively.
Moreover, $c$ fermion operators commute with the $\alpha\nu$
fermion operators and $\alpha\nu$ and $\alpha'\nu'$ fermion
operators belonging to different $\alpha\nu\neq\alpha'\nu'$ 
branches also commute with each other.

Interestingly, in 1D our description may be used in the study
of spectral functions involving matrix elements between Bethe 
states belonging to different subspaces with integer and half-integer 
character, respectively, for some of the quantum numbers of 
Eq. (\ref{q-j-f-repr}). In that case the anticommutation relations
of the operators $f^{\dag}_{x_{j},c}$ and $f^{\dag}_{x_{j},\alpha\nu}$
remain as given in Eqs. (\ref{albegra-cf-1D}) 
and (\ref{1D-anti-com-1D}) yet effective anticommutation relations,
which take into account the shifts of the discrete momentum
values upon the subspace transitions, must be used for
the operators $f^{\dag}_{q_{j},c}$ and $f^{\dag}_{q_{j},\alpha\nu}$.

Finally, for the model on the 1D lattice in the one- and two-electron
subspace as defined in this paper one may introduce a unitary transformation that 
under the ground-state - excited-state transitions, shifts
the discrete momentum values $q_{j}\rightarrow q_{j}+Q_c (q_j)/L$
of the $c$ fermions and $q_{j}\rightarrow q_{j}+Q_{s1} (q_j)/L$
of the $s1$ fermions. Here $Q_c (q_j)/2$ and $Q_{s1} (q_j)/2$
are well-defined overall phase shifts generated by such transitions. 
That allows the evaluation for finite excitation energy
of one- and two-electron matrix elements between the ground state and excited states 
and corresponding spectral-weight distributions for finite
excitation energy $\omega$ \cite{V}. In the $\omega\rightarrow 0$ limit that
method provides the well known low-energy expressions 
predicted long ago by other techniques \cite{LE}
and for finite $\omega$ values successfully describes the
unusual finite-energy one-electron spectral-weight distributions observed
by photo-emission experiments in quasi-one-dimensional
organic metals \cite{TTF}. 

Fortunately, the 1D dynamical theory
of such papers refers to the one- and two-electron subspace
where only the $c$ and $s1$ fermions have an active and
explicit role. The anti-commuting algebra of 
the $c$ and $s1$ fermion operators and the expressions of Eqs. (\ref{fc+}) 
and (\ref{f-an-operators}) apply to both the model in the square
and 1D lattices and hence together with the consistency with
the 1D exact solution found in this Appendix confirms the 
validity of the 1D dynamical theory. 
The general $c$ and $s1$ fermions considered here for
the model on the square and 1D lattices correspond to the $c$ and $s1$
pseudoparticles of the 1D problem of Ref. \cite{V}, respectively. In turn, the above
shifted momentum values ${\tilde{q}}_{j}\equiv q_{j}+Q_c (q_j)/L$
and ${\tilde{q}}_{j}\equiv q_{j}+Q_{\alpha\nu} (q_j)/L$ 
refer to the canonical momentum values of that reference and
in its language label the $c$ and $s1$ pseudofermions, respectively,
of the 1D {\it pseudofermion dynamical theory}.   
          
\section{Subspace-dimension summation} 

In this Appendix we perform the subspace-dimension summation of Eq. (\ref{Ntot}) that runs over $S_c$, $S_\eta$, and  $S_s$ 
integer and half-odd-integer values. Here we consider the square lattice so that $D=2$ in Eq. (\ref{Ntot}), yet the derivation 
proceeds in a similar way for $D=1$. The subspace dimensions have the form
$d_r\cdot\prod_{\alpha=\eta,s}{\cal{N}}(S_{\alpha} ,M_{\alpha})$ given in Eq. (\ref{dimension})
where $d_r$ and ${\cal{N}} (S_{\alpha},M_{\alpha})$ are provided in Eq. (\ref{N-Sa-Ma}). Recounting 
the terms of Eq. (\ref{Ntot}), one may choose $S_\eta$ to be the independent summation variable
what gives,
\begin{equation}\begin{split}
\sum_{S_c=0}^{N_a^2/2}\sum_{S_\eta=0}^{[N_a^2/2-S_c]}\sum_{S_s=0}^{S_c} &\frac{1+(-1)^{2(S_\eta + S_c)}}{2}  \;  \frac{1+(-1)^{2(S_s + S_c)}}{2}  \cdots = \\
 & = \sum_{S_\eta=0}^{N_a^2/2}\sum_{S_s=0}^{[\frac {N_a^2}{2}-S_\eta]}\sum_{S_c=S_s}^{[\frac {N_a^2}{2}-S_\eta]}\frac{1+(-1)^{2(S_\eta + S_s)}}{2} \; \frac{1+(-1)^{2(S_s + S_c)}}{2} \cdots \, .
\end{split}
\end{equation}
One can then rewrite the summation (\ref{Ntot}) in the form,
\begin{equation}
\mathcal{N}_{tot} = \sum_{S_\eta=0}^{N_a^2/2}\sum_{S_s=0}^{[N_a^2/2-S_\eta]}\frac{1+(-1)^{2(S_\eta + S_s)}}{2}(2S_\eta+1)(2S_s+1)\:\times \mathnormal{\Sigma}(S_\eta,S_s),
\label{Ntot2}
\end{equation}
where $\mathnormal{\Sigma}(S_\eta,S_s)$ denotes the $S_\eta$ and $S_s$ dependent summation over $S_c$
as follows,
\begin{eqnarray}
\mathnormal{\Sigma}(S_{\eta},S_s) & = &
\sum_{S_c=S_s}^{\frac {N_a^2}{2}-S_{\eta}} 
\frac{1+(-1)^{2(S_s + S_c)}}{2}  \binom{N_a^2}{2S_c} \times 
\nonumber \\ 
& & \Bigg[\binom{N_a^2-2S_c}{\frac {N_a^2}{2}-S_c-S_{\eta}} - \binom{N_a^2-2S_c}{\frac{N_a^2}{2}-S_c-S_{\eta}-1} \Bigg]
\Bigg[\binom{2S_c}{S_c-S_s}-\binom{2S_c}{S_c-S_s-1}\Bigg] 
\nonumber \\
&  & \sum_{S_c=S_s}^{\frac {N_a^2}{2}-S_\eta} 
\frac{1+(-1)^{2(S_s + S_c)}}{2}   N_a^2! \Bigg[\frac{1}{\displaystyle \left(S_c-S_s\right)! \left(S_c+S_s\right)!}-\frac{1}{\displaystyle \left( S_c-S_s-1\right)! \left( S_c+S_s+1\right)!}\Bigg] \times  
\nonumber \\ 
& & \Bigg[\frac{1}{\left( {N_a^2}/{2}-S_c-S_{\eta}\right)! \left( {N_a^2}/{2}-S_c+S_{\eta}\right)!} 
- \frac{1}{ \left( {N_a^2}/{2}-S_c-S_{\eta}-1\right))! \left( {N_a^2}/{2}-S_c+S_{\eta}+1\right)!}\Bigg] \, .
\label{sum-2}
\end{eqnarray}

In order to evaluate $\mathnormal{\Sigma}(S_\eta,S_s)$ it is useful to replace the variable $S_c$ by $k=S_c-S_s$.
To simplify the notation we then introduce,
\begin{equation}
\mathcal S = S_{\eta}+S_s = \mathcal S(S_\eta,S_s)
\, ; \hspace{0.35cm}
\mathcal D = S_{\eta}-S_s = \mathcal D(S_\eta,S_s) \, .
\label{SD}
\end{equation} 
Due to the parity factor, in the summation over $k$ only the terms with $k$ integer survive so that,
\begin{eqnarray}
\mathnormal{\Sigma} 
 & = &
\displaystyle \sum_{k=0 }^{\frac {N_a^2}{2}-\mathcal S}
N_a^2! \left[ \frac{1}{k! \left(\mathcal {S-D}+k\right)!} - \frac{1}{(k-1)! \left(\mathcal {S-D}+k+1\right)!} \right] \times
\nonumber \\
&& \hspace{1.5cm} \left[ \frac{1}{\left({N_a^2}/{2}-\mathcal S-k\right)! \left({N_a^2}/{2}+\mathcal D-k\right)!} - 
\frac{1}{ \left( {N_a^2}/{2}-\mathcal S-k-1\right)! \left({N_a^2}/{2}+\mathcal D-k+1\right)!} \right]
\nonumber \\
 & = &
\displaystyle \sum_{k=0}^{\frac {N_a^2}{2}-\mathcal S} 
N_a^2!  \Bigg \{ \frac{1}{k! \left(\mathcal {S-D}+k\right)!}\;\frac{1}{\left({N_a^2}/{2}-\mathcal S-k\right)! \left({N_a^2}/{2}+\mathcal D-k\right)!} \; -
\nonumber \\
&& \hspace{1.cm} - \;  
\frac{1}{k! \left(\mathcal {S-D}+k\right)!} \; \frac{1}{ \left( {N_a^2}/{2}-\mathcal S-k-1\right)! \left({N_a^2}/{2}+\mathcal D-k+1\right)!}   \; -
\nonumber \\
&& \hspace{1.cm} - \;  
\frac{1}{(k-1)! \left(\mathcal {S-D}+k+1\right)!} \; \frac{1}{\left({N_a^2}/{2}-\mathcal S-k\right)! \left({N_a^2}/{2}+\mathcal D-k\right)!}   \; +
\nonumber \\
&& \hspace{1.cm} + \; 
\frac{1}{(k-1)! \left(\mathcal {S-D}+k+1\right)!} \; \frac{1}{ \left( {N_a^2}/{2}-\mathcal S-k-1\right)! \left({N_a^2}/{2}+\mathcal D-k+1\right)!} \Bigg \} \, ,
\label{sum-3}
\end{eqnarray}
where now the $k$ summation runs over integers only.

In order to perform the summation (\ref{sum-3}) we rearrange the terms as follows,
\begin{eqnarray}
\mathnormal{\Sigma}
& = &  
\displaystyle \sum_{k=0}^{\frac {N_a^2}{2}-\mathcal S} 
\Bigg \{   \frac{1}{\left( {N_a^2}/{2}+\mathcal {D}\right)!\left( {N_a^2}/{2}-\mathcal {D}\right)!} 
\Bigg [\binom {N_a^2/2+\mathcal {D}}{k} \binom {N_a^2/2-\mathcal {D}}{N_a^2/2-\mathcal S-k} 
+\binom {N_a^2/2+\mathcal {D}}{k-1} \binom {N_a^2/2-\mathcal {D}}{N_a^2/2-\mathcal S-k-1} \Bigg ] -
\nonumber \\
&& - \; \frac{1}{\left( {N_a^2}/{2}+\mathcal {D}+1\right)! \left( {N_a^2}/{2}-\mathcal {D}-1\right)!} 
\binom {N_a^2/2+\mathcal {D}+1}{k} \binom {N_a^2/2-\mathcal {D}-1}{N_a^2/2-\mathcal S-k-1} \; -
\nonumber \\ 
&& - \; \frac{1}{\left( {N_a^2}/{2}+\mathcal {D}-1\right)! \left( {N_a^2}/{2}-\mathcal {D}+1\right)!} 
\binom {N_a^2/2+\mathcal {D}-1}{k-1} \binom {N_a^2/2-\mathcal {D}+1}{N_a^2/2-\mathcal S-k} \Bigg \}  N_a^2!   \, ,
\label{ARF2}
\end{eqnarray}
or
\begin{eqnarray}
\mathnormal{\Sigma}
& = &  
\displaystyle \binom {N_a^2}{N_a^2/2-\mathcal {D}} \sum_{k=0}^{\frac {N_a^2}{2}-\mathcal S} 
\Bigg [ \binom {N_a^2/2+\mathcal {D}}{k} \binom {N_a^2/2-\mathcal {D}}{N_a^2/2-\mathcal S-k} + 
\binom {N_a^2/2+\mathcal {D}}{k-1} \binom {N_a^2/2-\mathcal {D}}{N_a^2/2-\mathcal S-k-1} \Bigg ] \; -
\nonumber \\
&&  - \displaystyle \binom {N_a^2}{N_a^2/2-\mathcal {D}-1}  \sum_{k=0}^{\frac {N_a^2}{2}-\mathcal S}\binom {N_a^2/2+\mathcal {D}+1}{k} \binom {N_a^2/2-\mathcal {D}-1}{N_a^2/2-\mathcal S-k-1} \; -
\nonumber \\
&&  - \displaystyle \binom {N_a^2}{N_a^2/2-\mathcal {D}+1}  \sum_{k=0}^{\frac {N_a^2}{2}-\mathcal S} \binom {N_a^2/2+\mathcal {D}-1}{k-1} \binom {N_a^2/2-\mathcal {D}+1}{N_a^2/2-\mathcal S-k} \, .
\label{ARF}
\end{eqnarray}

Next, by using the identity,
\begin{equation}
\sum_{k=0}^{N} \binom {A}{k} \binom {B}{N-k}=\binom {A+B}{N} \, ,
\label{MJ}
\end{equation}
we carry out separately the summations in expression (\ref{ARF}), what gives,
\begin{equation}
\sum_{k=0}^{{N_a^2}/{2}-\mathcal S} \binom {N_a^2/2+\mathcal {D}}{k} \binom {N_a^2/2-\mathcal {D}}{N_a^2/2-\mathcal S-k} = \binom {N_a^2}{N_a^2/2-\mathcal S} \, ,
\label{MJ1}
\end{equation}
\begin{eqnarray}
\sum_{k=0}^{{N_a^2}/{2}-\mathcal S}\binom {N_a^2/2+\mathcal {D}}{k-1} \binom {N_a^2/2-\mathcal {D}}{N_a^2/2-\mathcal S-k-1}
& = &
\sum_{k=1}^{{N_a^2}/{2}-\mathcal S -1}\binom {N_a^2/2+\mathcal {D}}{k-1} \binom {N_a^2/2-\mathcal {D}}{N_a^2/2-\mathcal S-k-1}
\nonumber \\
& = &
\sum_{k'=0}^{ {N_a^2}/{2}-\mathcal S-2} \binom {N_a^2/2+\mathcal {D}}{k'} \binom {N_a^2/2-\mathcal {D}}{N_a^2/2-\mathcal S-2-k'} 
\nonumber \\
& = &
\binom {N_a^2}{N_a^2/2-\mathcal S-2} \, ,
\label{MJ2}
\end{eqnarray}
\begin{eqnarray}
\sum_{k=0}^{{N_a^2}/{2}-\mathcal S}\binom {N_a^2/2+\mathcal {D}+1}{k} \binom {N_a^2/2-\mathcal {D}-1}{N_a^2/2-\mathcal S-k-1} 
& = & 
\sum_{k=0}^{{N_a^2}/{2}-\mathcal S -1} \binom {N_a^2/2+\mathcal {D}+1}{k} \binom {N_a^2/2-\mathcal {D}-1}{N_a^2/2-\mathcal S-1-k}
\nonumber \\
& = &
\binom {N_a^2}{N_a^2/2-\mathcal S-1} \, ,
\label{MJ3}
\end{eqnarray}
and
\begin{eqnarray}
\sum_{k=0}^{{N_a^2}/{2}-\mathcal S}\binom {N_a^2/2+\mathcal {D}-1}{k-1} \binom {N_a^2/2-\mathcal {D}+1}{N_a^2/2-\mathcal S-k} 
& = & 
\sum_{k=1}^{{N_a^2}/{2}-\mathcal S}\binom {N_a^2/2+\mathcal {D}-1}{k-1} \binom {N_a^2/2-\mathcal {D}+1}{N_a^2/2-\mathcal S-k} 
\nonumber \\
& = &
\sum_{k'=0}^{{N_a^2}/{2}-\mathcal S-1}\binom {N_a^2/2+\mathcal {D}-1}{k'} \binom {N_a^2/2-\mathcal {D}+1}{N_a^2/2-\mathcal S-1-k'} 
\nonumber \\
& = &
\binom {N_a^2}{N_a^2/2-\mathcal S-1} \, .
\label{MJ4}
\end{eqnarray}

Introducing these results in expression (\ref{ARF}) for $\mathnormal{\Sigma}$ leads to,
\begin{eqnarray}
\mathnormal{\Sigma}(S_\eta,S_s)
& = & 
\displaystyle \binom {N_a^2}{N_a^2/2-\mathcal {D}} \left [\binom {N_a^2}{N_a^2/2-\mathcal S} + \binom {N_a^2}{N_a^2/2-\mathcal S-2} \right ] -
\nonumber \\
& & 
- \binom {N_a^2}{N_a^2/2-\mathcal S-1}   \left [ \displaystyle \binom {N_a^2}{N_a^2/2-\mathcal {D}+1} +  \displaystyle \binom {N_a^2}{N_a^2/2-\mathcal {D}-1}\right ]
\nonumber \\
&  \equiv  &  
\mathbf{\Sigma}(\mathcal S,\mathcal {D}) \, .
\label{sigma}
\end{eqnarray}

Expression (\ref {Ntot2}) for $\mathcal{N}_{tot}$ can now be rewritten as,
\begin{eqnarray}
\mathcal{N}_{tot} 
& = & 
\sum_{S_\eta=0}^{N_a^2/2}\sum_{S_s=0}^{[N_a^2/2-S_\eta]}\frac{1+(-1)^{2(S_\eta + S_s)}}{2}(2S_\eta+1)(2S_s+1) \times 
\nonumber \\
& & 
\Bigg \{ \displaystyle \binom {N_a^2}{N_a^2/2-(S_\eta-S_s)} \Bigg [\binom {N_a^2}{N_a^2/2-(S_\eta + S_s)} + \binom {N_a^2}{N_a^2/2-(S_\eta + S_s)-2} \Bigg ] -
\nonumber \\
& & 
- \binom {N_a^2}{N_a^2/2-(S_\eta + S_s)-1}   \Bigg [ \displaystyle \binom {N_a^2}{N_a^2/2-(S_\eta-S_s)+1} 
+  \displaystyle \binom {N_a^2}{N_a^2/2-(S_\eta-S_s)-1}\Bigg ] \Bigg \} \, ,
\label{N-tot-2}
\end{eqnarray}
where the summations run over both integers and half-odd integers. The use of the notation (\ref{SD}) then allows 
rewriting (\ref{N-tot-2}) in compact form,
\begin{equation}
\mathcal{N}_{tot} = \sum_{S_\eta=0}^{N_a^2/2}\sum_{S_s=0}^{[N_a^2/2-S_\eta]}\frac{1+(-1)^{2 \mathcal S}}{2}(\mathcal S + \mathcal {D}+1)(\mathcal S - \mathcal {D}+1)\:\times \mathbf{\Sigma}(\mathcal S,\mathcal {D}) \, ,
\label{Ntot*}
\end{equation}
where the summations run again over both integers and half-odd integers.

We can perform the summations of Eq. (\ref{Ntot*}) in the integers $\mathcal S$ and $ \mathcal {D}$ instead of in $S_\eta$  and $S_s$. 
Indeed, the first factor cancels all the terms with $\mathcal S$ and $ \mathcal {D}$ non-integer so that,
\begin{equation}\begin{split}
\sum_{S_\eta=0}^{N_a^2/2}\sum_{S_s=0}^{[N_a^2/2-S_\eta]}\frac{1+(-1)^{2 ( S_\eta+  S_s)}}{2} \cdots &  
\mbox {($S_\eta$ and $S_s$ both either integers or half odd integers)} = \\
& \qquad \qquad =\sum_{\mathcal S=0}^{N_a^2/2}\sum_{\mathcal {D}=-\mathcal S}^{+\mathcal S} \cdots \mbox {($\mathcal S$ 
and $\mathcal {D}$ integers)} \, .
\nonumber
\end{split}\end{equation}

Thus we find,
\begin{equation}
\mathcal{N}_{tot} = \sum_{\mathcal S=0}^{N_a^2/2}\sum_{\mathcal {D}=-\mathcal S}^{+\mathcal S} \left((\mathcal S +1)^2 - \mathcal {D}^2\right) \:\times \mathbf{\Sigma}(\mathcal S,\mathcal {D}) \, .
\nonumber
\end{equation}
The use of the result (\ref{sigma}) then leads to,
\begin{equation}\begin{split}
\mathcal{N}_{tot} = \sum_{\mathcal S=0}^{N_a^2/2}\sum_{\mathcal {D}=-\mathcal S}^{\mathcal S} &\left((\mathcal S +1)^2 - \mathcal {D}^2\right) \Bigg \{ \displaystyle \binom {N_a^2}{N_a^2/2-\mathcal {D}} \left [\binom {N_a^2}{N_a^2/2-\mathcal S} + \binom {N_a^2}{N_a^2/2-\mathcal S-2} \right ] -
 \\
 & \qquad \qquad - \binom {N_a^2}{N_a^2/2-\mathcal S-1}   \left [ \displaystyle \binom {N_a^2}{N_a^2/2-\mathcal {D}+1} +  \displaystyle \binom {N_a^2}{N_a^2/2-\mathcal {D}-1}\right ]\Bigg \} \, .
\end{split}\label{NtotSD}
\end{equation}

Replacing the variable $\mathcal S$ by $\mathcal S' = \mathcal S +1$ we reach a more tractable expression for $\mathcal{N}_{tot}$,
\begin{equation}
\mathcal{N}_{tot} = \sum_{\mathcal S'=1}^{N_a^2/2+1}\sum_{\mathcal {D}=-\mathcal S'+1}^{\mathcal S'-1} \mathcal T(S',D) \, ,
\label{NtotS'D2}
\end{equation}
where
\begin{equation}\begin{split}
\mathcal {T(S',D)} &=  \left( (\mathcal S')^2 - \mathcal {D}^2\right)\:\times \mathbf{\Sigma}(\mathcal S'-1,\mathcal {D}) \\
 &= \left(\mathcal S'^2 - \mathcal {D}^2\right) \Bigg \{\binom {N_a^2}{N_a^2/2-\mathcal {D}}\left [\binom {N_a^2}{N_a^2/2-\mathcal S'+1} + \binom {N_a^2}{N_a^2/2-\mathcal S'-1} \right ] - \\
  & \hspace{2.2cm}  - \binom {N_a^2}{N_a^2/2-\mathcal S'}   \left [ \displaystyle \binom {N_a^2}{N_a^2/2-\mathcal {D}+1} +  \displaystyle \binom {N_a^2}{N_a^2/2-\mathcal {D}-1}\right ] \Bigg \} \, ,
\end{split}\label{TermS'D}
\end{equation}
is completely symmetric in the summation variables.

Since $\mathcal {T(S',D=\pm S')}=0$, we can extend the summation over $\mathcal {D}$ of Eq.(\ref{NtotS'D2}) to $\mathcal {D} = \pm \mathcal S'$. 
We then formally extend the summation over $ \mathcal S' $ to $ \mathcal S'= 0 $ because the corresponding term vanishes:  
$\mathcal T(\mathcal S'=0,\mathcal D=0)=0 $. Futhermore, $\mathcal {T(\pm S',D)}=\mathcal {T(S',\pm D)}=\mathcal {T(S',D)}$, and  due to the symmetry 
$\mathcal S' \leftrightarrow \mathcal {D}$ we can write,
\begin{equation}
\sum_{\mathcal S'=1}^{N_a^2/2}\sum_{\mathcal {D}=-\mathcal S'+1}^{\mathcal S'-1} \mathcal {T(S',D)} = \frac{1}{4}\sum_{\mathcal S',\mathcal {D}=-(N_a^2/2+1)}^{N_a^2/2+1} \mathcal {T(S',D)} \, .
\label{ah}
\end{equation}

Let us introduce the numbers $p$ and $q$ such that,
\begin{eqnarray*}
\mathcal S'+N_a^2/2+1=p &\Leftrightarrow &  \mathcal S'=p-(N_a^2/2+1)   \\ 
\mathcal {D}+N_a^2/2+1=q &\Leftrightarrow & \mathcal {D}=q-(N_a^2/2+1) \, .
\label{pq}
\end{eqnarray*}
The use of (\ref{ah}) then allows rewriting (\ref{NtotS'D2}) as,
\begin{equation}\begin{split}
\mathcal{N}_{tot}  = \frac{1}{4}\sum_{p,q=0}^{N_a^2+2} & \left [ q(N_a^2+2-q)-p(N_a^2+2-p) \right ] \\
& \times \left \{ \binom {N_a^2}{q-1}\left [\binom {N_a^2}{p}+\binom {N_a^2}{p-2}\right ]- \binom {N_a^2}{p-1}\left [\binom {N_a^2}{q}+\binom {N_a^2}{q-2}\right ]\right \} \, .
\end{split}
\label {uff}
\end{equation}

This expression can be simplified noticing that,
\begin{equation}
\binom {N}{x} + \binom {N}{x-2} = -2 \binom {N}{x-1} + \binom {N+2}{x} \, .
\nonumber
\end{equation}
Replacing in Eq.(\ref {uff}) one then finds,
\begin{equation}\begin{split}
\mathcal{N}_{tot} 
& =   \frac{1}{4}\sum_{p,q=0}^{N_a^2+2} \Bigg [ q(N_a^2+2-q)-p(N_a^2+2-p) \Bigg ] \left \{ \binom {N_a^2}{q-1}\binom {N_a^2+2}{p}- \binom {N_a^2}{p-1} \binom {N_a^2+2}{q}\right \}  \\
& = \frac{1}{4}\sum_{p,q=0}^{N_a^2+2} \left \{q(N_a^2+2-q)\left [ \binom {N_a^2}{q-1}\binom {N_a^2+2}{p}- \binom {N_a^2}{p-1} \binom {N_a^2+2}{q}\right ] + (q \leftrightarrow p)\right \}
\\
& =  \frac{1}{4} 2  \Bigg \{ \; \sum_{q=0}^{N_a^2+2} q(N_a^2+2-q) \binom {N_a^2}{q-1} \sum_{p=0}^{N_a^2+2} \binom {N_a^2+2}{p} -  \sum_{q=0}^{N_a^2+2} q(N_a^2+2-q)\binom {N_a^2+2}{q} \sum_{p=0}^{N_a^2+2} \binom {N_a^2}{p-1} \Bigg \} \, .
\end{split}\end{equation}

Finally, the use of the identities,
\begin{equation}
\sum_{k=0}^{N} \binom {N}{k} = 2^{N} \, ,
\nonumber
\end{equation}
\begin{equation}
\sum_{k=0}^{N+2} \binom {N}{k-1} =\sum_{k=1}^{N+1} \binom {N}{k-1}  =\sum_{k'=0}^{N} \binom {N}{k'} = 2^N \, ,
\nonumber
\end{equation}
\begin{eqnarray*}
\sum_{k=0}^{N} k(N-k)\binom {N}{k}  &=& \sum_{k=1}^{N-1} \frac {N!}{(k-1)!(N-k-1)!}=  N(N-1) \sum_{k-1=0}^{N-2} \binom {N-2}{k-1} \\
& =& N(N-1) 2^{N-2} \, ,
\nonumber
\end{eqnarray*}
and
\begin{eqnarray*}
\sum_{k=0}^{N+2} k(N+2-k) \binom {N}{k-1} 
& = & 
\sum_{k=1}^{N+1}  k(N+2-k)\binom {N}{k-1} = \sum_{k-1=0}^{N} k(N+2-k)\binom {N}{k-1} \\ 
& = & \sum_{k'=0}^{N} (k'+1)(N-k'+1)\binom {N}{k'} = \sum_{k'=0}^{N} \left[ k'(N-k') + (N+1) \right ]\binom {N}{k'} \\
& = & N(N-1) 2^{N-2} +(N+1) 2^N = 2^{N-2} \left[  N(N-1)+4(N+1)\right ] \\
& = & \left[  N^2 + 3N + 4 \right ]2^{N-2} \, ,
\end{eqnarray*}
leads to,
\begin{equation}
\mathcal{N}_{tot} =  \frac{1}{2}  \left \{  \left[  N_a^2 + 3N_a^2 + 4 \right ]2^{N_a^2-2} \times 2^{N_a^2+2}  -  (N_a^2+2)(N_a^2+1) 2^{N_a^2} \times 2^{N_a^2}\right \} = \frac{1}{2}2^{2N_a^2} \times 2  = 4^{N_a^2} \, ,
\end{equation}
which is the desired result.
                                                            

\end{document}